\documentclass[12pt]{article}
\usepackage{amssymb,amsmath}
\usepackage{amsthm}
\usepackage{bm}
\usepackage{enumerate}
\usepackage{booktabs}
\usepackage{longtable}
\usepackage{array}
\usepackage{ytableau}
\usepackage{tikz-cd}
\usepackage{placeins}
\usepackage[hypertexnames=false]{hyperref}

\newtheorem{proposition}{Proposition}[section]
\newtheorem{theorem}[proposition]{Theorem}

\renewcommand{\baselinestretch}{1.1}

\newcommand{\Zb}{\mathbb{Z}}

\newcommand{\Cb}{\mathbb{C}}
\newcommand{\Acal}{\mathcal{A}}
\newcommand{\Dcal}{\mathcal{D}}
\newcommand{\Ical}{\mathcal{I}}
\newcommand{\Ncal}{\mathcal{N}}
\newcommand{\II}{\mathbb{II}}
\DeclareMathOperator*{\Tr}{{\rm Tr}}
\newcommand{\poch}[1]{(#1;q)_{\infty}}
\newcommand{\Tb}{\overline{\mathcal{T}}}

\newcommand{\lin}[3]{\langle #1\rangle^{#2}_{#3}}
\newcommand{\Qm}[1]{\left(\begin{smallmatrix}#1\end{smallmatrix}\right)}
\DeclareRobustCommand{\ysmall}[1]{{\ytableausetup{boxsize=0.42em,aligntableaux=top}\ydiagram{#1}}}
\newcommand{\yf}{\ysmall{1}}
\newcommand{\yaa}{\ysmall{1,1}}
\newcommand{\yb}{\ysmall{2}}
\newcommand{\yba}{\ysmall{2,1}}
\newcommand{\ybbd}{\ysmall{2,2}}

\numberwithin{equation}{section}
\allowdisplaybreaks[1]

\begin{document}

\thispagestyle{empty}
\begin{flushright}
YITP-26-158
\end{flushright}
\vskip0.5cm
\begin{center}
{\Large \bf
Boundary dualities of symplectic \\
\bigskip
Chern-Simons theories and minimal models
}

\vskip0.7cm
Yu Nakayama\footnote{yu.nakayama@yukawa.kyoto-u.ac.jp}

\medskip
{\it Center for Gravitational Physics and Quantum Information,\\
Yukawa Institute for Theoretical Physics, Kyoto University,\\
Kitashirakawa Oiwakecho, Sakyo-ku, Kyoto 606-8502, Japan
}

\bigskip
Tadashi Okazaki\footnote{tokazaki@seu.edu.cn}

\medskip
{\it School of Physics and Shing-Tung Yau Center, Southeast University,\\
Yifu Architecture Building, No.2 Sipailou, Xuanwu district, \\
Nanjing, Jiangsu, 210096, China
}
\end{center}

\vskip0.6cm
\begin{abstract}
We propose that a Neumann-type boundary condition of the 3d $\mathcal{N}=2$ $USp(2n)$ Chern-Simons theory at the level $-(n+\frac32)$
with a fundamental chiral multiplet is dual to a Dirichlet-type boundary condition of an Abelian rank-$0$ theory.
In a topological twist we propose that it supports the affine vertex operator algebra (VOA) $L_n(\mathfrak{osp}(1|2))$.
With the corresponding specialization of the fugacities, the line defect half-indices of the antisymmetric Wilson lines give, up to
monomials, the supercharacters of all its modules, which coincide up to signs with the characters of the Virasoro minimal model
$M(2,2n+3)$.
We extend the duality to the level $-(n+\ell+\frac12)$ and propose that the boundary VOA is
$L_n(\mathfrak{osp}(1|2\ell))$.
Together with the Giveon-Kutasov type duality and the mirror symmetry, the proposed dualities form a web of dualities of the boundary conditions and the
line operators.
For other gauge groups with real matter we propose that the boundary VOA in the holomorphic twist is the commutant of the gauge currents
in free real fermions, which is the Ising model for $Spin(7)$ with the spinor and the $\mathcal N=1$ superconformal minimal model
$SM(3,5)$ for $G_2$ with $\mathbf7$.
\end{abstract}

\newpage
\setcounter{tocdepth}{2}
\tableofcontents
\bigskip

\section{Introduction and conclusion}
\label{sec_intro}
The Virasoro minimal models $M(p,p')$, labeled by coprime integers $2\le p<p'$, are the simplest rational conformal field theories in
two dimensions \cite{Belavin:1984vu}.
They have finitely many irreducible modules, labeled by the Kac labels, whose characters are known
in closed form \cite{MR781391}.
The unitary series $M(p,p+1)$ was identified with the critical points of statistical models such as the
Ising model \cite{Friedan:1983xq}.
The modular invariant partition functions of all $M(p,p')$ were classified in \cite{Cappelli:1987xt}.
The non-unitary series $M(2,2n+3)$ is of particular interest.
Its simplest member $M(2,5)$, the Yang-Lee model, describes the Yang-Lee edge singularity
\cite{Lee:1952ig,Fisher:1978pf,Cardy:1985yy}.
The characters of $M(2,2n+3)$ are given by infinite products, and the Andrews-Gordon identities \cite{MR123484,MR351985}, which
generalize those of Rogers and Ramanujan \cite{Rogers:1894,MR1634067,MR3752624}, express them as multiple sums.
The relation of these identities to the representation theory of affine Lie algebras goes back to \cite{MR491467}.
The fermionic sum representations of these characters \cite{Kedem:1993ze} are a basic example of the relation
between rational conformal field theories and $q$-series identities of the Nahm sum type
\cite{Nahm:2004ch,MR2290758}.

More recently, the Virasoro vertex operator algebras (VOAs) of the minimal models, i.e.\ their chiral algebras, have appeared in
supersymmetric field theories in several settings.
In four dimensions, every $\mathcal{N}=2$ superconformal field theory has an associated VOA
\cite{Beem:2013sza}, e.g.\ the Virasoro VOA of $M(2,2n+3)$ for the Argyres-Douglas theory of type $(A_1,A_{2n})$
\cite{Beem:2017ooy,Cordova:2015nma}.
In three dimensions, holomorphic boundary conditions of topologically and holomorphically twisted supersymmetric theories support
VOAs \cite{Gaiotto:2016wcv,Gaiotto:2017euk,Costello:2018fnz,Costello:2018swh,Costello:2020ndc}, extending the classic relation
between Chern-Simons theories and Wess-Zumino-Witten models \cite{Witten:1988hf,Elitzur:1989nr}.
For topologically twisted 3d $\mathcal N=4$ theories the bulk line operators are conjectured to correspond to modules of these
VOAs \cite{Costello:2018swh}.
The three-dimensional $\mathcal{N}=4$ rank-$0$ theories \cite{Gang:2018huc,Gang:2021hrd} have trivial Higgs and
Coulomb branches, and their topological twists were proposed to be non-unitary semisimple topological field theories \cite{Gang:2021hrd}.
Their proposed boundary VOAs include the affine VOAs $L_n(\mathfrak{osp}(1|2))$ \cite{Ferrari:2023fez} and the Virasoro VOAs of
$M(2,2n+3)$ \cite{Gang:2023rei}, with a boundary condition for $M(2,5)$ in \cite{Ferrari:2023fez}, and, more generally, of the non-unitary minimal models
\cite{Gang:2024tlp}.
Creutzig, Garner and Kim \cite{Creutzig:2024ljv} further proposed $L_n(\mathfrak{osp}(1|2\ell))$ and the minimal $W$-algebras
$W^{\min}_{n-\frac12}(\mathfrak{usp}(2\ell))$ as boundary VOAs of the Abelian rank-$0$ theories $\mathcal{T}_{\ell,n}$, which are
$U(1)^{n(2\ell-1)}$ Chern-Simons matter theories (section \ref{sec_setup}).
These results for the Abelian theories raise the question of which boundary conditions of non-Abelian Lagrangian gauge theories in
three dimensions realize the Virasoro VOAs of the minimal models or VOAs related to them, and how these boundary conditions are related to those of the Abelian theories.

A family of non-Abelian Lagrangian gauge theories appears in a proposed bulk duality.
In \cite{Okazaki:2026efy} it was proposed that the 3d $\mathcal{N}=2$ $USp(2n)$ Chern-Simons theory with a single fundamental
chiral multiplet $Q$ at the level $n+\ell+\frac12$ is dual to the Abelian rank-$0$ theory $\mathcal{T}_{\ell,n}$.\footnote{The full indices
of $USp(2n)_{\pm(n+\ell+\frac12)}+Q$ are equal, so that the bulk analysis of \cite{Okazaki:2026efy} does not fix the relative sign of
the symplectic and Abelian levels. The boundary anomalies fix it, and so do the bulk line defect indices of the Wilson lines, for
$n=\ell$ only up to the Giveon-Kutasov type duality discussed below (sections \ref{sec_signs} and \ref{sec_bulksign}). In the conventions of \cite{Okazaki:2023kpq,Okazaki:2024paq}, which we use, the
dual pair is $USp(2n)_{-(n+\ell+\frac12)}+Q\leftrightarrow\mathcal{T}_{\ell,n}$, in agreement with the case $n=\ell=1$ found in
\cite{Okazaki:2024paq}.}
Both sides have trivial Higgs and Coulomb branches.
The supersymmetry is enhanced to $\mathcal{N}=4$ on both sides, and on the symplectic side the $\mathcal{N}=4$ theory is of the
Gaiotto-Witten type \cite{Gaiotto:2008sd}.
The symplectic gauge theories for $(n,\ell)$ and $(\ell,n)$ are related by the Giveon-Kutasov duality
\cite{Giveon:2008zn,Kapustin:2011vz,Willett:2011gp} continued to a single fundamental chiral multiplet, which we call the Giveon-Kutasov type duality.
The theories $\mathcal{T}_{\ell,n}$ and the parity image $\Tb_{n,\ell}$ of $\mathcal{T}_{n,\ell}$ are related by the mirror symmetry of
\cite{Creutzig:2024ljv}.
The evidence in \cite{Okazaki:2026efy} consists of the counting of vacua under real mass deformations and of the matching of the full
indices, i.e.\ the supersymmetric indices on $S^1\times S^2$, in a number of cases.
As emphasized there, the two sides look very different, with a non-Abelian gauge group and a single fundamental chiral multiplet on one side
and $n(2\ell-1)$ Abelian nodes on the other.
The boundary realization of the duality was left as an open problem.

More generally, this is the question of how infrared dualities of three-dimensional supersymmetric gauge theories act on boundary
conditions and line operators.
Such dualities are common in three dimensions, including mirror symmetry
\cite{Intriligator:1996ex,Aharony:1997bx} and the Seiberg-like dualities
\cite{Aharony:1997gp,Giveon:2008zn,Niarchos:2008jb,Kapustin:2010mh,Willett:2011gp,Benini:2011mf}.
They have been tested by exact results of supersymmetric localization such as the supersymmetric indices
\cite{Bhattacharya:2008zy,Kim:2009wb,Imamura:2011su,Kapustin:2011jm}.
A bulk duality is expected to give dualities of the boundary conditions on the two sides.
For the Abelian duality of the Maxwell theory the action of the duality on boundary conditions was studied with duality walls in
\cite{Kapustin:2009av}.
For the half-BPS boundary conditions of 3d $\mathcal{N}=2$ theories, both $\mathcal{N}=(1,1)$ and $\mathcal{N}=(0,2)$, dual pairs of
boundary conditions were proposed for the duality of the Maxwell theory, and a dual pair for that of SQED \cite{Okazaki:2013kaa}.
For the $\mathcal{N}=(0,2)$ boundary conditions the matching of boundary 't Hooft anomalies and of half-indices provides a sharp
test \cite{Gadde:2013wq,Yoshida:2014ssa,Dimofte:2017tpi}.
Dimofte, Gaiotto and Paquette constructed dual pairs of such boundary conditions for a variety of 3d $\mathcal{N}=2$ dualities and
matched their half-indices and boundary anomalies \cite{Dimofte:2017tpi}.
Bullimore, Dimofte, Gaiotto and Hilburn studied $\mathcal{N}=(2,2)$ boundary conditions of 3d $\mathcal{N}=4$ gauge theories and their
images under mirror symmetry \cite{Bullimore:2016nji}.
The matching of half-indices and boundary anomalies has been applied to $\mathcal{N}=(0,4)$ and $\mathcal{N}=(2,2)$ boundary conditions of Abelian theories
\cite{Okazaki:2019bok,Okazaki:2020lfy,Okazaki:2025rjl}.
It has also been applied to $\mathcal{N}=(0,2)$ boundary conditions of theories with orthogonal and symplectic gauge groups
\cite{Okazaki:2021pnc}, of quivers \cite{Okazaki:2021gkk}, of theories with exceptional gauge groups \cite{Okazaki:2023kpq} and of
Chern-Simons theories decorated by line defects \cite{Okazaki:2024paq,Okazaki:2024kzo}.
The half-indices can be decorated by line operators perpendicular to the boundary, and the
resulting line defect half-indices count the BPS local operators at the junction of the line
and the boundary \cite{Dimofte:2017tpi}.

The purpose of this paper is to address the boundary realization of the duality of \cite{Okazaki:2026efy} for all $(n,\ell)$, together
with the line operators that end on the boundaries and the boundary VOAs.
Assuming the duality, it was anticipated in \cite{Okazaki:2026efy} that the symplectic theory inherits the algebras
$L_n(\mathfrak{osp}(1|2\ell))$ and $W^{\min}_{n-\frac12}(\mathfrak{usp}(2\ell))$ of $\mathcal{T}_{\ell,n}$ \cite{Creutzig:2024ljv}.
We also study the boundary VOAs of the same boundary conditions for Chern-Simons matter theories with other gauge groups.
We find the following.
\begin{enumerate}
\item \textbf{Boundary dualities}\newline
With $k=n+\ell+\frac12$, we consider two boundary conditions of the symplectic gauge theory, (I) of $USp(2n)_{-k}+Q$ with Neumann
boundary conditions for the vector multiplet and $Q$ and with $\ell$ boundary Fermi multiplets, and (II) of the parity image
$USp(2n)_{k}+Q$ with Dirichlet boundary conditions for the vector multiplet and $Q$.
The dual of (I) is a Dirichlet-type boundary condition of $\mathcal{T}_{\ell,n}$.
The dual of (II) is a Neumann-type boundary condition of the parity image $\Tb_{\ell,n}$ with boundary Fermi multiplets, whose charges
come from a factorization of the level matrix (sections \ref{sec_gen_I} and \ref{sec_gen_II}).
We denote these two dualities of boundary conditions also by (I) and (II).

\item \textbf{Line operators}\newline
We also study the line operators which end on these boundary conditions, the Wilson lines of $USp(2n)$ on (I) and the
Wilson lines combined with powers of a vortex line on (II).
For some of them the dual is a single Wilson line of the Abelian theory, and we propose the equalities of their line defect
half-indices and line defect indices as the dictionary of the lines across the duality.
For the antisymmetric Wilson lines on (I) and their images on (II) under the Giveon-Kutasov type duality, the line defect half-index
of each line without a single dual is a combination of those of two Wilson lines.

\item \textbf{Web of dualities}\newline
The Giveon-Kutasov type duality and the mirror symmetry also act on the boundary conditions and on the line operators.
Together with the dualities (I) and (II) they form commutative squares (section \ref{sec_square}),
\begin{equation}
\label{intro_square}
\begin{tikzcd}[column sep=6.5em,row sep=3.2em]
USp(2n)_{-k}+Q\ [\Ncal] \arrow[r,leftrightarrow,"\textrm{(I)}"] \arrow[d,leftrightarrow,"\textrm{GK}"'] & \mathcal{T}_{\ell,n}\ [\Dcal] \arrow[d,leftrightarrow,"\textrm{mirror}"]\\
USp(2\ell)_{k}+Q\ [\Dcal] \arrow[r,leftrightarrow,"\textrm{(II)}"'] & \Tb_{n,\ell}\ [\Ncal]
\end{tikzcd}
\end{equation}
where $[\Ncal]$ and $[\Dcal]$ denote the Neumann-type and Dirichlet-type boundary conditions and GK the Giveon-Kutasov type
duality.
The bottom row is (II) for $(\ell,n)$.
The four vertices of a square describe the same boundary condition of the infrared theory in the four dual bulk theories, and the
antisymmetric Wilson lines and their duals are identified by all four dualities.
For the lines with single duals these identifications also hold for the bulk line defect indices.

\item \textbf{Boundary VOAs}\newline
In the H-twist (section \ref{sec_twist}) the boundary condition (I) supports free symplectic bosons and real fermions, which carry $\mathfrak{usp}(2n)$ currents
at the level $\ell-\frac12$.
We propose that its boundary VOA is the commutant of these gauge currents, which is the affine VOA $L_n(\mathfrak{osp}(1|2\ell))$
\cite{Linshaw:2021hoj,Creutzig:2024ljv}, and that the boundary VOA of (II) in the C-twist is $L_\ell(\mathfrak{osp}(1|2n))$.
In the twist limits the line defect half-indices of $W_\lambda$ on (I) and of $V\otimes W_\lambda$ on (II) with $\lambda\subseteq(\ell^n)$
give, up to monomials, the supercharacters of all the modules of these VOAs, and the four vertices of a square give the same
supercharacter (section \ref{sec_voa_usp}).
For $\ell=1$ the H-twist limits of the line defect half-indices of the antisymmetric Wilson lines give the supercharacters of all the modules of
$L_n(\mathfrak{osp}(1|2))$, which coincide, up to signs, with the characters of $M(2,2n+3)$.
We prove this for all $n$ with the Macdonald identity \cite{MR357528,MR1104219}, so that at the H-twist point the equalities of the
line defect half-indices of the antisymmetric Wilson lines on (I) are the Andrews-Gordon identities \cite{MR123484,MR351985} and
hold for all $n$.
The two VOAs are not isomorphic, since their central charges differ and the vacuum supercharacter of $L_n(\mathfrak{osp}(1|2))$ is the character of the module of
$M(2,2n+3)$ with the lowest conformal weight.
In a labeling which preserves the fusion rules and for a suitable choice of the parities of the modules, however, their modular data are Galois conjugate
\cite{MR1120140,MR1266785,Gannon:2001uy} (section \ref{sec_twist_osp}).
These results suggest a Lagrangian realization of $L_n(\mathfrak{osp}(1|2))$ as a boundary VOA, whose modules correspond to the antisymmetric
Wilson lines.

\item \textbf{Other gauge groups}\newline
We take the same boundary condition for Chern-Simons theories with other gauge groups and a 3d chiral multiplet in a real
representation.
We propose that the boundary VOA in the holomorphic twist \cite{Costello:2020ndc}, at the H-twist point, is the commutant of the gauge currents in the free
boundary fermions.
Its even part is a coset of Goddard, Kent and Olive \cite{Goddard:1984vk,Goddard:1986ee}, and it is unitary
\cite{Goddard:1986ee,MR3119224}.
By the Weyl-Kac character formula \cite{MR1104219} the H-twist limits of the line defect half-indices of all the Wilson lines vanish or
are given, up to monomials, by the supercharacters of its modules.
The commutant is the Ising model for $Spin(7)$ with the spinor \cite{MR1354601} and the $\mathcal N=1$ superconformal minimal
model $SM(3,5)$ \cite{MR1354601,Friedan:1984rv,Bershadsky:1985dq,Eichenherr:1985cx,Qiu:1986if} for $G_2$ with $\mathbf7$
\cite{Okazaki:2024kzo} and for $SU(2)$ with the spin-$3$ representation.
It is a lattice VOA \cite{MR843307,MR996026} for $SU(N)$ with $\mathbf N\oplus\overline{\mathbf N}$,
$SO(N)$ with $2\times\mathbf N$ and $E_6$ with $\mathbf{27}\oplus\overline{\mathbf{27}}$ (section \ref{sec_other_lat}).
\end{enumerate}

\subsection{Structure}
Section \ref{sec_setup} reviews the bulk duality and the conventions, fixes the signs of the levels by the boundary anomalies and
recalls the twist limits, the commutants, the VOAs and the bulk line defect indices.
Sections \ref{sec_n1} and \ref{sec_general} treat $\ell=1$ and general $\ell$ with examples.
Section \ref{sec_square} discusses the web of dualities and the boundary VOAs $L_n(\mathfrak{osp}(1|2\ell))$.
Section \ref{sec_other} treats the theories with other gauge groups and real matter, their boundary VOAs and the characters given by
the H-twist limits of their line defect half-indices.
Appendix \ref{app_proof} contains the proofs of \eqref{linechi} and \eqref{sun_result}, and appendix \ref{app_qseries} the
$q$-series expansions for small $(n,\ell)$.
Appendix \ref{app_pseudo} presents the characters given by the H-twist limits of the line defect half-indices of other Chern-Simons
theories with pseudo-real matter, including sums of pseudo-real and real representations of $SU(2)$, and appendix \ref{app_char} the
character formulae of the gauge groups of section \ref{sec_other} and appendix \ref{app_pseudo}.
Unless stated to be derived or proved, the equalities among half-indices, line defect half-indices and line defect indices and
between their twist limits and (super)characters are found in $q$-series expansions, at generic fugacities where stated
(appendix \ref{app_qseries}).

\subsection{Future work}
\label{sec_future}
\begin{itemize}
\item \textbf{Closed-form formulae}\newline
It is an interesting problem to derive exact closed-form formulae for the half-indices of other Chern-Simons matter theories.
The closed-form formulae for the H-twist limits of the half-indices and the line defect half-indices found in \cite{Okazaki:2024kzo} and in this paper are expressed by Ramanujan's theta
functions \eqref{ramanujan_f} and by characters of minimal models, and it would be interesting to clarify how they generalize and how
the gauge group, the representation of the matter and the Chern-Simons level determine them.
It would also be interesting to understand through the gauge theories the $q$-series identities of the Rogers-Ramanujan type given
by the dualities, such as the Andrews-Gordon identities (sections \ref{sec_n1_H} and \ref{sec_voa_closed}).

\item \textbf{General line operators}\newline
It remains to show that the $\lambda\subseteq(\ell^n)$ of \eqref{singlerule} are the only ones with a single
dual Wilson line.
For some lines we have written the line defect half-indices as combinations of those of two Wilson lines or as sums of those of
single lines, but the general rule which maps the lines $W_\lambda$ and $V^j\otimes W_\lambda$ to lines of the Abelian theories remains
to be found.
Such a rule may require more general line operators, defined by coupling a one-dimensional quantum mechanics to the bulk theory.
For example, the combination \eqref{antisym_two} has the form of the line defect half-index of a single Wilson line coupled to a
one-dimensional Fermi multiplet.
It would be interesting to study the line defect half-indices of such line operators and their transformations under the
dualities, which may extend the dictionary of the line operators found in this paper.
It would also be interesting to identify in general the boundary fields of the symplectic gauge theory which correspond to the
endpoints of the Abelian Wilson lines, such as the junction fields of sections \ref{sec_n1_I} and \ref{sec_gen_I}.
Another question is whether the Wilson lines $W_\lambda$, the vortex-Wilson lines $V^j\otimes W_\lambda$ and their fusion exhaust the
line operators of these rank-$0$ theories.

\item \textbf{String theory constructions}\newline
The $\mathcal N=4$ Chern-Simons matter theories of the Gaiotto-Witten type are realized by D3-branes ending on five-branes
\cite{Gaiotto:2008sd,Hosomichi:2008jd}, and topological field theories and rank-$0$ theories related to the minimal models arise from M5-branes, through the Argyres-Douglas
theories \cite{Dedushenko:2018bpp,Gaiotto:2024ioj} and through the 3d-3d correspondence \cite{Gang:2024tlp}.
Brane setups for half-BPS boundary conditions of 3d $\mathcal N=4$ theories and their dualities were constructed in
\cite{Chung:2016pgt,Hanany:2018hlz,Gaiotto:2019jvo,Okazaki:2019bok,Okazaki:2025rjl}.
It would be interesting to find string theory constructions of the dual pairs of this paper, with orientifold planes for the
symplectic gauge groups, and to realize their boundary conditions and line operators and the dualities among them in terms of branes.
\end{itemize}

\section{Setup}
\label{sec_setup}

\subsection{Bulk theories}
\label{sec_bulkduality}
In sections \ref{sec_setup}--\ref{sec_square}, $n$ and $\ell$ are positive integers and
\begin{align}
\label{k_def}
k\equiv n+\ell+\frac12 .
\end{align}

\subsubsection{Symplectic gauge theory}
\label{sec_bulk_usp}
The symplectic gauge theory is the 3d $\mathcal{N}=2$ $USp(2n)$ gauge theory with a single chiral
multiplet $Q$ in the fundamental representation and vanishing superpotential,
\begin{align}
\label{USp_charges}
\begin{array}{c|c|c|c}
& USp(2n) & U(1)_a & U(1)_R \\ \hline
\textrm{VM}& {\bf Adj} & 0 & 0 \\[2pt]
Q & {\bf 2n} & 1 & r_Q
\end{array}
\end{align}
at Chern-Simons level $\pm k$.
The sign of the level is fixed in section \ref{sec_signs}.
Where the sign is left open, as in sections \ref{sec_halfindex}, \ref{sec_signs} and \ref{sec_bulk}, we write the level as $K$ with
$|K|=k$.
We denote by $a$ the fugacity for $U(1)_a$.
The superconformal R-charge is $r_Q=\frac12$ \cite{Okazaki:2026efy}.

\subsubsection{Abelian gauge theory}
\label{sec_bulk_ab}
The Abelian theory $\mathcal{T}_{\ell,n}$ is a $U(1)^r$ Chern-Simons matter theory with $r=n(2\ell-1)$ nodes.
We label them by $I=(i,b)$ in $n$ rows $i=1,\dots,n$ and $2\ell-1$ columns $b=1,\dots,2\ell-1$, ordered as $(1,1),\dots,(1,2\ell-1),(2,1),\dots$.
The tensor product $\otimes$ of a vector over the rows and a vector over the columns is the vector with the products of their
$i$-th and $b$-th entries at the nodes $(i,b)$, and that of two matrices is defined in the same way.
With the unit vectors $\mathbf e_i$ of $\Zb^n$ over the rows and $\mathbf e_b$ of $\Zb^{2\ell-1}$ over the columns, the unit vector at
the node $I=(i,b)$ is
\begin{align}
\label{unit_node}
\mathbf e_I=\mathbf e_{(i,b)}=\mathbf e_i\otimes\mathbf e_b,\qquad (\mathbf e_{(i,b)})_{(i',b')}=\delta_{ii'}\delta_{bb'} .
\end{align}
Each node carries a chiral multiplet $\Phi_I$ of charge $+1$ under $U(1)_I$.
The theory has a monopole superpotential.
The mixed Chern-Simons levels $(\mathbf A)_I$ between $U(1)_I$ and the topological symmetry $U(1)_J$ give a monopole operator of flux
$m\in\Zb^r$ the charge $\mathbf A\cdot m$ under $U(1)_J$.
The level matrix $\kappa$ and $\mathbf A$ are \cite{Creutzig:2024ljv,Okazaki:2026efy}
\begin{align}
\label{kappa_A}
\kappa=C(T_n)^{-1}\otimes C(A_{2\ell-1}),\qquad
\bigl(C(T_n)^{-1}\bigr)_{ii'}=\min(i,i'),\qquad
(\mathbf A)_{(i,b)}=(-1)^{b+1}\,i .
\end{align}
Here $C(A_{2\ell-1})_{bb'}=2\delta_{bb'}-\delta_{|b-b'|,1}$ is the Cartan matrix of $A_{2\ell-1}$, and
$C(T_n)$ is the Cartan matrix of the tadpole graph $T_n$, obtained from $A_{2n}$ by folding,
\begin{align}
\label{CTn}
C(T_n)_{ii'}=2\delta_{ii'}-\delta_{|i-i'|,1}-\delta_{in}\delta_{i'n},\qquad
C(T_2)=\begin{pmatrix}2&-1\\-1&1\end{pmatrix}.
\end{align}
Here $\kappa$ is the level matrix in the convention where each chiral multiplet contributes
$+\frac12$, i.e.\ $\kappa=\tilde\kappa+\frac12\mathrm{id}$ with $\tilde\kappa$ the
parity-symmetric level matrix.
For $\ell=1$, $\kappa=2C(T_n)^{-1}$ and $\mathcal{T}_{1,n}$ is the tadpole theory of
\cite{Gang:2023rei,Okazaki:2026efy}.
In particular $\mathcal{T}_{1,1}$ is $U(1)_{3/2}$ with one chiral multiplet of charge $+1$, which is related by parity to the theory of
\cite{Gang:2018huc}, whose level is $-\frac32$ in the conventions there.
We denote by
\begin{align}
\label{S_def}
S=\{(n,2\beta-1)\ |\ \beta=1,\dots,\ell\}
\end{align}
the set of nodes with odd $b$ in the last row $i=n$.
It satisfies
\begin{align}
\label{kinvA}
\kappa^{-1}\mathbf A=\frac12\sum_{\beta=1}^{\ell}\mathbf e_{(n,2\beta-1)} .
\end{align}

\subsubsection{Duality}
\label{sec_bulk_dual}
In \cite{Okazaki:2026efy} it was proposed that the symplectic gauge theory is dual to $\mathcal{T}_{\ell,n}$.
With the sign of the level fixed in section \ref{sec_signs}, the dual pair is
\begin{align}
\label{bulk_dual}
USp(2n)_{-k}+Q\ \leftrightarrow\ \mathcal{T}_{\ell,n}.
\end{align}
Both sides have trivial Higgs and Coulomb branches, and their supersymmetry is enhanced to $\mathcal{N}=4$.
The evidence consists of the counting of vacua under real mass deformations and of the matching of the full indices in a number
of cases \cite{Okazaki:2026efy}.
Under the duality the R-charges of the chiral multiplets of $\mathcal{T}_{\ell,n}$ are
\begin{align}
\label{Rassign}
R_I=\begin{cases}
R_S & I\in S\\
0 & I\notin S
\end{cases},
\qquad
R_S=\frac12-r_Q,
\end{align}
and the fugacity $y$ for $U(1)_J$ is identified as \cite{Okazaki:2026efy}
\begin{align}
\label{y_a2}
y=a^2 .
\end{align}
The monopole superpotential has $n(2\ell-1)-1$ terms, and the condition that each of them has R-charge $2$ fixes the R-charges of the
chiral multiplets to \eqref{Rassign} up to the one parameter $R_S$ \cite{Okazaki:2026efy}.
At the superconformal point $r_Q=\frac12$ all the Abelian chiral multiplets have vanishing R-charge.
For the boundary conditions (I) and (II) of section \ref{sec_intro} the two sides share the bulk R-charge $r_Q$ and the bulk
dictionary $y=a^2$.

Two further bulk relations will be used.
The first is the Giveon-Kutasov type duality, i.e.\ the Giveon-Kutasov duality \cite{Giveon:2008zn,Kapustin:2011vz,Willett:2011gp} continued
to a single fundamental chiral multiplet \cite{Okazaki:2026efy}.
It relates $USp(2n)_{-k}$ to $USp(2\ell)_{k}$ with a dual quark of $U(1)_a$ charge $-1$ and R-charge $1-r_Q$.
The second is the mirror symmetry of \cite{Creutzig:2024ljv} between $\mathcal{T}_{\ell,n}$ and the parity image $\Tb_{n,\ell}$ of
$\mathcal{T}_{n,\ell}$, which inverts the topological fugacity (section \ref{sec_square}).

\subsection{Half-indices and line defect half-indices}
\label{sec_halfindex}
We follow the conventions of \cite{Okazaki:2023kpq,Okazaki:2024paq}.
The half-index of a 3d $\mathcal{N}=2$ theory $\mathcal{T}$ with an $\mathcal{N}=(0,2)$ half-BPS
boundary condition $\mathcal{B}$ is
\begin{align}
\label{halfindex_def}
\II^{\mathcal{T}}_{\mathcal{B}}
=\Tr(-1)^Fq^{J+\frac R2}\prod_\alpha x_\alpha^{f_\alpha},
\end{align}
where $F$ is the fermion number, $J$ the generator of the rotation of the plane supporting the
boundary operators, $R$ the R-charge and $f_\alpha$ the Cartan generators of the global
symmetries with fugacities $x_\alpha$ \cite{Gadde:2013wq,Dimofte:2017tpi}.\footnote{Here the half-index is weighted by the fermion number
$(-1)^F$, whereas in \cite{Dimofte:2017tpi} it is weighted by $(-1)^R$ with the R-charge $R$. When the R-charges can be chosen integral, the two are related by $q^{1/2}\to-q^{1/2}$.}
We use the $q$-Pochhammer symbol $\poch{z}=\prod_{j\ge0}(1-zq^j)$, $(q)_\infty=\poch{q}$,
$(q)_m=\prod_{j=1}^m(1-q^j)$, and write $\poch{zs^{\pm}}=\poch{zs}\poch{zs^{-1}}$.
A subscript lists the boundary conditions of the vector multiplet ($\Ncal$ or $\Dcal$) and of the 3d chiral multiplets ($N$ or $D$), and the
boundary Fermi multiplets $\Gamma$ that are present.
We write $D_c$ for a Dirichlet boundary condition of a chiral multiplet with a non-zero boundary value \cite{Dimofte:2017tpi} and keep $D$ in
the subscript when some of the chiral multiplets obey $D_c$.
We call a boundary condition Neumann-type, resp.\ Dirichlet-type, when the vector multiplets obey Neumann, resp.\ Dirichlet,
boundary conditions, whatever the boundary conditions of the 3d chiral multiplets and whichever boundary Fermi multiplets are present.
For example, $(\Ncal,N,\ell\Gamma)$ is Neumann-type.

We consider half-BPS line operators perpendicular to the boundary and ending on it
\cite{Dimofte:2017tpi}.
Following \cite{Okazaki:2024paq} we denote the half-index decorated by a line operator
$\mathcal{L}$ by $\lin{\mathcal{L}}{\mathcal{T}}{\mathcal{B}}$.
Under a duality we say that a line $\mathcal L$ maps to a line $\mathcal L'$ of the dual theory when their line defect half-indices
agree up to a monomial under the identification of the fugacities.
Abelian lines are labeled by their charge vectors, written as subscripts, without brackets when the charges are listed,
e.g.\ $W_{-1,1,0}$, while symplectic Wilson lines are labeled by Young diagrams, e.g.\
$W_{\yf}$, $W_{\yaa}$, $W_{(1^p)}$, $W_{(p)}$, $W_\lambda$.

\subsubsection{Symplectic gauge theory}
\paragraph{Neumann-type boundary conditions}
With Neumann boundary conditions for the vector multiplet and $Q$ and with $\ell$ boundary Fermi
multiplets $\Gamma_\beta$ in the fundamental of $USp(2n)$ with flavor fugacities $x_\beta$, the
half-index is
\begin{align}
\label{USpN_general}
\II^{USp(2n)}_{\Ncal,N,\ell\Gamma}(a,x)
&=\frac{(q)_\infty^n}{2^nn!}\oint\prod_{i=1}^n\frac{ds_i}{2\pi is_i}
\prod_{i\neq j}\poch{s_is_j^{-1}}\prod_{i\le j}\poch{s_is_j}\poch{s_i^{-1}s_j^{-1}}
\nonumber\\
&\qquad\times
\prod_{i=1}^n\frac{\prod_{\beta=1}^{\ell}\poch{q^{\frac12}x_\beta s_i^{\pm}}\poch{q^{\frac12}x_\beta^{-1}s_i^{\pm}}}{\poch{q^{\frac{r_Q}{2}}as_i^{\pm}}} .
\end{align}
The first line is the contribution of the vector multiplet, the numerator in the second line
that of the Fermi multiplets and the denominator that of $Q$.

\paragraph{Dirichlet-type boundary conditions}
With Dirichlet boundary conditions for the vector multiplet and $Q$ and no boundary degrees of
freedom, the half-index is a sum over boundary monopole fluxes $m\in\Zb^n$,
\begin{align}
\label{USpD_general}
\II^{USp(2n)}_{\Dcal,D}(u;a)
=\frac{1}{(q)_\infty^{n}}\sum_{m\in\Zb^n}
\frac{q^{k_{\rm eff}\sum_im_i^2}\prod_iu_i^{2k_{\rm eff}m_i}}{\prod_{\alpha\in\Delta}\poch{q^{1+\alpha(m)}u^\alpha}}
\prod_{i=1}^n\poch{q^{1-\frac{r_Q}{2}\pm m_i}a^{-1}u_i^{\pm1}} ,
\end{align}
where $u_i$ are the fugacities for the boundary global symmetry $USp(2n)_\partial$ and $\Delta=\{\pm2\mathbf e_i,\ \pm \mathbf e_i\pm \mathbf e_j\ (i<j)\}$ is
the set of roots.
The effective level $k_{\rm eff}=K-n-\frac12$ is the boundary anomaly of $USp(2n)_\partial$, and it equals $\ell$ for $K=k$
(section \ref{sec_signs}).
The weight $q^{\frac12k_{\rm eff}\Tr(m^2)}u^{k_{\rm eff}m}$ of \cite{Dimofte:2017tpi,Okazaki:2024paq},
with the trace in $\mathbf{2n}$, gives $q^{k_{\rm eff}\sum_im_i^2}\prod_iu_i^{2k_{\rm eff}m_i}$.

\paragraph{Wilson lines}
\label{sec_characters}
With the Neumann-type boundary conditions, a Wilson line in the representation $\lambda$ is introduced by
inserting the character $\chi_\lambda(s)$ into the integrand of \eqref{USpN_general}
\cite{Dimofte:2017tpi}.
With the Dirichlet-type boundary conditions, the endpoint of a Wilson line transforms under the boundary global symmetry.
In the flux sector $m$ the fugacities are shifted as $u\to q^mu$ \cite{Dimofte:2017tpi}, so that $W_\lambda$ inserts
$\chi_\lambda(q^mu)$ into the summand of \eqref{USpD_general}.
Following \cite{Hatsuda:2025yzp,Hatsuda:2025jze}, we write the characters of $USp(2n)$ \cite{MR1153249} with the
elementary and complete symmetric polynomials
\begin{align}
E_p=e_p(s_1,\dots,s_n,s_1^{-1},\dots,s_n^{-1}),\qquad
H_p=h_p(s_1,\dots,s_n,s_1^{-1},\dots,s_n^{-1}),
\end{align}
and by \cite{MR885807} the character of the irreducible representation labeled by the Young diagram $\lambda$ is
\begin{align}
\label{detchar}
\chi_\lambda=\det\bigl(E_{\lambda^T_i-i+j}-E_{\lambda^T_i-i-j}\bigr)_{1\le i,j\le l(\lambda^T)},
\end{align}
where $\lambda^T$ is the transpose, $E_0=1$ and $E_{p<0}=0$.
In particular the traceless rank-$p$ antisymmetric representation $(1^p)$, $p=1,\dots,n$, and the
rank-$p$ symmetric representation $(p)$ have
\begin{align}
\label{antisym_sym}
\chi_{(1^p)}=E_p-E_{p-2},\qquad \chi_{(p)}=H_p .
\end{align}
For $USp(2)$, $\chi_{(p)}(s)=\sum_{t=0}^ps^{p-2t}$ is the character of the spin-$\frac p2$ representation.

\paragraph{Vortex lines}
With the Dirichlet-type boundary conditions of the symplectic gauge theory the boundary global symmetry is
$(USp(2n)_\partial\times U(1)_a)/\Zb_2$.
Here $\Zb_2$ is generated by $(-\mathbf 1_{2n},-1)$, the central element $-\mathbf 1_{2n}$ of $USp(2n)$ combined with $-1\in U(1)_a$, which acts trivially
on $Q$.
This global structure allows a vortex line $V$ with holonomy $(-\mathbf 1_{2n},-1)$, namely gauge flux
$\frac12\mathbf{1}$, with $\mathbf 1=(1,\dots,1)$, together with $U(1)_a$ flux $\frac12$.
The total flux seen by each component of $Q$ is integral, so that $V$ satisfies the Dirac quantization condition.
We implement $V$ ending on the Dirichlet-type boundary following the prescription of \cite{Dimofte:2017tpi} for vortex lines of the
boundary global symmetry.
It (i) shifts every boundary flux $m\to m+\frac12\mathbf{1}$ in the summand, with the sum still over $m\in\Zb^n$, and (ii) shifts
$a\to q^{1/2}a$ in the factor contributed by $Q$.
More generally, $V^j$ ($j\in\Zb$, $V^{-1}=\bar V$) shifts $m\to m+\frac j2\mathbf 1$ and
$a\to q^{j/2}a$.
For $j$ even $V^j$ is a flavor vortex line for $U(1)_a$ alone, while for $j$ odd the fractional
$U(1)_a$ flux requires the central vortex of the gauge group.

\paragraph{Vortex-Wilson lines}
The vortex line $V^j$ combined with a Wilson line $W_\lambda$ ending at the same point of the
boundary, $V^j\otimes W_\lambda$, further inserts $\chi_\lambda(q^{m+\frac j2\mathbf 1}u)$,
\begin{align}
\label{Vj_general}
\lin{V^j\otimes W_\lambda}{USp(2n)}{\Dcal,D}(u;a)
&=\frac{1}{(q)_\infty^{n}}\sum_{m\in\Zb^n}
\frac{q^{k_{\rm eff}\sum_i(m_i+\frac j2)^2}\prod_iu_i^{2k_{\rm eff}(m_i+\frac j2)}}{\prod_{\alpha\in\Delta}\poch{q^{1+\alpha(m+\frac j2\mathbf 1)}u^\alpha}}
\nonumber\\
&\qquad\times
\prod_{i=1}^n\poch{q^{1-\frac{r_Q}{2}\pm(m_i+\frac j2)}(q^{\frac j2}a)^{-1}u_i^{\pm1}}\
\chi_\lambda(q^{m+\frac j2\mathbf{1}}u) .
\end{align}
Collecting the factors of $q^{\pm\frac j2}$ into shifts of the fugacities, one finds
\begin{align}
\label{Vj_shift}
\lin{V^j\otimes W_\lambda}{USp(2n)}{\Dcal,D}(u;a)
=q^{\frac{nj^2}{4}k_{\rm eff}}\prod_iu_i^{jk_{\rm eff}}\
\lin{W_\lambda}{USp(2n)}{\Dcal,D}(q^{j/2}u;q^{j/2}a) ,
\end{align}
which exhibits $V$ as a vortex line for the boundary symmetry with the anomaly prefactor of
\cite{Dimofte:2017tpi}.
Through the factor $\prod_iu_i^{jk_{\rm eff}}$, $V^j$ carries the charge $jk_{\rm eff}(1,\dots,1)$ under the Cartan of
$USp(2n)_\partial$, which for $j=1$ is the highest weight of $(k_{\rm eff}^n)$.
For $K=k$, where $(k_{\rm eff}^n)=(\ell^n)$ is a Young diagram, the Wilson line in this rectangular representation cancels $V$ in the
sense that
\begin{align}
\label{specflow}
\lin{V\otimes W_{(k_{\rm eff}^n)}}{USp(2n)}{\Dcal,D}(u;a)
=q^{-\frac{nk_{\rm eff}}{4}}\ \II^{USp(2n)}_{\Dcal,D}(u;a) .
\end{align}
We have checked \eqref{specflow} at generic fugacities.
Equivalently, by \eqref{Vj_shift}, the line defect half-index of the Wilson line $W_{(k_{\rm eff}^n)}$ alone equals that of the
anti-vortex line $\bar V$ up to a monomial.
Similarly, that of $V^2\otimes W_{(k_{\rm eff}^n)}$, with the flavor vortex line $V^2$ of $U(1)_a$, equals that of $V$ up to a
monomial.

\subsubsection{Abelian gauge theory}
\paragraph{Dirichlet-type boundary conditions}
\label{sec_invariants}
With Dirichlet boundary conditions for the vector multiplets and the chiral multiplets, the half-index is
\begin{align}
\label{AbD_general}
\II^{U(1)^r}_{\Dcal,D}(u,y)
=\frac{1}{(q)_\infty^r}\sum_{m\in\Zb^r}q^{\frac12m\kappa m}\prod_Iu_I^{(\kappa m)_I}\,
(q^{-\frac12}y)^{\mathbf A\cdot m}\prod_I\poch{q^{1-\frac{R_I}{2}-m_I}u_I^{-1}} .
\end{align}
The factor $(q^{-\frac12}y)^{\mathbf A\cdot m}$ encodes the topological charge of the boundary monopole together with the universal
shift of its R-charge.
It is normalized as in the bulk index of \cite{Okazaki:2026efy}, and the dictionary \eqref{y_a2} refers to this normalization.
Let $X_I=q^{R_I/2}$ and define
\begin{align}
\label{zw_def}
z_I=X_Iu_I,\qquad
w_J=(q^{-\frac12}y)^{(\mathbf A)_J}\prod_IX_I^{-\kappa_{IJ}} .
\end{align}
Then \eqref{AbD_general} depends on the R-charges and the fugacities only through $(z,w)$,
\begin{align}
\label{AbD_zw}
\II^{U(1)^r}_{\Dcal,D}
=\frac{1}{(q)_\infty^r}\sum_{m\in\Zb^r}q^{\frac12m\kappa m}\prod_Iz_I^{(\kappa m)_I}\prod_Jw_J^{m_J}
\prod_I\poch{q^{1-m_I}z_I^{-1}} .
\end{align}
Imposing the bulk dictionary \eqref{Rassign}, \eqref{y_a2} and using \eqref{kinvA}, one finds
\begin{align}
\label{w_value}
w_J=(q^{-\frac12}\hat y)^{(\mathbf A)_J},\qquad \hat y\equiv yq^{-R_S}=q^{r_Q-\frac12}a^2,
\end{align}
where $\hat y$ is the combination of \cite{Okazaki:2026efy} into which the trial R-charge is absorbed.
With the R-charges \eqref{Rassign}, $z$ and $w$ are invariant under the change $R_S\to R_S+\delta$ of the trial R-charge combined
with $u_I\to q^{-\delta/2}u_I$ for $I\in S$ and $y\to q^{\delta}y$, and we call them the invariants.
The boundary condition $D_c$ of $\Phi_I$ breaks the boundary $U(1)_I$ symmetry and sets $z_I=1$.
At a node with $z_I=1$ the factor $\poch{q^{1-m_I}}$ vanishes for $m_I\ge1$, and if $z_I=1$ at all nodes, \eqref{AbD_zw} becomes, for
$\mathcal T_{\ell,n}$, the fermionic sum
\begin{align}
\label{fermionic}
\II^{\mathcal T_{\ell,n}}_{\Dcal,D}\Big|_{z=1}=\sum_{N\in\Zb^{n(2\ell-1)}_{\ge0}}\frac{q^{\frac12N\kappa N}\prod_Jw_J^{-N_J}}{\prod_I(q)_{N_I}} ,
\end{align}
where $N_I$ is the entry of $N$ at the node $I=(i,b)$.

\paragraph{Neumann-type boundary conditions}
With Neumann boundary conditions for the vector multiplets and the 3d chiral multiplets, and boundary Fermi
multiplets $\Gamma_F$ of gauge charges $\mathbf g_F\in\Zb^r$ and fugacities $\nu_F$ (including the
powers of $q$ due to R-charges), the half-index is
\begin{align}
\label{AbN_general}
\II^{U(1)^r}_{\Ncal,N,\Gamma}(\nu)
=(q)_\infty^r\oint\prod_I\frac{ds_I}{2\pi is_I}\prod_I\frac{1}{\poch{q^{\frac{R_I}{2}}s_I}}
\prod_F\poch{q^{\frac12}\nu_Fs^{\mathbf g_F}}\poch{q^{\frac12}\nu_F^{-1}s^{-\mathbf g_F}} ,
\end{align}
where $s^{g}=\prod_Is_I^{g_I}$.
The contour $|s_I|=\rho_I<\min(1,|q|^{-R_I/2})$ leaves outside the poles $s_I=q^{-R_I/2-j}$, $j\ge0$, of the factors contributed by
the 3d chiral multiplets.
Since the factors contributed by the Fermi multiplets are regular away from $s_I=0,\infty$, the integral is the constant term of the
Laurent expansion of the integrand in the region $0<|s_I|<|q|^{-R_I/2}$, which contains the contour.
For $R_I=0$ the first pole lies on the unit circle, reflecting the boundary zero mode of the Neumann chiral multiplet, and the contour runs
just inside it.

\paragraph{Wilson lines}
The Wilson line $W_{\mathbf Q}$ of charge vector $\mathbf Q\in\Zb^r$ carries the flux $\kappa^{-1}\mathbf Q$ attached by the Gauss law.
Hence the names Wilson line and vortex line are not intrinsic to it (see \eqref{WQ_vortex} below), and we label it by its charge.
With the Neumann-type boundary conditions, $W_{\mathbf Q}$ inserts $\prod_I(q^{R_I/2}s_I)^{(\mathbf Q)_I}$ into the integrand of
\eqref{AbN_general}.
The power of $q$ is the R-charge carried by the endpoint through the mixing of the R-symmetry with the gauge group.
With the Dirichlet-type boundary conditions, $W_{\mathbf Q}$ inserts $\prod_I(u_Iq^{m_I})^{(\mathbf Q)_I}$ into the summand of
\eqref{AbD_general},
\begin{align}
\label{WQ_D}
\lin{W_{\mathbf Q}}{U(1)^r}{\Dcal,D}(u,y)
&=\frac{1}{(q)_\infty^r}\sum_{m\in\Zb^r}q^{\frac12m\kappa m}\prod_Iu_I^{(\kappa m)_I}\,
(q^{-\frac12}y)^{\mathbf A\cdot m}
\nonumber\\
&\qquad\times\prod_I(u_Iq^{m_I})^{(\mathbf Q)_I}\poch{q^{1-\frac{R_I}{2}-m_I}u_I^{-1}} .
\end{align}
At $z=1$ the sum is restricted to $m\le0$ as in \eqref{fermionic}, and with $N=-m$ the line defect half-index at $z=w=1$ is, up to a
monomial, the Nahm sum
\begin{align}
\label{nahm_Q}
\sum_{N_I\ge0}\frac{q^{\frac12N\kappa N-\mathbf Q\cdot N}}{\prod_I(q)_{N_I}} .
\end{align}

\paragraph{Vortex lines}
The vortex lines \cite{Dimofte:2019zzj} of the topological symmetry $U(1)_J$ appear on the Abelian side as Wilson lines.
In the Dirichlet-type line defect half-index \eqref{WQ_D} the insertion of $W_{-j\mathbf A}$ is absorbed into the shift of the
topological fugacity,
\begin{align}
\label{WQ_vortex}
\lin{W_{\mathbf Q-j\mathbf A}}{U(1)^r}{\Dcal,D}(u,y)=\prod_Iu_I^{-j(\mathbf A)_I}\ \lin{W_{\mathbf Q}}{U(1)^r}{\Dcal,D}(u,q^{-j}y),
\end{align}
which is the analogue of \eqref{Vj_shift}.
By \eqref{WQ_vortex} with $j=\pm1$, the vortex line and the anti-vortex line of $U(1)_J$ are the
Wilson lines $W_{-\mathbf A}$ and $W_{\mathbf A}$.
On (II) the Abelian theory $\Tb_{\ell,n}$ has the mixed levels $-\mathbf A$.
There we find that the vortex-Wilson lines $V^j\otimes W_\lambda$ of the symplectic gauge theory, which shift $a\to q^{j/2}a$, i.e.\
$y\to q^jy$, map to Abelian lines whose charges are shifted by $-j\mathbf A$ (section \ref{sec_gen_II}).
This is the shift expected from the Gauss law, as in \eqref{WQ_vortex} on the Dirichlet-type boundary with $\mathbf A$ replaced by $-\mathbf A$.

\subsection{Boundary anomalies and the signs of the levels}
\label{sec_signs}
We write the boundary anomaly polynomials in the conventions of
\cite{Okazaki:2023kpq,Okazaki:2024paq}.
The coefficients of $\Tr(s^2)$ are normalized with the index $T(\mathbf R)$ of a representation $\mathbf R$ equal to twice its
Dynkin index, so that $T(\mathbf{2n})=1$ for $USp(2n)$.
The vector multiplet contributes $+h^\vee$ with Neumann and $-h^\vee$ with Dirichlet boundary conditions, where $h^\vee=n+1$ for
$USp(2n)$.
A 3d chiral multiplet in $\mathbf R$ contributes $-\frac12T(\mathbf R)$ with Neumann and $+\frac12T(\mathbf R)$ with Dirichlet boundary conditions.
A boundary Fermi multiplet in $\mathbf R$ contributes $+T(\mathbf R)$, and a Chern-Simons term at the level $K$ contributes $+K$
\cite{Okazaki:2024paq}.

For the Neumann-type boundary conditions of the symplectic gauge theory at the level $K$ with $\ell$ fundamental
Fermi multiplets, the gauge anomaly is
\begin{align}
\label{anomaly_I}
\Acal=\underbrace{K\Tr(s^2)}_{\rm CS}
+\underbrace{(n+1)\Tr(s^2)}_{\textrm{VM},\,\Ncal}
-\underbrace{\tfrac12\Tr(s^2)}_{Q,\,N}
+\underbrace{\ell\Tr(s^2)}_{\Gamma}
=(K+k)\Tr(s^2),
\end{align}
which vanishes for $K=-k$.
With Dirichlet boundary conditions for the vector multiplet and $Q$, the anomaly of
$USp(2n)_\partial$ is $K-(n+1)+\frac12$, which is the effective level in \eqref{USpD_general}.
On the Abelian side of (II) the $2\ell n$ boundary Fermi multiplets form $\ell$ fundamental Fermi multiplets of $USp(2n)_\partial$
(section \ref{sec_gen_II}).
The matching of the boundary anomalies then requires $k_{\rm eff}=\ell$, i.e.\ $K=+k$.
For the Abelian theory with Dirichlet boundary conditions for the vector multiplets and the
chiral multiplets, the anomaly of $U(1)^r_\partial$ is $\tilde\kappa+\frac12\mathrm{id}=\kappa$, which is the
matrix in \eqref{AbD_general}.
With Neumann boundary conditions for the vector multiplets and the 3d chiral multiplets and with boundary Fermi multiplets of charges $\mathbf g_F$, the
gauge anomaly cancels if $\tilde\kappa_{\rm CS}-\frac12\mathrm{id}+\sum_F\mathbf g_F\mathbf g_F^T=0$.
The Fermi multiplets of section \ref{sec_gen_II} satisfy $\sum_F\mathbf g_F\mathbf g_F^T=\kappa$, so that $\tilde\kappa_{\rm CS}=-\tilde\kappa$.
Thus, in these conventions, the pair of boundary conditions (I) involves
\begin{align}
\label{pair_P}
USp(2n)_{-k}+Q
\quad\textrm{and}\quad
\mathcal{T}_{\ell,n},
\end{align}
while the pair (II) involves their parity images $USp(2n)_{k}+Q$ and $\Tb_{\ell,n}$, the latter having the parity-symmetric level
matrix $-\tilde\kappa$ and the mixed levels $-\mathbf A$.
For $n=\ell=1$ the pair \eqref{pair_P} is $USp(2)_{-5/2}+Q\leftrightarrow U(1)_{3/2}+\Phi$,
which is the duality of boundary conditions found in \cite{Okazaki:2024paq}.
The anomalies of $U(1)_a$ of the two sides differ by $\pm n(2\ell-1)$ for (I) and (II), respectively.
This corresponds to a relative background Chern-Simons term of $U(1)_a$, which the half-indices do not see.

In \cite{Okazaki:2026efy} the duality was written as $USp(2n)_{k}+Q\leftrightarrow\mathcal{T}_{\ell,n}$ with $k>0$.
The full indices used there are invariant under $K\to-K$ (section \ref{sec_bulk}), so that they do not fix the relative sign of the
symplectic and Abelian levels.
Accordingly, \cite{Okazaki:2026efy} noted that the case $n=\ell=1$ found in \cite{Okazaki:2024paq} from the half-indices agrees with
it up to a parity transformation.
The boundary anomalies fix this sign, and in our conventions the dual pair is \eqref{pair_P}.
With the identification $y=a^2$, the bulk line defect indices of the Wilson lines also single out
\eqref{pair_P}, for $n=\ell$ only up to the Giveon-Kutasov type duality (section \ref{sec_bulksign}).\footnote{The sign of a Chern-Simons level is meaningful only relative to an orientation.
In the presence of a boundary it is fixed relative to the chirality of the preserved
$\mathcal{N}=(0,2)$ supercharges, as in \eqref{anomaly_I}.}
We write $USp(2n)_{-k}$ for (I) and $USp(2n)_{k}$ for (II).

\subsection{Twist limits and VOAs}
\label{sec_twist}
\subsubsection{Twist limits}
\label{sec_twist_lim}
The chiral multiplet $Q$ is the bifundamental half-hypermultiplet of the Gaiotto-Witten description
$USp(2n)\times O(1)$ \cite{Gaiotto:2008sd,Okazaki:2026efy}, in which only $USp(2n)$ is gauged.
The flavor symmetry $U(1)_a$ is the axial combination of the Cartans of $SU(2)_H\times SU(2)_C$.
In the conventions of \cite{Okazaki:2019bok,Okazaki:2020lfy}, the half-indices of an $\mathcal{N}=4$
hypermultiplet with the boundary condition $N$ or $D$ for both of its chiral multiplets are
\begin{align}
\II^{\rm HM}_{N}(t,x)=\frac{1}{\poch{q^{1/4}tx^{\pm}}},\qquad
\II^{\rm HM}_{D}(t,x)=\poch{q^{3/4}t^{-1}x^{\pm}},
\end{align}
and in the H-twist and the C-twist \cite{Costello:2018fnz} their limits give vacuum (super)characters.
The limit $t=q^{1/4}$ of $\II^{\rm HM}_{N}$ gives the vacuum character of the symplectic bosons, and the limit $t=q^{-1/4}$ of
$\II^{\rm HM}_{D}$ the vacuum supercharacter of the fermionic currents \cite{Gaiotto:2019jvo,Okazaki:2019bok}.
Comparing with the factors contributed by $Q$ in \eqref{USpN_general} and \eqref{USpD_general} at the
superconformal point, we identify
\begin{align}
\label{a_t}
a=t\qquad\bigl(r_Q=\tfrac12\bigr).
\end{align}
At the H-twist point $a=q^{1/4}$ the boundary value of $Q$ becomes weight-$\frac12$ symplectic bosons, and
at the C-twist point $a=q^{-1/4}$ the boundary fermions of $Q$ become weight-one currents.\footnote{These labels refer to the
$\mathcal{N}=4$ structure of the Gaiotto-Witten description and are transported to the dual sides through the duality. They
depend on the description, since the Giveon-Kutasov type duality exchanges them (section \ref{sec_square}). Under this duality the
H-twist point of (I) is the C-twist point of $USp(2\ell)_k$. On the Abelian side it is the B-twist of $\mathcal{T}_{\ell,n}$ in
\cite{Creutzig:2024ljv,Gang:2023rei}, called the C-twist in the terminology recalled in \cite{Creutzig:2024ljv}.}
In the Neumann-type half-index \eqref{USpN_general} the factor contributed by $Q$ has the form of $\II^{\rm HM}_N$, and in the
Dirichlet-type half-index \eqref{USpD_general} that of $\II^{\rm HM}_D$.
We therefore consider the H-twist point for the former and the C-twist point for the latter, together with the corresponding points
on the dual sides.
These combinations need no flips, i.e.\ no boundary Fermi multiplets coupled to the boundary values of $Q$ by a boundary
superpotential, which exchange its Neumann and Dirichlet boundary conditions \cite{Dimofte:2017tpi}.
The twist limits with flips are not considered in this paper.
At the twist points we set $r_Q=\frac12$ and all the flavor fugacities other than $a$ to one, which on the Dirichlet-type boundary of the symplectic
side is the limit $u\to1$ (section \ref{sec_gen_II}).
For a half-index or a line defect half-index $\II(a)$ of the symplectic side we write
\begin{align}
\label{HC_def}
\II\big|_{\rm H}=\II(a)\big|_{a=q^{\frac14}},\qquad \II\big|_{\rm C}=\II(a)\big|_{a=q^{-\frac14}}.
\end{align}
At these points the Abelian fugacities follow from the identifications of the fugacities.
For example, $y=a^2=q^{1/2}$ at the H-twist point on the Dirichlet-type boundary of (I).
On the Neumann-type boundary of (II), $h=q^{(r_Q-1)/2}a=q^{-1/2}$ at the C-twist point, where $h$ is the fugacity of the flavor
symmetry $U(1)_h$ which acts with the same charge on all the boundary Fermi multiplets (section \ref{sec_n1_II}).
We use the term VOA also for vertex operator superalgebras, such as $L_n(\mathfrak{osp}(1|2\ell))$ below, and the parity enters only
through the supercharacters.
We normalize the (super)character $\chi^{\mathcal{V}}_M(q)$ of a module $M$ of a VOA $\mathcal{V}$ by removing
$q^{h_M-c/24}$, with the conformal weight $h_M$ of $M$, and by choosing the overall sign so that the leading coefficient is positive.
In these limits the half-indices and the line defect half-indices are expected to give the (super)characters of the boundary VOAs and
of their modules (section \ref{sec_twist_com}).

\subsubsection{Commutants}
\label{sec_twist_com}
At the H-twist point the boundary VOA of the Neumann-type boundary conditions is expected to be a commutant.
With the Neumann boundary condition for the $\mathcal N=2$ vector multiplet and the Chern-Simons term, the boundary VOA is given by
the derived invariants of the boundary fields under the group $G_{\Cb}[[z]]$ of boundary gauge transformations \cite{Costello:2020ndc},
where $\mathfrak g$ is the Lie algebra of the gauge group $G$.\footnote{Here $z$ is a holomorphic coordinate on the boundary centered at
the point of the boundary local operators, and $G_{\Cb}[[z]]$ consists of the gauge transformations given by formal power series in
$z$ with values in the complexified gauge group $G_{\Cb}$. Its Lie algebra is $\mathfrak g[[z]]$, and $z\mathfrak g[[z]]$ is the ideal of
the elements which vanish at $z=0$, so that every element of $G_{\Cb}[[z]]$ is a gauge transformation independent of $z$ times one generated
by $z\mathfrak g[[z]]$. The ghost for $z^m$ has the conformal weight $m$.}
The invariants under the gauge transformations independent of $z$ are taken directly, and those under $z\mathfrak g[[z]]$ by ghosts in the
adjoint representation, one for each power $z^m$, $m\ge1$.
In the half-index the gauge transformations independent of $z$ give the Haar measure, and the ghosts give the remaining factor
$(q)_\infty^r\prod_{\alpha\in\Delta}\poch{qs^\alpha}$ of the vector multiplet, where $r$ is the rank of $G$, $\Delta$ is the set of its
roots and $s^\alpha$ is the monomial in the fugacities $s$ of the maximal torus for the root $\alpha$.
So the H-twist limit of the half-index is the supercharacter of the complex of the boundary fields and the ghosts.
Its cohomology of degree zero consists of the states annihilated by the non-negative modes of the gauge currents, i.e.\ the commutant
of the gauge currents in the boundary fields.

We assume that the derived invariants are concentrated in degree zero, so that the boundary VOA is this commutant.
The H-twist limits of the line defect half-indices of the Wilson lines $W_\lambda$ are then expected to give the (super)characters of
the modules of the commutant which appear together with the conjugate of $\lambda$ in the decomposition of the boundary fields.
We apply this construction to $USp(2n)$ below and to other gauge groups with real matter in section \ref{sec_other}, where, given
the singular operator product of real fermions, the assumption follows from the theorem of Garland and Lepowsky (section
\ref{sec_other_unit_real}).

For the boundary conditions (I) of $USp(2n)_{-k}$, with $k$ of \eqref{k_def}, the boundary fields at the H-twist point are the
symplectic bosons $\beta\gamma^n$ in $\mathbf{2n}$ from the boundary value of $Q$ and $4n\ell$ real fermions $F^{4n\ell}$ from the
Fermi multiplets.
They carry $\mathfrak{usp}(2n)$ currents at the level $\ell-\frac12$, in agreement with the anomaly $k-(n+1)=\ell-\frac12$.
The boundary VOA is expected to be the commutant
\begin{align}
\label{H_coset}
\mathrm{Com}\bigl(L_{\ell-\frac12}(\mathfrak{usp}(2n)),\,F^{4n\ell}\otimes\beta\gamma^n\bigr)\cong L_n(\mathfrak{osp}(1|2\ell)),
\end{align}
where $\mathrm{Com}(\mathcal{A},\mathcal{V})$ is the commutant of $\mathcal{A}$ in $\mathcal{V}$, and $L_{\ell-\frac12}(\mathfrak{usp}(2n))$ and
$L_n(\mathfrak{osp}(1|2\ell))$ are the simple affine VOAs of $\mathfrak{usp}(2n)$ at the level $\ell-\frac12$ and of
$\mathfrak{osp}(1|2\ell)$ at the level $n$.
The isomorphism is proved in \cite{Linshaw:2021hoj} for the vertex subalgebras generated by the currents.
The one generated by the currents of $\mathfrak{osp}(1|2\ell)$ is the simple affine VOA $L_n(\mathfrak{osp}(1|2\ell))$ \cite{MR4806362},
and we assume that the one generated by the currents of $\mathfrak{usp}(2n)$ is $L_{\ell-\frac12}(\mathfrak{usp}(2n))$.
By \eqref{Hmodule} with $\lambda=\varnothing$ and the isomorphism \eqref{H_coset}, the supercharacter of the complex equals that of its
cohomology of degree zero, for $\ell=1$ by \eqref{linechi} with \eqref{osp12M} and for $\ell\ge2$ in $q$-series expansions.
This is consistent with the vanishing of the cohomology in the other degrees.

At the C-twist point the Dirichlet-type boundary conditions (II) of $USp(2n)_k$ support the
$\mathfrak{usp}(2n)$ currents at the level $k_{\rm eff}=\ell$ together with the weight-one currents from
the boundary fermions of $Q$.
We expect the boundary VOA to be $L_\ell(\mathfrak{osp}(1|2n))$.
This is supported by the Giveon-Kutasov type duality, which maps this twist limit to the H-twist limit of (I) for $(\ell,n)$
(section \ref{sec_square}).
It is also supported by the C-twist limits of the line defect half-indices, which give the
supercharacters of all the modules of $L_\ell(\mathfrak{osp}(1|2n))$ \eqref{CVW}.

In the rest of this section we recall the VOAs whose characters appear in this paper.

\subsubsection{Virasoro minimal models}
\label{sec_twist_vir}
The Virasoro minimal model $M(p,p')$ \cite{Belavin:1984vu}, with coprime $2\le p<p'$, has central charge $c=1-\frac{6(p-p')^2}{pp'}$.
Its modules are labeled by the Kac labels $(r,s)$, $1\le r\le p-1$, $1\le s\le p'-1$, with $(r,s)\sim(p-r,p'-s)$, and have the
conformal weights $h_{r,s}=\frac{(p'r-ps)^2-(p'-p)^2}{4pp'}$ \cite{DiFrancesco:1997nk}.
The vacuum module is $(1,1)$.
Their characters, normalized as in section \ref{sec_twist_lim}, are given by the formula of Rocha-Caridi \cite{MR781391}.
The non-unitary model $M(2,2n+3)$ has the $n+1$ modules $(1,s)$, $s=1,\dots,n+1$, with the lowest conformal weight
$h_{1,n+1}=-\frac{n(n+1)}{2(2n+3)}$.
The formula of Rocha-Caridi and the Jacobi triple product give
\begin{align}
\label{chiM}
\chi^{M(2,2n+3)}_{(1,s)}(q)=\prod_{\substack{m\ge1\\ m\not\equiv0,\pm s\ (\mathrm{mod}\ 2n+3)}}\frac{1}{1-q^m},
\qquad s=1,\dots,n+1 .
\end{align}
We also use Ramanujan's general theta function
\begin{align}
\label{ramanujan_f}
f(a,b)=\sum_{m\in\Zb}a^{\frac{m(m+1)}{2}}b^{\frac{m(m-1)}{2}}=(-a;ab)_\infty(-b;ab)_\infty(ab;ab)_\infty ,
\end{align}
where the second equality is the Jacobi triple product, and $f(-q)=f(-q,-q^2)=(q)_\infty$.
We also write $\theta(x;q)=\poch{x}\poch{q/x}$.

\subsubsection{\texorpdfstring{$\mathcal N=1$}{N=1} superconformal minimal models}
\label{sec_twist_svir}
The $\mathcal N=1$ superconformal minimal model $SM(p,p')$ \cite{Friedan:1984rv,Bershadsky:1985dq,Eichenherr:1985cx,Qiu:1986if}, with
$p<p'$, $p'-p$ even and $\gcd(\frac{p'-p}2,p)=1$, has central charge $c=\frac32\bigl(1-\frac{2(p'-p)^2}{pp'}\bigr)$.
Its modules $(r,s)$, $1\le r\le p-1$, $1\le s\le p'-1$, have the conformal weights
$h_{r,s}=\frac{(p'r-ps)^2-(p'-p)^2}{8pp'}+\frac{1-(-1)^{r-s}}{32}$, and they are in the Neveu-Schwarz sector for even $r-s$ and in the
Ramond sector for odd $r-s$.
For the Neveu-Schwarz modules $(r,s)$, $r-s$ even, and the Ramond modules $(r,s)$, $r-s$ odd, we write
\begin{align}
&\chi^{SM(p,p')}_{{\rm NS},(r,s)}=q^{\frac{c}{24}-h_{r,s}}\,\mathrm{Tr}_{(r,s)}\,q^{L_0-\frac{c}{24}},\qquad
\chi^{SM(p,p')}_{\widetilde{\rm NS},(r,s)}=q^{\frac{c}{24}-h_{r,s}}\,\mathrm{Tr}_{(r,s)}(-1)^Fq^{L_0-\frac{c}{24}},
\nonumber\\
&\chi^{SM(p,p')}_{{\rm R},(r,s)}=\frac12\,q^{\frac{c}{24}-h_{r,s}}\,\mathrm{Tr}_{(r,s)}\,q^{L_0-\frac{c}{24}},
\end{align}
with $(r,s)\sim(p-r,p'-s)$.
The lowest states of the Neveu-Schwarz modules are even, so that the $\widetilde{\rm NS}$ characters are the supercharacters.
For $h_{r,s}\neq\frac c{24}$ the zero mode $G_0$ pairs the lowest states of a Ramond module into an even and an odd state, which gives
the factor $\frac12$.

\subsubsection{Lattice VOAs}
\label{sec_twist_lat}
The lattice VOA $V_{\sqrt N\Zb}$ \cite{MR843307,MR996026} of the lattice $\sqrt N\Zb$, with $(\alpha,\alpha)=N$ for the generator
$\alpha$, is the chiral algebra of a compact boson, of central charge $1$, e.g.\ $\widehat{\mathfrak{su}}(2)_1$ for $N=2$.
It is rational, and its modules $V_{\sqrt N\Zb+\frac pN\alpha}$, $p\in\Zb_N$, have the fusion rules of $\Zb_N$ \cite{MR1245855} and
the conformal weights $h_p=\frac{p^2}{2N}$ for $|p|\le\frac N2$.
For even $N$ the normalized characters are
\begin{align}
\label{lattice_chi}
\chi^{V_{\sqrt N\Zb}}_p=\frac{1}{(q)_\infty}\sum_{m\in\Zb}q^{\frac N2m^2+pm},\qquad \chi^{V_{\sqrt N\Zb}}_{-p}=\chi^{V_{\sqrt N\Zb}}_p,
\end{align}
while for odd $N$ the lattice is odd and $V_{\sqrt N\Zb}$ is a vertex operator superalgebra, whose supercharacters have the sign
$(-1)^m$ in the sum.

\subsubsection{\texorpdfstring{The affine VOA $L_n(\mathfrak{osp}(1|2\ell))$}{The affine VOA L(osp(1|2l))}}
\label{sec_twist_osp}
The Lie superalgebra $\mathfrak{osp}(1|2\ell)$ \cite{Kac:1977em} has the even part $\mathfrak{usp}(2\ell)$, the odd part
$\mathbf{2\ell}$ and the superdimension\footnote{The superdimension of a $\Zb_2$-graded vector space is the dimension of its
even part minus that of its odd part. For a module of a Lie superalgebra it is the value of its supercharacter at the identity.}
$\ell(2\ell-1)$.
With the long roots of length squared two, its dual Coxeter number is $h^\vee=\ell+\frac12$, so that $n+h^\vee=k$.
Its simple affine VOA $L_n(\mathfrak{osp}(1|2\ell))$ is generated by the currents of $\mathfrak{usp}(2\ell)$ and the fermionic
currents in $\mathbf{2\ell}$, of conformal weight one, and has the central charge $c=\frac{n\ell(2\ell-1)}{k}$.
At positive integer level $n$ its category of ordinary modules is semisimple with finitely many simple objects
\cite{Creutzig:2022osp}.
Although $c>0$, $L_n(\mathfrak{osp}(1|2\ell))$ is not unitary.
The commutant of its subalgebra $L_n(\mathfrak{usp}(2\ell))$ is the principal $\mathcal W$-algebra of $\mathfrak{usp}(2\ell)$ at the level
$-(\ell+1)+\frac{n+\ell+1}{2n+2\ell+1}$ \cite{Creutzig:2022osp}, whose Virasoro algebra, generated by the difference of the Sugawara
fields of $L_n(\mathfrak{osp}(1|2\ell))$ and $L_n(\mathfrak{usp}(2\ell))$, has the central charge
\begin{align}
\label{osp_com_c}
c-\frac{n\ell(2\ell+1)}{n+\ell+1}=-\frac{n\ell(4n+2\ell+3)}{2k(n+\ell+1)}<0 .
\end{align}
Hence in any invariant Hermitian form the state $L_{-2}|0\rangle$ of this Virasoro algebra has a negative norm squared, one half of
\eqref{osp_com_c}, so that no such form is positive definite.
For $\ell=1$ the commutant is the minimal model $M(n+2,2n+3)$ \cite{MR3869280,MR3856129}.
The simple ordinary modules of $L_n(\mathfrak{osp}(1|2\ell))$ are the integrable highest weight modules \cite{MR1327543,Creutzig:2022osp},
and $L_n(\mathfrak{osp}(1|2\ell))$ is rational \cite{Creutzig:2021trial}.
With the fundamental weights $\omega_i=\mathbf e_1+\dots+\mathbf e_i$ of $\mathfrak{usp}(2\ell)$ they are labeled by the dominant integral
weights $\mu=\sum_i(\mu_i-\mu_{i+1})\omega_i$ with $\mu_1\le n$, i.e.\ by the $\binom{\ell+n}{\ell}$ Young diagrams
$\mu\subseteq(n^\ell)$.
The top component of the module $\mu$ is the irreducible $\mathfrak{osp}(1|2\ell)$-module with highest weight $\mu$.
Its supercharacter, in the fugacities $z$ of \eqref{KWz} below, is the character of the irreducible $\mathfrak{so}(2\ell+1)$-module with
the same highest weight at $-z$ \cite{MR648354}, where a weight $\sum_im_i\mathbf e_i$ has the parity $\sum_im_i$ mod $2$.
Thus the superdimension of the $\mathfrak{osp}(1|2\ell)$-module with highest weight $\omega_p$, which corresponds to the antisymmetric
representation $(1^p)$ of $\mathfrak{so}(2\ell+1)$, is $(-1)^p\bigl(\binom{2\ell}{p}-\binom{2\ell}{p-1}\bigr)$.
The conformal weight of the module $\mu$ is
\begin{align}
\label{hmu}
h_\mu=\frac{(\mu,\mu+2\rho)}{2(n+h^\vee)}=\sum_{i=1}^\ell\frac{\mu_i(\mu_i+2\ell-2i+1)}{4k} ,
\end{align}
with the super Weyl vector $\rho=\sum_i(\ell-i+\frac12)\mathbf e_i$.\footnote{The super Weyl vector is half the sum of the even
positive roots minus half the sum of the odd positive roots. For $\mathfrak{osp}(1|2\ell)$ the even positive roots are
$\mathbf e_i\pm\mathbf e_j$, $i<j$, and $2\mathbf e_i$, and the odd positive roots are $\mathbf e_i$. The super Weyl vector equals
the Weyl vector of $\mathfrak{so}(2\ell+1)$.}
The twist limits of the half-indices compute the supercharacters $\mathrm{Tr}\,(-1)^Fq^{L_0-\frac{c}{24}}$.
The character formula of Kac and Wakimoto \cite{MR949675,MR1327543} gives the refined supercharacter \eqref{KWz} below, and at $z=1$
the supercharacter of the module $\mu$ is
\begin{align}
\label{KWmu}
\chi^{L_n(\mathfrak{osp}(1|2\ell))}_\mu&=\frac{1}{(q)_\infty^{\ell(2\ell-1)}}\sum_{t\in\Zb^\ell}(-1)^{\sum_mt_m}
\prod_{i<m}\frac{v_i^2-v_m^2}{\rho_i^2-\rho_m^2}\,q^{k\sum_mt_m^2+\sum_mb_mt_m},
\nonumber\\
v_m&=b_m+2kt_m ,
\end{align}
with the components $\rho_m=\ell-m+\frac12$ of the super Weyl vector $\rho$ and $b_m=\mu_m+\rho_m$.
Its leading coefficient $\prod_{i<m}(b_i^2-b_m^2)/(\rho_i^2-\rho_m^2)$ is the absolute value of the superdimension of the top component.
With the fugacities $z=(z_1,\dots,z_\ell)$ of the Cartan subalgebra of $\mathfrak{usp}(2\ell)$ the refined supercharacter
$\chi^{L_n(\mathfrak{osp}(1|2\ell))}_\mu(z)$, normalized in the same way, is
\begin{align}
\label{KWz}
\chi^{L_n(\mathfrak{osp}(1|2\ell))}_\mu(z)&=\sum_{t\in\Zb^\ell}(-1)^{\sum_mt_m}q^{k\sum_mt_m^2+\sum_mb_mt_m}\,
\frac{\mathsf P_{b+2kt}(z)}{\mathsf P_{\rho}(z)\,\Xi_\ell(z)},
\nonumber\\
\mathsf P_v(z)&=\det\bigl(z_i^{v_j}+z_i^{-v_j}\bigr)_{i,j=1}^\ell,\qquad
\Xi_\ell(z)=(q)_\infty^\ell\,\frac{\prod_{\alpha\in\Delta}(qz^\alpha;q)_\infty}{\prod_{i=1}^\ell(qz_i^{\pm1};q)_\infty},
\end{align}
with the roots $\Delta$ of $\mathfrak{usp}(2\ell)$.
At $z=1$ it reduces to \eqref{KWmu}.
The dualities of this paper give further forms of these supercharacters, integrals over $USp(2n)$ and Nahm sums
(\eqref{Hmodule} and \eqref{chinahmmu}), collected in section \ref{sec_voa_closed}.
For $\ell=1$ the sum \eqref{KWmu} is the Jacobi triple product form of \eqref{chiM}, and the supercharacters coincide with the
characters of the minimal models \cite{Ferrari:2023fez},
\begin{align}
\label{osp12M}
\chi^{L_n(\mathfrak{osp}(1|2))}_{(p)}=\chi^{M(2,2n+3)}_{(1,n+1-p)},\qquad p=0,\dots,n,
\end{align}
so that the vacuum module of $L_n(\mathfrak{osp}(1|2))$ corresponds to the module $(1,n+1)$ of $M(2,2n+3)$ with the
lowest conformal weight.
The Virasoro vacuum module $(1,1)$ corresponds to the module $(n)$ with the highest conformal weight.
Without the choice of the overall signs in section \ref{sec_twist_lim}, the two sides agree up to signs, which depend on the choice of
the parity of the modules, e.g.\ through the parity shifts in \cite{Ferrari:2023fez} and the additional R-symmetry Wilson lines in
\cite{Ferrari:2023fez,Creutzig:2024ljv}.
The equality \eqref{osp12M} also holds with the factors $q^{h-\frac c{24}}$.
Indeed, the central charge $c=\frac{2n}{2n+3}$ of $L_n(\mathfrak{osp}(1|2))$ equals $c_M-24h_{1,n+1}$, the effective central charge of
$M(2,2n+3)$, where $c_M=1-\frac{3(2n+1)^2}{2n+3}$, and the conformal weights are $h_{(p)}=h_{1,n+1-p}-h_{1,n+1}$.
Thus the supercharacters of $L_n(\mathfrak{osp}(1|2))$ and the characters of $M(2,2n+3)$ form the same vector-valued modular form,
with the same modular matrices $S$ and $T$ in the labeling \eqref{osp12M}, which act on the (super)characters
$q^{h_M-\frac c{24}}\chi^{\mathcal V}_M$, $q=e^{2\pi i\tau}$, under $\tau\to-\frac1\tau$ and $\tau\to\tau+1$,
as found in \cite{Ferrari:2023fez}.
The two VOAs are nevertheless different, since their central charges differ and this labeling does not map the vacuum module to
the vacuum module.

In a labeling which preserves the fusion rules, the modular data of the two VOAs, their matrices $S$ and $T$, are instead Galois
conjugate \cite{MR1120140,MR1266785,Gannon:2001uy}, i.e.\ related by replacing a primitive root of unity by another
one of the same order in all their entries.
We identify the module $(p)$ of $L_n(\mathfrak{osp}(1|2))$ with the module $(1,2p+1)=(1,2n+2-2p)$ of $M(2,2n+3)$, which preserves the
fusion rules \cite{Ferrari:2023fez}, those of the Wilson lines of spin $p$ in $SU(2)_{2n+1}$ Chern-Simons theory \cite{MR3856129,Harvey:2019qzs}.
The conformal weights $h_{(p)}=\frac{p(p+1)}{2(2n+3)}$ and $h_{1,2p+1}=\frac{p(2p-2n-1)}{2n+3}$ have the denominator $2n+3$, so that the
topological spins\footnote{The topological spin $e^{2\pi ih_M}$ of a module $M$ is the phase of the corresponding line of the 3d topological field
theory under a $2\pi$ rotation. It is the eigenvalue of $T$ on the character $q^{h_M-\frac c{24}}\chi^{\mathcal V}_M$ up to the common factor $e^{-\frac{2\pi ic}{24}}$.}
$e^{2\pi ih_{(p)}}$ and $e^{2\pi ih_{1,2p+1}}$ are powers of $e^{\frac{2\pi i}{2n+3}}$.
The entries of $S$ of $M(2,2n+3)$ for the modules $(1,s)$ and $(1,s')$ are $\pm\frac{2}{\sqrt{2n+3}}\sin\frac{2\pi ss'}{2n+3}$
\cite{Cardy:1986ie,DiFrancesco:1997nk}, which are polynomials in $e^{\frac{2\pi i}{2n+3}}$ and $i$ with rational coefficients.
The Galois conjugation we consider replaces $e^{\frac{2\pi i}{2n+3}}$ by $e^{\frac{8\pi i}{2n+3}}$ in these polynomials and keeps
$i$.\footnote{It is the automorphism of the field generated by $i$ and $e^{\frac{2\pi i}{2n+3}}$ over the rational numbers which fixes $i$ and
maps $e^{\frac{2\pi i}{2n+3}}$ to its fourth power, which exists since $4$ is coprime to $2n+3$. It maps $\sin\frac{2\pi x}{2n+3}$ to $\sin\frac{8\pi x}{2n+3}$. It leaves $\sqrt{2n+3}$ invariant, since up to a power of $i$ this is the Gauss sum
$\sum_te^{\frac{2\pi it^2}{2n+3}}$ over the residues $t$ modulo $2n+3$, which $t\to2t$ permutes. It extends to an automorphism of the field generated by $e^{\frac{2\pi i}{24(2n+3)}}$ which also maps the factor
$e^{-\frac{2\pi ic}{24}}$ of $T$ to $e^{-\frac{2\pi ic_M}{24}}$.}
Since $4h_{(p)}-h_{1,2p+1}=p$, it maps $e^{2\pi ih_{(p)}}$ to $e^{2\pi ih_{1,2p+1}}$, and by \eqref{osp12M} it maps the entry $S_{(p),(p')}$
of $S$ of the supercharacters to $(-1)^{p+p'}S_{(1,2p+1),(1,2p'+1)}$.
The signs disappear if the supercharacter of the module $(p)$ is multiplied by $(-1)^p$, the sign of the superdimension of its top
component, and with this choice of the parity of the modules the modular data are Galois conjugate.\footnote{The modular data of
both also appear in \cite{Go:2025ixu} for small $n$, among those of the 3d topological field theories obtained from the Argyres-Douglas
theory of type $(A_1,A_{2n})$ with powers of its monodromy operator and the A- or B-twist. There the modular data for
different powers are observed to be related by Galois transformations, which raise $T$ to the same power. Changing the holonomy in the construction of
\cite{Dedushenko:2018bpp} also acts by a Galois transformation, and the modular representation of $M(2,k+2)$ is, up to a
one-dimensional representation, a Galois conjugate of that of $(A_1,k)_{\frac12}$ \cite{Harvey:2019qzs}.}

At the C-twist point of (II) for $\ell=1$ the boundary VOA is expected to be $L_1(\mathfrak{osp}(1|2n))$ (section \ref{sec_twist_com}).
By the Giveon-Kutasov type duality it is also the commutant \eqref{H_coset} with $(n,\ell)$ replaced by $(1,n)$.
Its $n+1$ modules $(1^p)$ have by \eqref{hmu} the conformal weights $h_{(1^p)}=\frac{p(2n+1-p)}{2(2n+3)}=-h_{1,p+1}$, those of
$M(2,2n+3)$ with the opposite signs, so that their topological spins are the complex conjugates of those of $M(2,2n+3)$,
in accordance with the relation between the $T$ matrices of mirror pairs in \cite{Creutzig:2024ljv}.
In \cite{Creutzig:2024ljv} $L_1(\mathfrak{osp}(1|2n))$ and $M(2,2n+3)=W^{\min}_{n-\frac12}(\mathfrak{usp}(2))$ \cite{Arakawa:2005krw} are
related by a level-rank duality in $2n$ $bc$ systems, as $L_1(\mathfrak{osp}(1|2))$ and $M(2,5)$ for $n=1$ \cite{Ferrari:2023fez}.

\subsection{Bulk line defect indices}
\label{sec_bulk}
The supersymmetric index on $S^1\times S^2$ \cite{Bhattacharya:2008zy,Kim:2009wb,Imamura:2011su,Kapustin:2011jm}
can be decorated by line operators which wrap $S^1$ at the north ($N$) or south ($S$) pole of $S^2$.
These line defect indices test the map of the line operators independently of the boundary conditions and detect the
sign of the Chern-Simons level (section \ref{sec_bulksign}).

\subsubsection{Insertions}
By the localization of \cite{Drukker:2012sr}, a Wilson line of charge $\eta$ at the north pole
and one of charge $\eta'$ at the south pole contribute $s^{\eta+\eta'}q^{-(\eta-\eta')m/2}$ at the
saddle point with flux $m$.
Hence the line defect index of the Wilson line $W_\lambda$ in the representation $\lambda$ at the
north or south pole is
\begin{align}
\label{bulk_ins}
\Ical^{USp(2n)_K}[W^{N/S}_\lambda](a)
&=\frac{1}{2^nn!}\sum_{m\in\Zb^n}\oint\prod_{i=1}^n\frac{ds_i}{2\pi is_i}(-s_i)^{2Km_i}\
q^{-\frac12\sum_{\alpha>0}|\alpha(m)|}\bigl(q^{\frac{1-r_Q}{2}}a^{-1}\bigr)^{\sum_i|m_i|}
\nonumber\\
&\qquad\times\prod_{\alpha\in\Delta}\bigl(1-q^{\frac{|\alpha(m)|}{2}}s^\alpha\bigr)
\prod_{i=1}^n\frac{\poch{q^{1-\frac{r_Q}{2}+\frac{|m_i|}{2}}a^{-1}s_i^{\mp}}}{\poch{q^{\frac{r_Q}{2}+\frac{|m_i|}{2}}as_i^{\pm}}}
\nonumber\\
&\qquad\times\chi_\lambda\bigl(s_1q^{-\epsilon\frac{m_1}{2}},\dots,s_nq^{-\epsilon\frac{m_n}{2}}\bigr),
\end{align}
with $\epsilon=+1$ for $W^N_\lambda$ and $\epsilon=-1$ for $W^S_\lambda$, and $\lambda=\varnothing$ gives
the full index $\Ical^{USp(2n)_K}$ of $USp(2n)_K+Q$.
The Weyl group maps each insertion to itself, and the two insertions give different indices in
general.
The symmetric insertion $\chi_\lambda(s)$, which for an Abelian Wilson line of even charge
corresponds to splitting the charge between the two poles, does not describe a BPS Wilson line
of $USp(2n)$.
We have checked these prescriptions in the duality between $U(1)_{-1/2}+\Phi$ and a free chiral multiplet, and in
$USp(2)_1+2Q$, which is dual to a free meson.
We have also checked them in $USp(2)_{3/2}+Q$, whose index is $1$ and whose line defect indices of the fundamental Wilson line are
monomials.

For $\mathcal{T}_{\ell,n}$ the line defect index of the Wilson line $W_{\mathbf Q}$ of charge $\mathbf Q$ at the north or
south pole is the index of \cite{Okazaki:2026efy} with the Wilson line inserted,
\begin{align}
\label{Ab_fullindex}
\Ical^{\mathcal{T}_{\ell,n}}[W^{N/S}_{\mathbf Q}](y)
&=\sum_{m\in\Zb^r}\oint\prod_I\frac{ds_I}{2\pi is_I}\,s_I^{(\kappa m)_I}\,(q^{-\frac12}y)^{\mathbf A\cdot m}
\nonumber\\
&\qquad\times\prod_I\Bigl(\frac{-q^{\frac{1-R_I}{2}}}{s_I}\Bigr)^{m_I^+}
\frac{\poch{q^{1-\frac{R_I}{2}+\frac{|m_I|}{2}}s_I^{-1}}}{\poch{q^{\frac{R_I}{2}+\frac{|m_I|}{2}}s_I}}\
\prod_I(s_Iq^{-\epsilon\frac{m_I}{2}})^{(\mathbf Q)_I},
\end{align}
with $m^+=\max(m,0)$ and, as in \eqref{bulk_ins}, $\epsilon=+1$ for $W^N_{\mathbf Q}$ and $\epsilon=-1$ for $W^S_{\mathbf Q}$.
Here $\kappa$ includes the contribution $+\frac12$ of each chiral multiplet (section \ref{sec_bulk_ab}), and the factors
$(-q^{\frac{1-R_I}{2}}/s_I)^{m_I^+}$ are the remaining contributions of the chiral multiplets to the bare monopoles, including
their signs, so that the exponent of $s_I$ is $(\kappa m)_I-m_I^+=(\tilde\kappa m)_I-\frac{|m_I|}{2}$.
At nodes with $R_I=0$ the contour is taken slightly inside the unit circle, by continuity from
$R_I>0$.
As for the half-indices, the line defect indices of the two theories are compared under the
identification $y=a^2$ of the fugacities of the topological symmetry $U(1)_J$ of $\mathcal{T}_{\ell,n}$
and of $U(1)_a$, which we indicate by $\big|_{y=a^2}$.

\subsubsection{Vortex lines}
The line defect index of the vortex line $V$ of the symplectic gauge theory for the global structure
$(USp(2n)\times U(1)_a)/\Zb_2$ is computed by extending the prescription of \cite{Drukker:2012sr}.
The gauge fluxes are shifted by the
center, $m\in(\Zb+\frac12)^n$, together with the background flux $\pm\frac12$ of $U(1)_a$ and the
shift $a\to aq^{1/4}$ of its fugacity, while the gauge holonomies are not shifted.
A Wilson line
combined with $V$ is put at the same pole.
The vortex line $V$ at the south pole of $S^2$, possibly combined with $W_\lambda$, is inserted as
\begin{align}
\label{bulkV}
\Ical^{USp(2n)_K}[(V\otimes W_\lambda)^{S}]&=\frac{1}{2^nn!}\sum_{m\in(\Zb+\frac12)^n}\oint\prod_{i=1}^n\frac{ds_i}{2\pi is_i}\,s_i^{2Km_i}
\prod_{\alpha\in\Delta}q^{-\frac{|\alpha(m)|}{4}}\bigl(1-s^\alpha q^{\frac{|\alpha(m)|}{2}}\bigr)
\nonumber\\
&\qquad\times\prod_{w=\pm \mathbf e_i}Z_{\rm chi}\bigl(w(m)+\tfrac12,\ aq^{\frac14}s^w\bigr)\ \chi_\lambda(sq^{\frac m2}),
\nonumber\\
Z_{\rm chi}(f,X)&=q^{-\frac{f(1-r_Q)}{4}}X^{\frac f2}\frac{\poch{q^{1-\frac{r_Q}{2}-\frac f2}X^{-1}}}{\poch{q^{\frac{r_Q}{2}-\frac f2}X}},
\end{align}
while $\bar V$ at the north pole has $f=w(m)-\frac12$, the same $X=aq^{\frac14}s^w$ and $\chi_\lambda(sq^{-m/2})$.
With $m\in\Zb^n$, $f=w(m)$ and $X=as^w$ this reduces to \eqref{bulk_ins} without lines, including the signs.

\subsubsection{The two poles}
\paragraph{Symplectic gauge theory}
The full index is invariant under $K\to-K$.
Indeed, the pseudo-reality of $\mathbf{2n}$ makes the one-loop determinant in \eqref{bulk_ins} depend only on $|m_i|$, and the change of
the Chern-Simons factor is undone by $m\to-m$.
With the Wilson lines, the change of variables $m\to-m$ in \eqref{bulk_ins} gives
\begin{align}
\label{pole_flip}
\Ical^{USp(2n)_{K}}[W^S_\lambda]=\Ical^{USp(2n)_{-K}}[W^N_\lambda].
\end{align}

\paragraph{Abelian gauge theory}
The substitution $m\to-m$ in \eqref{Ab_fullindex} maps $W^S_{\mathbf Q}$ to $W^N_{\mathbf Q}$.
It gives \eqref{Ab_fullindex} with $\kappa$, $\mathbf A$ and $m_I^+$ replaced by $-\kappa$, $-\mathbf A$ and $m_I^-=\max(-m_I,0)$.
In terms of $m_I^+$ the gauge levels become $\mathrm{id}-\kappa$, i.e.\ the parity-symmetric levels $\kappa-\frac12\mathrm{id}$ change
sign, so that this is the index of the parity image $\Tb$.
Hence
\begin{align}
\label{pole_flip_Ab}
\Ical^{\mathcal T}[W^S_{\mathbf Q}](y)=\Ical^{\Tb}[W^N_{\mathbf Q}](y),
\end{align}
which, like \eqref{pole_flip}, expresses that the reflection of $S^2$ exchanging the two poles
reverses the orientation.
By \eqref{pole_flip_Ab} with $\mathbf Q=0$ the full indices of $\mathcal{T}$ and $\Tb$ coincide, as observed in
\cite{Creutzig:2024ljv}, so that only the line defect indices distinguish them.

\section{\texorpdfstring{$USp(2n)_{-(n+\frac32)}$ and $\mathcal{T}_{1,n}$}{USp(2n) and T(1,n)}}
\label{sec_n1}
For $\ell=1$ the dual pair \eqref{pair_P} consists of $USp(2n)_{-(n+\frac32)}+Q$ and the tadpole
theory $\mathcal{T}_{1,n}$ of \cite{Gang:2023rei,Okazaki:2026efy}, the $U(1)^n$ Chern-Simons matter theory with
\begin{align}
\label{n1_data}
\kappa=2C(T_n)^{-1}=\bigl(2\min(I,J)\bigr)_{I,J=1}^n,\qquad \mathbf A=(1,2,\dots,n),\qquad R=(0,\dots,0,R_S),
\end{align}
and $S=\{n\}$.
In sections \ref{sec_n1_I} and \ref{sec_n1_II} we give the results for general $n$, and in section
\ref{sec_n1_ex} we illustrate them for $n=1,2,3$.
For $\ell=1$ we find that all the Wilson lines in the representations inside the rectangle $(1^n)$, i.e.\ the
antisymmetric Wilson lines $W_{(1^p)}$, map to single Wilson lines.
We prove for all $n$ that the H-twist limits of their line defect half-indices with the Neumann-type boundary condition of
$USp(2n)_{-(n+\frac32)}$ give, up to monomials, the characters of all the modules of the Virasoro minimal model $M(2,2n+3)$
(appendix \ref{app_proof}).
By \eqref{osp12M} these are the supercharacters of all the modules of $L_n(\mathfrak{osp}(1|2))$.
With the Andrews-Gordon identities it follows that the equalities \eqref{n1_I_lines} of these line defect half-indices hold at the
H-twist point for all $n$ (section \ref{sec_n1_H}).

\subsection{Boundary conditions (I)}
\label{sec_n1_I}
\subsubsection{Symplectic gauge theory}
We take Neumann boundary conditions for the vector multiplet and $Q$ together with a single
fundamental Fermi multiplet $\Gamma$ with flavor fugacity $x$.
The line defect half-index of the Wilson line $W_\lambda$ is
\begin{align}
\label{n1_I_A}
\lin{W_\lambda}{USp(2n)_{-(n+\frac32)}}{\Ncal,N,\Gamma}(a,x)
&=\frac{(q)_\infty^n}{2^nn!}\oint\prod_{i=1}^n\frac{ds_i}{2\pi is_i}
\prod_{i\neq j}\poch{s_is_j^{-1}}\prod_{i\le j}\poch{s_is_j}\poch{s_i^{-1}s_j^{-1}}
\nonumber\\
&\qquad\times\prod_{i=1}^n\frac{\poch{q^{\frac12}xs_i^{\pm}}\poch{q^{\frac12}x^{-1}s_i^{\pm}}}{\poch{q^{\frac{r_Q}{2}}as_i^{\pm}}}\,\chi_\lambda(s),
\end{align}
with $\chi_{(1^p)}=E_p-E_{p-2}$ for the antisymmetric representations (section \ref{sec_characters}).

\subsubsection{Abelian gauge theory}
We take Dirichlet boundary conditions for the vector multiplets and the chiral multiplets.
With $M_j=m_j+m_{j+1}+\cdots+m_n$, the line defect half-index \eqref{WQ_D} reads
\begin{align}
\label{n1_I_B}
\lin{W_{\mathbf Q}}{\mathcal{T}_{1,n}}{\Dcal,D}(u,y)
&=\frac{1}{(q)_\infty^n}\sum_{m\in\Zb^n}q^{\sum_{j=1}^nM_j^2}\,(q^{-\frac12}y)^{\sum_{j=1}^njm_j}
\prod_{I=1}^nu_I^{2\sum_{J}\min(I,J)m_J}
\nonumber\\
&\qquad\times\prod_{I=1}^n(u_Iq^{m_I})^{(\mathbf Q)_I}\poch{q^{1-\frac{R_I}{2}-m_I}u_I^{-1}} .
\end{align}

\subsubsection{Identification of the fugacities}
The identification of the fugacities is
\begin{align}
\label{n1_I_C}
u_I=1,\quad R_I=0\quad(I<n),\qquad u_n=q^{1/4}x\,a^{-1},\quad R_n=\tfrac12-r_Q,\qquad y=a^2 .
\end{align}
It corresponds to $D$ for $\Phi_n$, whose boundary $U(1)$ is identified with the Cartan of the flavor symmetry of $\Gamma$ combined with
$U(1)_a$, and to $D_c$ for the
other chiral multiplets, i.e.\ to non-zero boundary values which break their boundary $U(1)$ symmetries.
With this identification the half-indices agree,
\begin{align}
\label{n1_I_D}
\II^{USp(2n)_{-(n+\frac32)}}_{\Ncal,N,\Gamma}(a,x)=\II^{\mathcal{T}_{1,n}}_{\Dcal,D}(u,y)\Big|_{\eqref{n1_I_C}} .
\end{align}
This is the case $\ell=1$ of \eqref{gen_I_claim} (section \ref{sec_gen_I}), and we have checked it at
generic fugacities.

\subsubsection{Line operators}
We find at generic fugacities that the rank-$p$ antisymmetric Wilson lines map to single Wilson lines,
\begin{align}
\label{n1_I_lines}
\lin{W_{(1^p)}}{USp(2n)_{-(n+\frac32)}}{\Ncal,N,\Gamma}(a,x)=(-q^{\frac34}a^{-1})^p\,
\lin{W_{\mathbf Q_{(1^p)}}}{\mathcal{T}_{1,n}}{\Dcal,D}(u,y)\Big|_{\eqref{n1_I_C}},\qquad p=0,1,\dots,n,
\end{align}
with the charges $\mathbf Q_{(1^p)}=-(0,\dots,0,1,2,\dots,p)$, e.g.\ $\mathbf Q_{\yf}=(0,\dots,0,-1)$.
The largest one $W_{\mathbf Q_{(1^n)}}=W_{-\mathbf A}$ is the unit vortex line of $U(1)_J$.
The flux $v=\kappa^{-1}\mathbf Q_{(1^p)}$ attached by the Gauss law gives the endpoint of $W_{\mathbf Q_{(1^p)}}$ the topological charge
$\mathbf A\cdot v=-\frac p2$.
Under \eqref{n1_I_C} the coefficient in \eqref{n1_I_lines} times the fugacity monomial $u_n^{-p}$ of
$W_{\mathbf Q_{(1^p)}}$ is $(-1)^p(q^{1/2}x^{-1})^p$, the sign and the weight of $\Gamma^p$.
We therefore call $\Gamma$ the junction field.
The equality \eqref{n1_I_lines} is the case $\ell=1$ of \eqref{singlerule} (section \ref{sec_gen_I}).

\subsubsection{H-twist limit and VOA}
\label{sec_n1_H}
At the H-twist point the boundary VOA is expected to be the commutant \eqref{H_coset}, $L_n(\mathfrak{osp}(1|2))$ with
$c=\frac{2n}{2n+3}$ (section \ref{sec_twist_com}), whose $n+1$ modules $(p)$, $p=0,\dots,n$, have as
supercharacters the characters of the modules $(1,n+1-p)$ of $M(2,2n+3)$ \cite{Ferrari:2023fez}.
At the H-twist point \eqref{HC_def}, where $a=q^{1/4}$, $r_Q=\frac12$ and $x=1$, the contribution of the Neumann 3d chiral multiplet
$Q$ cancels one of the two factors contributed by the Fermi multiplet in \eqref{n1_I_A}.
The line defect half-index becomes
\begin{align}
\label{n1_H_int}
\lin{W_\lambda}{USp(2n)_{-(n+\frac32)}}{\Ncal,N,\Gamma}\Big|_{\rm H}
&=\frac{(q)_\infty^n}{2^nn!}\oint\prod_{i=1}^n\frac{ds_i}{2\pi is_i}
\prod_{i\neq j}\poch{s_is_j^{-1}}\prod_{i\le j}\poch{s_is_j}\poch{s_i^{-1}s_j^{-1}}
\nonumber\\
&\qquad\times\prod_{i=1}^n\poch{q^{\frac12}s_i^{\pm}}\,\chi_\lambda(s).
\end{align}
It coincides with the line defect half-index considered in \cite{Okazaki:2024kzo} for the Neumann boundary condition of the vector multiplet
and the Dirichlet boundary condition of a fundamental chiral multiplet of R-charge $1$.
Indeed, at the H-twist point the factors contributed by $\Gamma$ and $Q$ combine into that of a Dirichlet chiral multiplet
(section \ref{sec_gen_I}).
On the Abelian side the invariants \eqref{zw_def} are $z_I=w_J=1$, and \eqref{nahm_Q} gives the
Andrews-Gordon sums \cite{MR123484,MR351985}
\begin{align}
\label{n1_H_AG}
\lin{W_{\mathbf Q_{(1^p)}}}{\mathcal{T}_{1,n}}{\Dcal,D}\Big|_{\rm H}
=\sum_{N\in\Zb_{\ge0}^n}\frac{q^{\mathcal{N}_1^2+\cdots+\mathcal{N}_n^2+\mathcal{N}_{n-p+1}+\cdots+\mathcal{N}_n}}
{(q)_{N_1}\cdots(q)_{N_n}},\qquad \mathcal{N}_j=N_j+N_{j+1}+\cdots+N_n,
\end{align}
where the linear term $-\mathbf Q_{(1^p)}\cdot N$ comes from the Wilson line.

By \eqref{ramanujan_f} and \eqref{chiM}, for $1\le s\le n+1$,
\begin{align}
\label{n1_chi}
\frac{f(-q^s,-q^{2n+3-s})}{f(-q)}=\prod_{\substack{m\ge1\\ m\not\equiv0,\pm s\ ({\rm mod}\ 2n+3)}}\frac{1}{1-q^m}
=\chi^{M(2,2n+3)}_{(1,s)},
\end{align}
the normalized character of the module $(1,s)$ of $M(2,2n+3)$ \cite{MR781391}.

For every $n\ge1$ and $p=0,1,\dots,n$, the H-twist limits of the line defect half-indices of the antisymmetric Wilson lines are
\begin{align}
\label{linechi}
\lin{W_{(1^p)}}{USp(2n)_{-(n+\frac32)}}{\Ncal,N,\Gamma}\Big|_{\rm H}
=(-q^{\frac12})^p\,\frac{f(-q^{n+1-p},-q^{n+2+p})}{f(-q)}
=(-q^{\frac12})^p\,\chi^{M(2,2n+3)}_{(1,n+1-p)} ,
\end{align}
as we prove in appendix \ref{app_proof}.
The modulus $2n+3$ of the characters \eqref{n1_chi} is twice the absolute value $n+\frac32$ of the level and equals $2h^\vee+1$
with the dual Coxeter number $h^\vee=n+1$ of $USp(2n)$.
The H-twist limits of the line defect half-indices of the $n+1$ antisymmetric Wilson lines $W_{(1^p)}$, $p=0,\dots,n$, give each of the $n+1$ characters
of $M(2,2n+3)$ exactly once.
The H-twist limit of the half-index gives the character $\chi^{M(2,2n+3)}_{(1,n+1)}$ of the module with the
lowest conformal weight, and that for $p=n$ gives the vacuum character $\chi^{M(2,2n+3)}_{(1,1)}$.
The cases $p=0,1$ were conjectured in \cite{Okazaki:2024kzo}, and the case $n=1$ gives the pair of Rogers-Ramanujan identities, found for this theory in \cite{Okazaki:2024paq}.

The module of the VOA $L_{\frac12}(\mathfrak{usp}(2n))$ of the gauge currents with top component $(1^p)$ has the conformal weight
$\frac{p(2n+2-p)}{2(2n+3)}$, and multiplying \eqref{linechi} by $q^{-\frac c{24}}$ gives
\begin{align}
\label{usp_full}
q^{-\frac c{24}}\lin{W_{(1^p)}}{USp(2n)_{-(n+\frac32)}}{\Ncal,N,\Gamma}\Big|_{\rm H}=(-1)^pq^{\frac{p(2n+2-p)}{2(2n+3)}}\,q^{h_{1,n+1-p}-\frac{c_M}{24}}
\chi^{M(2,2n+3)}_{(1,n+1-p)},
\end{align}
where $c_M=c+24h_{1,n+1}$ is the central charge of $M(2,2n+3)$.
The last two factors form the character of the module $(1,n+1-p)$ of $M(2,2n+3)$ including $q^{h_{1,n+1-p}-\frac{c_M}{24}}$.

On the Abelian side the duality fixes the twist point of the Dirichlet-type half-index of $\mathcal{T}_{1,n}$ to $y=q^{1/2}$.
The same half-index appears in \cite{Gang:2023rei} at the point $\nu=-1$, i.e.\ $y=q^{-1/2}$, regarded there as the
A-twist, where it gives the vacuum character of $M(2,2n+3)$.
By \eqref{WQ_vortex} with $j=1$, the shift $y\to q^{-1}y$ at $u_I=1$ is the insertion of $W_{-\mathbf A}$.
Hence the half-index at $y=q^{-1/2}$ equals the line defect half-index of $W_{-\mathbf A}$ at our twist point $y=q^{1/2}$.
Since $W_{-\mathbf A}$ is dual to $W_{(1^n)}$, this is the vacuum character, in accordance with \eqref{linechi}.
For $n=1$ the case $p=0$ of \eqref{linechi} agrees with \cite{Okazaki:2024paq}, where the Dirichlet-type half-index of $U(1)_{3/2}$ without lines was identified with
$\chi^{M(2,5)}_{(1,2)}$.

At the H-twist point the line defect half-indices of the Wilson lines in the other representations reduce to those of the
$W_{(1^p)}$.
Let $\rho=(n,n-1,\dots,1)$ be the Weyl vector of $USp(2n)$ and $\Lambda=\rho+\lambda$ for a dominant weight $\lambda$.
Then the H-twist limit of the line defect half-index of $W_\lambda$ is either zero or
$\pm q^{\sum_i(\Lambda_i^2-(\Lambda^{(p)}_i)^2)/(4n+6)}$ times that of $W_{(1^p)}$ in \eqref{linechi} for a single $p$, where
$\Lambda^{(p)}=\rho+(1^p)$ (see
appendix \ref{app_proof} for the derivation).
In particular, the H-twist limit of the line defect half-index of the Wilson line in the rank-$m$ symmetric representation either
vanishes or is a monomial times the H-twist limit of the half-index or of the line defect half-index of the fundamental Wilson line.
This was observed for $USp(4)$ and conjectured for $USp(6)$ in \cite{Okazaki:2024kzo}.\footnote{In \cite{Okazaki:2024kzo} the relations for the rank-$2$ and rank-$3$ symmetric Wilson
lines of $USp(4)_{-7/2}$ should read $\langle W_{(2)}\rangle=q^{\frac12}\langle W_{(1)}\rangle$ and $\langle W_{(3)}\rangle=q^{\frac32}\II$.}
For example, for the rank-$2$ symmetric representation of $USp(4)$ we have
\begin{align}
\label{usp4_sym}
\lin{W_{\yb}}{USp(4)_{-7/2}}{\Ncal,N,\Gamma}\Big|_{\rm H}=q^{\frac12}
\lin{W_{\yf}}{USp(4)_{-7/2}}{\Ncal,N,\Gamma}\Big|_{\rm H}.
\end{align}

At the H-twist point the Abelian side of \eqref{n1_I_lines} is the Andrews-Gordon sum \eqref{n1_H_AG}.
By \eqref{linechi}, the equality \eqref{n1_I_lines} at this point is the Andrews-Gordon identity \cite{MR123484,MR351985}, i.e.\ the
equality of the sum \eqref{n1_H_AG} and the product \eqref{n1_chi}.
Hence at the H-twist point \eqref{n1_I_lines} holds for all $n$ and $p$.
In \cite{Creutzig:2024ljv} the B-twisted lines $\mathcal{P}^B_\mu$ in the Dirichlet-type half-indices of the Abelian theories were
proposed to give the supercharacters of the modules $\mu$ of $L_n(\mathfrak{osp}(1|2\ell))$.
The charges $\mathbf Q_{(1^p)}$ and the signs $(-1)^p$ coincide, up to the convention for the signs of the
charges, with those of $\mathcal{P}^B_\mu=(-1)^\mu\prod_{i=0}^{\mu-1}z_{n-i}^{\mu-i}$ for $\mu=p$, in the variables $z$ of \cite{Creutzig:2024ljv}.

\subsection{Boundary conditions (II)}
\label{sec_n1_II}
\subsubsection{Symplectic gauge theory}
We take Dirichlet boundary conditions for the vector multiplet and $Q$.
This is a boundary condition of the parity image $USp(2n)_{n+\frac32}+Q$ with $k_{\rm eff}=1$ (section \ref{sec_signs}).
With $\tilde m=m+\frac j2\mathbf 1$, the line defect half-index \eqref{Vj_general} of $V^j\otimes W_\lambda$ is
\begin{align}
\label{n1_II_A}
\lin{V^j\otimes W_{\lambda}}{USp(2n)_{n+\frac32}}{\Dcal,D}(u;a)
&=\frac{1}{(q)_\infty^n}\sum_{m\in\Zb^n}q^{\sum_i\tilde m_i^2}\prod_iu_i^{2\tilde m_i}\ \chi_{\lambda}(q^{\tilde m}u)
\nonumber\\
&\qquad\times\prod_{i=1}^{n}\frac{\poch{q^{1-\frac{r_Q}{2}\pm\tilde m_i}(q^{\frac j2}a)^{-1}u_i^{\pm1}}}{\poch{q^{1\pm2\tilde m_i}u_i^{\pm2}}}
\nonumber\\
&\qquad\times\prod_{i<i'}\frac{1}{\poch{q^{1\pm(\tilde m_i-\tilde m_{i'})}u_i^{\pm}u_{i'}^{\mp}}\poch{q^{1\pm(\tilde m_i+\tilde m_{i'})}u_i^{\pm}u_{i'}^{\pm}}} .
\end{align}

\subsubsection{Abelian gauge theory}
We take Neumann boundary conditions for the vector multiplets and the 3d chiral multiplets of the parity image $\Tb_{1,n}$ and add $2n$ boundary
Fermi multiplets charged under its gauge group $U(1)^n$, two of each of the charges $\mathbf f_j=\sum_{i\ge j}\mathbf e_i$, $j=1,\dots,n$, where $\mathbf e_i$ is the
unit charge under the $i$-th $U(1)$.
The two Fermi multiplets of charge $\mathbf f_j$ have the fugacities $q^{R_S/2}hx_j^{\pm1}$, where $R_S$ is the R-charge of $\Phi_n$.
Here $h$ is the fugacity of the boundary flavor symmetry $U(1)_h$, which acts on all the Fermi multiplets with the same charge.
The fugacity $x_j$ is that of $U(1)_{x_j}$, which acts on the two Fermi multiplets of charge $\mathbf f_j$ with the charges $\pm1$.
Since $\sum_F\mathbf g_F\mathbf g_F^T=2\sum_j\mathbf f_j\mathbf f_j^T=2C(T_n)^{-1}=\kappa$, the gauge anomaly cancels (section \ref{sec_signs}).
As shown in section \ref{sec_gen_II}, $U(1)_h$ is $U(1)_J$ acting with the charge $\frac12$.
With $\sigma_j=s^{\mathbf f_j}=\prod_{i\ge j}s_i$, the line defect half-index of the Wilson line $W_{\mathbf Q}$ is
\begin{align}
\label{n1_II_B}
\lin{W_{\mathbf Q}}{\Tb_{1,n}}{\Ncal,N,2n\Gamma}(h,x)
&=(q)_\infty^n\oint\prod_{I=1}^n\frac{ds_I}{2\pi is_I}\,
\prod_{I=1}^{n-1}\frac{s_I^{(\mathbf Q)_I}}{\poch{s_I}}\ \frac{(q^{\frac{R_S}2}s_n)^{(\mathbf Q)_n}}{\poch{q^{\frac{R_S}2}s_n}}
\nonumber\\
&\qquad\times\prod_{j=1}^n\prod_{\epsilon=\pm1}\poch{q^{\frac{1+R_S}{2}}hx_j^{\epsilon}\sigma_j}\poch{q^{\frac{1-R_S}{2}}h^{-1}x_j^{-\epsilon}\sigma_j^{-1}} .
\end{align}

\subsubsection{Identification of the fugacities}
With $u_j=x_j$, $h=q^{(r_Q-1)/2}a$ and $R_S=\frac12-r_Q$ the half-indices agree,
\begin{align}
\label{n1_II_D}
\II^{\Tb_{1,n}}_{\Ncal,N,2n\Gamma}(h,x)=\II^{USp(2n)_{n+\frac32}}_{\Dcal,D}(x;a)\Big|_{h=q^{(r_Q-1)/2}a}.
\end{align}
This is the case $\ell=1$ of \eqref{gen_II_claim}, and we have checked it at generic fugacities.
The equality \eqref{n1_II_D} indicates that the boundary flavor symmetry $U(1)^n$ of the Abelian side is enhanced to $USp(2n)$.

\subsubsection{Line operators}
For $p=0,\dots,n$ we find at generic fugacities that the lines map as
\begin{align}
\label{n1_II_rule}
\lin{V^j\otimes W_{(1^p)}}{USp(2n)_{n+\frac32}}{\Dcal,D}
=(-1)^{p+nj}\,q^{-\frac n4j^2+\frac{p(j-1)}{2}}\,h^{\,p-nj}\,
\lin{W_{-j\mathbf A-\mathbf Q_{(1^p)}}}{\Tb_{1,n}}{\Ncal,N,2n\Gamma}\Big|_{h=q^{(r_Q-1)/2}a},
\end{align}
with the identification of the fugacities of \eqref{n1_II_D}.
We have checked \eqref{n1_II_rule} for $|j|\le3$ with $n\le4$ and for $|j|\le8$, $6$ and $5$ with $n=1$, $2$ and $3$, respectively, and we expect it
to hold for all $j\in\Zb$.
For $V^j\otimes W_{(1^n)}$ and $V^j\otimes W_{(1^{n-1})}$, \eqref{n1_II_rule} is the case $\ell=1$ of \eqref{II_two} (section
\ref{sec_gen_II}).

By the Giveon-Kutasov type duality the line $V\otimes W_{(1^p)}$ is related to the Wilson line $W_{(n-p)}$ of $USp(2)_{-(n+\frac32)}$ with $n$
Fermi multiplets on (I) (see \eqref{GKlines}), which is a symmetric Wilson line for $p\le n-2$.

\subsubsection{C-twist limit and VOA}
At the C-twist point the boundary VOA is expected to be $L_1(\mathfrak{osp}(1|2n))$ with
$c=\frac{2n(2n-1)}{2n+3}$ (sections \ref{sec_twist_com} and \ref{sec_twist_osp}), whose $n+1$ modules $(1^p)$ have the highest
weights $\omega_p$.
We find that their supercharacters are given by the C-twist limits of the line defect half-indices of the lines
$V\otimes W_{(1^{n-p})}$,
\begin{align}
\label{n1_II_twist}
\lin{V\otimes W_{(1^{n-p})}}{USp(2n)_{n+\frac32}}{\Dcal,D}\Big|_{\rm C}=q^{\frac p2-\frac n4}\,\chi^{L_1(\mathfrak{osp}(1|2n))}_{(1^p)},
\end{align}
the case $\ell=1$ of \eqref{CVW}.
For $n\ge2$ the individual flux sectors of \eqref{n1_II_A} are singular at $u=1$, but their sum is finite.
By the Giveon-Kutasov type duality these supercharacters are also given by the H-twist limits of the line defect half-indices of the
Wilson lines $W_{(p)}$ of $USp(2)_{-(n+\frac32)}$ with $n$ Fermi multiplets on (I), i.e.\ of (I) with $(n,\ell)\to(1,n)$ (see
\eqref{GKlines}).
Section \ref{sec_voa_usp} collects their closed forms.

\subsection{Examples}
\label{sec_n1_ex}
The equalities of the half-indices and of the line defect half-indices of the antisymmetric Wilson lines on (I) and of
$V^j\otimes W_{(1^p)}$, $p=0,\dots,n$, on (II) below are the cases $n=1,2,3$ of \eqref{n1_I_D}, \eqref{n1_I_lines},
\eqref{n1_II_D} and \eqref{n1_II_rule}.
The proposed boundary VOAs at the two twist points are listed in table \ref{tab_n1_voa}.
On the Abelian side the Neumann-type line defect half-indices of the dual lines at the C-twist point are \eqref{n1_II_B} at $h=q^{-1/2}$
and $x=1$, which \eqref{NNfsum} with $\ell=1$ writes as alternating fermionic sums.

\begin{table}[t]
\begin{center}\footnotesize\setlength{\tabcolsep}{3pt}
\begin{tabular}{l|llll|lll}
\toprule
& \multicolumn{4}{c|}{H-twist point, boundary conditions (I)} & \multicolumn{3}{c}{C-twist point, boundary conditions (II)}\\
$n$ & VOA & $c$ & $c$ of $M(2,2n+3)$ & $h_{1,n+1},\dots,h_{1,1}$ & VOA & $c$ & superdimensions\\
\midrule
$1$ & $L_1(\mathfrak{osp}(1|2))$ & $\frac25$ & $-\frac{22}{5}$ & $-\frac15,\ 0$ & $L_1(\mathfrak{osp}(1|2))$ & $\frac25$ & $1,\ -1$\\
$2$ & $L_2(\mathfrak{osp}(1|2))$ & $\frac47$ & $-\frac{68}{7}$ & $-\frac37,\ -\frac27,\ 0$ & $L_1(\mathfrak{osp}(1|4))$ & $\frac{12}{7}$ & $1,\ -3,\ 2$\\
$3$ & $L_3(\mathfrak{osp}(1|2))$ & $\frac23$ & $-\frac{46}{3}$ & $-\frac23,\ -\frac59,\ -\frac13,\ 0$ & $L_1(\mathfrak{osp}(1|6))$ & $\frac{10}{3}$ & $1,\ -5,\ 9,\ -5$\\
\bottomrule
\end{tabular}
\caption{The proposed boundary VOAs of the examples with $\ell=1$, where the columns $c$ give their central charges.
At the H-twist point the modules $(p)$ of $L_n(\mathfrak{osp}(1|2))$ have as supercharacters the characters of the modules
$(1,n+1-p)$ of the Virasoro minimal model $M(2,2n+3)$, whose central charges and conformal weights are listed.
The central charge $\frac{2n}{2n+3}$ of $L_n(\mathfrak{osp}(1|2))$ equals the effective central charge $c_M-24h_{1,n+1}$ of
$M(2,2n+3)$, where $c_M$ is the central charge of $M(2,2n+3)$.
At the C-twist point the modules $(1^p)$ of $L_1(\mathfrak{osp}(1|2n))$, $p=0,\dots,n$, have the highest weights
$\omega_p$, and their top components have the listed superdimensions.}
\label{tab_n1_voa}
\end{center}
\end{table}

\subsubsection{\texorpdfstring{$n=1$}{n=1}}
\label{sec_n1_ex_1}
For $n=1$ the symplectic side is $USp(2)\cong SU(2)$ with one doublet $Q$, and $\mathcal{T}_{1,1}$ is
$U(1)_{3/2}$ with one chiral multiplet $\Phi$ of charge $+1$, with $\kappa=2$ and $\mathbf A=1$.

\paragraph{Boundary conditions (I)}
The identification \eqref{n1_I_C} is $y=a^2$, $R=\frac12-r_Q$ and $u=q^{1/4}x\,a^{-1}$.
Since the left-hand side of \eqref{n1_I_D} is invariant under $x\to x^{-1}$, the equality \eqref{n1_I_D} implies the non-trivial
symmetry of \eqref{AbD_zw} under $(z,w)\to((zw)^{-1},w)$.
By \eqref{n1_I_lines} the Wilson line $W_{\yf}$ maps to the Abelian Wilson line $W_{-1}$, whose line defect half-index comes with the
coefficient $-q^{\frac34}a^{-1}$.
The charge $-1=-\mathbf A$ carries the flux $v=-\frac12$ attached by the Gauss law (section \ref{sec_n1_I}).
Hence the flux sectors $m$ have the fluxes $m+v$, and the endpoint has the half-integral topological charge $\mathbf A\cdot v=-\frac12$,
consistent with the fundamental representation being odd under the center $\Zb_2$.
The lowest junction operators match as follows.
\begin{center}\small
\begin{tabular}{@{}lll@{}}
\toprule
endpoint of $W_{-1}$ & term of $-q^{3/4}a^{-1}\lin{W_{-1}}{}{}$ & junction on the symplectic side\\
\midrule
flux $-\frac12$ ($m=0$) & $-q^{1/2}x^{-1}$ & $\Gamma$\\
flux $+\frac12$ ($m=1$) & $-q^{1/2}x$ & $\Gamma$\\
flux $+\frac12$ with the fermion of $\Phi$ & $q^{r_Q/2}a$ & $Q$\\
\bottomrule
\end{tabular}
\end{center}
At $a=q^{1/4}$ and $r_Q=\frac12$, with generic $x$, the Abelian half-index coincides with the expression of the refined vacuum
supercharacter of
$L_1(\mathfrak{osp}(1|2))$ as a Dirichlet-type half-index \cite{Ferrari:2023fez,Creutzig:2024ljv}, $x$ being the fugacity
for the Cartan of $\mathfrak{osp}(1|2)$.
At $x=1$ the sums \eqref{n1_H_AG} are the Rogers-Ramanujan sums
\begin{align}
\label{RR_sums}
\chi^{M(2,5)}_{(1,2)}=\sum_{N\ge0}\frac{q^{N^2}}{(q)_N},\qquad \chi^{M(2,5)}_{(1,1)}=\sum_{N\ge0}\frac{q^{N^2+N}}{(q)_N}.
\end{align}

\paragraph{Boundary conditions (II)}
On the Abelian side of (II) the two boundary Fermi multiplets $\Gamma_{1,2}$ of gauge charge $+1$ cancel the gauge
anomaly $-\frac32-\frac12=-2$ of the parity image $\Tb_{1,1}=U(1)_{-3/2}+\Phi$ (section \ref{sec_signs}).
Of the symmetries acting on the Fermi multiplets, $SU(2)_x$ rotates them and has no mixed anomaly with the gauge group, while
$U(1)_h$ is $U(1)_J$ acting with the charge $\frac12$, as in section \ref{sec_n1_II}.
By \eqref{n1_II_D}, $SU(2)_x$ is identified with $USp(2)_\partial$, whose boundary anomaly $k_{\rm eff}=1$ matches
that of the doublet of Fermi multiplets, and $U(1)_h$ with $U(1)_a$, where $h^2=q^{-1/2}\hat y$ in accordance with
\eqref{w_value}.
By \eqref{n1_II_rule} the vortex lines and the fundamental Wilson line map to Wilson lines,
\begin{align}
\label{11_II_line}
\lin{V^j}{USp(2)_{5/2}}{\Dcal,D}&=(-1)^jq^{-\frac{j^2}{4}}h^{-j}\ \lin{W_{-j}}{U(1)_{-3/2}}{\Ncal,N,2\Gamma}\Big|_{h=q^{(r_Q-1)/2}a},
\nonumber\\
\lin{W_{\yf}}{USp(2)_{5/2}}{\Dcal,D}&=-q^{-\frac12}h\ \lin{W_{1}}{U(1)_{-3/2}}{\Ncal,N,2\Gamma}\Big|_{h=q^{(r_Q-1)/2}a}.
\end{align}
Since $V^j$ shifts $y\to q^jy$, the first relation is the familiar relation between a gauge Wilson line and a vortex
line of $U(1)_J$ through the mixed Chern-Simons coupling.
At the C-twist point the supercharacters in \eqref{n1_II_twist} are $\chi^{M(2,5)}_{(1,2)}$ and $\chi^{M(2,5)}_{(1,1)}$.

\subsubsection{\texorpdfstring{$n=2,3$}{n=2,3}}
\label{sec_n1_ex_23}
For $n=2,3$ the Abelian theories are the tadpole theories \eqref{n1_data} of rank two and three, e.g.\
$\kappa=\bigl(\begin{smallmatrix}2&2\\2&4\end{smallmatrix}\bigr)$, $\mathbf A=(1,2)$ and $S=\{2\}$ for $n=2$.

\paragraph{Boundary conditions (I)}
By \eqref{n1_I_lines} the antisymmetric Wilson lines map to the single Wilson lines of the charges $\mathbf Q_{(1^p)}$.
For $n=2$, $W_{\yf}$ and $W_{\yaa}$ map to $W_{0,-1}$ and $W_{-1,-2}=W_{-\mathbf A}$, and for $n=3$, $W_{\yf}$, $W_{\yaa}$ and
$W_{\ysmall{1,1,1}}$ map to $W_{0,0,-1}$, $W_{0,-1,-2}$ and $W_{-1,-2,-3}=W_{-\mathbf A}$.
As for $n=1$, the invariance of the left-hand side of \eqref{n1_I_D} under $x\to x^{-1}$ implies a non-trivial symmetry of the
Abelian side.
Under \eqref{n1_I_C} the invariants \eqref{zw_def} are $z_I=1$ for $I<n$ and $w_J=w_1^J$, and \eqref{AbD_zw} is invariant under
$z_n\to(z_nw_1)^{-1}$.
We find that the line defect half-index of $W_{\yb}$ of the symmetric representation is $q^{r_Q/2}a$ times that of $W_{\yf}$, so that
the two lines map to the same Abelian Wilson line.
More generally, $W_{(2,1^{p-1})}$ maps in this way to the same Abelian Wilson line as $W_{(1^p)}$, while the line defect half-indices
of the $W_\lambda$ with $\lambda_1=\lambda_2=2$ vanish.
These two statements are the case $\ell=1$ of \eqref{out_N} (section \ref{sec_gen_I}), and for $n=2$ at the H-twist point the first gives
\eqref{usp4_sym}.
At the H-twist point \eqref{n1_I_lines} gives the Andrews-Gordon identities for the characters of $M(2,7)$ and $M(2,9)$
(section \ref{sec_n1_H}).

\paragraph{Boundary conditions (II)}
For $n=2$ the Neumann-type boundary condition of the parity image $\Tb_{1,2}$ has four Fermi multiplets, two of charge $(1,1)$ with the flavor fugacities
$x_1^{\pm1}$ and two of charge $(0,1)$ with $x_2^{\pm1}$.
Although $x_1$ and $x_2$ enter differently, we find that the half-index is invariant under
$x_1\leftrightarrow x_2$.
The coefficient of $q^1h^0$ is the adjoint character $x_1^{\pm2}+x_2^{\pm2}+x_1^{\pm1}x_2^{\pm1}+2$ of $\mathfrak{usp}(4)$, as expected
from the enhancement of $U(1)^2$ to $USp(4)$.
By \eqref{n1_II_rule} the lines $V^j\otimes W_{(1^p)}$, $p=0,1,2$, map to single Wilson lines.
We find that the line defect half-index of $V^j\otimes W_{\yb}$ of the symmetric representation is $q^{-1/2}(q^{j/2}h)^{-1}$ times
that of $V^j\otimes W_{\yf}$, so that the two lines map to the same Abelian Wilson line.
More generally, $V^j\otimes W_{(2,1^{p-1})}$ maps in this way to the same Abelian Wilson line as $V^j\otimes W_{(1^p)}$, while the line
defect half-indices of the $V^j\otimes W_\lambda$ with $\lambda_1=\lambda_2=2$ vanish.
These two statements are the case $\ell=1$ of \eqref{out_D} (section \ref{sec_gen_II}).

\paragraph{C-twist limit}
We find that the supercharacters in \eqref{n1_II_twist} are the Nahm sums \eqref{nahm_Q} of $\mathcal{T}_{n,1}$ with the charges
$(-1,1,-1,\dots,0,\dots,0)$, whose first $\min(2p,2n-1)$ entries are non-vanishing and alternate in sign, e.g.\ for $n=2$
\begin{align}
\label{osp14chars}
\chi^{L_1(\mathfrak{osp}(1|4))}_{(1^p)}=\sum_{N\in\Zb_{\ge0}^3}\frac{q^{N_1^2+N_2^2+N_3^2-N_1N_2-N_2N_3+\sum_{b=1}^{\min(2p,3)}(-1)^{b+1}N_b}}{(q)_{N_1}(q)_{N_2}(q)_{N_3}} .
\end{align}
The cases $p=0,1,2$ also appear in \cite{Gang:2024loa}, where they are identified with the supercharacters of
$L_1(\mathfrak{osp}(1|4))$.

\section{\texorpdfstring{General $(n,\ell)$}{General (n,l)}}
\label{sec_general}
In sections \ref{sec_gen_I} and \ref{sec_gen_II} we state the results for general $(n,\ell)$, which extend
those of section \ref{sec_n1} for $\ell=1$, and in section \ref{sec_gen_ex} we illustrate them for
$(n,\ell)=(1,2)$, $(1,3)$ and $(2,2)$.

\subsection{Boundary conditions (I)}
\label{sec_gen_I}
\subsubsection{Symplectic gauge theory}
On the symplectic side we take Neumann boundary conditions for the vector multiplet and $Q$,
together with $\ell$ fundamental Fermi multiplets $\Gamma_\beta$ with flavor fugacities
$x_\beta$, $\beta=1,\dots,\ell$, so that the gauge anomaly \eqref{anomaly_I} cancels at the level
$-k$.
The flavor symmetry rotating the Fermi multiplets is $USp(2\ell)$, with the boundary anomaly $\frac12\cdot2n=n$, and the $x_\beta$
parametrize its Cartan subalgebra.
The line defect half-index of the Wilson line $W_\lambda$ is
\begin{align}
\label{gen_I_A}
\lin{W_\lambda}{USp(2n)_{-k}}{\Ncal,N,\ell\Gamma}(a,x)
&=\frac{(q)_\infty^n}{2^nn!}\oint\prod_{i=1}^n\frac{ds_i}{2\pi is_i}
\prod_{i\neq j}\poch{s_is_j^{-1}}\prod_{i\le j}\poch{s_is_j}\poch{s_i^{-1}s_j^{-1}}
\nonumber\\
&\qquad\times
\prod_{i=1}^n\frac{\prod_{\beta=1}^{\ell}\poch{q^{\frac12}x_\beta s_i^{\pm}}\poch{q^{\frac12}x_\beta^{-1}s_i^{\pm}}}{\poch{q^{\frac{r_Q}{2}}as_i^{\pm}}}\
\chi_\lambda(s).
\end{align}

\subsubsection{Abelian gauge theory}
On the Abelian side we take Dirichlet boundary conditions for all the vector multiplets.
The chiral multiplets at the nodes of $S$ obey the ordinary Dirichlet boundary conditions $D$ and keep
their boundary $U(1)$ symmetries, while those at the other nodes obey $D_c$ as in section \ref{sec_n1_I}.
The latter is compatible with their vanishing R-charge, which is required by the monopole
superpotential (section \ref{sec_bulk_dual}).
With the labels $I=(i,b)$ of the nodes (section \ref{sec_bulk_ab}) we write $\mathbf m_i=(m_{(i,1)},\dots,m_{(i,2\ell-1)})$ and
$\mathbf M_j=\sum_{i\ge j}\mathbf m_i$.
Then $\frac12m\kappa m=\frac12\sum_j\mathbf M_jC(A_{2\ell-1})\mathbf M_j$, and the line defect half-index of the Wilson line $W_{\mathbf Q}$ is
\begin{align}
\label{gen_I_B}
\lin{W_{\mathbf Q}}{\mathcal{T}_{\ell,n}}{\Dcal,D}(u,y)
&=\frac{1}{(q)_\infty^{n(2\ell-1)}}\sum_{m\in\Zb^{n(2\ell-1)}}q^{\frac12\sum_{j=1}^n\mathbf M_jC(A_{2\ell-1})\mathbf M_j}\,
(q^{-\frac12}y)^{\mathbf A\cdot m}\prod_{i,b}u_{(i,b)}^{(\kappa m)_{(i,b)}}
\nonumber\\
&\qquad\times\prod_{i,b}(u_{(i,b)}q^{m_{(i,b)}})^{(\mathbf Q)_{(i,b)}}
\poch{q^{1-\frac{R_{(i,b)}}{2}-m_{(i,b)}}u_{(i,b)}^{-1}} .
\end{align}

\subsubsection{Identification of the fugacities}
The identification of the fugacities is
\begin{align}
\label{gen_I_id}
u_{(n,2\beta-1)}=q^{1/4}x_\beta\,a^{-1}\quad(\beta=1,\dots,\ell),\qquad
u_I=1\quad(I\notin S),\qquad y=a^2,
\end{align}
with the R-charges \eqref{Rassign}.
In terms of the invariants \eqref{zw_def},
\begin{align}
z_{(n,2\beta-1)}=q^{\frac{1-r_Q}{2}}a^{-1}x_\beta,\qquad z_I=1\quad(I\notin S),\qquad w_J=(q^{-\frac12}\hat y)^{(\mathbf A)_J}.
\end{align}
The half-indices, i.e.\ \eqref{gen_I_A} with $\lambda=\varnothing$ and \eqref{gen_I_B}
with $\mathbf Q=0$, agree,
\begin{align}
\label{gen_I_claim}
\II^{USp(2n)_{-k}}_{\Ncal,N,\ell\Gamma}(a,x)=\II^{\mathcal{T}_{\ell,n}}_{\Dcal,D}(u,y)\Big|_{\eqref{gen_I_id}} .
\end{align}
We have checked \eqref{gen_I_claim} at generic fugacities.
Under \eqref{gen_I_id} the boundary $U(1)$ symmetries of the chiral multiplets at $S$ are identified one-to-one with the Cartan generators of
$USp(2\ell)$ combined with $U(1)_a$.
The factor $q^{1/4}$ in \eqref{gen_I_id} describes the mixing of this identification with the R-symmetry and does not depend on $r_Q$.
On the Abelian side $\kappa$ restricted to $S$ is $2n\,\mathrm{id}$, which gives the same anomaly $n$ of $USp(2\ell)$.

If we set $x_\ell=q^{(r_Q-1)/2}a$ in \eqref{gen_I_A}, the factor $\poch{q^{1/2}x_\ell s_i^{\pm}}$ cancels the contribution of $Q$,
and the remaining $\poch{q^{1/2}x_\ell^{-1}s_i^{\pm}}$ becomes that of a Dirichlet 3d chiral multiplet, i.e.\ the Fermi multiplet $\Gamma_\ell$
flips $Q$.
On the Abelian side this gives $z_{(n,2\ell-1)}=1$, so that the chiral multiplet at that node acquires a
boundary value.
Flipping on the Neumann-type side corresponds to Higgsing on the Dirichlet-type side.
Neumann-type boundary conditions with the Dirichlet boundary condition for $Q$ and $\ell-1$ Fermi multiplets are thus included
in (I) as specializations and will not be treated separately.

\subsubsection{Line operators}
\paragraph{Single Wilson lines}
Let $\lambda\subseteq(\ell^n)$ be a Young diagram inside the $n\times\ell$ rectangle, whose columns have the heights
$\lambda^T_1\ge\lambda^T_2\ge\cdots$, the entries of the transpose $\lambda^T$, and let
$\lambda^c=(\ell-\lambda_n,\dots,\ell-\lambda_1)$ be its complement in the rectangle.
We define
\begin{align}
\mathbf w_p=\bigl(\max(0,i-n+p)\bigr)_{i=1,\dots,n},\qquad
\boldsymbol\eta_j=\mathbf e_{2j-1}-\mathbf e_{2j}\ (j<\ell),\qquad \boldsymbol\eta_\ell=\mathbf e_{2\ell-1},
\end{align}
where $\mathbf e_b$, $\mathbf e_{(i,b)}$ and $\otimes$ are those of \eqref{unit_node}.

Let $\lambda\subseteq(\ell^n)$ with $\lambda=(v^n)$ for some $0\le v\le\ell$, or $\lambda^c=(1^p)$ for some $0\le p\le n$.
For these $\lambda$ we find at generic fugacities that the line defect half-index \eqref{gen_I_A} of $W_\lambda$
equals, up to a monomial, that of the single Wilson line $W_{\mathbf Q_\lambda}$ of $\mathcal{T}_{\ell,n}$ in \eqref{gen_I_B},
\begin{align}
\label{singlerule}
\lin{W_\lambda}{USp(2n)_{-k}}{\Ncal,N,\ell\Gamma}(a,x)
=(-q^{\frac34}a^{-1})^{|\lambda|}\ \lin{W_{\mathbf Q_\lambda}}{\mathcal{T}_{\ell,n}}{\Dcal,D}\Big|_{\eqref{gen_I_id}},\qquad
\mathbf Q_\lambda=-\sum_j\mathbf w_{\lambda^T_j}\otimes\boldsymbol\eta_j.
\end{align}
The same holds with $\mathbf Q_\lambda$ replaced by its image under the reflection $b\to2\ell-b$ of $A_{2\ell-1}$.
The equality \eqref{singlerule} is evidence that $W_\lambda$ is dual to the single Wilson line $W_{\mathbf Q_\lambda}$ of
$\mathcal{T}_{\ell,n}$.\footnote{We have also found that, for $v+v'\le\ell$, the line defect half-index of the product
$W_{(v^n)}W_{(v'^n)}$ of Wilson lines agrees, up to the factor $(-q^{\frac34}a^{-1})^{(v+v')n}$, with that of a single Wilson line of
$\mathcal T_{\ell,n}$. The charge of this Wilson line is $\mathbf Q_{(v^n)}$ plus the image of $\mathbf Q_{(v'^n)}$ under the reflection
$b\to2\ell-b$.}

For $\ell=1$ and for $n=1$, \eqref{singlerule} covers all $\lambda\subseteq(\ell^n)$.
For $\ell=1$ the complement of $(1^p)$ is $(1^{n-p})$, and \eqref{singlerule} is \eqref{n1_I_lines}, with $\mathbf Q_{(1^p)}=-\mathbf w_p$.
For $n=1$ every $\lambda=(p)$ is a rectangle $(v^n)$ with $v=p$, and
\begin{align}
\label{IWp}
\lin{W_{(p)}}{USp(2)_{-(\ell+\frac32)}}{\Ncal,N,\ell\Gamma}=(-q^{\frac34}a^{-1})^p\ \lin{W_{-1,1,-1,\dots}}{\mathcal{T}_{\ell,1}}{\Dcal,D}\Big|_{\eqref{gen_I_id}},
\end{align}
for $p=0,\dots,\ell$, where the charge $(-1,1,-1,\dots,0,\dots,0)$ has $\min(2p,2\ell-1)$ non-vanishing entries, which come first
and alternate in sign, so that it is $-\mathbf A$ for $p=\ell$.
In the examples we have studied, the line defect half-index of the product of two of the dual Wilson lines of
\eqref{singlerule}, i.e.\ of the Wilson line with the sum of their charges, is, up to a power of $q$ in each term, the sum of those
of the Wilson lines of $USp(2n)$ in the corresponding fusion.
For example, for $(n,\ell)=(1,2)$ the Wilson line $W_{-2,2,0}$, whose charge is twice that of the line $W_{-1,1,0}$ dual to $W_{\yf}$ in
\eqref{IWp}, satisfies, under \eqref{gen_I_id},
\begin{align}
\label{fusion_ex}
(-q^{\frac34}a^{-1})^2\ \lin{W_{-2,2,0}}{\mathcal{T}_{2,1}}{\Dcal,D}\Big|_{\eqref{gen_I_id}}
=q^{-1}\lin{W_{\yb}}{USp(2)_{-\frac72}}{\Ncal,N,2\Gamma}+\II^{USp(2)_{-\frac72}}_{\Ncal,N,2\Gamma},
\end{align}
in accordance with $W_{\yf}W_{\yf}=W_{\yb}+1$.
We have found similar relations for $(n,\ell)=(1,3)$, and for $W_{\yaa}W_{\yaa}=W_{\yb}+W_{\ybbd}+1$ with $(n,\ell)=(2,2)$.
For $\ell\ge2$ the complement of $(1^p)$ is $(\ell^{n-p},(\ell-1)^p)$, so that $(1^p)$ is of the type of \eqref{singlerule} only for
$p=0$ and $p=n$, with $\mathbf Q_{(1^n)}=-(1,2,\dots,n)\otimes(\mathbf e_1-\mathbf e_2)$.
For the antisymmetric Wilson lines with $0<p<n$ we give below their line defect half-indices as combinations of those of single
Wilson lines.
Some non-antisymmetric Wilson lines, such as $W_{\yb}$ for $(n,\ell)=(1,2)$ or $W_{\yba}$ for $(2,2)$, map to single lines.

\paragraph{Sums of Wilson lines}
We write the line defect half-index of $W_\lambda$ as a sum or a combination of those of single Wilson lines in the following cases:
\begin{enumerate}
\item $\lambda=\yf$ with $n,\ell\ge2$ (table \ref{tab_sums}),
\item $\lambda=(1^p)$ with $0\le p\le n$, as a combination of those of at most two single Wilson lines \eqref{antisym_two},
\item $\lambda=\yb$ with $\ell=2$ and $n\ge2$ \eqref{symdual}.
\end{enumerate}
In the first case the coefficient of the line defect half-index of each single Wilson line in the sum is determined by
boundary fields of the symplectic side, which we call
junction fields as for $\Gamma$ in section \ref{sec_n1_I}.
They are assigned to the nodes $(n,b)$ of the last row.
The junction field $\Gamma_\beta$, of weight $q^{1/2}x_\beta^{-1}$, is assigned to the odd node $(n,2\beta-1)$, and $Q$, of weight
$q^{r_Q/2}a$, to every even node.
The coefficient of the line defect half-index of $W_{\mathbf Q}$ is
\begin{align}
\label{junctionrule}
(-1)^{|\lambda|}\,\frac{\prod\,(\textrm{weights of the junction fields})}{u^{\mathbf Q}},
\end{align}
where $u^{\mathbf Q}=\prod_Iu_I^{(\mathbf Q)_I}$ is the fugacity monomial of
$W_{\mathbf Q}$ under \eqref{gen_I_id}.
For a charge $-1$ this monomial is $q^{-1/4}a\,x_\beta^{-1}$ at a node of $S$ and $1$ at an even node, so that each $\Gamma_\beta$
contributes $-q^{3/4}a^{-1}$ and each $Q$ contributes $-q^{r_Q/2}a$.
The sum is listed in table \ref{tab_sums}.
Its terms are labeled by the $2\ell-1$ nodes $b$ of the last row, and the term labeled by $b$ is the line defect half-index of the
Wilson line of charge $-\mathbf e_{(n,b)}$.

For all $p$ the line defect half-index of $W_{(1^p)}$ is a combination of those of two single Wilson lines with charges in
the first column.
For $p=0,1,\dots,n$ the charge $(\mathbf w_n-\mathbf w_{n-p})\otimes \mathbf e_1$ is $\min(i,p)$ at the node $(i,1)$ and zero elsewhere.
We find at generic fugacities
\begin{align}
\label{antisym_two}
\lin{W_{(1^p)}}{USp(2n)_{-k}}{\Ncal,N,\ell\Gamma}(a,x)&=(-q^{\frac14}a)^p\Big[\lin{W_{(\mathbf w_n-\mathbf w_{n-p})\otimes \mathbf e_1}}{\mathcal{T}_{\ell,n}}{\Dcal,D}
\nonumber\\
&\qquad-q^{\frac{r_Q}{2}-\frac14}\lin{W_{(\mathbf w_n-\mathbf w_{n-p+1})\otimes \mathbf e_1}}{\mathcal{T}_{\ell,n}}{\Dcal,D}\Big]_{\eqref{gen_I_id}},
\end{align}
where the second term is absent for $p=0$.
These are equalities of line defect half-indices, and for a general $\lambda\subseteq(\ell^n)$ we do not know the rule which maps
$W_\lambda$ to lines of $\mathcal{T}_{\ell,n}$.
For the symmetric representation $\yb$ with $\ell=2$ and $n\ge2$ we find
\begin{align}
\label{symdual}
\lin{W_{\yb}}{USp(2n)_{-k}}{\Ncal,N,2\Gamma}
=\Big[q\,\lin{W_{-\mathbf e_{(n-1,2)}-\mathbf e_{(n,2)}}}{\mathcal{T}_{2,n}}{\Dcal,D}
+q^{\frac32}a^{-2}\lin{W_{-\mathbf e_{(n,1)}+\mathbf e_{(n,2)}-\mathbf e_{(n,3)}}}{\mathcal{T}_{2,n}}{\Dcal,D}\Big]_{\eqref{gen_I_id}},
\end{align}
whose second term is the line defect half-index of the Wilson line of the charge $\mathbf Q_\lambda$ defined in \eqref{singlerule},
evaluated at $\lambda=\yb$, with the
coefficient $(-q^{\frac34}a^{-1})^2$.
For $n=1$ only the second term remains, the case $p=2$ of \eqref{IWp}.

\paragraph{Quantum numbers of the endpoints}
In \eqref{WQ_D} the factor contributed by $W_{\mathbf Q}$ combines with the Chern-Simons and topological terms into the same terms at
the flux $m+\kappa^{-1}\mathbf Q$, up to an $m$-independent factor, where $\kappa^{-1}\mathbf Q$ is the flux attached by the Gauss law.
The factors contributed by the chiral multiplets still depend on $m$, so that the endpoint carries the topological charge $\mathbf A\cdot\kappa^{-1}\mathbf Q$.
For $W_{\mathbf Q_\lambda}$ this charge is $\mathbf A\cdot\kappa^{-1}\mathbf Q_\lambda=\frac12\sum_{I\in S}(\mathbf Q_\lambda)_I=-\frac{|\lambda|}{2}$, since
$\kappa^{-1}\mathbf A=\frac12\sum_{\beta=1}^{\ell}\mathbf e_{(n,2\beta-1)}$ by \eqref{kinvA} and $\boldsymbol\eta_j$ is $+1$ at the odd node $2j-1$.
The corresponding weight $(q^{-\frac12}y)^{-\frac{|\lambda|}{2}}$ is the second factor of
$(-q^{\frac34}a^{-1})^{|\lambda|}=(-q^{\frac12})^{|\lambda|}(q^{-\frac12}y)^{-\frac{|\lambda|}{2}}|_{y=a^2}$ in \eqref{singlerule}.
This charge matches the charge $(-1)^{|\lambda|}$ of $W_\lambda$ under the center $\Zb_2$.
Indeed, by \eqref{gen_I_id} the $U(1)_J$ charge of a boundary operator is half the sum of its charges under $U(1)_a$ and the Cartan
of $USp(2\ell)$.
It therefore lies in $\frac12+\Zb$ for the components of $Q$ and $\Gamma_\beta$ and vanishes for the vector multiplet.
The charge $-\lambda^T_\beta$ of $W_{\mathbf Q_\lambda}$ at the node $(n,2\beta-1)$ gives the endpoint the $\mathfrak{usp}(2\ell)$
weight $-\lambda^T$ through \eqref{gen_I_id}.

\paragraph{Representations outside the rectangle}
Here we consider only the $\lambda$ whose first row extends one box beyond the rectangle $(\ell^n)$, i.e.\ $\lambda_1=\ell+1$.
For $\lambda_2\le\ell$ we find at generic fugacities
\begin{align}
\label{out_N}
\lin{W_\lambda}{USp(2n)_{-k}}{\Ncal,N,\ell\Gamma}(a,x)=q^{\frac{r_Q}{2}}a\;\lin{W_{\lambda-\mathbf e_1}}{USp(2n)_{-k}}{\Ncal,N,\ell\Gamma}(a,x),
\end{align}
and for $\lambda_1=\lambda_2=\ell+1$ the line defect half-index of $W_\lambda$ vanishes.
Hence in the former case $W_\lambda$ maps to the same Abelian lines as $W_{\lambda-\mathbf e_1}$.

\paragraph{Non-uniqueness}
The expression of a line defect half-index by those of single Wilson lines is not unique, since the line defect half-indices of
different Wilson lines obey linear relations with monomial coefficients, such as \eqref{shiftrelmain} on the Neumann-type boundary.
For example, for $n,\ell\ge2$ the line defect half-index of $W_{\yf}$ is given both by the sum in table \ref{tab_sums} and by
\eqref{antisym_two} with $p=1$.

\begin{table}[t]
\begin{center}\small
\begin{tabular}{lll}
\toprule
$(n,\ell)$ & $p$ & Wilson lines of $\mathcal{T}_{\ell,n}$\\
\midrule
$(n,1)$ & $0\le p\le n$ & $W_{\mathbf Q_{(1^p)}}$,\quad $\mathbf Q_{(1^p)}=-(0,\dots,0,1,2,\dots,p)\otimes(1)$\\[2pt]
$(n,\ell)$, $\ell\ge2$ & $n$ & $W_{\mathbf Q_{(1^n)}}$,\quad $\mathbf Q_{(1^n)}=-(1,2,\dots,n)\otimes(\mathbf e_1-\mathbf e_2)$\\[2pt]
$(n,\ell)$, $n,\ell\ge2$ & $1$ & $\displaystyle\sum_{b=1}^{2\ell-1}W_{-\mathbf e_{(n,b)}}$\\[4pt]
\bottomrule
\end{tabular}
\caption{The line defect half-indices of the antisymmetric Wilson lines $W_{(1^p)}$ on the Neumann-type
boundary (I), as those of single Wilson lines or a sum of those of single Wilson lines of $\mathcal{T}_{\ell,n}$ on the Dirichlet-type
boundary, where $\mathbf e_{(i,b)}$
is the unit vector \eqref{unit_node} at the node $(i,b)$.
The line defect half-indices of the single lines come with the coefficient $(-q^{\frac34}a^{-1})^p$ and those in the sum
with the coefficients \eqref{junctionrule}, i.e.\ $-q^{\frac34}a^{-1}$ for $b$ odd and $-q^{\frac{r_Q}{2}}a$ for $b$ even.}
\label{tab_sums}
\end{center}
\end{table}

\subsubsection{H-twist limit and VOA}
\label{sec_gen_I_H}
For all $\ell$, at $a=q^{1/4}$ and $r_Q=\frac12$ with the flavor fugacities kept, the invariants \eqref{zw_def} are
$z_{(n,2\beta-1)}=x_\beta$, $z_I=1$ at the other nodes and $w_J=1$.
Then the right-hand side of \eqref{gen_I_claim} coincides identically with the Dirichlet-type half-index which is conjectured in \cite{Creutzig:2024ljv} to
be the refined vacuum supercharacter of $L_n(\mathfrak{osp}(1|2\ell))$, with all the $\ell$ flavor fugacities.
At $x_\beta=1$ it is the Nahm sum on the sum side of a conjecture of Warnaar and Zudilin \cite{MR2881319}, with $p=k=n+1$ in the
notation there.
The other side is a Macdonald-type theta sum of rank $\ell$ and modulus $2n+2\ell+1$, and the identity is proved for $n=1$ in
\cite{MR2881319} and in general in \cite{Creutzig:2025wz}.
The boundary VOA is expected to be the commutant \eqref{H_coset}, $L_n(\mathfrak{osp}(1|2\ell))$ (section \ref{sec_twist_com}).
For general $\lambda\subseteq(\ell^n)$ the H-twist limit of the line defect half-index of $W_\lambda$ gives the supercharacter of the module $\lambda^T$ of
$L_n(\mathfrak{osp}(1|2\ell))$ in the
labeling of section \ref{sec_twist_osp}.
We find that at $a=q^{1/4}$ and $r_Q=\frac12$, with the flavor fugacities $x$ kept,
\begin{align}
\label{Hl_refined}
\lin{W_\lambda}{USp(2n)_{-k}}{\Ncal,N,\ell\Gamma}(q^{\frac14},x)=(-q^{\frac12})^{|\lambda|}\,\chi^{L_n(\mathfrak{osp}(1|2\ell))}_{\lambda^T}(x),
\end{align}
with the refined supercharacter \eqref{KWz}.
Hence at the H-twist point
\begin{align}
\label{Hmodule}
\lin{W_\lambda}{USp(2n)_{-k}}{\Ncal,N,\ell\Gamma}\Big|_{\rm H}=(-q^{\frac12})^{|\lambda|}\,\chi^{L_n(\mathfrak{osp}(1|2\ell))}_{\lambda^T}.
\end{align}
We have checked \eqref{Hl_refined}, and hence \eqref{Hmodule}, for all $\lambda\subseteq(\ell^n)$ with $n,\ell\le4$.
These include the Wilson lines whose line defect half-indices are not those of single Wilson lines.
For $\ell=1$, \eqref{Hmodule} is \eqref{linechi}, which we prove in appendix \ref{app_proof}.
The leading coefficient of the supercharacter in \eqref{Hmodule} is the absolute value of the superdimension of the
$\mathfrak{osp}(1|2\ell)$ representation with highest weight $\lambda^T$.

At the H-twist point the line defect half-index of a single Wilson line of $\mathcal{T}_{\ell,n}$ is a single Nahm sum \eqref{nahm_Q}.
We have searched for single Wilson lines dual to the $W_\lambda$ by comparing the H-twist limits of their line defect half-indices, up
to monomials, with these Nahm sums.
For $(n,\ell)=(2,2)$ the search includes all the charges whose six entries are integers from $-3$ to $2$.
For $(n,\ell)=(3,2)$, $(2,3)$, $(4,2)$, $(2,4)$, $(3,3)$ and $(5,2)$ it includes the charges
$\mathbf Q=-\sum_j\mathbf w^{(j)}\otimes\boldsymbol\eta_j$ and their reflections, where each $\mathbf w^{(j)}$ is a non-decreasing sequence
of $n$ integers from $0$ to $n$.
Only the representations of \eqref{singlerule} are reproduced, and only by $\mathbf Q_\lambda$ up to the reflection.
We therefore expect that among the $\lambda\subseteq(\ell^n)$ only those of \eqref{singlerule} have a single dual Wilson
line.
For $\lambda=\yf$ and $n,\ell\ge2$ the coefficients of the sum in table \ref{tab_sums} all become $-q^{\frac12}$ at the H-twist point,
so that by \eqref{Hmodule} the supercharacter $\chi^{L_n(\mathfrak{osp}(1|2\ell))}_{\yf}$ is the sum of the $2\ell-1$ Nahm sums
\eqref{nahm_Q} with the charges $-\mathbf e_{(n,b)}$.

\subsection{Boundary conditions (II)}
\label{sec_gen_II}
\subsubsection{Symplectic gauge theory}
On the symplectic side we take Dirichlet boundary conditions for the vector multiplet and $Q$.
As shown in section \ref{sec_signs}, this is a boundary condition of $USp(2n)_{k}+Q$ with $k_{\rm eff}=\ell$.
With $\tilde m=m+\frac j2\mathbf 1$, the line defect half-index \eqref{Vj_general} of $V^j\otimes W_\lambda$ is
\begin{align}
\label{gen_II_A}
\lin{V^j\otimes W_\lambda}{USp(2n)_{k}}{\Dcal,D}(u;a)
&=\frac{1}{(q)_\infty^n}\sum_{m\in\Zb^n}q^{\ell\sum_i\tilde m_i^2}\prod_iu_i^{2\ell\tilde m_i}\ \chi_{\lambda}(q^{\tilde m}u)
\nonumber\\
&\qquad\times\prod_{i=1}^{n}\frac{\poch{q^{1-\frac{r_Q}{2}\pm\tilde m_i}(q^{\frac j2}a)^{-1}u_i^{\pm1}}}{\poch{q^{1\pm2\tilde m_i}u_i^{\pm2}}}
\nonumber\\
&\qquad\times\prod_{i<i'}\frac{1}{\poch{q^{1\pm(\tilde m_i-\tilde m_{i'})}u_i^{\pm}u_{i'}^{\mp}}\poch{q^{1\pm(\tilde m_i+\tilde m_{i'})}u_i^{\pm}u_{i'}^{\pm}}} .
\end{align}

\subsubsection{Abelian gauge theory}
On the Abelian side we take Neumann boundary conditions for all the vector multiplets and 3d chiral multiplets
of $\Tb_{\ell,n}$ and add $2\ell n$ boundary Fermi multiplets charged under its gauge group $U(1)^{n(2\ell-1)}$.
Their charges are obtained from the factorizations
\begin{align}
\label{LBfact}
C(T_n)^{-1}=LL^T,\qquad C(A_{2\ell-1})=BB^T,
\end{align}
where $L$ is the lower triangular matrix with all the entries on and below the diagonal equal to $1$, whose columns are
$\mathbf f_j=\sum_{i\ge j}\mathbf e_i$, $j=1,\dots,n$, and $B$ is the $(2\ell-1)\times2\ell$ matrix with columns
$\mathbf d_b=\mathbf e_b-\mathbf e_{b-1}$, $b=1,\dots,2\ell$, with $\mathbf e_0=\mathbf e_{2\ell}=0$.
The Fermi multiplet $(j,b)$ has the charge
\begin{align}
\label{NNcharges}
\mathbf g_{(j,b)}=(-1)^{b+1}\mathbf f_j\otimes \mathbf d_b,\qquad j=1,\dots,n,\quad b=1,\dots,2\ell ,
\end{align}
so that $\sum_F\mathbf g_F\mathbf g_F^T=\kappa$ and the gauge anomaly cancels for $\Tb_{\ell,n}$ (section
\ref{sec_signs}).
Since $\sum_j\mathbf f_j=(1,2,\dots,n)$ and $\sum_b(-1)^{b+1}\mathbf d_b=2(1,-1,\dots,-1,1)$, the mixed
anomaly between the gauge group and $U(1)_J$ cancels if every Fermi multiplet has charge
$\frac12$ under $U(1)_J$.
The boundary flavor symmetries whose mixed anomalies with the gauge group cancel are of two kinds.
One is $U(1)_{x_j}$, which acts on the Fermi multiplets $(j,b)$, $b=1,\dots,2\ell$, with charge $(-1)^{b+1}$ and has no mixed anomaly since
$\sum_b\mathbf d_b=0$.
The other is $U(1)_h$, which acts on all the Fermi multiplets with the same charge and whose mixed anomaly cancels against that of
$U(1)_J$, so that $U(1)_h$ is $U(1)_J$ acting with the charge $\frac12$.
The fugacities are
\begin{align}
\nu_{(j,b)}=q^{\frac{R_S}{2}}h\,x_j^{(-1)^{b+1}},
\end{align}
where $R_S$ is the R-charge of the chiral multiplets at $S$, while the chiral multiplets at the other nodes have $R_I=0$, as in \eqref{Rassign}.
With $\sigma_{j,b}=s^{\mathbf g_{(j,b)}}=\prod_{i\ge j}(s_{(i,b)}/s_{(i,b-1)})^{(-1)^{b+1}}$ and
$s_{(i,0)}=s_{(i,2\ell)}=1$, the line defect half-index of the Wilson line $W_{\mathbf Q}$ is
\begin{align}
\label{gen_II_B}
\lin{W_{\mathbf Q}}{\Tb_{\ell,n}}{\Ncal,N,2\ell n\Gamma}(h,x)
&=(q)_\infty^{n(2\ell-1)}\oint\prod_{I}\frac{ds_I}{2\pi is_I}\prod_I\frac{1}{\poch{q^{\frac{R_I}{2}}s_I}}
\nonumber\\
&\qquad\times\prod_{j=1}^n\prod_{b=1}^{2\ell}\poch{q^{\frac{1+R_S}{2}}hx_j^{(-1)^{b+1}}\sigma_{j,b}}
\poch{q^{\frac{1-R_S}{2}}h^{-1}x_j^{(-1)^b}\sigma_{j,b}^{-1}}
\nonumber\\
&\qquad\times\prod_I(q^{\frac{R_I}{2}}s_I)^{(\mathbf Q)_I}.
\end{align}
Every Fermi multiplet has total charge $1$ at the nodes of $S$, so that the shift $s_I\to q^{-R_I/2}s_I$ removes $q^{R_S/2}$ from all
$\nu_F$ and from the Wilson line insertions.
Then the Jacobi triple product and Euler's identity give the fermionic sum
\begin{align}
\label{NNfsum}
\lin{W_{\mathbf Q}}{\Tb_{\ell,n}}{\Ncal,N,2\ell n\Gamma}(h,x)
&=\frac{(q)_\infty^{n(2\ell-1)}}{(q)_\infty^{2\ell n}}\sum_{\substack{t\in\Zb^{2\ell n}\\ d_I\ge0}}
(-1)^{\sum_Ft_F}q^{\sum_F\frac{t_F^2}{2}}\prod_F(hx_{j_F}^{\pm1})^{t_F}\prod_I\frac{1}{(q)_{d_I}},
\nonumber\\
d&=-\sum_Ft_F\mathbf g_F-\mathbf Q ,
\end{align}
where $x_{j_F}^{\pm1}$ is the flavor fugacity of $F$.

\subsubsection{Identification of the fugacities}
The half-indices, i.e.\ \eqref{gen_II_A} with $j=0$, $\lambda=\varnothing$ and
\eqref{gen_II_B} with $\mathbf Q=0$, agree with the identification
\begin{align}
\label{gen_II_id}
u_j=x_j\quad(j=1,\dots,n),\qquad h=q^{\frac{r_Q-1}{2}}a,\qquad R_S=\tfrac12-r_Q ,
\end{align}
i.e.
\begin{align}
\label{gen_II_claim}
\II^{\Tb_{\ell,n}}_{\Ncal,N,2\ell n\Gamma}(h,x)=\II^{USp(2n)_{k}}_{\Dcal,D}(x;a)\Big|_{\eqref{gen_II_id}} .
\end{align}
We have checked \eqref{gen_II_claim} at generic fugacities.
The flavor symmetries $U(1)_{x_j}$ are identified with the Cartan of
$USp(2n)_\partial$, and $U(1)_h$ with $U(1)_a$.
In terms of the Abelian variables, $h^2=q^{-1/2}\hat y$,
in accordance with \eqref{w_value}.
The Abelian half-index is invariant under the signed permutations of the $x_j$, although the rows are not treated symmetrically in
\eqref{NNcharges}.
This indicates that the boundary flavor symmetry $U(1)^n$ of $\Tb_{\ell,n}$ is enhanced to $USp(2n)$.
Under $U(1)^n$ the $2\ell n$ Fermi multiplets have the weights $\pm\mathbf e_j$, each with multiplicity $\ell$, i.e.\ those of $\ell$
fundamental Fermi multiplets of $USp(2n)_\partial$.

At $x_n=q^{-1/2}h^{-1}$ the Fermi multiplet $\Gamma_{(n,1)}$ of charge $\mathbf e_{(n,1)}$ flips the 3d chiral multiplet $\Phi_{(n,1)}$.
On the symplectic side this corresponds to $u_n=q^{-r_Q/2}a^{-1}$, where one of the factors contributed by $Q$
becomes $\poch{q^{1-m_n}}$.
Thus a component of $Q$ acquires a boundary value which breaks $USp(2n)_\partial$ to $USp(2n-2)_\partial$, in accordance with the
change of only the last row on the Abelian side.
Under the Giveon-Kutasov type duality and the mirror symmetry, whose identifications of the fugacities are given in section
\ref{sec_square}, this flip is exchanged with that of section \ref{sec_gen_I}.

\subsubsection{Line operators}
\paragraph{Vortex-Wilson lines}
We consider the vortex-Wilson lines $V^j\otimes W_\lambda$ of section \ref{sec_halfindex} on the Dirichlet-type boundary, with
$\lambda=(\ell^{n-1},\ell-p)$, $p=0,\dots,\ell$, the complement of $(p)$ in the rectangle.
Here and in the examples of (II) in section \ref{sec_gen_ex} the label $p$ refers to this complement, as in \eqref{n1_II_twist}.
On (I) in sections \ref{sec_n1} and \ref{sec_gen_I} and in \eqref{n1_II_rule} on (II), in contrast, $p$ labels the antisymmetric
representation $(1^p)$, which is $\lambda$ itself except in the condition $\lambda^c=(1^p)$ of \eqref{singlerule}.
We arrange the charge $\mathbf Q$ of a Wilson line of $\Tb_{\ell,n}$ as a matrix with $n$ rows and $2\ell-1$ columns, e.g.\
$W_{\Qm{-1&1&-1\\-1&1&-1}}$ for $(n,\ell)=(2,2)$.
The entry in the row $i$ and the column $b$ is the charge at the node $(i,b)$.
For $1\le b\le2\ell-1$ the charge $-\mathbf 1\otimes \mathbf e_b$, with $\mathbf 1=(1,\dots,1)$, is $-1$ at the $n$ nodes $(i,b)$ of the column $b$ and
zero elsewhere, and it vanishes for $b=0$ since $\mathbf e_0=0$.
We find that the line defect half-index of each of these vortex-Wilson lines is a linear combination of those of at most two Wilson
lines $W_{-\mathbf 1\otimes \mathbf e_b+(1-j)\mathbf A}$, i.e.\ of the Wilson lines $W_{-\mathbf 1\otimes \mathbf e_b}$ of the columns shifted by vortex lines of
$U(1)_J$.
More precisely, we find at generic fugacities that the line defect half-index \eqref{gen_II_A} of
$V^j\otimes W_\lambda$ is related to the line defect half-indices \eqref{gen_II_B} under \eqref{gen_II_id} by
\begin{align}
\label{II_two}
&\lin{V^j\otimes W_{(\ell^{n-1},\ell-p)}}{USp(2n)_{k}}{\Dcal,D}(x;a)
=(-1)^{p+n\ell(j-1)}\,q^{-\frac{n\ell}{4}j^2+\frac{(n\ell-\epsilon_p)(j-1)}{2}}\,h^{\,n\ell(1-j)-\epsilon_p}
\nonumber\\
&\qquad\times\Big(\lin{W_{-\mathbf 1\otimes \mathbf e_p+(1-j)\mathbf A}}{\Tb_{\ell,n}}{\Ncal,N,2\ell n\Gamma}(h,x)
-\lin{W_{-\mathbf 1\otimes \mathbf e_{p-2}+(1-j)\mathbf A}}{\Tb_{\ell,n}}{\Ncal,N,2\ell n\Gamma}(h,x)\Big)\Big|_{\eqref{gen_II_id}},
\end{align}
where $\epsilon_p=1$ for odd $p$ and $\epsilon_p=0$ for even $p$, and the second term is absent for $p<2$.
For $p\ge2$ and $n\ge2$, \eqref{II_two} expresses the line defect half-index of the single line $V^j\otimes W_\lambda$ as a difference
of those of two Wilson lines.\footnote{For $p\ge2$ and $(n,\ell)=(2,2)$, $(3,2)$ and $(2,3)$ we have found that the right-hand side of
\eqref{II_two} can also be written as a monomial times a sum, with all the coefficients $+1$, of the line
defect half-indices of single lines, some of which are Wilson lines combined with gauge vortex lines on which the chiral multiplets
vanish, as in the last equality of \eqref{22_II_lines} for $(n,\ell,p)=(2,2,2)$ and $j=1$. These equalities also hold for the line defect
indices.}
The relation \eqref{II_two} covers only $\lambda=(\ell^{n-1},\ell-p)$, and as for (I), the rule which maps $V^j\otimes W_\lambda$ with a
general $\lambda\subseteq(\ell^n)$ to lines of $\Tb_{\ell,n}$ remains to be found.
We have checked \eqref{II_two} for $j=-1,0,1,2$ with $n+\ell\le5$, for $|j|\le5$ with $(n,\ell)=(1,2)$, for $|j|\le4$ with $(1,3)$ and
for $-2\le j\le4$ with $(2,2)$, and we expect it to hold for all $j\in\Zb$.
For $j=1$ it reads
\begin{align}
\label{NNlinetwo}
&\lin{V\otimes W_{(\ell^{n-1},\ell-p)}}{USp(2n)_{k}}{\Dcal,D}(x;a)
\nonumber\\
&=(-1)^pq^{-\frac{n\ell}{4}}h^{-\epsilon_p}
\Big(\lin{W_{-\mathbf 1\otimes \mathbf e_p}}{\Tb_{\ell,n}}{\Ncal,N,2\ell n\Gamma}(h,x)-\lin{W_{-\mathbf 1\otimes \mathbf e_{p-2}}}{\Tb_{\ell,n}}{\Ncal,N,2\ell n\Gamma}(h,x)\Big)\Big|_{\eqref{gen_II_id}},
\end{align}
where $p=0$ is \eqref{gen_II_claim} by \eqref{specflow}.

\paragraph{Single Wilson lines}
For $p=0$ the lines $V^j\otimes W_{(\ell^n)}$ map to the vortex lines $W_{(1-j)\mathbf A}$ of $U(1)_J$, which are trivial for $j=1$.
For $p=1$ the second term of \eqref{II_two} is absent, and the line $V^j\otimes W_{(\ell^{n-1},\ell-1)}$ maps to the single Wilson line
$W_{-\mathbf 1\otimes \mathbf e_1+(1-j)\mathbf A}$.
For $n=1$ we also find that the combination in \eqref{II_two} is that of a single Wilson line,
\begin{align}
\label{hc_two}
&\lin{W_{-\mathbf e_p+(1-j)\mathbf A}}{\Tb_{\ell,1}}{\Ncal,N,2\ell\Gamma}(h,x)
-\lin{W_{-\mathbf e_{p-2}+(1-j)\mathbf A}}{\Tb_{\ell,1}}{\Ncal,N,2\ell\Gamma}(h,x)
\nonumber\\
&=(q^{\frac{j-1}{2}}h)^{-2\lfloor p/2\rfloor}\lin{W_{(-1,1,\dots,1,-1,0,\dots,0)+(1-j)\mathbf A}}{\Tb_{\ell,1}}{\Ncal,N,2\ell\Gamma}(h,x),
\end{align}
where the second term on the left-hand side is absent for $p<2$ and the charge $(-1,1,\dots,1,-1,0,\dots,0)$ has $2p-1$ non-vanishing entries (the trivial line for $p=0$).
Hence for $n=1$ the lines $V^j\otimes W_{(\ell-p)}$ map to single Wilson lines for all $p$.
For $\ell=1$ the representations are $\lambda=(1^{n-p})$ with $p=0,1$, and the charges are $-j\mathbf A-\mathbf Q_{(1^{n-p})}$.
The dictionary \eqref{n1_II_rule} of section \ref{sec_n1_II} extends this to all the antisymmetric
representations.

\paragraph{Quantum numbers of the endpoints}
In \eqref{II_two} the powers of $V$ shift the charges by multiples of the charge $-\mathbf A$ of the vortex line $W_{-\mathbf A}$ of
$U(1)_J$ (section \ref{sec_halfindex}).
Every Fermi multiplet has charge $\frac12$ under $U(1)_J$ and total charge $1$ at the nodes of $S$.
Hence the power of $h$ in \eqref{II_two} equals the sum of the charges at $S$ of each of the two Wilson lines.
With $h^2=q^{-1/2}\hat y$ it gives the topological charge $\frac12(n\ell(1-j)-\epsilon_p)$ of the endpoint.
As for (I), this is fractional when $|\lambda|-n\ell j=n\ell(1-j)-p$ is odd, which reflects the charge of $V^j\otimes W_\lambda$ under
the center $\Zb_2$ through the global structure $(USp(2n)\times U(1)_a)/\Zb_2$.

\paragraph{Representations outside the rectangle}
Here we consider only the $\lambda$ whose first row extends one box beyond the rectangle $(\ell^n)$, i.e.\ $\lambda_1=\ell+1$.
For $\lambda_2\le\ell$ we find at generic fugacities
\begin{align}
\label{out_D}
\lin{V^j\otimes W_\lambda}{USp(2n)_k}{\Dcal,D}(x;a)=q^{-\frac12}(q^{\frac j2}h)^{-1}\,\lin{V^j\otimes W_{\lambda-\mathbf e_1}}{USp(2n)_k}{\Dcal,D}(x;a),
\end{align}
with $h=q^{(r_Q-1)/2}a$, and for $\lambda_1=\lambda_2=\ell+1$ the line defect half-index of $V^j\otimes W_\lambda$ vanishes.
We have checked both statements for $j=-1,0,1,2$ with $n+\ell\le5$ and for $-4\le j\le5$ with $(n,\ell)=(1,2)$, $(1,3)$, $(2,1)$ and
$(2,2)$, and we expect them to hold for all $j\in\Zb$.
Hence in the former case the line $V^j\otimes W_\lambda$ maps to the same lines as $V^j\otimes W_{\lambda-\mathbf e_1}$.

\paragraph{Non-uniqueness}
The expression \eqref{II_two} by the line defect half-indices of Wilson lines is not unique.
Indeed, for each node $I$ the line defect half-indices \eqref{gen_II_B} satisfy
\begin{align}
\label{shiftrelmain}
\lin{W_{\mathbf Q}}{\Tb_{\ell,n}}{\Ncal,N,2\ell n\Gamma}
&=\xi_I(\mathbf Q)\Bigl(\lin{W_{\mathbf Q-\kappa \mathbf e_I}}{\Tb_{\ell,n}}{\Ncal,N,2\ell n\Gamma}
-\lin{W_{\mathbf Q-\kappa \mathbf e_I+\mathbf e_I}}{\Tb_{\ell,n}}{\Ncal,N,2\ell n\Gamma}\Bigr),
\nonumber\\
\xi_I(\mathbf Q)&=q^{(\mathbf Q)_I-\frac{\kappa_{II}}{2}}(-1)^{\sum_F(\mathbf g_F)_I}\prod_F(hx_{j_F}^{\pm1})^{-(\mathbf g_F)_I},
\end{align}
with the flavor fugacities $x_{j_F}^{\pm1}$ of \eqref{NNfsum}.
This relation follows from the quasi-periodicity of the factors contributed by the Fermi multiplets under $s_I\to qs_I$ in
\eqref{gen_II_B}.
For $(n,\ell)=(1,1)$ it reads
\begin{align}
\label{shift11}
\lin{W_{\mathbf Q}}{\Tb_{1,1}}{\Ncal,N,2\Gamma}=q^{(\mathbf Q)_1-1}h^{-2}\Bigl(\lin{W_{\mathbf Q-2\mathbf e_1}}{\Tb_{1,1}}{\Ncal,N,2\Gamma}
-\lin{W_{\mathbf Q-\mathbf e_1}}{\Tb_{1,1}}{\Ncal,N,2\Gamma}\Bigr),
\end{align}
which becomes the recursion of the Rogers-Ramanujan functions at the C-twist point.

\subsubsection{C-twist limit and VOA}
The C-twist point \eqref{HC_def} is $r_Q=\frac12$ and $a=q^{-1/4}$ in the limit $u\to1$, i.e.\ $h=q^{-1/2}$ and $x_j=1$ on the
Abelian side through \eqref{gen_II_id}.
There the boundary VOA is expected to be $L_\ell(\mathfrak{osp}(1|2n))$ (section \ref{sec_twist_com}), with the rank and the level
exchanged relative to $L_n(\mathfrak{osp}(1|2\ell))$ of (I).
We find that the C-twist limits of the line defect half-indices of the lines $V\otimes W_\lambda$ give the supercharacters of the
modules $\mu=\lambda^c$ of $L_\ell(\mathfrak{osp}(1|2n))$.
More precisely, at $r_Q=\frac12$ and $a=q^{-1/4}$, for generic $u$,
\begin{align}
\label{Cl_refined}
\lin{V\otimes W_\lambda}{USp(2n)_{k}}{\Dcal,D}(u;q^{-\frac14})=q^{\frac{|\mu|}{2}-\frac{n\ell}{4}}\,\chi^{L_\ell(\mathfrak{osp}(1|2n))}_{\mu}(u),
\qquad\mu=\lambda^c,
\end{align}
with the refined supercharacter \eqref{KWz} of $L_\ell(\mathfrak{osp}(1|2n))$, i.e.\ \eqref{KWz} with $n$ and $\ell$ exchanged.
We have checked \eqref{Cl_refined} for all $\lambda\subseteq(\ell^n)$ with $n,\ell\le4$.
For $n\ge2$ the individual flux sectors of \eqref{gen_II_A} are singular at $u=1$, while the right-hand side of \eqref{Cl_refined} is holomorphic there.
Hence the limit $u\to1$ in \eqref{HC_def} exists, and
\begin{align}
\label{CVW}
\lin{V\otimes W_\lambda}{USp(2n)_{k}}{\Dcal,D}\Big|_{\rm C}=q^{\frac{|\mu|}{2}-\frac{n\ell}{4}}\,\chi^{L_\ell(\mathfrak{osp}(1|2n))}_{\mu}.
\end{align}
For $\lambda=(\ell^n)$, i.e.\ $\mu=\varnothing$, \eqref{CVW} and \eqref{specflow} imply that the half-index is the vacuum supercharacter.
For $\lambda$ outside the rectangle we find that the line defect half-index of $V\otimes W_\lambda$ at $a=q^{-1/4}$ vanishes or is a
signed power of $q$ times that of $V\otimes W_{\lambda'}$ for a unique $\lambda'\subseteq(\ell^n)$.
The line defect half-index vanishes if $b^\lambda_i\equiv\pm b^\lambda_j$ modulo $2n+2\ell+1$ for some $i\neq j$ or $2n+2\ell+1$
divides $2b^\lambda_i$ for some $i$, where $b^\lambda_i=\ell+i-\lambda_i-\frac12$, $i=1,\dots,n$.
Otherwise $\lambda'$ is fixed by the condition that the entries of $b^{\lambda'}$ are, up to signs and modulo $2n+2\ell+1$, those of
$b^\lambda$ in some order.

\subsection{Examples}
\label{sec_gen_ex}
On (I) the charges related by the reflection $b\to2\ell-b$ of $A_{2\ell-1}$ give the same line defect half-indices up to
$x_\beta\to x_{\ell+1-\beta}$.
Since the symplectic side is invariant under the permutations of the $x_\beta$, the reflected charges of the lines dual to the
$W_\lambda$ give the same line defect half-indices as the original charges under \eqref{gen_I_id}.
On (II) the reflection maps the Fermi multiplets $(j,b)$ of $\Tb_{\ell,n}$ to $(j,2\ell+1-b)$, so that the reflected charges give the
same line defect half-indices up to $x_j\to x_j^{-1}$, under which the symplectic side is also invariant.
The equalities of the half-indices below are cases of \eqref{gen_I_claim} and \eqref{gen_II_claim}.
The statements on the representations outside the rectangle are cases of \eqref{out_N} and \eqref{out_D}, and the twist limits are
cases of \eqref{Hmodule} and \eqref{CVW}.

\subsubsection{\texorpdfstring{$n=1$}{n=1}}
\label{sec_gen_ex_n1}
For $n=1$ the Abelian theory $\mathcal{T}_{\ell,1}$ is $U(1)^{2\ell-1}$ with $\kappa=C(A_{2\ell-1})$,
$\mathbf A=(1,-1,1,\dots,1)$ and $S$ the set of odd nodes.
Its mirror is the parity image $\Tb_{1,\ell}$ of the theory $\mathcal{T}_{1,\ell}$ of section \ref{sec_n1}.
The Neumann-type boundary condition (I) of $USp(2)_{-(\ell+\frac32)}$ was proposed in \cite{Okazaki:2024paq} to be dual also to
a Dirichlet-type boundary condition of $U(\ell)_{\ell+\frac32,\frac32}$, as checked there for $\ell=2,3$.
With $\mathcal{T}_{\ell,1}$ and the Giveon-Kutasov dual $USp(2\ell)_{\ell+\frac32}$ (section \ref{sec_square}) it thus has three dual
descriptions, and the relation of the $U(\ell)$ description to the other two is left open.
The proposed boundary VOAs are those of table \ref{tab_n1_voa} with $n\to\ell$ and the two twist points exchanged, i.e.\
$L_1(\mathfrak{osp}(1|2\ell))$ at the H-twist point and $L_\ell(\mathfrak{osp}(1|2))$ at the C-twist point.
The equalities of the line defect half-indices of $W_{(p)}$ on (I) and of $V^j\otimes W_{(\ell-p)}$ on (II),
$p=0,\dots,\ell$, are cases of \eqref{singlerule}, \eqref{II_two} and \eqref{hc_two}.

\paragraph{Boundary conditions (I)}
By \eqref{IWp}, the Wilson lines $W_{(p)}$, $p=1,\dots,\ell$, map to the single Wilson lines whose charges consist
of the first $\min(2p,2\ell-1)$ components of $-\mathbf A$.
These are $W_{-1,1,0}$ and $W_{-\mathbf A}$ for $\ell=2$, and $W_{-1,1,0,0,0}$, $W_{-1,1,-1,1,0}$ and $W_{-\mathbf A}$ for $\ell=3$.
For $\ell=2$, under \eqref{gen_I_id}, the coefficient in \eqref{IWp} times the fugacity monomial of $W_{-1,1,0}$ is $-q^{1/2}x_1^{-1}$,
the sign
and the weight of $\Gamma_1$.
The line $W_{\yf}$ corresponds to the B-twisted line $\mathcal{P}^B_{\mu=1}=-z_1z_2^{-1}$ of \cite{Creutzig:2024ljv}, in the notation
there.

\paragraph{Boundary conditions (II)}
By \eqref{II_two} and \eqref{hc_two}, the lines $V^j\otimes W_{(\ell-p)}$ map to single Wilson lines.
For example, $V\otimes W_{(\ell-p)}$ maps to the Wilson line whose charge is $-\mathbf A$ restricted to the first $2p-1$ nodes, i.e.\ to
$W_{-1,0,0,0,0}$, $W_{-1,1,-1,0,0}$ and $W_{-\mathbf A}$ for $\ell=3$ and $p=1,2,3$.
For $\ell=2$ the lines $V^j\otimes W_\lambda$, $\lambda=\varnothing,\yf,\yb$, map to $W_{\mathbf Q+(1-j)\mathbf A}$ with
$\mathbf Q=-\mathbf A$, $(-1,0,0)$ and $0$, up to the reflection, with the coefficients of their line defect half-indices
given by \eqref{II_two} and \eqref{hc_two}.
For example, $W_{\yb}$ and $\bar V$ map to $W_{\mathbf A}$, and $W_{\yf}$ maps to $W_{0,-1,1}$, or by the reflection to $W_{1,-1,0}$, the
reversal of the line $W_{-1,1,0}$ of (I).
By \eqref{out_D} the line $V^j\otimes W_{(\ell+1)}$ maps to the same Abelian Wilson line as $V^j\otimes W_{(\ell)}$, with the extra
factor $q^{-1/2}(q^{j/2}h)^{-1}$, as for $(n,\ell)=(2,1)$ in section \ref{sec_n1_ex_23}.

\paragraph{Twist limits}
For $p=0,\dots,\ell$ the H-twist limits of the line defect half-indices of the Wilson lines $W_{(p)}$ give the
supercharacters of the $\ell+1$ modules $(1^p)$ of $L_1(\mathfrak{osp}(1|2\ell))$, by \eqref{Hmodule},
\begin{align}
\label{n1_l_H}
\lin{W_{(p)}}{USp(2)_{-(\ell+\frac32)}}{\Ncal,N,\ell\Gamma}\Big|_{\rm H}=(-q^{\frac12})^p\,\chi^{L_1(\mathfrak{osp}(1|2\ell))}_{(1^p)}.
\end{align}
The dual Dirichlet-type line defect half-indices of $\mathcal{T}_{\ell,1}$ become the Nahm sums \eqref{nahm_Q} with the charges of
\eqref{IWp}.
For $\ell=2,3$ these supercharacters also appear in section \ref{sec_n1_ex_23} as the C-twist limits of the line defect half-indices of
$V\otimes W_{(1^{\ell-p})}$ of $USp(2\ell)_{\ell+\frac32}$, which are related to $W_{(p)}$ by the Giveon-Kutasov type duality (see
\eqref{GKlines}).
The C-twist limits of the line defect half-indices of the lines $V\otimes W_{(\ell-p)}$ give the supercharacters of the modules $(p)$
of $L_\ell(\mathfrak{osp}(1|2))$, by \eqref{CVW} and \eqref{osp12M},
\begin{align}
\label{n1_l_C}
\lin{V\otimes W_{(\ell-p)}}{USp(2)_{\ell+\frac32}}{\Dcal,D}\Big|_{\rm C}=q^{\frac p2-\frac\ell4}\,\chi^{M(2,2\ell+3)}_{(1,\ell+1-p)}.
\end{align}
On the Abelian side the line defect half-indices of the dual lines give $(-1)^p\chi^{M(2,2\ell+3)}_{(1,\ell+1-p)}$ as the alternating
fermionic sums \eqref{NNfsum} at $h=q^{-1/2}$ and $x_j=1$.

\subsubsection{\texorpdfstring{$(n,\ell)=(2,2)$}{(n,l)=(2,2)}}
\label{sec_gen_ex_22}
For $(n,\ell)=(2,2)$, the simplest case with both the tadpole and the $A_{2\ell-1}$ structures, $\mathcal{T}_{2,2}$ is the $U(1)^6$
theory with $\kappa=\bigl(\begin{smallmatrix}1&1\\1&2\end{smallmatrix}\bigr)\otimes C(A_3)$, $\mathbf A=\Qm{1&-1&1\\2&-2&2}$ and
$S=\{(2,1),(2,3)\}$, where the charges are written as matrices with the rows $i$ and the columns $b$.
Its mirror is its parity image $\Tb_{2,2}$.
The proposed boundary VOA at both twist points is $L_2(\mathfrak{osp}(1|4))$ with $c=\frac83$, which has the six
modules $\mu\subseteq\ybbd$.
The charges \eqref{22_I_single} and the equalities \eqref{22_II_lines} below are cases of \eqref{singlerule} and \eqref{II_two}.

\paragraph{Boundary conditions (I)}
We find that the line defect half-index of $W_{\yf}$ is the sum of those of the Wilson lines at the
three nodes of the last row,
\begin{align}
\label{22_I_fund}
\lin{W_{\yf}}{USp(4)_{-9/2}}{\Ncal,N,2\Gamma}
&=\Big[-q^{\frac34}a^{-1}\Bigl(\lin{W_{\Qm{0&0&0\\-1&0&0}}}{\mathcal{T}_{2,2}}{\Dcal,D}+\lin{W_{\Qm{0&0&0\\0&0&-1}}}{\mathcal{T}_{2,2}}{\Dcal,D}\Bigr)
\nonumber\\
&\qquad-q^{\frac{r_Q}{2}}a\ \lin{W_{\Qm{0&0&0\\0&-1&0}}}{\mathcal{T}_{2,2}}{\Dcal,D}\Big]_{\eqref{gen_I_id}}.
\end{align}
The coefficients of the three terms of \eqref{22_I_fund} are given by \eqref{junctionrule}, with the junction fields $\Gamma_1$
and $\Gamma_2$ for the
nodes $(2,1)$ and $(2,3)$ of $S$ and $Q$ for the even node $(2,2)$.
The structure of \eqref{22_I_fund} matches the non-monomial B-twisted line
$\mathcal{P}^B_{(1,0)}=-(z_{21}+z_{22}+z_{23})$ found in \cite{Creutzig:2024ljv} for $N=k=2$ in the notation there.
As noted in section \ref{sec_gen_I}, the expression as a sum of line defect half-indices is not unique.
The case $p=1$ of \eqref{antisym_two} gives the two-term expression with the line defect half-indices of the Wilson lines of the
charges $0$ and $\Qm{1&0&0\\1&0&0}$, and a
search among the pairs of Wilson lines whose charges have all their six entries equal to $-1$, $0$ or $1$ finds two more.
Of these expressions, \eqref{22_I_fund} is the one whose coefficients follow \eqref{junctionrule}.
By \eqref{singlerule} the Wilson lines $W_\lambda$, $\lambda\subseteq\ybbd$, with two rows map to the single Wilson
lines of the charges
\begin{align}
\label{22_I_single}
\mathbf Q_{\yaa}=\Qm{-1&1&0\\-2&2&0},\qquad \mathbf Q_{\yba}=\Qm{-1&1&0\\-2&2&-1},\qquad
\mathbf Q_{\ybbd}=\Qm{-1&1&-1\\-2&2&-2}=-\mathbf A,
\end{align}
the last one being the vortex line of $U(1)_J$.
The line defect half-index of $W_{\yb}$ is the two-term sum \eqref{symdual} with $n=2$.
By \eqref{out_N}, the line defect half-indices for the representations $\ysmall{3}$, $\ysmall{3,1}$ and $\ysmall{3,2}$ outside the rectangle are
$q^{r_Q/2}a$ times those for $\yb$, $\yba$ and $\ybbd$.
Hence these Wilson lines map to the same Abelian lines as $W_{\yb}$, $W_{\yba}$ and $W_{\ybbd}$, while the line defect half-index for
$\ysmall{3,3}$ vanishes.

\paragraph{Boundary conditions (II)}
Here the symplectic side is a boundary condition of $USp(4)_{9/2}+Q$ with $k_{\rm eff}=2$, and the Neumann-type boundary
condition of $\Tb_{2,2}$ has the eight Fermi multiplets of \eqref{NNcharges}.
By \eqref{II_two} with $p=1,2$,
\begin{align}
\label{22_II_lines}
\lin{V\otimes W_{\yba}}{USp(4)_{9/2}}{\Dcal,D}&=-q^{-1}h^{-1}\lin{W_{\Qm{-1&0&0\\-1&0&0}}}{\Tb_{2,2}}{\Ncal,N,8\Gamma}\Big|_{\eqref{gen_II_id}},
\nonumber\\
\lin{V\otimes W_{\yb}}{USp(4)_{9/2}}{\Dcal,D}&=q^{-1}\Bigl(\lin{W_{\Qm{0&-1&0\\0&-1&0}}}{\Tb_{2,2}}{\Ncal,N,8\Gamma}
-\II^{\Tb_{2,2}}_{\Ncal,N,8\Gamma}\Bigr)\Big|_{\eqref{gen_II_id}}
\nonumber\\
&=q^{-1}h^{-2}\Bigl(\lin{W_{\Qm{-1&1&-1\\-1&1&-1}}}{\Tb_{2,2}}{\Ncal,N,8\Gamma}+\lin{W_{\Qm{0&0&0\\-1&1&-1}}}{\Tb_{2,2}}{\Ncal,N,8\Gamma}
\nonumber\\
&\qquad-\lin{W_{\Qm{0&1&0\\-1&1&-1}}}{\Tb_{2,2}}{\Ncal,N,8\Gamma}\Bigr)\Big|_{\eqref{gen_II_id}},
\end{align}
while the line defect half-index of $V\otimes W_{\ybbd}$ is $q^{-1}$ times the half-index without lines by \eqref{specflow}.
In the last equality the last two terms form the line defect half-index of $W_{\Qm{0&0&0\\-1&1&-1}}$ combined with a gauge vortex line at the
node $(1,2)$ on which the chiral multiplet vanishes, so that the bracket is a sum of the line defect half-indices of two single lines.

\paragraph{Twist limits}
At $a=q^{1/4}$ and $r_Q=\frac12$ the invariants \eqref{zw_def} become $z_{(2,1)}=x_1$, $z_{(2,3)}=x_2$, $z_I=1$ at the other nodes and
$w_J=1$, and the Abelian half-index is the refined vacuum supercharacter of $L_2(\mathfrak{osp}(1|4))$ (section \ref{sec_gen_I_H}).
The H-twist limits of the line defect half-indices of the lines $W_{(1^p)}$, $p=0,1,2$, give the supercharacters of the modules $\mu=(p)$,
\begin{align}
\label{22_I_lines}
\lin{W_{(1^p)}}{USp(4)_{-9/2}}{\Ncal,N,2\Gamma}\Big|_{\rm H}=(-q^{\frac12})^p\,\chi^{L_2(\mathfrak{osp}(1|4))}_{(p)},
\end{align}
with
\begin{align}
\label{22_I_twist}
\chi^{L_2(\mathfrak{osp}(1|4))}_{(p)}=\sum_{N\in\Zb_{\ge0}^6}\frac{q^{\frac12N\kappa N}}{\prod_I(q)_{N_I}}\times
\begin{cases}
1 & p=0\\
q^{N_{(2,1)}}+q^{N_{(2,2)}}+q^{N_{(2,3)}} & p=1\\
q^{N_{(1,1)}-N_{(1,2)}+2N_{(2,1)}-2N_{(2,2)}} & p=2
\end{cases},
\end{align}
where the three coefficients of \eqref{22_I_fund} all become $-q^{1/2}$ at this point.
By \eqref{Hmodule}, the supercharacters of the remaining modules $\mu=\yaa$, $\yba$, $\ybbd$ are given by the H-twist limits of the line defect
half-indices of
$W_{\yb}$, $W_{\yba}$ and $W_{\ybbd}$, and for the latter two they are the Nahm sums \eqref{nahm_Q} with the charges
\eqref{22_I_single}.
The H-twist limits of the line defect half-indices of all the six lines $W_\lambda$ are thus $(-q^{\frac12})^{|\lambda|}$ times
\eqref{KWmu} for $\mu=\lambda^T$.
The supercharacters of $\mu=\yb$, $\yaa$, $\yba$, $\ybbd$ start with $6$, $2$, $5$ and $3$, the absolute values of the superdimensions of the
corresponding representations of $\mathfrak{osp}(1|4)$.
At the C-twist point the half-index of $\Tb_{2,2}$ is the vacuum supercharacter $\chi^{L_2(\mathfrak{osp}(1|4))}_\varnothing$.
The C-twist limits of the line defect half-indices of the lines $V\otimes W_{(2,2-p)}$ give the supercharacters of the modules $\mu=(p)$,
\begin{align}
\label{22_II_Clines}
\lin{V\otimes W_{(2,2-p)}}{USp(4)_{9/2}}{\Dcal,D}\Big|_{\rm C}=q^{\frac p2-1}\,\chi^{L_2(\mathfrak{osp}(1|4))}_{(p)}.
\end{align}
By \eqref{CVW} the C-twist limits of the line defect half-indices of $V\otimes W_{\yf}$, $V\otimes W_{\yaa}$ and $V$ give the supercharacters of the modules
$\yba$, $\yaa$ and $\ybbd$.
We have not determined the Abelian duals of the first two.
By \eqref{Vj_shift}, \eqref{specflow} and the case $j=2$, $p=0$ of \eqref{II_two}, the line $V$ maps to the vortex line $W_{-\mathbf A}$
of $U(1)_J$,
\begin{align}
\lin{V}{USp(4)_{9/2}}{\Dcal,D}(x;a)=q^{-1}h^{-4}\,\lin{W_{-\mathbf A}}{\Tb_{2,2}}{\Ncal,N,8\Gamma}(h,x)\Big|_{\eqref{gen_II_id}}.
\end{align}

\section{Web of dualities and boundary VOAs}
\label{sec_square}
The duality between $USp(2n)$ and $\mathcal{T}_{\ell,n}$ of sections \ref{sec_n1} and \ref{sec_general}, the mirror symmetry
between $\mathcal{T}_{\ell,n}$ and $\Tb_{n,\ell}$ and the Giveon-Kutasov (GK) type duality between $USp(2n)$ and $USp(2\ell)$ form a
square.
The first of them gives two opposite sides of the square, for $(n,\ell)$ and for the parity images of the theories for $(\ell,n)$.
This square is consistent at the level of the full indices \cite{Okazaki:2026efy} (sections \ref{sec_bulkduality} and \ref{sec_bulk}).
With the signs of the levels of section \ref{sec_signs}, the Giveon-Kutasov type duality relates
$USp(2n)_{-k}$ and $USp(2\ell)_{k}$.
The mirror symmetry between $\mathcal{T}_{\ell,n}$ and $\Tb_{n,\ell}$ agrees with that of \cite{Creutzig:2024ljv}, which relates
$\mathcal{T}_{\ell,n}$ to the theory with the levels $\mathrm{id}-\kappa$, where $\kappa$ is the level matrix of $\mathcal{T}_{n,\ell}$.
This theory is $\Tb_{n,\ell}$, whose index is obtained from that of $\mathcal T_{n,\ell}$ by $m\to-m$, as in \eqref{pole_flip_Ab}.

The purpose of this section is to test the Giveon-Kutasov type duality and the mirror symmetry
for the boundary conditions and line operators.
We find that they map the boundary conditions (I) for $(n,\ell)$ and the lines ending on them to
those of (II) for $(\ell,n)$, and vice versa.
Together with the duality between $USp(2n)$ and $\mathcal{T}_{\ell,n}$, they form commutative squares of boundary conditions
(section \ref{sec_bc_square}) and of the lines ending on them (section \ref{sec_bline_square}).
We also check the corresponding relations of the bulk line defect indices (section \ref{sec_bulkid}).
Finally we read the twist limits of the squares as the supercharacters of the modules of the proposed boundary VOAs
$L_n(\mathfrak{osp}(1|2\ell))$ and $L_\ell(\mathfrak{osp}(1|2n))$ (section \ref{sec_voa_usp}).

\subsection{Giveon-Kutasov type duality}
The Neumann-type boundary condition of $USp(2n)_{-k}$ with $\ell$ Fermi multiplets is dual to the
Dirichlet-type boundary condition of $USp(2\ell)_{k}$,
\begin{align}
\label{GKbdry}
\II^{USp(2n)_{-k}}_{\Ncal,N,\ell\Gamma}(a,x)=\II^{USp(2\ell)_{k}}_{\Dcal,D}(x;a^{-1})\Big|_{r_Q\to1-r_Q},
\end{align}
where the right-hand side is \eqref{USpD_general} with $n$ and $\ell$ exchanged and $k_{\rm eff}=n$.
The dual quark has the fugacity $a^{-1}$ and the R-charge $1-r_Q$.
This is the case of a single flavor of the boundary duality of \cite{Okazaki:2021pnc}, which allows half-integral levels and
odd numbers of flavors.\footnote{In our conventions it reads as follows. Consider $N_f$ fundamental chiral multiplets $Q_r$ of fugacities $a_r$ and R-charges $r_r$ on the Neumann-type
side. For $N_f\le2n+2\ell+1$ the half-index equals the Dirichlet-type half-index of $USp(2\ell)$ with $N_f$ dual quarks of
fugacities $a_r^{-1}$ and R-charges $1-r_r$, times the factors $1/(q^{\frac{r_r+r_{r'}}{2}}a_ra_{r'};q)_\infty$, $r<r'$, contributed
by the gauge-singlet mesons. For $N_f=2n+2\ell+2$ the same holds with one more factor
$(q^{-\ell}\prod_rq^{\frac{r_r}{2}}a_r;q)_\infty$. For $N_f\le2n+2\ell+1$ this is the form expected from the Giveon-Kutasov duality
with $N_f$ fundamental chiral multiplets, in which the level $n+\ell+1-\frac{N_f}2$ keeps the rank $\ell$ of the dual gauge group. For
$N_f=2n+2\ell+2$ the level vanishes, and the form is that expected from the Aharony duality \cite{Aharony:1997gp} with the
monopole singlet. For $\ell=0$ and $N_f=2n+2$ the equality is
Gustafson's generalization of the Askey-Wilson integral to $USp(2n)$ \cite{MR1139492}, which gives the boundary confining duality of
\cite{Okazaki:2023hiv}, and for $N_f<2n+2$ it follows as a limit of the fugacities, as for $n=1$ in \cite{Okazaki:2026rbd}. We have checked the general
equality for $(n,\ell)=(1,1)$, $(1,2)$, $(2,1)$ and $(2,2)$ with $1\le N_f\le2n+2\ell+1$, and with
$N_f=2n+2\ell+2$ at $r_r=1$, where the Dirichlet-type sum converges as a $q$-series.}
The flavor symmetry $USp(2\ell)$ of the Fermi multiplets becomes the boundary symmetry
$USp(2\ell)_\partial$, with the matching anomaly $n=k_{\rm eff}$.
The condition $k_{\rm eff}=K-(\ell+1)+\frac12=n$ then fixes the level $K=k$ of the dual gauge group.
The lines ending on the boundary conditions, which were not considered in \cite{Okazaki:2021pnc}, map as
\begin{align}
\label{GKlines}
\lin{V\otimes W_{(\lambda^T)^c}}{USp(2\ell)_{k}}{\Dcal,D}(x;a^{-1})\Big|_{r_Q\to1-r_Q}=(-1)^{|\lambda|}q^{-\frac{n\ell}{4}}\
\lin{W_{\lambda}}{USp(2n)_{-k}}{\Ncal,N,\ell\Gamma}(a,x),
\end{align}
for all $\lambda\subseteq(\ell^n)$, where $(\lambda^T)^c$ is the complement of the transpose $\lambda^T$ in the $\ell\times n$
rectangle $(n^\ell)$.
For $\lambda=(1^p)$ the line on the left is $V\otimes W_{(n^{\ell-1},n-p)}$, and the case $\lambda=\varnothing$ is \eqref{GKbdry}
by \eqref{specflow}.

We have checked \eqref{GKbdry} and \eqref{GKlines} at generic fugacities for all $\lambda\subseteq(\ell^n)$
with $n+\ell\le6$.

\subsection{Mirror symmetry}
Composing (I) for $(n,\ell)$, \eqref{GKbdry} and (II) for $(\ell,n)$, we obtain the equality in the first line of
\begin{align}
\label{mirrorchain}
\begin{array}{>{\displaystyle}c>{\displaystyle}c>{\displaystyle}c}
\II^{\mathcal{T}_{\ell,n}}_{\Dcal,D}(a)&=&\II^{\Tb_{n,\ell}}_{\Ncal,N,2n\ell\Gamma}(a^{-1})\\[2pt]
\Vert&&\Vert\\[1pt]
\II^{USp(2n)_{-k}}_{\Ncal,N,\ell\Gamma}(a)&=&\II^{USp(2\ell)_{k}}_{\Dcal,D}(a^{-1}),
\end{array}
\end{align}
where the second line is \eqref{GKbdry} and the left-hand sides are equal by (I) and the right-hand sides by (II).
As in \eqref{GKbdry}, the right-hand sides are at $r_Q\to1-r_Q$.
This realizes the mirror symmetry as a duality which exchanges the Dirichlet-type and Neumann-type boundary conditions and
inverts the topological fugacity $y=a^2$.
Since \eqref{gen_II_claim} enters \eqref{mirrorchain} with $n$ and $\ell$ exchanged, $a\to a^{-1}$ and $r_Q\to1-r_Q$, the fugacity of
$U(1)_h$ on the Neumann-type side is $h'=q^{-r_Q/2}a^{-1}$.
The chiral multiplets of $\Tb_{n,\ell}$ at the nodes of $S$, i.e.\ of \eqref{S_def} with $n$ and $\ell$ exchanged, have R-charge $r_Q-\frac12$.
For $\ell=1$, and for $n=1$ with $p=1$, the duals of $W_{(1^p)}$ of $USp(2n)_{-k}$ on both Abelian sides are single Wilson
lines, $W_{\mathbf Q}$ of $\mathcal{T}_{\ell,n}$ and $W_{\mathbf Q'}$ of $\Tb_{n,\ell}$.
Their line defect half-indices satisfy
\begin{align}
\label{mirrorlines}
\lin{W_{\mathbf Q'}}{\Tb_{n,\ell}}{\Ncal,N,2n\ell\Gamma}(h',x)=(-q^{\frac34}a^{-1}h')^p\
\lin{W_{\mathbf Q}}{\mathcal{T}_{\ell,n}}{\Dcal,D}\Big|_{\eqref{gen_I_id}}.
\end{align}
The charges, which follow from \eqref{singlerule}, \eqref{II_two}, \eqref{hc_two} and \eqref{GKlines}, are given in the following table.
\begin{center}\small
\setlength{\tabcolsep}{4pt}
\begin{tabular}{llll}
\toprule
$(n,\ell)$ & $\mathcal{T}_{\ell,n}$, Dirichlet-type, $W_{\mathbf Q}$ & $\Tb_{n,\ell}$, Neumann-type, $W_{\mathbf Q'}$ & $p$\\
\midrule
$(1,1)$ & $W_{-1}$ & $W_{-1}$ & $1$\\
$(2,1)$ & $W_{0,-1}$, $W_{-1,-2}$ & $W_{-1,0,0}$, $W_{-1,1,-1}$ & $1,2$\\
$(3,1)$ & $W_{0,0,-1}$, $W_{0,-1,-2}$, $W_{-1,-2,-3}$ & $W_{-1,0,0,0,0}$, $W_{-1,1,-1,0,0}$, $W_{-1,1,-1,1,-1}$ & $1,2,3$\\
$(1,2)$ & $W_{-1,1,0}$ & $W_{-1,-1}$ & $1$\\
$(1,3)$ & $W_{-1,1,0,0,0}$ & $W_{-1,-1,-1}$ & $1$\\
\bottomrule
\end{tabular}
\end{center}
On the Dirichlet-type side the flux attached by the Gauss law carries the topological charge $-\frac p2$, which is fractional for
odd $p$.
On the Neumann-type side this topological charge becomes the charge $p$ under $U(1)_h$, whose fugacity is $h'$ here.
In the other cases the line defect half-index on at least one side is a combination of those of two Wilson lines, and the mirror
relation equates the combination \eqref{antisym_two} with the combination \eqref{NNlinetwo} for $(\ell,n)$ (section
\ref{sec_bline_square}).

\subsection{Squares of boundary conditions}
\label{sec_bc_square}
Starting from the boundary condition (I) or (II) of $USp(2n)$, we obtain two squares of boundary conditions, the first of which is
\eqref{intro_square} with the boundary conditions specified in full,
\begin{equation}
\label{square_A}
\tag{A}
\begin{tikzcd}[column sep=5.5em,row sep=3.2em]
USp(2n)_{-k}\,[\Ncal,N,\ell\Gamma] \arrow[r,leftrightarrow,"\textrm{(I)}"] \arrow[d,leftrightarrow,"\textrm{GK}"'] & \mathcal{T}_{\ell,n}\,[\Dcal,D] \arrow[d,leftrightarrow,"\textrm{mirror}"]\\
USp(2\ell)_{k}\,[\Dcal,D] \arrow[r,leftrightarrow,"\textrm{(II)}"'] & \Tb_{n,\ell}\,[\Ncal,N,2n\ell\Gamma]
\end{tikzcd}
\end{equation}
\begin{equation}
\label{square_B}
\tag{B}
\begin{tikzcd}[column sep=5.5em,row sep=3.2em]
USp(2n)_{k}\,[\Dcal,D] \arrow[r,leftrightarrow,"\textrm{(II)}"] \arrow[d,leftrightarrow,"\textrm{GK}"'] & \Tb_{\ell,n}\,[\Ncal,N,2\ell n\Gamma] \arrow[d,leftrightarrow,"\textrm{mirror}"]\\
USp(2\ell)_{-k}\,[\Ncal,N,n\Gamma] \arrow[r,leftrightarrow,"\textrm{(I)}"'] & \mathcal{T}_{n,\ell}\,[\Dcal,D]
\end{tikzcd}
\end{equation}
The square \eqref{square_B} is \eqref{square_A} for $(\ell,n)$ with its two rows exchanged, and its bulk theories are the parity images of those of \eqref{square_A}.
Each of the four dualities exchanges a Neumann-type boundary condition with a Dirichlet-type one.
The half-indices agree along all of them by \eqref{gen_I_claim}, \eqref{GKbdry}, \eqref{gen_II_claim} and \eqref{mirrorchain}.
The dualities (I) and (II) preserve the twist, while the Giveon-Kutasov type duality and the mirror symmetry exchange the H-twist and
C-twist points through
$a\to a^{-1}$, so that each square carries a single VOA (section \ref{sec_voa_usp}).

The diagonal of the square \eqref{square_B} is the composition of (II) for $(n,\ell)$ with \eqref{GKbdry} for $(\ell,n)$.
It shows that the Neumann-type boundary condition of the Abelian theory is also dual to a Neumann-type boundary condition of the
symplectic gauge theory,
\begin{align}
\label{NNmirrorGK}
\II^{\Tb_{\ell,n}}_{\Ncal,N,2\ell n\Gamma}(h,x)=\II^{USp(2\ell)_{-k}}_{\Ncal,N,n\Gamma}(a^{-1},x)\Big|_{r_Q\to1-r_Q},
\end{align}
where $h=q^{\frac{r_Q-1}{2}}a$ as in \eqref{gen_II_id}, and the flavor fugacities $x_j$ of the rows become those of the $n$ Fermi
multiplets.
We have also checked \eqref{NNmirrorGK} directly for $n+\ell\le5$.
The C-twist point of the left-hand side corresponds to the H-twist point of the right-hand side,
where the VOA $L_\ell(\mathfrak{osp}(1|2n))$ appears as in section \ref{sec_gen_I} with $n$ and $\ell$
exchanged.

\subsection{Squares of boundary lines}
\label{sec_bline_square}
\subsubsection{Line defect half-indices}
The antisymmetric Wilson lines of the symplectic gauge theories and their duals form squares of lines ending on the boundary
conditions of \eqref{square_A} and \eqref{square_B}.
For the square \eqref{square_A} the four dualities give the chain
\begin{align}
\label{squarerel}
&\lin{W_{(1^p)}}{USp(2n)_{-k}}{\Ncal,N,\ell\Gamma}(a,x)
\nonumber\\
&=(-q^{\frac14}a)^p\Big[\lin{W_{(\mathbf w_n-\mathbf w_{n-p})\otimes \mathbf e_1}}{\mathcal{T}_{\ell,n}}{\Dcal,D}
-q^{\frac{r_Q}{2}-\frac14}\lin{W_{(\mathbf w_n-\mathbf w_{n-p+1})\otimes \mathbf e_1}}{\mathcal{T}_{\ell,n}}{\Dcal,D}\Big]_{\eqref{gen_I_id}}
\nonumber\\
&=(-1)^pq^{\frac{n\ell}{4}}\lin{V\otimes W_{(n^{\ell-1},n-p)}}{USp(2\ell)_{k}}{\Dcal,D}(x;a^{-1})\Big|_{r_Q\to1-r_Q}
\nonumber\\
&=h'^{-\epsilon_p}\Big(\lin{W_{-\mathbf 1\otimes \mathbf e_p}}{\Tb_{n,\ell}}{\Ncal,N,2n\ell\Gamma}(h',x)-\lin{W_{-\mathbf 1\otimes \mathbf e_{p-2}}}{\Tb_{n,\ell}}{\Ncal,N,2n\ell\Gamma}(h',x)\Big),
\nonumber\\
&\qquad h'=q^{-\frac{r_Q}{2}}a^{-1},
\end{align}
where the first line is the duality (I), \eqref{antisym_two}.
The equality of its left-hand side with the second line is the Giveon-Kutasov type duality, \eqref{GKlines}, and the equality of
the second and third lines is the duality (II), \eqref{NNlinetwo} for $(\ell,n)$.
The equality of the right-hand side of the first line with the third line is the mirror symmetry.
The second terms of the first and third lines are absent for $p=0$ and for $p<2$, respectively, and $\epsilon_p$ is $1$ for odd $p$ and $0$ for even $p$, as in
\eqref{II_two}.
For $\ell=1$ the left-hand side and the second line are the line defect half-indices of the single lines $W_{(1^p)}$ and
$V\otimes W_{(n-p)}$.
By \eqref{singlerule}, the left-hand side is also $(-q^{\frac34}a^{-1})^p$ times the line defect half-index of the single line
$W_{\mathbf Q_{(1^p)}}$ of $\mathcal T_{1,n}$.
By \eqref{hc_two} with $\ell$ replaced by $n$, the third line is $h'^{-p}$ times the line defect half-index of the single line
$W_{-1,1,\dots,-1,0,\dots,0}$ with $2p-1$ non-vanishing entries (the trivial line for $p=0$, section \ref{sec_gen_II}).
The dualities of \eqref{square_B} give the same equalities with $n$ and $\ell$ exchanged.
The factor $q^{n\ell/4}$ is the inverse of the factor of \eqref{specflow} for $USp(2\ell)$, where $k_{\rm eff}=n$.
The powers of $h'$ reflect the quantum numbers of the endpoints discussed in section \ref{sec_gen_II}.
On the Abelian sides of \eqref{square_A} the duals are single Wilson lines on the Dirichlet-type boundary of $\mathcal{T}_{\ell,n}$
for $\ell=1$ or $p\in\{0,n\}$ and on the Neumann-type boundary of $\Tb_{n,\ell}$ for $\ell=1$ or $p\in\{0,1\}$, and for
\eqref{square_B} the same holds with $n$ and $\ell$ exchanged.
In the other cases their line defect half-indices are combinations of those of two Wilson lines or sums of those of single lines.
On the diagonal of \eqref{square_B} the lines map as
\begin{align}
\label{NNmirrorGKlines}
&\lin{W_{-\mathbf 1\otimes \mathbf e_p}}{\Tb_{\ell,n}}{\Ncal,N,2\ell n\Gamma}(h,x)-\lin{W_{-\mathbf 1\otimes \mathbf e_{p-2}}}{\Tb_{\ell,n}}{\Ncal,N,2\ell n\Gamma}(h,x)
\nonumber\\
&=h^{\epsilon_p}\Bigl(\lin{W_{(1^p)}}{USp(2\ell)_{-k}}{\Ncal,N,n\Gamma}(a^{-1},x)\Big|_{r_Q\to1-r_Q}\Bigr),
\end{align}
where $p=0,\dots,\ell$ and the second term on the left-hand side is absent for $p<2$.
For $p=0$ and $p=1$ the left-hand side is the line defect half-index of the trivial line and of a single Wilson line, respectively, so that
\eqref{NNmirrorGKlines} relates those of single lines, as it does for $n=1$ by \eqref{hc_two}.
We have also checked \eqref{NNmirrorGKlines} directly for $n+\ell\le5$.
The Giveon-Kutasov type duality \eqref{GKlines} applies to all the lines $W_\lambda$ and $V\otimes W_\lambda$ with $\lambda\subseteq(\ell^n)$,
and the twist limits of their line defect half-indices are given by \eqref{Hmodule} and \eqref{CVW}.
Whether the dual of a line is a single line, however, depends on the line and on the boundary condition.

\subsubsection{Line defect indices}
\label{sec_bulkid}
\label{sec_bulksign}
We also test the dictionaries of the lines with the line defect indices of section \ref{sec_bulk}, which do not depend on the
boundary conditions.
For the single lines of \eqref{singlerule} we find
\begin{align}
\label{bulk_id}
\Ical^{USp(2n)_{-k}}[W^N_\lambda](a)&=(-q^{-\frac34}a)^{|\lambda|}\
\Ical^{\mathcal{T}_{\ell,n}}[W^N_{-\mathbf Q_\lambda}](y)\big|_{y=a^2},
\nonumber\\
\Ical^{USp(2n)_{-k}}[W^S_\lambda](a)&=(-q^{\frac34}a^{-1})^{|\lambda|}\
\Ical^{\mathcal{T}_{\ell,n}}[W^S_{\mathbf Q_\lambda}](y)\big|_{y=a^2},
\end{align}
with the charges $\mathbf Q_\lambda$ of \eqref{singlerule}, so that the north pole line $W^N_{-\mathbf Q_\lambda}$ carries the flux
$v_\lambda=-\kappa^{-1}\mathbf Q_\lambda$ with the topological charge $\mathbf A\cdot v_\lambda=\frac{|\lambda|}{2}$.
The coefficients do not depend on $r_Q$.
We have checked \eqref{bulk_id} at both poles for $\ell=1$ with $n\le5$ and all $p$, and for all the $\lambda$ of
\eqref{singlerule} with $(n,\ell)=(1,2)$, $(1,3)$, $(1,4)$, $(2,2)$, $(3,2)$, $(2,3)$, $(4,2)$ and $(2,4)$.
The sums for $W_{\yf}$ (table \ref{tab_sums}) with $(n,\ell)=(2,2)$, $(3,2)$, $(2,3)$, $(4,2)$ and $(2,4)$, and
\eqref{symdual} with $n=2,3,4$, also hold in the bulk.
At the south pole they have the same charges and coefficients as on the boundary, and at the north pole the charges are reversed and the coefficients
are replaced by their images under $(q,a)\to(q^{-1},a^{-1})$.
The south pole equalities \eqref{bulk_id} have the charges and the coefficients of \eqref{singlerule} on (I).
By \eqref{pole_flip} and \eqref{pole_flip_Ab}, the north pole equalities are the south pole equalities of $USp(2n)_{k}$ and
$\Tb_{\ell,n}$.
For $\lambda=(\ell^{n-1},\ell-p)$ with $p\le1$ or $n=1$ these equalities have the charges and the coefficients of \eqref{II_two} with $j=0$
on (II), together with \eqref{hc_two} for $n=1$, at $h=q^{-1/4}a$, and up to the reflection $b\to2\ell-b$ these charges are those of
(I) reversed.
Thus the two poles of the bulk correspond to the two boundary conditions (I) and (II).

For the lines $V\otimes W_{(\ell^{n-1},\ell-p)}$ with single Abelian duals, i.e.\ for $n=1$ or $p\le1$, we insert $V$ as in \eqref{bulkV} and find
at the south pole with $r_Q=\frac12$
\begin{align}
\label{bulkII_gen}
&\Ical^{USp(2n)_{k}}[(V\otimes W_{(\ell^{n-1},\ell-p)})^S](a)
\nonumber\\
&=(-1)^{\frac{n(n+1)}{2}}a^{\frac n2}\,(-1)^pq^{-\frac{n\ell}{4}}h^{-p}\Big|_{h=q^{-1/4}a}\,
\Ical^{\Tb_{\ell,n}}[W^S_{\mathbf Q}](y)\big|_{y=a^2},
\end{align}
where $W_{\mathbf Q}$ is the single Wilson line dual to $V\otimes W_{(\ell^{n-1},\ell-p)}$ on (II) (section \ref{sec_gen_II}).
This is the equality on (II) with the coefficient multiplied by $(-1)^{n(n+1)/2}a^{n/2}$.
This factor does not depend on $\ell$ and $p$, and we call it the normalization factor of $V$.
Its dependence on $a$ is that of the product of the factors $X^{f/2}$ of $Z_{\rm chi}$ in \eqref{bulkV},
$\prod_{w=\pm\mathbf e_i}(aq^{\frac14}s^w)^{\frac{w(m)}{2}+\frac14}=(aq^{\frac14})^{\frac n2}\prod_is_i^{m_i}$, whose power of $a$ does not
depend on $m$, so that it reflects the choice of the background Chern-Simons contact term of $U(1)_a$ in $Z_{\rm chi}$.
We have checked \eqref{bulkII_gen} for all $p$ for $n=1$ with $\ell\le3$ and for $p\le1$ with $(n,\ell)=(2,2)$, $(3,2)$ and $(2,3)$.
For $\ell=1$, $n\le5$ and $j=0,1$, the south pole line defect indices of $V^j\otimes W_{(1^p)}$ of $USp(2n)_{n+\frac32}$
equal those of the Wilson lines of $\Tb_{1,n}$ with the charges and the coefficients of \eqref{n1_II_rule} at $h=q^{-1/4}a$.
For $j=1$ the coefficient is multiplied by the same factor $(-1)^{n(n+1)/2}a^{n/2}$.
For $(n,\ell,p)=(2,2,2)$ we have also checked that \eqref{NNlinetwo}, with the coefficient multiplied by the normalization factor of
$V$, holds for the line defect indices at the south pole.

By \eqref{pole_flip}, the equalities \eqref{bulk_id} written for $USp(2n)_{+k}$ would map a Wilson
line at one pole to a Wilson line at the opposite pole.
A duality maps a line at a point of $S^2$ to a line at the same point.
Hence, with the identification $y=a^2$, the line defect indices of the Wilson lines identify the dual of $\mathcal{T}_{\ell,n}$ as
$USp(2n)_{-k}+Q$, in agreement with the boundary analysis of section \ref{sec_signs}.
For $n=\ell$ the Giveon-Kutasov type duality relates $USp(2n)_{-k}$ at $a$ to $USp(2n)_{k}$ at $a^{-1}$, so that the
bulk analysis fixes the sign only up to this duality.
With the levels assigned in this way, all the dualities of \eqref{square_A} and \eqref{square_B} map a line at a given pole to a line at
the same pole.
For $(n,\ell)=(1,1)$ and $(2,1)$ we have checked that the line defect indices of all the single lines of the two squares, including
those with $V$, agree along the squares at both poles up to monomials with coefficient $\pm1$, whose product along each square is one.
These monomials agree with those of the boundary squares at $r_Q=\frac12$, including the powers of $q$, except for the dualities
involving $V$, where the bulk coefficient is the boundary one times the normalization factor of $V$.
These results suggest that the single-line dictionaries come from identifications of bulk line operators.

\subsection{\texorpdfstring{Boundary VOAs $L_n(\mathfrak{osp}(1|2\ell))$}{Boundary VOAs L(osp(1|2l))}}
\label{sec_voa_usp}
As discussed in section \ref{sec_twist_com}, the boundary VOAs of (I) at the H-twist point and of (II) at the C-twist point are expected to be
$L_n(\mathfrak{osp}(1|2\ell))$ and $L_\ell(\mathfrak{osp}(1|2n))$, respectively.
We now read the twist limits of the squares of section \ref{sec_bline_square} as the supercharacters of the modules of these VOAs.

\subsubsection{Squares of modules}
\label{sec_voa_squares}
The twist limits of the squares \eqref{square_A} and \eqref{square_B} are taken at $a=q^{1/4}$ and $a=q^{-1/4}$, respectively, with $r_Q=\frac12$ and all
the flavor fugacities set to one.
Evaluating \eqref{squarerel} at $a=q^{1/4}$ and $r_Q=\frac12$ for $(n,\ell)$ and for $(\ell,n)$ and using \eqref{Hmodule}, we obtain
\begin{equation}
\label{square_Ap}
\tag{A$'$}
\begin{tikzcd}[column sep=1.6em,row sep=3.2em]
\bigl[USp(2n)_{-k},\Ncal,\textrm{H}\bigr]\ (-q^{\frac12})^p\,\chi^{L_n(\mathfrak{osp}(1|2\ell))}_{(p)} \arrow[r,leftrightarrow] \arrow[d,leftrightarrow] & \bigl[\mathcal{T}_{\ell,n},\Dcal,\textrm{H}\bigr]\ \chi^{L_n(\mathfrak{osp}(1|2\ell))}_{(p)} \arrow[d,leftrightarrow]\\
\bigl[USp(2\ell)_{k},\Dcal,\textrm{C}\bigr]\ q^{\frac p2-\frac{n\ell}{4}}\,\chi^{L_n(\mathfrak{osp}(1|2\ell))}_{(p)} \arrow[r,leftrightarrow] & \bigl[\Tb_{n,\ell},\Ncal,\textrm{C}\bigr]\ (-1)^p\,\chi^{L_n(\mathfrak{osp}(1|2\ell))}_{(p)}
\end{tikzcd}
\end{equation}
\begin{equation}
\label{square_Bp}
\tag{B$'$}
\begin{tikzcd}[column sep=1.6em,row sep=3.2em]
\bigl[USp(2n)_{k},\Dcal,\textrm{C}\bigr]\ q^{\frac p2-\frac{n\ell}{4}}\,\chi^{L_\ell(\mathfrak{osp}(1|2n))}_{(p)} \arrow[r,leftrightarrow] \arrow[d,leftrightarrow] & \bigl[\Tb_{\ell,n},\Ncal,\textrm{C}\bigr]\ (-1)^p\,\chi^{L_\ell(\mathfrak{osp}(1|2n))}_{(p)} \arrow[d,leftrightarrow]\\
\bigl[USp(2\ell)_{-k},\Ncal,\textrm{H}\bigr]\ (-q^{\frac12})^p\,\chi^{L_\ell(\mathfrak{osp}(1|2n))}_{(p)} \arrow[r,leftrightarrow] & \bigl[\mathcal{T}_{n,\ell},\Dcal,\textrm{H}\bigr]\ \chi^{L_\ell(\mathfrak{osp}(1|2n))}_{(p)}
\end{tikzcd}
\end{equation}
Here $\chi^{L_n(\mathfrak{osp}(1|2\ell))}_{(p)}$ and $\chi^{L_\ell(\mathfrak{osp}(1|2n))}_{(p)}$ are the supercharacters of the modules
$\mu=(p)$ of $L_n(\mathfrak{osp}(1|2\ell))$ and $L_\ell(\mathfrak{osp}(1|2n))$, and H and C mark the H-twist and C-twist points.
Each vertex carries the twist limit of the line defect half-index of the corresponding line in \eqref{squarerel}.
At the Dirichlet-type Abelian vertex it is that of the combination of those of two Wilson lines in the first line of
\eqref{squarerel}, and at the Neumann-type Abelian vertex it is $q^{-\lfloor p/2\rfloor}$ times that of the combination in the third
line, as in \eqref{chiferm}.
By the equalities of the four dualities the four vertices give the same supercharacter once one of them is identified, and by
\eqref{Hmodule} and \eqref{CVW} the symplectic vertices give the stated supercharacters.
For $\ell=1$ in \eqref{square_Ap} all the characters of $M(2,2n+3)$ appear at every vertex,
\begin{equation}
\label{square_Ap1}
\tag{A$'_{\ell=1}$}
\begin{tikzcd}[column sep=2.2em,row sep=3.6em]
{\begin{array}{c}USp(2n)_{-(n+\frac32)},\ \Ncal\\ \lin{W_{(1^p)}}{}{}=(-q^{\frac12})^p\,\chi^{M(2,2n+3)}_{(1,n+1-p)}\end{array}} \arrow[r,leftrightarrow] \arrow[d,leftrightarrow] & {\begin{array}{c}\mathcal{T}_{1,n},\ \Dcal\\ \lin{W_{\mathbf Q_{(1^p)}}}{}{}=\chi^{M(2,2n+3)}_{(1,n+1-p)}\end{array}} \arrow[d,leftrightarrow]\\
{\begin{array}{c}USp(2)_{n+\frac32},\ \Dcal\\ \lin{V\otimes W_{(n-p)}}{}{}=q^{\frac p2-\frac n4}\,\chi^{M(2,2n+3)}_{(1,n+1-p)}\end{array}} \arrow[r,leftrightarrow] & {\begin{array}{c}\Tb_{n,1},\ \Ncal\\ \lin{W_{(p)}^{\rm ab}}{}{}=(-1)^p\,\chi^{M(2,2n+3)}_{(1,n+1-p)}\end{array}}
\end{tikzcd}
\end{equation}
where $W_{(p)}^{\rm ab}=W_{-1,1,\dots,-1,0,\dots,0}$ is the single Wilson line of $\Tb_{n,1}$ with $2p-1$
non-vanishing alternating entries, the trivial line for $p=0$.
The square \eqref{square_Bp} for $n=1$ is \eqref{square_Ap1} with $n$ replaced by $\ell$ and its two rows exchanged.
The squares collect the twist limits in which the line defect half-indices give the supercharacters of single modules.
The opposite twist points are the C-twist point of the Neumann-type and the H-twist point of the Dirichlet-type boundary
conditions of the symplectic gauge theory.
There these boundary conditions require flips, which we do not consider (section \ref{sec_twist_lim}).\footnote{In
\cite{Creutzig:2024ljv} the boundary VOAs of $\mathcal{T}_{\ell,n}$ were proposed to be $L_n(\mathfrak{osp}(1|2\ell))$ in the B-twist
and $W^{\min}_{n-\frac12}(\mathfrak{usp}(2\ell))$ in the A-twist. Here the C-twist point of the square \eqref{square_B}, at the vertex $\Tb_{\ell,n}$, gives the
supercharacters of $L_\ell(\mathfrak{osp}(1|2n))$, the B-twisted algebra of its mirror $\mathcal T_{n,\ell}$. For $\ell\ge2$ the supercharacters of $W^{\min}$ do not appear at
the twist points without flips considered here. For $\ell=1$ the characters of $W^{\min}_{n-\frac12}(\mathfrak{usp}(2))=M(2,2n+3)$ are
the supercharacters of $L_n(\mathfrak{osp}(1|2))$ by \eqref{osp12M}.}

\subsubsection{Closed forms}
\label{sec_voa_closed}
The four vertices of \eqref{square_Ap} give four expressions of the same supercharacter.
For $\ell=1$ the first two are also equal by \eqref{linechi} and the Andrews-Gordon identities.
\begin{enumerate}
\item For the Neumann-type boundary of $USp(2n)$, at $a=q^{1/4}$ the factor contributed by $Q$ cancels one of the factors contributed
by the Fermi multiplets, and
\begin{align}
\label{chiint}
\chi^{L_n(\mathfrak{osp}(1|2\ell))}_{(p)}&=\frac{(-q^{\frac12})^{-p}(q)_\infty^n}{2^nn!}\oint\prod_{i=1}^n\frac{ds_i}{2\pi is_i}
\prod_{i\neq j}\poch{s_is_j^{-1}}\prod_{i\le j}\poch{s_is_j}\poch{s_i^{-1}s_j^{-1}}
\nonumber\\
&\qquad\times\prod_{i=1}^n\poch{q^{\frac12}s_i^{\pm}}^{2\ell-1}\,\chi_{(1^p)}(s).
\end{align}
\item For the Dirichlet-type boundary of $\mathcal{T}_{\ell,n}$, the invariants \eqref{zw_def} are $z_I=w_J=1$, and \eqref{nahm_Q} gives the
Nahm sum
\begin{align}
\label{chinahm}
\chi^{L_n(\mathfrak{osp}(1|2\ell))}_{(p)}=\sum_{N\in\Zb_{\ge0}^{n(2\ell-1)}}\frac{q^{\frac12N\kappa N-\mathbf Q_{(1^p)}\cdot N}}{\prod_I(q)_{N_I}},
\end{align}
where the linear term $-\mathbf Q_{(1^p)}\cdot N=-\sum_{i,b}(\mathbf Q_{(1^p)})_{(i,b)}N_{(i,b)}$
comes from the single Wilson line $W_{\mathbf Q_{(1^p)}}$ dual to $W_{(1^p)}$.
By \eqref{singlerule}, $W_{(1^p)}$ has this single dual Wilson line for $\ell=1$ or $p\in\{0,n\}$.
For all $p$, \eqref{antisym_two} gives the supercharacter as the difference of the Nahm sums of its two Wilson lines,
\begin{align}
\label{Hl_two}
\chi^{L_n(\mathfrak{osp}(1|2\ell))}_{(p)}=\sum_{N\in\Zb_{\ge0}^{n(2\ell-1)}}\frac{q^{\frac12N\kappa N}\bigl(q^{-\sum_i\min(i,p)N_{(i,1)}}-q^{-\sum_i\min(i,p-1)N_{(i,1)}}\bigr)}{\prod_I(q)_{N_I}},
\end{align}
where the second term is absent for $p=0$.
\item For the Neumann-type boundary of $\Tb_{n,\ell}$, \eqref{NNfsum} at $h=q^{-1/2}$ and $x_j=1$ gives the alternating
fermionic sum
\begin{align}
\label{chiferm}
(-1)^p\chi^{L_n(\mathfrak{osp}(1|2\ell))}_{(p)}&=\frac{(q)_\infty^{\ell(2n-1)}}{(q)_\infty^{2n\ell}}\sum_{\substack{t\in\Zb^{2n\ell}\\ d_I\ge0}}(-1)^{\sum_Ft_F}q^{\sum_F\frac{t_F^2-t_F}{2}}\prod_I\frac{1}{(q)_{d_I}},
\nonumber\\
d&=-\sum_Ft_F\mathbf g_F-\mathbf Q,
\end{align}
with the data of $\Tb_{n,\ell}$.
For $p=0$, $p=1$ or $\ell=1$, $\mathbf Q$ is the charge of the single Wilson line dual to $V\otimes W_{(n^{\ell-1},n-p)}$.
Otherwise, by \eqref{II_two}, the right-hand side is replaced by $q^{-\lfloor p/2\rfloor}$ times the difference of the sums for
$\mathbf Q=-\mathbf 1\otimes \mathbf e_p$ and $\mathbf Q=-\mathbf 1\otimes \mathbf e_{p-2}$.
\item For the Dirichlet-type boundary of $USp(2\ell)$, the supercharacter times $q^{\frac p2-\frac{n\ell}{4}}$ is given by \eqref{Vj_general}
for $V\otimes W_{(n^{\ell-1},n-p)}$ at $a=q^{-1/4}$ in the limit $u\to1$ (\eqref{CVW} with $n$ and $\ell$ exchanged).
\end{enumerate}
For $\ell=1$ these give four expressions of the characters of $M(2,2n+3)$.
The first, the integral \eqref{chiint}, equals the Andrews-Gordon product \eqref{chiM} by \eqref{linechi}, and the second is the
Andrews-Gordon sum
\eqref{n1_H_AG} \cite{MR123484,MR351985}, which for $n=1$ is the Rogers-Ramanujan sum \cite{Rogers:1894}.
The third is $(-1)^p$ times the alternating sum \eqref{chiferm} with $\ell=1$.
The fourth is the bosonic sum
\begin{align}
\label{chibos}
\chi^{M(2,2n+3)}_{(1,n+1-p)}=\frac{q^{\frac{n-p}{2}}}{2(q)_\infty}\sum_{m\in\Zb}(-1)^mq^{\frac{(2n+3)m^2+(2n+1)m}{2}}\bigl(1-q^{2m+1}\bigr)\,\chi_{(n-p)}(q^{m+\frac12})
\end{align}
of the Rocha-Caridi type \cite{MR781391}, where $\chi_{(n-p)}$ is the character of the spin-$\frac{n-p}{2}$
representation of $USp(2)$ (section \ref{sec_characters}).
The bosonic sum is obtained from the fourth expression, the Dirichlet-type line defect half-index of $V\otimes W_{(n-p)}$ of
$USp(2)_{n+\frac32}$.
In its flux sum the quasi-periodicity of $\theta$ reduces the factors contributed by $Q$ and by the vector multiplet to the
$m$-independent ratio $\theta(u^{-1};q)/\theta(u^{-2};q)$, which tends to $\frac12$ as $u\to1$, times signed powers of $q$ and $u$ and
the factor $1-q^{2m+1}u^2$.
Hence the limit can be taken term by term.
With $(1-q^{2m+1})\chi_{(n-p)}(q^{m+\frac12})=q^{-(n-p)(m+\frac12)}\bigl(1-q^{(n-p+1)(2m+1)}\bigr)$ and $m\to-m-1$ in the second term,
the sum becomes $f(-q^{n+2+p},-q^{n+1-p})/f(-q)$, which gives \eqref{chibos} by \eqref{n1_chi}.
This proves \eqref{CVW} for $USp(2)$, i.e.\ \eqref{n1_l_C}, for all $\ell$.
Together with \eqref{linechi} it also proves \eqref{GKlines} at the twist point for $\ell=1$ and all $n$.

For a general module $\mu\subseteq(n^\ell)$ of $L_n(\mathfrak{osp}(1|2\ell))$, \eqref{Hmodule} gives the supercharacter as the integral
\eqref{chiint} with $(-q^{\frac12})^{-p}$ and $\chi_{(1^p)}(s)$ replaced by $(-q^{\frac12})^{-|\mu|}$ and $\chi_{\mu^T}(s)$.
When $\mu^T$ is of the type of \eqref{singlerule}, the supercharacter is also given by the Nahm sum
\begin{align}
\label{chinahmmu}
\chi^{L_n(\mathfrak{osp}(1|2\ell))}_\mu=\sum_{N\in\Zb_{\ge0}^{n(2\ell-1)}}\frac{q^{\frac12N\kappa N-\mathbf Q_{\mu^T}\cdot N}}{\prod_I(q)_{N_I}},
\end{align}
whose linear term is minus the charge $\mathbf Q_{\mu^T}$ of the dual Wilson line in \eqref{singlerule}.
This Nahm sum is the H-twist limit of the Dirichlet-type line defect half-index of this Wilson line.
For $n=1$ all the modules $\mu=(1^p)$, $p=0,\dots,\ell$, are obtained, with the linear terms
\begin{align}
\label{Lp}
-\mathbf Q_{(p)}\cdot N=\sum_{b=1}^{\min(2p,2\ell-1)}(-1)^{b+1}N_b ,
\end{align}
by \eqref{IWp}.

As noted in section \ref{sec_gen_I_H}, the Nahm sums \eqref{chinahmmu} with $\mu=\varnothing$ are the sum sides of the identities of Warnaar and
Zudilin \cite{MR2881319}, proved in general in \cite{Creutzig:2025wz}, and for $(n,\ell)=(1,2)$ the supercharacters \eqref{KWmu} equal the three Nahm sums
\eqref{osp14chars}.
The integrals of the form \eqref{chiint} for $\mu\neq\varnothing$ do not appear in \cite{Creutzig:2025wz} or in \cite{Creutzig:2024ljv},
where the B-twisted lines
$\mathcal P^B_\mu$ were proposed to give these supercharacters (section \ref{sec_n1_H}).

\section{Other gauge groups}
\label{sec_other}
In this section we take the boundary conditions (I) for Chern-Simons theories with other gauge groups and a 3d chiral multiplet in a
real representation and compute the H-twist limits of their line defect half-indices.
These H-twist limits follow from the Weyl-Kac character formula (section \ref{sec_other_unit_real}).
We identify the commutants, the proposed boundary VOAs, in sections \ref{sec_other_real} and \ref{sec_other_lat}, and table
\ref{tab_other_sum} lists them.

\subsection{Theories and boundary conditions}
\label{sec_other_setup}
The theories are 3d $\mathcal N=2$ Chern-Simons matter theories at the level $-k$ with vanishing superpotential.
Their gauge group $G$, with the Lie algebra $\mathfrak g$, is a compact simple Lie group of rank $r$, or $SO(4)$,\footnote{We include
$SO(4)$, whose Lie algebra $\mathfrak{so}(4)=\mathfrak{su}(2)\oplus\mathfrak{su}(2)$ is not simple, for the family $SO(N)$, $N\ge3$, of
section \ref{sec_voa_lattice}.} and the chiral multiplet $Q$ is in a real representation
$\mathbf R$ of dimension $d$, which may be reducible.
We write them as $G_{-k}+Q$, as $USp(2n)_{-k}+Q$ in \eqref{pair_P}.
We put them on a half-space with the $\mathcal N=(0,2)$ Neumann boundary conditions for the vector multiplet and for $Q$ and with a
boundary Fermi
multiplet $\Gamma$ in $\mathbf R$, as for the boundary conditions (I) of the $USp(2n)$ theories.
Here $Q$ has the charge $1$ under $U(1)_a$ and the R-charge $r_Q$, and $\Gamma$ has the charge $1$ under $U(1)_x$, with the fugacities
$a$ and $x$.
With the Neumann boundary condition for the vector multiplet the gauge group $G$ is preserved on the boundary, and its anomaly there
must vanish.
In the conventions of section \ref{sec_signs} the gauge anomaly of the Neumann-type boundary condition is
\begin{align}
\label{other_anomaly}
\Acal=\underbrace{K\Tr(s^2)}_{\rm CS}+\underbrace{h^\vee\Tr(s^2)}_{\textrm{VM},\,\Ncal}-\underbrace{\tfrac12T(\mathbf R)\Tr(s^2)}_{Q,\,N}
+\underbrace{T(\mathbf R)\Tr(s^2)}_{\Gamma}=(K+k)\Tr(s^2),
\end{align}
as in \eqref{anomaly_I}, which vanishes only for $K=-k$ with
\begin{align}
\label{other_level}
k=h^\vee+j_{\mathbf R},\qquad j_{\mathbf R}=\tfrac12T(\mathbf R),
\end{align}
where $T(\mathbf R)$ is the index of section \ref{sec_signs}, e.g.\ $T=2$ for the vector representation of $SO(N)$.
We write $\rho$ for the Weyl vector of $G$ and $(\cdot,\cdot)$ for the invariant form with $(\alpha,\alpha)=2$ for the long roots.
We also write $Q_{\rm long}$ for the lattice spanned by the long roots, $\theta$ for the highest root and $w_0$ for the longest element
of the Weyl group $W$.
Let $\Delta$ be the set of roots of $G$, and let $s^\beta$ be the monomial in the fugacities $s=(s_1,\dots,s_r)$ of the maximal
torus for a weight $\beta$.
The line defect half-index of the Wilson line $W_\lambda$ is
\begin{align}
\label{other_general}
\lin{W_\lambda}{G_{-k}}{\Ncal,N,\Gamma}(a,x)=\frac{(q)_\infty^r}{|W|}\oint\prod_{i=1}^r\frac{ds_i}{2\pi is_i}
\prod_{\alpha\in\Delta}\poch{s^\alpha}
\prod_{\mu\in\mathbf R}\frac{\poch{q^{\frac12}xs^\mu}\poch{q^{\frac12}x^{-1}s^{-\mu}}}{\poch{q^{\frac{r_Q}{2}}as^\mu}}
\,\chi_\lambda(s),
\end{align}
where $\mu$ runs over the weights of $\mathbf R$ with multiplicity.
For $USp(2n)$ and $\mathbf{2n}$, which is pseudo-real, \eqref{other_general} is \eqref{gen_I_A} with $\ell=1$.
The roots, the weights of $\mathbf R$ and the characters $\chi_\lambda(s)$ of the gauge groups of this section are given in appendix
\ref{app_char}.
As in \eqref{HC_def}, the H-twist point is
\begin{align}
\label{other_HHm}
\lin{W_\lambda}{G_{-k}}{\Ncal,N,\Gamma}\Big|_{\rm H}=\lin{W_\lambda}{G_{-k}}{\Ncal,N,\Gamma}(a,x)\Big|_{r_Q=\frac12,\ a=q^{\frac14},\ x=1}.
\end{align}
At the H-twist point the contribution of $Q$ cancels half of the factors contributed by $\Gamma$, as in \eqref{n1_H_int}, and
\begin{align}
\label{other_int}
\lin{W_\lambda}{G_{-k}}{\Ncal,N,\Gamma}\Big|_{\rm H}=\frac{(q)_\infty^r}{|W|}\oint\prod_{i=1}^r\frac{ds_i}{2\pi is_i}
\prod_{\alpha\in\Delta}\poch{s^\alpha}\prod_{\mu\in\mathbf R}\poch{q^{\frac12}s^\mu}\,\chi_\lambda(s).
\end{align}
For $USp(2n)$ this is the H-twist of the theories with $\mathcal{N}=4$ supersymmetry.
For the other theories we do not expect $\mathcal N=4$ supersymmetry.
For $Spin(7)_{-6}$ the $q$-series expansion of the full index contains, at $r_Q=\frac12$, the tower $(a^2q^{\frac12})^j$ of the invariant chiral
operators $(QQ)^j$ and shows no sign of additional supercurrents.
For these theories we take \eqref{other_HHm} as the holomorphic twist \cite{Costello:2020ndc} in which the R-symmetry is
mixed with $U(1)_a$ so that $Q$ has R-charge one.
We call it the H-twist point also for these theories.
At this point the factors contributed by $\Gamma$ and $Q$ combine into that of a Dirichlet chiral multiplet in $\mathbf R$, as for
$USp(2n)$ in section \ref{sec_gen_I}, so that the H-twist limit is the half-index with the
Neumann boundary condition for the vector multiplet and the Dirichlet boundary condition for $Q$ \cite{Okazaki:2024kzo}.
For real $\mathbf R$ we assume that the boundary values of the fermions in $Q$ acquire the singular operator product of $d$ real
fermions $F^d$, so that the gauge currents have the level $j_{\mathbf R}$.
We have not derived this from the boundary condition.
Then the derived invariants of section \ref{sec_twist_com} are concentrated in degree zero (section \ref{sec_other_unit_real}), and the
boundary VOA is the commutant
\begin{align}
\label{other_coset}
\mathrm{Com}\bigl(L_{j_{\mathbf R}}(\mathfrak g),\,F^d\bigr),\qquad c=\frac d2-\frac{j_{\mathbf R}\dim G}{k},
\end{align}
where $L_{j_{\mathbf R}}(\mathfrak g)$ is the simple affine VOA of $\mathfrak g$ at the level $j_{\mathbf R}$.
Here $c$ is the central charge $\frac d2$ of $F^d$ minus the Sugawara central charge
$\frac{j_{\mathbf R}\dim G}{j_{\mathbf R}+h^\vee}=\frac{j_{\mathbf R}\dim G}{k}$ of $L_{j_{\mathbf R}}(\mathfrak g)$.
Since $F^d$ has the even part $\widehat{\mathfrak{so}}(d)_1$ \cite{Kac:1981ax,Goddard:1986ee}, the even part of the commutant is the
coset $\widehat{\mathfrak{so}}(d)_1/\widehat{\mathfrak g}_{j_{\mathbf R}}$ \cite{Goddard:1984vk,Goddard:1986ee}.
Independently of the assumption, the $G$-invariant polynomials in the fermions without derivatives are classes of degree zero of
the complex, e.g.\ the Cayley form for $Spin(7)$ and the associative and coassociative forms for $G_2$, which have the conformal
weights of the generators of the commutants.
The identification of the boundary VOA with the commutant is a proposal supported by the characters.
The ${\rm H}^-$-twist point is the image of the H-twist point under the $\Zb_2$ symmetry which acts by $-1$ on $Q$ and
$\Gamma$, i.e.\ $(a,x)\to(-a,-x)$,
\begin{align}
\label{other_Hm}
\lin{W_\lambda}{G_{-k}}{\Ncal,N,\Gamma}\Big|_{{\rm H}^-}=\lin{W_\lambda}{G_{-k}}{\Ncal,N,\Gamma}(a,x)\Big|_{r_Q=\frac12,\ a=-q^{\frac14},\ x=-1}.
\end{align}
The half-indices there are those at the H-twist point with this element of $\Zb_2$ inserted, and
\begin{align}
\label{other_int_m}
\lin{W_\lambda}{G_{-k}}{\Ncal,N,\Gamma}\Big|_{{\rm H}^-}=\frac{(q)_\infty^r}{|W|}\oint\prod_{i=1}^r\frac{ds_i}{2\pi is_i}
\prod_{\alpha\in\Delta}\poch{s^\alpha}\prod_{\mu\in\mathbf R}\poch{-q^{\frac12}s^\mu}\,\chi_\lambda(s),
\end{align}
which is \eqref{other_int} with $q^{\frac12}$ replaced by $-q^{\frac12}$ in the matter factor.
When an element of the center of $G$ acts as $-1$ on $\mathbf R$, this $\Zb_2$ is a gauge transformation.
Then the line defect half-indices at the H-twist point and at the ${\rm H}^-$-twist point agree up to signs.
Otherwise it is a global symmetry, which can be gauged, namely for $G_2$, $F_4$, $E_6$, $E_8$, for $SU(N)$ and $SO(N)$ with odd $N$ and for
$SU(2)$ with $\mathbf7$.
Gauging this $\Zb_2$ splits the line defect half-indices into the parts even and odd under it,
\begin{align}
\label{other_gauge}
\frac12\Bigl(\lin{W_\lambda}{G_{-k}}{\Ncal,N,\Gamma}\Big|_{\rm H}\pm\lin{W_\lambda}{G_{-k}}{\Ncal,N,\Gamma}\Big|_{{\rm H}^-}\Bigr),
\end{align}
and adds a twisted sector.
In the gauged theory the twist line of the $\Zb_2$ is a line operator, around which the fermions odd under the $\Zb_2$ are periodic.
The twisted sector consists of the boundary local operators at the end of this line.
We write $d_-=\dim\mathbf R_-$ for the subrepresentation $\mathbf R_-$ of $\mathbf R=\mathbf R_+\oplus\mathbf R_-$ on which the $\Zb_2$ acts
by $-1$.
For the $\Zb_2$ above we have $\mathbf R_-=\mathbf R$, and $d_-$ is odd in the cases in which we compute the twisted sector.
We write \eqref{other_tw} for general $\mathbf R_\pm$, since in appendix \ref{app_pseudo} a $\Zb_2$ acting only on a summand of $\mathbf R$
appears.
With the twist line the factors $\poch{q^{\frac12}s^\mu}$ of the weights $\mu$ of $\mathbf R_-$ in \eqref{other_int} are replaced by
those of the Ramond moding.
The line defect half-index of the twist line together with $W_\lambda$ at the H-twist point, with the $\Zb_2$ inserted, is then
\begin{align}
\label{other_tw}
\begin{split}
\lin{W_\lambda}{G_{-k}}{\Ncal,N,\Gamma}\Big|^{\rm tw}_{\rm H}&=\frac{(q)_\infty^r}{|W|}\oint\prod_{i=1}^r\frac{ds_i}{2\pi is_i}
\prod_{\alpha\in\Delta}\poch{s^\alpha}\prod_{\mu\in\mathbf R_+}\poch{q^{\frac12}s^\mu}\\
&\quad\times q^{\frac{d_-}{16}}\chi_{\rm sp}(s)\prod_{\mu\in\mathbf R_-}(-qs^\mu;q)_\infty\,\chi_\lambda(s).
\end{split}
\end{align}
Here $\frac{d_-}{16}$ is the conformal weight of the Ramond ground states of the $d_-$ boundary fermions in $\mathbf R_-$.
The zero modes of these fermions give the spinor representation of $Spin(d_-)$, and $\chi_{\rm sp}(s)$ is its character restricted
to $G$.

\subsection{Unitary VOAs as commutants}
\label{sec_other_unit_real}
The boundary fermions decompose as $F^d=\bigoplus_\lambda L(\lambda)\otimes M_\lambda$ into the integrable highest
weight modules $L(\lambda)$ of $\widehat{\mathfrak g}$ at the level $j_{\mathbf R}$, whose top components are the representations $\lambda$ of
$\mathfrak g$, and the modules $M_\lambda$ of the commutant.
In the notation of \cite{MR1104219}, where $\Lambda_0$ is the fundamental weight of the affine node, $L(\lambda)$ is
$L(j_{\mathbf R}\Lambda_0+\lambda)$.
This decomposition and the Weyl-Kac character formula give the H-twist limits of the line defect half-indices of all the Wilson lines.
We write $\|x\|^2=(x,x)$, $\Delta_+$ for the positive roots, $P$ for the weight lattice and $\mu^*=-w_0\mu$ for the conjugate of $\mu$.
An element $g=t_\beta w$ of $W\ltimes kQ_{\rm long}$, with $w\in W$ and the translation $t_\beta$ by $\beta\in kQ_{\rm long}$, acts by
$g(x)=w(x)+\beta$.
Then
\begin{align}
\label{real_wk}
\begin{split}
&(q)_\infty^r\prod_{\alpha\in\Delta}\poch{s^\alpha}=\prod_{\alpha\in\Delta_+}(1-s^\alpha)\,D(s;q),\\
&D(s;q)\,\chi^{L_{j_{\mathbf R}}(\mathfrak g)}_\lambda(s)=\sum_{g=t_\beta w\in W\ltimes kQ_{\rm long}}{\rm sgn}(w)\,q^{\frac{\|g(\lambda+\rho)\|^2-\|\lambda+\rho\|^2}{2k}}\,s^{g(\lambda+\rho)-\rho},\\
&\prod_{\alpha\in\Delta_+}(1-s^\alpha)\,\chi_\mu(s)=\sum_{w'\in W}{\rm sgn}(w')\,s^{\rho-w'(\mu^*+\rho)},
\end{split}
\end{align}
where the first line is the factor contributed by the vector multiplet in \eqref{other_int}, written with the affine Weyl denominator $D(s;q)$
of \eqref{app_Ddef}.
The second line is the Weyl-Kac formula, with the character $\chi^{L_{j_{\mathbf R}}(\mathfrak g)}_\lambda(s)$ of $L(\lambda)$ in $q$
and $s$, normalized as in section \ref{sec_twist_lim}, and the third line is the Weyl character formula.
The matter factor of \eqref{other_int} is $\sum_\lambda q^{h_\lambda}\chi^{L_{j_{\mathbf R}}(\mathfrak g)}_\lambda(s)\,
\mathrm{Tr}_{M_\lambda}(-1)^Fq^{L_0}$, with the conformal weight $h_\lambda$ of $L(\lambda)$ and the supercharacters of the
$M_\lambda$, where $L_0$ is the zero mode of the Virasoro algebra of the commutant, without the shift by its central charge.
For $W_\mu$, i.e.\ with $\chi_\mu(s)$ in \eqref{other_int}, the integral picks the terms with $g(\lambda+\rho)=w'(\mu^*+\rho)$, and the $|W|$ elements $w'$ give the same contribution.
The points $\lambda+\rho$ of the integrable weights $\lambda$ at the level $j_{\mathbf R}$ are the points of $\rho+P$ in the open
alcove, in which $(x,\theta)<k$ for the highest root $\theta$ of each simple factor of $G$.\footnote{The affine Weyl group
$W\ltimes kQ_{\rm long}$ is generated by the reflections in the hyperplanes $(x,\alpha)=mk$ with the roots $\alpha$ and $m\in\Zb$, and
the alcoves are the connected components of the complement of these hyperplanes. The open alcove is the alcove given by
$(x,\alpha_i)>0$ and $(x,\theta)<k$, with the simple roots $\alpha_i$ and the highest root $\theta$ of each simple factor of $G$. Every
$x$ not on any of the hyperplanes is $g(y)$ for a unique $g\in W\ltimes kQ_{\rm long}$ and a unique $y$ in the open alcove
\cite{MR1066460}. For a dominant weight $\lambda$ the inequalities $(\lambda+\rho,\alpha_i)>0$ always hold.}
We say that $\lambda$ is in the open alcove when $\lambda+\rho$ is.
Since $W\ltimes kQ_{\rm long}$ preserves $\rho+P$ for the integral level $k$, we obtain for every dominant weight $\mu$
\begin{align}
\label{real_reduce}
\lin{W_\mu}{G_{-k}}{\Ncal,N,\Gamma}\Big|_{\rm H}={\rm sgn}(w)\,q^{h_\mu}\,\mathrm{Tr}_{M_\lambda}(-1)^Fq^{L_0},\qquad h_\mu=\frac{(\mu,\mu+2\rho)}{2k},
\end{align}
when $\mu^*+\rho=g(\lambda+\rho)$ with $g=t_\beta w$ and $\lambda$ in the open alcove.
The H-twist limit of the line defect half-index of $W_\mu$ vanishes when $\mu^*+\rho$ lies on a reflecting hyperplane of $W\ltimes kQ_{\rm long}$, since then there
are no such $g$ and $\lambda$.
For $\lambda$ in the open alcove the conformal weight $h_\lambda$ of $L(\lambda)$ is given by the formula for $h_\mu$ in \eqref{real_reduce}.
The H-twist limit of the line defect half-index of $W_{\lambda^*}$ is $\pm q^h$ times the normalized supercharacter of $M_\lambda$, where $h$ is the lowest
conformal weight of $L(\lambda)\otimes M_\lambda$ in $F^d$, i.e.\ $h_\lambda$ plus the conformal weight of $M_\lambda$.
Those of the other Wilson lines are $0$ or these up to signs and integral powers of $q$.
Physically, the H-twist limit of the line defect half-index of $W_\mu$ is the supercharacter of the states of the complex of section
\ref{sec_twist_com} that transform in the representation $\mu^*$.
These are the states which can end the Wilson line on the boundary.
In this supercharacter the ghosts cancel most of the states which the gauge currents generate, and \eqref{real_reduce} gives what remains.
Given $F^d$, the derived invariants are concentrated in degree zero, as assumed in section \ref{sec_twist_com}.
By the theorem of Garland and Lepowsky \cite{MR414645}, in the cohomological form of \cite{MR543216}, the cohomology of
$z\mathfrak g[z]$ in $L(\lambda)$, which the ghosts compute, is the sum of the
irreducible $\mathfrak g$-modules of the dominant highest weights $g(\lambda+\rho)-\rho$ with $g\in W\ltimes kQ_{\rm long}$, each in the
degree given by the length of $g$.
It does not vanish in positive degrees, but its $G$-invariant part is non-zero only for $g(\lambda+\rho)=\rho$.
Since $\lambda+\rho$ and $\rho$ lie in the open alcove, this happens only for $g=1$ and $\lambda=0$.

Since $\widehat{\mathfrak g}$ acts unitarily at the integral level $j_{\mathbf R}$, the commutant is unitary, in the sense that its
Virasoro algebra acts unitarily on all the $M_\lambda$ and on the multiplicity spaces of the Ramond sector of
$F^d$ \cite{Goddard:1986ee}.
Its even part, the coset $\widehat{\mathfrak{so}}(d)_1/\widehat{\mathfrak g}_{j_{\mathbf R}}$ of section \ref{sec_other_setup}, is a unitary VOA
\cite{MR3119224}.
Unitarity has three consequences for these theories.
The conformal weights of the $M_\lambda$ are non-negative, so that the vacuum module, whose (super)character the H-twist limit of the half-index gives, has the
lowest conformal weight.
The central charge $c$ is non-negative, and it vanishes only when the commutant is $\Cb$.
This happens e.g.\ for $F_4$, $E_8$ and $Spin(9)$ with $\mathbf{26}$, $\mathbf{248}$ and the spinor $\mathbf{16}$, whose gauge currents form
conformal embeddings in the boundary fermions \cite{Goddard:1985jp}, and these theories give neither a non-trivial VOA nor
non-trivial characters.
For $0<c<1$ the unitary representations of the Virasoro algebra are those of the unitary minimal models
\cite{Friedan:1983xq,Goddard:1986ee}.
Hence the commutant is an extension of the unitary minimal model with central charge $c$, as for $Spin(7)$, $G_2$
and $SU(2)$ with $\mathbf7$.
For $SU(N)$, $SO(N)$ and $E_6$ the central charge is $c=1$, and the conformal embeddings of section \ref{sec_voa_lattice} show that the
commutant is a lattice VOA.

\subsection{\texorpdfstring{$Spin(7)_{-6}$ and $(G_2)_{-5}$}{Spin(7) and G2}}
\label{sec_other_real}
\begin{table}[!htbp]
\centering\footnotesize
\renewcommand{\arraystretch}{1.3}\setlength{\tabcolsep}{4pt}
\begin{tabular}{@{}llll>{\raggedright\arraybackslash}p{0.46\textwidth}@{}}
\toprule
theory & $\mathbf R$ & $c$ & boundary VOA & Wilson lines and modules\\
\midrule
$USp(2n)_{-(n+\frac32)}$ & $\mathbf{2n}$ & $\frac{2n}{2n+3}$ & $L_n(\mathfrak{osp}(1|2))$ & $W_{(1^p)}$, $0\le p\le n$, give all the modules, with the
characters of $(1,n+1-p)$ of $M(2,2n+3)$\\
\midrule
$Spin(7)_{-6}$ & $\mathbf8$ & $\frac12$ & $M(3,4)$ & $1$, $W_{(1)}$, $W_{(\frac12,\frac12,\frac12)}$ give $(1,1)$, $(1,3)$, $(1,2)$\\
$(G_2)_{-5}$ & $\mathbf7$ & $\frac7{10}$ & $SM(3,5)$ & $1$, $W_{(1)}$ give the $\widetilde{\rm NS}$ characters of $(1,1)$, $(1,3)$, and with the $\Zb_2$
gauged all the six modules of $M(4,5)$\\
$SU(2)_{-30}$ & $\mathbf7$ & $\frac7{10}$ & $SM(3,5)$ & as for $G_2$, from the $W_{(a)}$ with $a+1$ an exponent of $E_8$\\
$SU(N)_{-N-1}$ & $\mathbf N\oplus\overline{\mathbf N}$ & $1$ & $V_{\sqrt N\Zb}$ & $W_{(1^p)}$, $0\le p\le N-1$, give all the modules $p$\\
$SO(N)_{-N}$ & $2\times\mathbf N$ & $1$ & $V_{\sqrt N\Zb}$ & $W_{(1^p)}$ give the sums over $\pm p$, and for $O(N)$ the even labels of
$V_{2\sqrt N\Zb}$\\
$(E_6)_{-18}$ & $\mathbf{27}\oplus\overline{\mathbf{27}}$ & $1$ & $V_{\sqrt{27}\Zb}$ & $1$, $W_{\mathbf{27}}$, $W_{\mathbf{351}}$, $W_{\mathbf{2925}}$ give $0$, $1$,
$2$, $3$\\
\bottomrule
\end{tabular}
\caption{The theories with real $\mathbf R$ and $c>0$, with $USp(2n)$ for comparison, and their proposed boundary VOAs, the commutants \eqref{other_coset}
and, for $USp(2n)$, \eqref{H_coset} (sections \ref{sec_other_real} and \ref{sec_other_lat}).
The last column gives the Wilson lines, with $1$ the trivial line, and the modules whose characters, or supercharacters, the H-twist limits of
their line defect half-indices give up to monomials.}
\label{tab_other_sum}
\end{table}

\subsubsection{\texorpdfstring{$Spin(7)_{-6}$ and the Ising model}{Spin(7) and the Ising model}}
\label{sec_voa_spin7}
For $Spin(7)_{-6}$ with $Q$ and $\Gamma$ in the spinor $\mathbf8$ we have $j_{\mathbf R}=1$ and label the representations by
$\lambda=\sum_i\lambda_i\mathbf e_i$, so that $(1)$ is the vector, $(1,1,0)$ the adjoint and $(\frac12,\frac12,\frac12)$ the spinor.
The commutant of $\widehat{\mathfrak{so}}(7)_1$ in $F^8$ is the
coset $\widehat{\mathfrak{so}}(8)_1/\widehat{\mathfrak{so}}(7)_1$ with central charge $4-\frac72=\frac12$, the Ising model $M(3,4)$
\cite{MR1354601}.
Since the eight fermions are in the spinor of $Spin(7)$, $F^8$ decomposes under $\widehat{\mathfrak{so}}(7)_1\otimes M(3,4)$ as
\begin{align}
\label{spin7_branch}
F^8=L(\varnothing)\otimes(1,1)\oplus L((1))\otimes(1,3)\oplus L((\tfrac12,\tfrac12,\tfrac12))\otimes(1,2),
\end{align}
where $L(\varnothing)$, $L((1))$ and $L((\frac12,\frac12,\frac12))$ are the modules of $\widehat{\mathfrak{so}}(7)_1$ with conformal
weights $0$, $\frac12$ and $\frac7{16}$.
Here $(r,s)$ are the modules of $M(3,4)$, and the last summand is the odd part of $F^8$.
The multiplicity space of $L(\varnothing)$ is the vacuum module, so that the commutant is $M(3,4)$ itself.
By \eqref{real_reduce} with $k=6$ and \eqref{spin7_branch} we obtain
\begin{align}
\label{spin7_ising}
&\II^{Spin(7)_{-6}}_{\Ncal,N,\Gamma}\Big|_{\rm H}=\chi^{M(3,4)}_{(1,1)},\qquad
\lin{W_{(1)}}{Spin(7)_{-6}}{\Ncal,N,\Gamma}\Big|_{\rm H}=q\,\chi^{M(3,4)}_{(1,3)},
\nonumber\\
&\lin{W_{(\frac12,\frac12,\frac12)}}{Spin(7)_{-6}}{\Ncal,N,\Gamma}\Big|_{\rm H}=-q^{\frac12}\chi^{M(3,4)}_{(1,2)},
\end{align}
with the conformal weights $h_{1,1}=0$, $h_{1,2}=\frac1{16}$ and $h_{1,3}=\frac12$, and with the classical product forms \cite{MR49225}
\begin{align}
\label{ising_prod}
\begin{split}
&\chi^{M(3,4)}_{(1,1)}=\prod_{m\equiv\pm2,\pm3,\pm4,\pm5\,(16)}\frac{1}{1-q^m},\qquad
\chi^{M(3,4)}_{(1,3)}=\prod_{m\equiv\pm1,\pm4,\pm6,\pm7\,(16)}\frac{1}{1-q^m},\\
&\chi^{M(3,4)}_{(1,2)}=\prod_{m\ge1}(1+q^m).
\end{split}
\end{align}
So the H-twist limits of the half-index and of the line defect half-indices of $W_{(1)}$ and $W_{(\frac12,\frac12,\frac12)}$ give the characters of the three
modules of the Ising model.
These modules have the fusion rules of the modules of $\widehat{\mathfrak{so}}(7)_1$, in which
$(\frac12,\frac12,\frac12)\times(\frac12,\frac12,\frac12)=\varnothing+(1)$ matches $(1,2)\times(1,2)=(1,1)+(1,3)$.
The powers $q$ and $q^{\frac12}$ in \eqref{spin7_ising} are the conformal weights $\frac12+\frac12$ and $\frac7{16}+\frac1{16}$ of the
lowest states of the second and third summands of \eqref{spin7_branch}.
The sign of the H-twist limit of the line defect half-index of $W_{(\frac12,\frac12,\frac12)}$ is that of the odd part, as in section \ref{sec_other_unit_real}.
The H-twist limits of the line defect half-indices of the other Wilson lines are also given by \eqref{real_reduce}.
That of the adjoint Wilson line $W_{(1,1,0)}$ vanishes, since $(\theta+\rho,\theta)=6$, i.e.\ $\theta+\rho$ lies on the reflecting hyperplane
$(x,\theta)=6$.
Physically, none of the three modules of $\widehat{\mathfrak{so}}(7)_1$ in \eqref{spin7_branch} has the adjoint as its top component.
The adjoint states of $F^8$, which the gauge currents generate, cancel against those with the ghosts, e.g.\ at the order $q$ the gauge
currents against the ghosts for $z\mathfrak g$.
For $W_{(2,1,0)}$ and $W_{(2,2,0)}$ the reflection in the same hyperplane maps $\mu+\rho$ to $(1)+\rho$ and $\rho$, with ${\rm sgn}(w)=-1$,
$h_{(2,1,0)}=\frac32$ and $h_{(2,2,0)}=2$, so that \eqref{real_reduce} gives $-q^2\chi^{M(3,4)}_{(1,3)}$ and $-q^2\chi^{M(3,4)}_{(1,1)}$.

\subsubsection{\texorpdfstring{$(G_2)_{-5}$ and the tricritical Ising model}{G2 and the tricritical Ising model}}
\label{sec_voa_g2}
For $(G_2)_{-5}$ with $Q$ and $\Gamma$ in $\mathbf7$ we have $j_{\mathbf R}=1$.
The commutant of $(\widehat{\mathfrak g}_2)_1$ in $F^7$, with central charge $\frac72-\frac{14}5=\frac7{10}$, is the $\mathcal N=1$
superconformal minimal model $SM(3,5)$.
Its even part is the coset $\widehat{\mathfrak{so}}(7)_1/(\widehat{\mathfrak g}_2)_1$ \cite{Goddard:1984vk}, the tricritical Ising model
$M(4,5)$ \cite{MR1354601}.
The modules $(r,s)$ of $M(4,5)$ have the conformal weights $h_{1,1}=0$, $h_{1,2}=\frac1{10}$, $h_{1,3}=\frac35$, $h_{1,4}=\frac32$, $h_{2,1}=\frac7{16}$ and $h_{2,2}=\frac3{80}$.
The Neveu-Schwarz modules $(1,1)$ and $(1,3)$ of $SM(3,5)$, with $h=0$ and $h=\frac1{10}$, decompose into the Virasoro modules
$(1,1)\oplus(1,4)$ and $(1,2)\oplus(1,3)$ of $M(4,5)$ \cite{Friedan:1984rv,Goddard:1986ee}.
Here $(1,4)$ and $(1,3)$, with conformal weights $0+\frac32$ and $\frac1{10}+\frac12$, are generated by odd numbers of modes of the
supercurrent and are fermionic.
The Ramond modules $(1,4)$ and $(1,2)$ of $SM(3,5)$ have $h=\frac7{16}$ and $h=\frac3{80}$, and as Virasoro modules they are two copies of $(2,1)$ and of $(2,2)$, one even
and one odd, so that their $\widetilde{\rm R}$ characters $\mathrm{Tr}_{(r,s)}(-1)^Fq^{L_0-\frac{c}{24}}$ vanish.
Hence, in the notation of section \ref{sec_twist_svir},
\begin{align}
\label{NS_vir}
\chi^{SM(3,5)}_{{\rm NS},(1,1)}&=\chi^{M(4,5)}_{(1,1)}+q^{\frac32}\chi^{M(4,5)}_{(1,4)},&
\chi^{SM(3,5)}_{\widetilde{\rm NS},(1,1)}&=\chi^{M(4,5)}_{(1,1)}-q^{\frac32}\chi^{M(4,5)}_{(1,4)},
\nonumber\\
\chi^{SM(3,5)}_{{\rm NS},(1,3)}&=\chi^{M(4,5)}_{(1,2)}+q^{\frac12}\chi^{M(4,5)}_{(1,3)},&
\chi^{SM(3,5)}_{\widetilde{\rm NS},(1,3)}&=\chi^{M(4,5)}_{(1,2)}-q^{\frac12}\chi^{M(4,5)}_{(1,3)},
\nonumber\\
\chi^{SM(3,5)}_{{\rm R},(1,4)}&=\chi^{M(4,5)}_{(2,1)},&
\chi^{SM(3,5)}_{{\rm R},(1,2)}&=\chi^{M(4,5)}_{(2,2)}.
\end{align}
Let $L(\varnothing)$ and $L((1))$ be the modules of $(\widehat{\mathfrak g}_2)_1$ whose top components are the trivial representation and the
$7$-dimensional representation $(1)$, with conformal weights $0$ and $\frac25$.
The Neveu-Schwarz sector of $F^7$ and its Ramond sector, which is the module $(\frac12,\frac12,\frac12)$ of
$\widehat{\mathfrak{so}}(7)_1$, decompose under $(\widehat{\mathfrak g}_2)_1\otimes M(4,5)$ as
\begin{align}
\label{g2_branch}
\begin{split}
F^7_{\rm NS}&=L(\varnothing)\otimes\bigl((1,1)\oplus(1,4)\bigr)\oplus L((1))\otimes\bigl((1,2)\oplus(1,3)\bigr),\\
F^7_{\rm R}&=L(\varnothing)\otimes(2,1)\oplus L((1))\otimes(2,2),
\end{split}
\end{align}
where $(r,s)$ are the modules of $M(4,5)$.
The summands $L(\varnothing)\otimes(1,4)$ and $L((1))\otimes(1,2)$ form the odd part of $F^7_{\rm NS}$.
By \eqref{NS_vir} the multiplicity spaces of $L(\varnothing)$ and $L((1))$ in $F^7_{\rm NS}$ are the Neveu-Schwarz modules $(1,1)$ and
$(1,3)$ of $SM(3,5)$, the latter with the opposite parity, so that the commutant is $SM(3,5)$ itself.
The lowest states of $L((1))\otimes(1,2)$ have conformal weight $\frac25+\frac1{10}=\frac12$ and lie in the odd part, so that they
give the factor $-q^{\frac12}$ in \eqref{g2_theta}.
The twisted sector, discussed below, gives the two summands of $F^7_{\rm R}$.
It was found in \cite{Okazaki:2024kzo} that the H-twist limits of the half-index and of the line defect half-index of $W_{(1)}$ are
\begin{align}
\label{g2_theta}
\II^{(G_2)_{-5}}_{\Ncal,N,\Gamma}\Big|_{\rm H}&=\frac{f(-q^{\frac32},-q)}{f(-q)}=\frac{f(-q^{\frac12})\,\chi^{M(2,5)}_{(1,2)}(q^{\frac12})}{f(-q)},
\nonumber\\
\lin{W_{(1)}}{(G_2)_{-5}}{\Ncal,N,\Gamma}\Big|_{\rm H}&=-q^{\frac12}\,\frac{f(-q^{\frac12},-q^2)}{f(-q)}
=-q^{\frac12}\,\frac{f(-q^{\frac12})\,\chi^{M(2,5)}_{(1,1)}(q^{\frac12})}{f(-q)},
\end{align}
where the second equalities are \eqref{n1_chi} for $n=1$ at $q^{\frac12}$.
These are the $\widetilde{\rm NS}$ characters, i.e.\ the supercharacters, of $SM(3,5)$ of \eqref{NS_vir},
\begin{align}
\label{g2_tci}
\II^{(G_2)_{-5}}_{\Ncal,N,\Gamma}\Big|_{\rm H}=\chi^{SM(3,5)}_{\widetilde{\rm NS},(1,1)},\qquad
\lin{W_{(1)}}{(G_2)_{-5}}{\Ncal,N,\Gamma}\Big|_{\rm H}=-q^{\frac12}\chi^{SM(3,5)}_{\widetilde{\rm NS},(1,3)}.
\end{align}
The H-twist limit of the line defect half-index of the adjoint Wilson line vanishes, and those of the other Wilson lines vanish or give these two characters up to
signs and integral powers of $q$.

Since the center of $G_2$ is trivial, the $\Zb_2$ which acts by $-1$ on $Q$ and $\Gamma$ is a global symmetry, the element $-1$ of
$O(7)$.
It acts on the boundary VOA as the fermion parity of $SM(3,5)$.
At the ${\rm H}^-$-twist point the line defect half-indices of the same lines give the Neveu-Schwarz characters,
\begin{align}
\label{g2_NS}
\II^{(G_2)_{-5}}_{\Ncal,N,\Gamma}\Big|_{{\rm H}^-}=\chi^{SM(3,5)}_{{\rm NS},(1,1)},\qquad
\lin{W_{(1)}}{(G_2)_{-5}}{\Ncal,N,\Gamma}\Big|_{{\rm H}^-}=q^{\frac12}\chi^{SM(3,5)}_{{\rm NS},(1,3)}.
\end{align}
By \eqref{NS_vir} and \eqref{other_gauge}, the gauging of this $\Zb_2$ therefore gives the characters of the four modules $(1,1)$,
$(1,4)$, $(1,2)$ and $(1,3)$ of $M(4,5)$ individually.
In the twisted sector \eqref{other_tw} of this $\Zb_2$ we have $\mathbf R_-=\mathbf R=(1)$ and $d_-=7$.
The seven fermion zero modes give the spinor $\mathbf8=(1)\oplus\varnothing$ of $Spin(7)\supset G_2$, and the matter factor is
\begin{align}
\label{g2_ramond}
q^{\frac7{16}}(-q;q)_\infty\prod_{i=1}^3(-s_i;q)_\infty(-qs_i^{-1};q)_\infty,
\end{align}
with the fugacities $s_1,s_2,s_3$ of $SU(3)\subset G_2$, $s_1s_2s_3=1$.
Then the line defect half-indices \eqref{other_tw} of the twist line alone and together with $W_{(1)}$ give the Ramond characters
$\chi^{SM(3,5)}_{{\rm R},(1,4)}=\chi^{M(4,5)}_{(2,1)}$ and $\chi^{SM(3,5)}_{{\rm R},(1,2)}=\chi^{M(4,5)}_{(2,2)}$.
Without the $\Zb_2$ inserted, i.e.\ without $(a,x)\to(-a,-x)$, the integrand is odd under $s\to s^{-1}$, so that the line defect
half-indices of the twist line vanish.
At the H-twist point alone no finite sum of the line defect half-indices with coefficients in $\mathbb Q(q)$ gives an individual
character of $M(4,5)$.\footnote{Such a sum is a combination of $\chi^{SM(3,5)}_{\widetilde{\rm NS},(1,1)}$ and
$q^{\frac12}\chi^{SM(3,5)}_{\widetilde{\rm NS},(1,3)}$ with coefficients in $\mathbb Q(q)$. For an individual character its terms with
half-integral or those with integral powers of $q$ must cancel. By \eqref{NS_vir} the former cancel only if
$\chi^{M(4,5)}_{(1,4)}/\chi^{M(4,5)}_{(1,2)}$ is in $\mathbb Q(q)$, and the latter only if $\chi^{M(4,5)}_{(1,1)}/\chi^{M(4,5)}_{(1,3)}$ is in
$\mathbb Q(q)$. Neither is the case, since up to fractional powers of $q$ these quotients are non-constant modular functions.}

The representation $\mathbf7$ of spin $3$ of $SU(2)$ is the restriction of $\mathbf7$ of $G_2$ to the principal $SU(2)\subset G_2$.
For this representation $j_{\mathbf R}=28$ and $k=30$, and $\widehat{\mathfrak{su}}(2)_{28}\subset(\widehat{\mathfrak g}_2)_1$ is a conformal embedding, whose branching
rules give the $E_8$-type modular invariant of $\widehat{\mathfrak{su}}(2)_{28}$ \cite{Cappelli:1987xt,Altschuler:1989nm}.
So $SU(2)_{-30}$ has the same commutant $SM(3,5)$ as $(G_2)_{-5}$.
Let $W_{(a)}$ be the Wilson line of spin $\frac a2$ and $h_{(a)}=\frac{a(a+2)}{120}$.
Then \eqref{real_reduce} and these branching rules give for $0\le a\le28$
\begin{align}
\label{spin3_result}
\lin{W_{(a)}}{SU(2)_{-30}}{\Ncal,N,\Gamma}\Big|_{\rm H}=q^{h_{(a)}-h_\lambda}\lin{W_\lambda}{(G_2)_{-5}}{\Ncal,N,\Gamma}\Big|_{\rm H},
\end{align}
with $\lambda=\varnothing$ for $a\in\{0,10,18,28\}$ and $\lambda=(1)$ for $a\in\{6,12,16,22\}$, i.e.\ when $a+1$ is an exponent of
$E_8$.
The H-twist limits of the line defect half-indices of the other $W_{(a)}$ with $0\le a\le28$ vanish.
The same holds at the ${\rm H}^-$-twist point and in the twisted sector, and $SU(2)_{-30}$ gives the characters of the tricritical
Ising model in the same way as $G_2$.

Thus the H-twist limits of the line defect half-indices of the Wilson lines of the theories of $Spin(7)$, $G_2$ and $SU(2)$ with $\mathbf7$ give the
characters of the first two non-trivial members $M(3,4)$ and $M(4,5)$ of the series of unitary minimal models \cite{Belavin:1984vu,Friedan:1983xq}.
For $M(4,5)$ all the characters appear individually only after the gauging of the $\Zb_2$, which also adds the twisted sector.
The tricritical Ising model describes the tricritical point of the Ising model with vacancies \cite{Friedan:1983xq,Zamolodchikov:1986db}, and its
fermionization is $SM(3,5)$ \cite{Friedan:1984rv,Bershadsky:1985dq,Eichenherr:1985cx,Qiu:1986if,Hsieh:2020uwb}.
The dictionary between the Wilson lines and the modules is
\begin{align}
\label{voa_ising_dict}
\begin{array}{ll}
Spin(7)_{-6}: & 1\ \leftrightarrow\ (1,1),\quad W_{(1)}\ \leftrightarrow\ (1,3),\quad W_{(\frac12,\frac12,\frac12)}\ \leftrightarrow\ (1,2),\\[2pt]
(G_2)_{-5}: & 1\ \leftrightarrow\ (1,1),\ (1,4),\ (2,1),\quad W_{(1)}\ \leftrightarrow\ (1,2),\ (1,3),\ (2,2),
\end{array}
\end{align}
where for $G_2$ the modules $(1,s)$ come from the even and odd parts \eqref{other_gauge} and $(2,s)$ from the twisted sector.
For the Ising model every module is realized by a single Wilson line up to a monomial.
For both theories the commutant is the minimal model $M(3,4)$ or $SM(3,5)$ itself, and the H-twist limit of the half-index gives the (super)character of its vacuum module.

\subsection{\texorpdfstring{$SU(N)_{-N-1}$, $SO(N)_{-N}$ and $(E_6)_{-18}$}{SU(N), SO(N) and E6}}
\label{sec_other_lat}
\subsubsection{\texorpdfstring{$SU(N)$, $SO(N)$ and lattice VOAs}{SU(N), SO(N) and lattice VOAs}}
\label{sec_voa_lattice}
The representations $\mathbf N\oplus\overline{\mathbf N}$ of $SU(N)$, $2\times\mathbf N$ of $SO(N)$ and $\mathbf{27}\oplus\overline{\mathbf{27}}$ of $E_6$ are the
isotropy representations of compact symmetric spaces,\footnote{A compact symmetric space $G'/K$ is the quotient of a compact
Lie group $G'$ by a subgroup $K$ fixed by an involution of $G'$. The involution splits $\mathfrak g'=\mathfrak k\oplus\mathfrak p$ into
its eigenspaces with the eigenvalues $1$ and $-1$, where $\mathfrak k$, the isotropy algebra, is the Lie algebra of $K$ and $\mathfrak p$
is the tangent space at the point $K$. The isotropy representation is the representation of $K$ on $\mathfrak p$ \cite{MR514561}. In
\eqref{symm_lattice} $\mathfrak p$ is the representation of $Q$.}
\begin{align}
\label{symm_lattice}
\mathfrak{su}(N+1)&=\mathfrak{su}(N)\oplus\mathfrak u(1)\oplus\mathbf N\oplus\overline{\mathbf N}, & \mathfrak{so}(N+2)&=\mathfrak{so}(N)\oplus\mathfrak{so}(2)\oplus2\times\mathbf N,
\nonumber\\
\mathfrak e_7&=\mathfrak e_6\oplus\mathfrak u(1)\oplus\mathbf{27}\oplus\overline{\mathbf{27}},
\end{align}
whose isotropy algebras contain, besides $\mathfrak g$, the factor $\mathfrak u(1)$ or $\mathfrak{so}(2)$.
By the theorem of Goddard, Nahm and Olive \cite{Goddard:1985jp} the gauge currents and the current of this factor then form a
conformal embedding into $\widehat{\mathfrak{so}}(d)_1$.
Hence the commutant has central charge $1$ and is the lattice VOA of this current.
For $SU(N)_{-N-1}$ with $Q$ and $\Gamma$ in $\mathbf N\oplus\overline{\mathbf N}$ the commutant is $\widehat{\mathfrak u}(N)_1/\widehat{\mathfrak{su}}(N)_1$, the
lattice VOA $V_{\sqrt N\Zb}$ of section \ref{sec_twist_lat}, and its modules are given by the antisymmetric Wilson lines.

For every $N\ge2$ and $p=0,1,\dots,N-1$, at the H-twist point
\begin{align}
\label{sun_result}
\lin{W_{(1^p)}}{SU(N)_{-N-1}}{\Ncal,N,\Gamma}\Big|_{\rm H}=(-q^{\frac12})^p\,\frac{f\bigl((-1)^Nq^{\frac N2-p},(-1)^Nq^{\frac N2+p}\bigr)}{f(-q)}
=(-q^{\frac12})^{\min(p,N-p)}\chi^{V_{\sqrt N\Zb}}_p,
\end{align}
which also follows from \eqref{real_reduce} and which we prove directly in appendix \ref{app_proof} together with \eqref{linechi}.
With the (super)characters \eqref{lattice_chi}, the H-twist limits of the line defect half-indices of the antisymmetric Wilson lines give the
(super)characters of all the modules of $V_{\sqrt N\Zb}$, labeled by the $N$-ality $p$.
This parallels the $USp(2n)$ theories, in which the H-twist limits of the line defect half-indices of the antisymmetric Wilson lines give the characters
of the modules of $M(2,2n+3)$.
The cases $p=0,1$ were conjectured in \cite{Okazaki:2024paq} and proved, with a flavor fugacity, in \cite{Wang:2026kvh}.
For $N\ge4$ the lines $W_{(1^p)}$ with $2\le p\le N-2$ give characters which are not monomials times those for $p=0,1$, so
that the statement of \cite{Okazaki:2024kzo} that the other representations give, up to monomials, those for $p=0,1$ holds only for
$N\le3$.

For $SO(N)_{-N}$, $N\ge3$, with two copies of $Q$ and $\Gamma$ in the vector representation
\cite{Okazaki:2024kzo}, the commutant of $\widehat{\mathfrak{so}}(N)_2$ in $F^{2N}$ is $\widehat{\mathfrak{so}}(2)_N$, with central
charge $N-\frac{N(N-1)}N=1$, again $V_{\sqrt N\Zb}$.
For $N=3$ the level is written in the normalization of $SO(N)$, in which $SO(3)_K$ corresponds to $SU(2)_{2K}$.
With the discrete fugacity $\chi=\pm1$ of the charge conjugation $\Zb_2^{\mathcal C}$ \cite{Okazaki:2021pnc,Okazaki:2024kzo} (appendix \ref{app_char}) we find
\begin{align}
\label{intro_SO}
\lin{W_{(1^p)}}{SO(N)_{-N,\chi}}{\Ncal,N,\Gamma}\Big|_{\rm H}
=C_p\,\frac{f\bigl(\chi(-1)^Nq^{\frac N2-p},\chi(-1)^Nq^{\frac N2+p}\bigr)}{f(-q)},
\end{align}
with $C_0=1$ and $C_p=2(-q^{\frac12})^p$ for $1\le p\le\lfloor\frac N2\rfloor$, where for even $N$ the representation $(1^{N/2})$ is the sum of the self-dual and anti-self-dual
parts.
For $\chi=1$ and $1\le p\le\lfloor\frac N2\rfloor$ these are the sums of the (super)characters of the conjugate modules $p$ and $-p$,
which for even $N$ and $p=\frac N2$ are the same module,
\begin{align}
\lin{W_{(1^p)}}{SO(N)_{-N,1}}{\Ncal,N,\Gamma}\Big|_{\rm H}=(-q^{\frac12})^p\bigl(\chi^{V_{\sqrt N\Zb}}_p+\chi^{V_{\sqrt N\Zb}}_{-p}\bigr),
\end{align}
since $SO(N)$ has no $\Zb_N$ center.
For odd $N$ the element $-\mathbf 1_N$ of $O(N)$ has determinant $-1$, so that $O(N)=SO(N)\times\Zb_2$, and the $\Zb_2$ is the one of
section \ref{sec_other_setup} acting by $-1$ on $Q$ and $\Gamma$.
Hence $\chi=-1$ is the ${\rm H}^-$-twist point, up to the sign $(-1)^p$ of $W_{(1^p)}$.
The charge conjugation acts on $V_{\sqrt N\Zb}$ by $e^{j\alpha}\mapsto(-1)^je^{j\alpha}$, and gauging it, which gives the $O(N)$ theories, projects
onto the lattice VOA $V_{2\sqrt N\Zb}$.
The H-twist limits of the line defect half-indices then give the characters of the modules of $V_{2\sqrt N\Zb}$ with even labels, with conjugate modules summed in
pairs, while the modules with odd labels are not obtained from the Wilson lines.
The $SO(N)$ theory also has the magnetic symmetry $\Zb_2^{\mathcal M}$, which acts on the monopole operators, and the two $O(N)$ theories
$O(N)_\pm$ differ by a discrete theta angle \cite{Aharony:2013kma}.
The Neumann-type half-indices, with or without Wilson lines, have no sum over magnetic fluxes.
So they do not depend on the discrete fugacity of $\Zb_2^{\mathcal M}$, which enters the indices through the fluxes
\cite{Aharony:2013kma,Okazaki:2021pnc}.
Hence $O(N)_+$ and $O(N)_-$ give the same line defect half-indices.

\subsubsection{\texorpdfstring{$(E_6)_{-18}$ and the lattice VOA $V_{\sqrt{27}\Zb}$}{E6 and the lattice VOA}}
\label{sec_voa_exc}
For $(E_6)_{-18}$ with $Q$ and $\Gamma$ in $\mathbf{27}\oplus\overline{\mathbf{27}}$ we have $j_{\mathbf R}=6$, and the boundary fermions are $27$ complex fermions.
As discussed in section \ref{sec_voa_lattice}, the gauge currents and the $\mathfrak u(1)$ current of these fermions, under which each of
the $27$ complex fermions has charge $1$, form the conformal embedding $(\widehat{\mathfrak e}_6)_6\oplus\widehat{\mathfrak u}(1)\subset\widehat{\mathfrak{so}}(54)_1$.
Hence the commutant is the lattice VOA $V_{\sqrt{27}\Zb}$ of this $\mathfrak u(1)$ current, as for $SU(27)$, and \eqref{real_reduce} gives
\begin{align}
\label{Re_e6}
&\II^{(E_6)_{-18}}_{\Ncal,N,\Gamma}\Big|_{\rm H}=\chi^{V_{\sqrt{27}\Zb}}_0,\qquad
\lin{W_{\mathbf{27}}}{(E_6)_{-18}}{\Ncal,N,\Gamma}\Big|_{\rm H}=-q^{\frac12}\chi^{V_{\sqrt{27}\Zb}}_1,
\nonumber\\
&\lin{W_{\mathbf{351}}}{(E_6)_{-18}}{\Ncal,N,\Gamma}\Big|_{\rm H}=q\,\chi^{V_{\sqrt{27}\Zb}}_2,\qquad
\lin{W_{\mathbf{2925}}}{(E_6)_{-18}}{\Ncal,N,\Gamma}\Big|_{\rm H}=-2q^{\frac32}\chi^{V_{\sqrt{27}\Zb}}_3,
\end{align}
with the supercharacters \eqref{lattice_chi}, $\mathbf{351}=\Lambda^2\mathbf{27}$ and $\mathbf{2925}=\Lambda^3\mathbf{27}$.
The same holds for $\overline{\mathbf{27}}$ and $\overline{\mathbf{351}}$, and
$\lin{W_{\mathbf{78}}}{(E_6)_{-18}}{\Ncal,N,\Gamma}\big|_{\rm H}=0$.
The factor $2$ arises since $\Lambda^3\mathbf{27}$ is self-conjugate under $E_6$ and appears with the $\mathfrak u(1)$ charges $3$ and $-3$.
The H-twist limits of the line defect half-indices of the other Wilson lines are sums of the supercharacters of the modules of $V_{\sqrt{27}\Zb}$, with the
multiplicities of the branching rules of $(\widehat{\mathfrak e}_6)_6\subset\widehat{\mathfrak{su}}(27)_1$.

\FloatBarrier
\subsection*{Acknowledgments}
We would like to thank Douglas J. Smith for useful discussions.
YN is supported in part by JSPS KAKENHI Grant Numbers JP21K03581 and JP26K00699.
The work of TO was supported by the Startup Funding no.\ 4007012317 of Southeast University.

\appendix

\section{Proofs of the product formulae}
\label{app_proof}
In this appendix we prove the following two theorems and treat the two cases together.

\begin{theorem}
\label{thm_usp}
For every $n\ge1$ and $p=0,1,\dots,n$, the H-twist limit of the line defect half-index of $W_{(1^p)}$ of $USp(2n)_{-(n+\frac32)}$ with the
Neumann-type boundary condition (I) is given by \eqref{linechi}, i.e.\ it equals $(-q^{\frac12})^p\chi^{M(2,2n+3)}_{(1,n+1-p)}$.
\end{theorem}

\begin{theorem}
\label{thm_sun}
For every $N\ge2$ and $p=0,1,\dots,N-1$, the H-twist limit of the line defect half-index of $W_{(1^p)}$ of $SU(N)_{-N-1}$ with $Q$ and
$\Gamma$ in $\mathbf N\oplus\overline{\mathbf N}$ and the Neumann-type boundary condition is given by \eqref{sun_result}, i.e.\ it equals
$(-q^{\frac12})^{\min(p,N-p)}\chi^{V_{\sqrt N\Zb}}_p$.
\end{theorem}

For $p=0,1$ theorem \ref{thm_sun} was proved, with a flavor fugacity, in \cite{Wang:2026kvh}.
In the first case $G=USp(2n)$ at the level $-(n+\frac32)$ with $\mathbf R=\mathbf{2n}$.
We use the coordinates $s_1,\dots,s_n$ of the maximal torus, the roots $\Delta=\{\pm \mathbf e_i\pm \mathbf e_j\ (i<j),\ \pm2\mathbf e_i\}$, the positive roots
$\Delta_+=\{\mathbf e_i\pm \mathbf e_j\ (i<j),\ 2\mathbf e_i\}$ and $\rho=(n,n-1,\dots,1)$.
In the second case $G=SU(N)$ at the level $-(N+1)$ with $\mathbf R=\mathbf N\oplus\overline{\mathbf N}$.
We use $s_1,\dots,s_N$ with $s_1\cdots s_N=1$, the roots $\Delta=\{\mathbf e_i-\mathbf e_j\ (i\neq j)\}$, the positive roots $\Delta_+=\{\mathbf e_i-\mathbf e_j\ (i<j)\}$ and
$\rho=(\frac{N-1}2,\frac{N-3}2,\dots,\frac{1-N}2)$, and the weights $\gamma\in\Zb^N$ and $\gamma+m\mathbf 1$, $m\in\Zb$, with $\mathbf 1=(1,\dots,1)$, give
the same monomial $s^\gamma=\prod_is_i^{\gamma_i}$.
In both cases $r$ is the rank and $W$ the Weyl group, and $\alpha\cdot\gamma$ and $\|\gamma\|^2$ are the Euclidean scalar product and norm in
these coordinates, with $|\gamma|=\sum_i\gamma_i$.
The Wilson line $W_\lambda$ is labeled by a Young diagram $\lambda$ with at most $n$, resp.\ $N-1$, rows, and $\Lambda=\lambda+\rho$.
For $SU(N)$ with even $N$ the entries of $\rho$ are half-integers, and $s^\rho$ is not a monomial in the $s_i$.
It enters below only through the quotient $\chi_\lambda=A_\Lambda/A_\rho$ and the products $A_\Lambda(s)A_\rho(s^{-1})$ and
$s^\Lambda A_\rho(s^{-1})$, which are Laurent polynomials in the $s_i$.
The weights of $\mathbf R$ are the $\pm \mathbf e_i$, so that the H-twist limit \eqref{other_int} is
\begin{align}
\label{app_int}
\lin{W_\lambda}{G}{\Ncal,N,\Gamma}\Big|_{\rm H}=\frac{(q)_\infty^r}{|W|}\oint\prod_{i=1}^r\frac{ds_i}{2\pi is_i}\prod_{\alpha\in\Delta}\poch{s^\alpha}
\prod_i\theta(q^{\frac12}s_i;q)\,\chi_\lambda(s),
\end{align}
where the last product runs over $i=1,\dots,n$, resp.\ $i=1,\dots,N$.
For $USp(2n)$ this is \eqref{n1_H_int}.

\paragraph{Two identities}
We use the Jacobi triple product \cite{MR2128719}
\begin{align}
\label{app_jtp}
(q)_\infty\,\theta(q^{\frac12}z;q)=\sum_{b\in\Zb}(-1)^bq^{\frac{b^2}2}z^b,
\end{align}
and the Macdonald identity for the affine root system of $G$ \cite{MR357528,MR1104219}.
For a nome $\mathfrak q$ with $|\mathfrak q|<1$ let
\begin{align}
\label{app_Ddef}
D(s;\mathfrak q)=(\mathfrak q;\mathfrak q)_\infty^r\prod_{\alpha\in\Delta_+}(1-s^{-\alpha})\prod_{j\ge1}\prod_{\alpha\in\Delta}(1-\mathfrak q^js^\alpha)
\end{align}
be the affine Weyl denominator.
The Macdonald identity is
\begin{align}
\label{app_macdonald}
D(s;\mathfrak q)=\sum_{w\in W}\sum_{t\in L}{\rm sgn}(w)\,\mathfrak q^{\frac{\|v\|^2-\|\rho\|^2}{2g}}\,s^{w(v)-\rho},\qquad v=\rho+gt,
\end{align}
with $g=2n+2$ and $L=\Zb^n$ for $USp(2n)$, and $g=N$ and $L=\{t\in\Zb^N\,|\,|t|=0\}$ for $SU(N)$.
Here $g$ is $2h^\vee$ for $USp(2n)$ and $h^\vee$ for $SU(N)$, since the long roots have the length squared $4$ and $2$ in these coordinates.
At $s=q^{-\Lambda}$, i.e.\ $s_i=q^{-\Lambda_i}$, the factors of \eqref{app_Ddef} combine into
\begin{align}
\label{app_Dprod}
D(q^{-\Lambda};\mathfrak q)=(\mathfrak q;\mathfrak q)_\infty^r\prod_{\alpha\in\Delta_+}(q^{\alpha\cdot\Lambda};\mathfrak q)_\infty
(\mathfrak q\,q^{-\alpha\cdot\Lambda};\mathfrak q)_\infty .
\end{align}

\begin{proposition}
\label{prop_app}
For every Young diagram $\lambda$ with at most $n$, resp.\ $N-1$, rows,
\begin{align}
\label{app_general}
\lin{W_\lambda}{USp(2n)_{-(n+\frac32)}}{\Ncal,N,\Gamma}\Big|_{\rm H}&=(-1)^{|\lambda|}q^{\frac{\|\lambda\|^2}2}\,
\frac{D(q^{-\Lambda};q^{2n+3})}{(q)_\infty^n},
\nonumber\\
\lin{W_\lambda}{SU(N)_{-N-1}}{\Ncal,N,\Gamma}\Big|_{\rm H}&=(-1)^{|\lambda|}q^{\frac{\|\lambda\|^2}2}\,
\frac{D(q^{-\Lambda};q^{N+1})}{(q)_\infty^N}\,f\bigl((-1)^Nq^{\frac N2+|\lambda|},(-1)^Nq^{\frac N2-|\lambda|}\bigr).
\end{align}
\end{proposition}

\begin{proof}
The integrand of \eqref{app_int} is holomorphic on $(\Cb^\times)^r$, and the integral is the constant term of its Laurent expansion in $s$,
which we denote by ${\rm CT}_s$.
By the Weyl character formula $\chi_\lambda=A_\Lambda/A_\rho$, with the alternants $A_\gamma(s)=\sum_{w\in W}{\rm sgn}(w)s^{w(\gamma)}$, and the Weyl
denominator formula $A_\rho(s)=s^\rho\prod_{\alpha\in\Delta_+}(1-s^{-\alpha})$ \cite{MR1153249},
\begin{align}
\prod_{\alpha\in\Delta}(1-s^\alpha)\,\chi_\lambda(s)=A_\Lambda(s)\,A_\rho(s^{-1}).
\end{align}
The other factors of the integrand are invariant under $W$, and $A_\rho(s^{-1})$ changes by ${\rm sgn}(w)$ under $s\to w(s)$.
So the substitution $s\to w^{-1}(s)$ in the term of $w$ in $A_\Lambda(s)$ turns it into the term of the identity, and the $|W|$ terms are equal.
With $s^\Lambda A_\rho(s^{-1})=s^\lambda\prod_{\alpha\in\Delta_+}(1-s^\alpha)$, \eqref{app_Ddef} and the substitution $s\to s^{-1}$, under which
$\theta(q^{\frac12}s_i;q)$ is invariant, we obtain
\begin{align}
\label{app_CT}
\lin{W_\lambda}{G}{\Ncal,N,\Gamma}\Big|_{\rm H}={\rm CT}_s\Bigl[s^{-\lambda}D(s;q)\prod_i\theta(q^{\frac12}s_i;q)\Bigr].
\end{align}

We insert \eqref{app_macdonald} with $\mathfrak q=q$ and \eqref{app_jtp} for every $s_i$.
The series converge absolutely and uniformly on the torus $|s_i|=1$, so that the constant term is the sum over the pairs of terms whose monomials
cancel.
For $USp(2n)$ the term $(w,t)$ of \eqref{app_macdonald} pairs with the term $b=\Lambda-w(v)$ of the product of the series \eqref{app_jtp}, and for
$SU(N)$ with the terms $b=\Lambda-w(v)+m\mathbf 1$, $m\in\Zb$.
The sign of this term is $(-1)^{|b|}=(-1)^{|\lambda|}$ for $USp(2n)$, since the sign changes of the coordinates change $|v|$ by even numbers and
$|v|=|\rho|+g|t|$ with $g$ even, and $(-1)^{|b|}=(-1)^{|\lambda|+Nm}$ for $SU(N)$, since $|\rho|=|v|=0$.
Since $\|w(v)\|=\|v\|$ and $\Lambda=\lambda+\rho$, a direct computation gives
\begin{align}
\label{app_square}
\frac{\|v\|^2-\|\rho\|^2}{2g}+\frac12\|\Lambda-w(v)\|^2=\frac12\|\lambda\|^2+(g+1)\,\frac{\|v\|^2-\|\rho\|^2}{2g}-\Lambda\cdot\bigl(w(v)-\rho\bigr),
\end{align}
and the last two terms are the exponent of $q$ in the term $(w,t)$ of \eqref{app_macdonald} with $\mathfrak q=q^{g+1}$ at $s=q^{-\Lambda}$.
For $SU(N)$ moreover $\|\Lambda-w(v)+m\mathbf 1\|^2=\|\Lambda-w(v)\|^2+2|\lambda|m+Nm^2$, so that the sum over $m$ factors out, and by
\eqref{ramanujan_f} it is
\begin{align}
\sum_{m\in\Zb}(-1)^{Nm}q^{\frac N2m^2+|\lambda|m}=f\bigl((-1)^Nq^{\frac N2+|\lambda|},(-1)^Nq^{\frac N2-|\lambda|}\bigr).
\end{align}
With $g+1=2n+3$ and $g+1=N+1$, and with the factors $1/(q)_\infty$ of \eqref{app_jtp}, this gives \eqref{app_general}.
\end{proof}

\begin{proof}[Proof of theorems \ref{thm_usp} and \ref{thm_sun}]
For $\lambda=(1^p)$ we have $|\lambda|=\|\lambda\|^2=p$, and it remains to evaluate \eqref{app_Dprod} at $\Lambda=\rho+(1^p)$.
Here $g+1=2n+3$ for $USp(2n)$ and $g+1=N+1$ for $SU(N)$.
For $\alpha\in\Delta_+$ we have $1\le\alpha\cdot\Lambda\le g$, and the negative roots give the residues $g+1-\alpha\cdot\Lambda$ modulo $g+1$, so that
\eqref{app_Dprod} is
\begin{align}
\label{app_Nm}
D(q^{-\Lambda};q^{g+1})=(q^{g+1};q^{g+1})_\infty^r\prod_{m=1}^{g}(q^m;q^{g+1})_\infty^{\nu_m},
\end{align}
where $\nu_m$ is the number of the roots $\alpha\in\Delta$ with $\alpha\cdot\Lambda\equiv m$ modulo $g+1$.
We count them with the polynomial $P(z)=\sum_{\alpha\in\Delta}z^{\alpha\cdot\Lambda}$ in $\Zb[z]/(z^{g+1}-1)$, in which $\Sigma=\sum_{m=0}^{g}z^m$
satisfies $z\Sigma=\Sigma$ and $\Sigma^2=(g+1)\Sigma$.

For $USp(2n)$ the entries of $\Lambda$ are the numbers $1,2,\dots,n+1$ except $n+1-p$, so that the $\pm\Lambda_i$ give all the non-zero residues
except $\pm(n+1-p)$, i.e.\ $X(z)=\sum_i(z^{\Lambda_i}+z^{-\Lambda_i})=\Sigma-1-z^{n+1-p}-z^{-(n+1-p)}$.
The roots $\pm \mathbf e_i\pm \mathbf e_j$ and $\pm2\mathbf e_i$ give
\begin{align}
P(z)=\frac12\bigl(X(z)^2+X(z^2)\bigr)-n=(n-1)(\Sigma-1)+z^{n+1-p}+z^{-(n+1-p)},
\end{align}
where we have used $\Sigma(z^2)=\Sigma$, since $g+1$ is odd.
So $\nu_m=n-1$ except $\nu_m=n$ for $m=n+1-p$ and $m=n+2+p$, and with $\prod_{m=1}^{g+1}(q^m;q^{g+1})_\infty=(q)_\infty$ and \eqref{ramanujan_f} we obtain
$D(q^{-\Lambda};q^{2n+3})=(q)_\infty^{n-1}f(-q^{n+1-p},-q^{n+2+p})$, which with \eqref{app_general} is \eqref{linechi}.

For $SU(N)$ the entries of $\Lambda-\Lambda_N\mathbf 1$ are the numbers $0,1,\dots,N$ except $N-p$, so that $X(z)=\sum_iz^{\Lambda_i-\Lambda_N}=\Sigma-z^{N-p}$
and
\begin{align}
P(z)=X(z)X(z^{-1})-N=(N-1)(\Sigma-1).
\end{align}
So $\nu_m=N-1$ for all $m$, and $D(q^{-\Lambda};q^{N+1})=(q)_\infty^{N-1}$ for all $p$.
With \eqref{app_general} and the symmetry $f(a,b)=f(b,a)$ this gives the first equality of \eqref{sun_result}.
By \eqref{ramanujan_f}, $f\bigl((-1)^Nq^{\frac N2-p},(-1)^Nq^{\frac N2+p}\bigr)/f(-q)$ is the series \eqref{lattice_chi}, which is
normalized for $p\le\frac N2$.
For $p>\frac N2$ a shift of the summation variable gives
\begin{align}
(-q^{\frac12})^pf\bigl((-1)^Nq^{\frac N2-p},(-1)^Nq^{\frac N2+p}\bigr)=(-q^{\frac12})^{N-p}f\bigl((-1)^Nq^{p-\frac N2},(-1)^Nq^{\frac{3N}2-p}\bigr),
\end{align}
whose right-hand side is $(-q^{\frac12})^{N-p}f(-q)$ times the series \eqref{lattice_chi} for the label $N-p<\frac N2$, i.e.\ the
normalized character $\chi^{V_{\sqrt N\Zb}}_{N-p}=\chi^{V_{\sqrt N\Zb}}_{-p}=\chi^{V_{\sqrt N\Zb}}_p$.
This gives the second equality.
\end{proof}

\paragraph{Remark}
By \eqref{app_Dprod}, $D(q^{-\Lambda};\mathfrak q)$ is $(\mathfrak q;\mathfrak q)_\infty^r$ times the product of the theta functions
$\theta(q^{\alpha\cdot\Lambda};\mathfrak q)$, $\alpha\in\Delta_+$.
For $USp(2n)$ and $\mathfrak q=q^{2n+3}$ it vanishes when $\alpha\cdot\Lambda\in(2n+3)\Zb$ for a root $\alpha$, and by the quasi-periodicity of
$\theta$ it changes by a sign and a power of $q$ under the affine Weyl group $W\ltimes(2n+3)\Zb^n$.
Since $\sum_{\alpha\in\Delta_+}(\alpha\cdot\Lambda)^2=(2n+2)\|\Lambda\|^2$, the right-hand side of \eqref{app_general} for $USp(2n)$, as a
function of $\Lambda$, then changes by a sign and by $q^{(\|w(\Lambda)+\beta\|^2-\|\Lambda\|^2)/(4n+6)}$ under $\Lambda\to w(\Lambda)+\beta$ with
$w\in W$ and $\beta\in(2n+3)\Zb^n$.
Every integral $\Lambda$ with $\alpha\cdot\Lambda\notin(2n+3)\Zb$ for all the roots $\alpha$ is mapped by a unique element of
$W\ltimes(2n+3)\Zb^n$ into the open alcove $n+\frac32>\Lambda_1>\dots>\Lambda_n>0$, whose integral points are $\Lambda=\rho+(1^p)$,
$p=0,\dots,n$.
These properties give the reduction of the H-twist limits of the line defect half-indices of all the Wilson lines to those of the
antisymmetric ones (section \ref{sec_n1_H}).

\section{\texorpdfstring{$q$}{q}-series expansions}
\label{app_qseries}
For small $(n,\ell)$ we list the $q$-series expansions of the half-indices and the line defect half-indices in appendix
\ref{app_qseries_N}, and of the full indices and the line defect indices in appendix \ref{app_qseries_bulk}.
In each entry the left column gives the identity of the two dual sides and the right column their common expansion.
We set $r_Q=\frac12$ and all the flavor fugacities to one, and keep the fugacity $a$ of $U(1)_a$.

\subsection{Line defect half-indices}
\label{app_qseries_N}
We take the boundary conditions (I) of $USp(2n)_{-k}$, $k=n+\ell+\frac12$, with $\ell$ Fermi multiplets and $(n,\ell)=(1,1)$, $(2,1)$,
$(1,2)$ and $(2,2)$.
Their half-indices and the line defect half-indices of all the Wilson lines $W_\lambda$ with $\lambda\subseteq(\ell^n)$ are as follows,
where the Abelian sides are evaluated with \eqref{gen_I_id}.
By \eqref{GKlines} the same expansions give, up to monomials, the line defect half-indices of the dual lines of $USp(2\ell)_k$ at the
fugacity $a^{-1}$.
By \eqref{squarerel} they also give, for $\lambda=(1^p)$, those of the dual combinations of Wilson lines of $\Tb_{n,\ell}$, i.e.\ of the
boundary conditions (II) for $(\ell,n)$.
Setting $a=q^{\frac14}$ gives the H-twist limits, which are the C-twist limits of the boundary conditions (II) for $(\ell,n)$.

\begin{longtable}{c|>{\raggedright\arraybackslash}p{0.47\textwidth}}
\multicolumn{2}{c}{$(n,\ell)=(1,1)$}\\[4pt]
\hline
line defect half-index & expansion\\
\hline
\endfirsthead
\multicolumn{2}{c}{$(n,\ell)=(1,1)$ (continued)}\\[4pt]
\hline
line defect half-index & expansion\\
\hline
\endhead
\hline
\endfoot
\begin{tabular}[t]{@{}c@{}}$\II^{USp(2)_{-5/2}}_{\Ncal,N,\Gamma}(a)$\\$=\II^{\mathcal{T}_{1,1}}_{\Dcal,D}$\end{tabular} & $1-2aq^{\frac{3}{4}}+3q+a^{2}q^{\frac{3}{2}}-4aq^{\frac{7}{4}}+4q^{2}+4a^{2}q^{\frac{5}{2}}-10aq^{\frac{11}{4}}+7q^{3}+\mathcal{O}(q^{\frac{13}{4}})$\\
\hline
\begin{tabular}[t]{@{}c@{}}$\lin{W_{\yf}}{USp(2)_{-5/2}}{\Ncal,N,\Gamma}(a)$\\$=-q^{\frac{3}{4}}a^{-1}\lin{W_{-1}}{\mathcal{T}_{1,1}}{\Dcal,D}$\end{tabular} & $aq^{\frac{1}{4}}-2q^{\frac{1}{2}}-2a^{2}q+4aq^{\frac{5}{4}}-2q^{\frac{3}{2}}+a^{3}q^{\frac{7}{4}}-4a^{2}q^{2}+8aq^{\frac{9}{4}}-6q^{\frac{5}{2}}+4a^{3}q^{\frac{11}{4}}-12a^{2}q^{3}+\mathcal{O}(q^{\frac{13}{4}})$\\
\end{longtable}
\addtocounter{table}{-1}

\begin{longtable}{c|>{\raggedright\arraybackslash}p{0.47\textwidth}}
\multicolumn{2}{c}{$(n,\ell)=(2,1)$}\\[4pt]
\hline
line defect half-index & expansion\\
\hline
\endfirsthead
\multicolumn{2}{c}{$(n,\ell)=(2,1)$ (continued)}\\[4pt]
\hline
line defect half-index & expansion\\
\hline
\endhead
\hline
\endfoot
\begin{tabular}[t]{@{}c@{}}$\II^{USp(4)_{-7/2}}_{\Ncal,N,\Gamma}(a)$\\$=\II^{\mathcal{T}_{1,2}}_{\Dcal,D}$\end{tabular} & $1-2aq^{\frac{3}{4}}+3q+a^{2}q^{\frac{3}{2}}-8aq^{\frac{7}{4}}+9q^{2}+7a^{2}q^{\frac{5}{2}}-20aq^{\frac{11}{4}}+15q^{3}+\mathcal{O}(q^{\frac{13}{4}})$\\
\hline
\begin{tabular}[t]{@{}c@{}}$\lin{W_{\yf}}{USp(4)_{-7/2}}{\Ncal,N,\Gamma}(a)$\\$=-q^{\frac{3}{4}}a^{-1}\lin{W_{0,-1}}{\mathcal{T}_{1,2}}{\Dcal,D}$\end{tabular} & $aq^{\frac{1}{4}}-2q^{\frac{1}{2}}-2a^{2}q+7aq^{\frac{5}{4}}-6q^{\frac{3}{2}}+a^{3}q^{\frac{7}{4}}-10a^{2}q^{2}+20aq^{\frac{9}{4}}-12q^{\frac{5}{2}}+7a^{3}q^{\frac{11}{4}}-30a^{2}q^{3}+\mathcal{O}(q^{\frac{13}{4}})$\\
\hline
\begin{tabular}[t]{@{}c@{}}$\lin{W_{\yaa}}{USp(4)_{-7/2}}{\Ncal,N,\Gamma}(a)$\\$=q^{\frac{3}{2}}a^{-2}\lin{W_{-1,-2}}{\mathcal{T}_{1,2}}{\Dcal,D}$\end{tabular} & $-2aq^{\frac{3}{4}}+3q+4a^{2}q^{\frac{3}{2}}-8aq^{\frac{7}{4}}+4q^{2}-2a^{3}q^{\frac{9}{4}}+11a^{2}q^{\frac{5}{2}}-20aq^{\frac{11}{4}}+12q^{3}+\mathcal{O}(q^{\frac{13}{4}})$\\
\end{longtable}
\addtocounter{table}{-1}

\begin{longtable}{c|>{\raggedright\arraybackslash}p{0.47\textwidth}}
\multicolumn{2}{c}{$(n,\ell)=(1,2)$}\\[4pt]
\hline
line defect half-index & expansion\\
\hline
\endfirsthead
\multicolumn{2}{c}{$(n,\ell)=(1,2)$ (continued)}\\[4pt]
\hline
line defect half-index & expansion\\
\hline
\endhead
\hline
\endfoot
\begin{tabular}[t]{@{}c@{}}$\II^{USp(2)_{-7/2}}_{\Ncal,N,2\Gamma}(a)$\\$=\II^{\mathcal{T}_{2,1}}_{\Dcal,D}$\end{tabular} & $1-4aq^{\frac{3}{4}}+10q+6a^{2}q^{\frac{3}{2}}-24aq^{\frac{7}{4}}+30q^{2}-4a^{3}q^{\frac{9}{4}}+31a^{2}q^{\frac{5}{2}}-84aq^{\frac{11}{4}}+a^{4}q^{3}+85q^{3}+\mathcal{O}(q^{\frac{13}{4}})$\\
\hline
\begin{tabular}[t]{@{}c@{}}$\lin{W_{\yf}}{USp(2)_{-7/2}}{\Ncal,N,2\Gamma}(a)$\\$=-q^{\frac{3}{4}}a^{-1}\lin{W_{-1,1,0}}{\mathcal{T}_{2,1}}{\Dcal,D}$\end{tabular} & $aq^{\frac{1}{4}}-4q^{\frac{1}{2}}-4a^{2}q+16aq^{\frac{5}{4}}-20q^{\frac{3}{2}}+6a^{3}q^{\frac{7}{4}}-28a^{2}q^{2}+61aq^{\frac{9}{4}}-4a^{4}q^{\frac{5}{2}}-60q^{\frac{5}{2}}+32a^{3}q^{\frac{11}{4}}-112a^{2}q^{3}+\mathcal{O}(q^{\frac{13}{4}})$\\
\hline
\begin{tabular}[t]{@{}c@{}}$\lin{W_{\yb}}{USp(2)_{-7/2}}{\Ncal,N,2\Gamma}(a)$\\$=q^{\frac{3}{2}}a^{-2}\lin{W_{-1,1,-1}}{\mathcal{T}_{2,1}}{\Dcal,D}$\end{tabular} & $a^{2}q^{\frac{1}{2}}-4aq^{\frac{3}{4}}+5q-4a^{3}q^{\frac{5}{4}}+16a^{2}q^{\frac{3}{2}}-24aq^{\frac{7}{4}}+6a^{4}q^{2}+15q^{2}-28a^{3}q^{\frac{9}{4}}+62a^{2}q^{\frac{5}{2}}-4a^{5}q^{\frac{11}{4}}-84aq^{\frac{11}{4}}+32a^{4}q^{3}+56q^{3}+\mathcal{O}(q^{\frac{13}{4}})$\\
\end{longtable}
\addtocounter{table}{-1}

\begin{longtable}{c|>{\raggedright\arraybackslash}p{0.47\textwidth}}
\multicolumn{2}{c}{$(n,\ell)=(2,2)$}\\[4pt]
\hline
line defect half-index & expansion\\
\hline
\endfirsthead
\multicolumn{2}{c}{$(n,\ell)=(2,2)$ (continued)}\\[4pt]
\hline
line defect half-index & expansion\\
\hline
\endhead
\hline
\endfoot
\begin{tabular}[t]{@{}c@{}}$\II^{USp(4)_{-9/2}}_{\Ncal,N,2\Gamma}(a)$\\$=\II^{\mathcal{T}_{2,2}}_{\Dcal,D}$\end{tabular} & $1-4aq^{\frac{3}{4}}+10q+6a^{2}q^{\frac{3}{2}}-44aq^{\frac{7}{4}}+65q^{2}-4a^{3}q^{\frac{9}{4}}+76a^{2}q^{\frac{5}{2}}-248aq^{\frac{11}{4}}+a^{4}q^{3}+246q^{3}+\mathcal{O}(q^{\frac{13}{4}})$\\
\hline
\begin{tabular}[t]{@{}c@{}}$\lin{W_{\yf}}{USp(4)_{-9/2}}{\Ncal,N,2\Gamma}(a)$\\$=-q^{\frac14}a\Big(\lin{W_{\Qm{1&0&0\\1&0&0}}}{\mathcal{T}_{2,2}}{\Dcal,D}-\II^{\mathcal{T}_{2,2}}_{\Dcal,D}\Big)$\end{tabular} & $aq^{\frac{1}{4}}-4q^{\frac{1}{2}}-4a^{2}q+26aq^{\frac{5}{4}}-40q^{\frac{3}{2}}+6a^{3}q^{\frac{7}{4}}-68a^{2}q^{2}+206aq^{\frac{9}{4}}-4a^{4}q^{\frac{5}{2}}-204q^{\frac{5}{2}}+92a^{3}q^{\frac{11}{4}}-468a^{2}q^{3}+\mathcal{O}(q^{\frac{13}{4}})$\\
\hline
\begin{tabular}[t]{@{}c@{}}$\lin{W_{\yaa}}{USp(4)_{-9/2}}{\Ncal,N,2\Gamma}(a)$\\$=q^{\frac{3}{2}}a^{-2}\lin{W_{\Qm{-1&1&0\\-2&2&0}}}{\mathcal{T}_{2,2}}{\Dcal,D}$\end{tabular} & $-4aq^{\frac{3}{4}}+10q+16a^{2}q^{\frac{3}{2}}-60aq^{\frac{7}{4}}+65q^{2}-24a^{3}q^{\frac{9}{4}}+156a^{2}q^{\frac{5}{2}}-360aq^{\frac{11}{4}}+16a^{4}q^{3}+300q^{3}+\mathcal{O}(q^{\frac{13}{4}})$\\
\hline
\begin{tabular}[t]{@{}c@{}}$\lin{W_{\yb}}{USp(4)_{-9/2}}{\Ncal,N,2\Gamma}(a)$\\$=q\,\lin{W_{\Qm{0&-1&0\\0&-1&0}}}{\mathcal{T}_{2,2}}{\Dcal,D}$\\$+q^{\frac32}a^{-2}\lin{W_{\Qm{0&0&0\\-1&1&-1}}}{\mathcal{T}_{2,2}}{\Dcal,D}$\end{tabular} & $a^{2}q^{\frac{1}{2}}-4aq^{\frac{3}{4}}+5q-4a^{3}q^{\frac{5}{4}}+26a^{2}q^{\frac{3}{2}}-60aq^{\frac{7}{4}}+6a^{4}q^{2}+50q^{2}-68a^{3}q^{\frac{9}{4}}+236a^{2}q^{\frac{5}{2}}-4a^{5}q^{\frac{11}{4}}-360aq^{\frac{11}{4}}+92a^{4}q^{3}+220q^{3}+\mathcal{O}(q^{\frac{13}{4}})$\\
\hline
\begin{tabular}[t]{@{}c@{}}$\lin{W_{\yba}}{USp(4)_{-9/2}}{\Ncal,N,2\Gamma}(a)$\\$=-q^{\frac{9}{4}}a^{-3}\lin{W_{\Qm{-1&1&0\\-2&2&-1}}}{\mathcal{T}_{2,2}}{\Dcal,D}$\end{tabular} & $-4a^{2}q+15aq^{\frac{5}{4}}-16q^{\frac{3}{2}}+16a^{3}q^{\frac{7}{4}}-80a^{2}q^{2}+144aq^{\frac{9}{4}}-24a^{4}q^{\frac{5}{2}}-96q^{\frac{5}{2}}+186a^{3}q^{\frac{11}{4}}-536a^{2}q^{3}+\mathcal{O}(q^{\frac{13}{4}})$\\
\hline
\begin{tabular}[t]{@{}c@{}}$\lin{W_{\ybbd}}{USp(4)_{-9/2}}{\Ncal,N,2\Gamma}(a)$\\$=q^{3}a^{-4}\lin{W_{\Qm{-1&1&-1\\-2&2&-2}}}{\mathcal{T}_{2,2}}{\Dcal,D}$\end{tabular} & $5a^{2}q^{\frac{3}{2}}-16aq^{\frac{7}{4}}+14q^{2}-20a^{3}q^{\frac{9}{4}}+79a^{2}q^{\frac{5}{2}}-112aq^{\frac{11}{4}}+30a^{4}q^{3}+59q^{3}+\mathcal{O}(q^{\frac{13}{4}})$\\
\end{longtable}
\addtocounter{table}{-1}

\subsection{Bulk line defect indices}
\label{app_qseries_bulk}
For $\ell=1$ and $n=1,2$ the full indices and the line defect indices \eqref{bulk_id} of the Wilson lines $W_{(1^p)}$ at the two
poles are as follows, where the Abelian sides are evaluated at $y=a^2$.
At $r_Q=\frac12$ all the chiral multiplets of $\mathcal T_{1,n}$ have vanishing R-charge.
By \eqref{pole_flip} and \eqref{pole_flip_Ab} the line defect indices at the north pole are also those at the south pole of $USp(2n)_k$
and $\Tb_{1,n}$, which correspond to the boundary conditions (II).

\begin{longtable}{c|>{\raggedright\arraybackslash}p{0.47\textwidth}}
\multicolumn{2}{c}{$(n,\ell)=(1,1)$}\\[4pt]
\hline
line defect index & expansion\\
\hline
\endfirsthead
\multicolumn{2}{c}{$(n,\ell)=(1,1)$ (continued)}\\[4pt]
\hline
line defect index & expansion\\
\hline
\endhead
\hline
\endfoot
\begin{tabular}[t]{@{}c@{}}$\Ical^{USp(2)_{-5/2}}(a)$\\$=\Ical^{\mathcal{T}_{1,1}}$\end{tabular} & $1-q+a^{2}q^{\frac{3}{2}}+a^{-2}q^{\frac{3}{2}}-2q^{2}+a^{2}q^{\frac{5}{2}}+a^{-2}q^{\frac{5}{2}}-2q^{3}+\mathcal{O}(q^{\frac{13}{4}})$\\
\hline
\begin{tabular}[t]{@{}c@{}}$\Ical^{USp(2)_{-5/2}}[W^S_{\yf}](a)$\\$=-q^{\frac{3}{4}}a^{-1}\,\Ical^{\mathcal{T}_{1,1}}[W^S_{-1}]$\end{tabular} & $aq^{\frac{1}{4}}-a^{-1}q^{\frac{3}{4}}+a^{3}q^{\frac{7}{4}}-aq^{\frac{9}{4}}+a^{3}q^{\frac{11}{4}}+a^{-1}q^{\frac{11}{4}}+\mathcal{O}(q^{\frac{13}{4}})$\\
\hline
\begin{tabular}[t]{@{}c@{}}$\Ical^{USp(2)_{-5/2}}[W^N_{\yf}](a)$\\$=-q^{-\frac{3}{4}}a\,\Ical^{\mathcal{T}_{1,1}}[W^N_{1}]$\end{tabular} & $aq^{\frac{1}{4}}-a^{-1}q^{\frac{3}{4}}+aq^{\frac{5}{4}}-a^{-1}q^{\frac{7}{4}}+a^{3}q^{\frac{11}{4}}+\mathcal{O}(q^{\frac{13}{4}})$\\
\end{longtable}
\addtocounter{table}{-1}

\begin{longtable}{c|>{\raggedright\arraybackslash}p{0.47\textwidth}}
\multicolumn{2}{c}{$(n,\ell)=(2,1)$}\\[4pt]
\hline
line defect index & expansion\\
\hline
\endfirsthead
\multicolumn{2}{c}{$(n,\ell)=(2,1)$ (continued)}\\[4pt]
\hline
line defect index & expansion\\
\hline
\endhead
\hline
\endfoot
\begin{tabular}[t]{@{}c@{}}$\Ical^{USp(4)_{-7/2}}(a)$\\$=\Ical^{\mathcal{T}_{1,2}}$\end{tabular} & $1-q+a^{2}q^{\frac{3}{2}}+a^{-2}q^{\frac{3}{2}}-2q^{2}+a^{2}q^{\frac{5}{2}}-q^{3}+a^{-4}q^{3}+\mathcal{O}(q^{\frac{13}{4}})$\\
\hline
\begin{tabular}[t]{@{}c@{}}$\Ical^{USp(4)_{-7/2}}[W^S_{\yf}](a)$\\$=-q^{\frac{3}{4}}a^{-1}\,\Ical^{\mathcal{T}_{1,2}}[W^S_{0,-1}]$\end{tabular} & $aq^{\frac{1}{4}}-a^{-1}q^{\frac{3}{4}}+a^{3}q^{\frac{7}{4}}+a^{-1}q^{\frac{7}{4}}-2aq^{\frac{9}{4}}-a^{-3}q^{\frac{9}{4}}+a^{3}q^{\frac{11}{4}}+2a^{-1}q^{\frac{11}{4}}+\mathcal{O}(q^{\frac{13}{4}})$\\
\hline
\begin{tabular}[t]{@{}c@{}}$\Ical^{USp(4)_{-7/2}}[W^N_{\yf}](a)$\\$=-q^{-\frac{3}{4}}a\,\Ical^{\mathcal{T}_{1,2}}[W^N_{0,1}]$\end{tabular} & $aq^{\frac{1}{4}}-a^{-1}q^{\frac{3}{4}}+aq^{\frac{5}{4}}-aq^{\frac{9}{4}}-a^{-3}q^{\frac{9}{4}}+a^{3}q^{\frac{11}{4}}+2a^{-1}q^{\frac{11}{4}}+\mathcal{O}(q^{\frac{13}{4}})$\\
\hline
\begin{tabular}[t]{@{}c@{}}$\Ical^{USp(4)_{-7/2}}[W^S_{\yaa}](a)$\\$=q^{\frac{3}{2}}a^{-2}\,\Ical^{\mathcal{T}_{1,2}}[W^S_{-1,-2}]$\end{tabular} & $-q+a^{2}q^{\frac{3}{2}}+a^{-2}q^{\frac{3}{2}}-q^{2}+\mathcal{O}(q^{\frac{13}{4}})$\\
\hline
\begin{tabular}[t]{@{}c@{}}$\Ical^{USp(4)_{-7/2}}[W^N_{\yaa}](a)$\\$=q^{-\frac{3}{2}}a^{2}\,\Ical^{\mathcal{T}_{1,2}}[W^N_{1,2}]$\end{tabular} & $-q+a^{2}q^{\frac{3}{2}}+a^{-2}q^{\frac{3}{2}}-2q^{2}+a^{2}q^{\frac{5}{2}}+a^{-2}q^{\frac{5}{2}}-2q^{3}+\mathcal{O}(q^{\frac{13}{4}})$\\
\end{longtable}
\addtocounter{table}{-1}

\FloatBarrier
\section{Characters from other Chern-Simons theories with pseudo-real matter}
\label{app_pseudo}
In this appendix we consider the theories $G_{-k}+Q$ as in section \ref{sec_other_setup} but with pseudo-real matter, other than
$USp(2n)$ with $\mathbf{2n}$.
Here $\mathbf R$ is pseudo-real or, for $SU(2)$, the sum of a pseudo-real and a real representation, and $G$ may also be $SU(2)^3$.
The boundary conditions and the line defect half-indices \eqref{other_general}--\eqref{other_tw} are those of section \ref{sec_other_setup}.
We find that the H-twist limits of their line defect half-indices are, up to monomials, characters, sums of two
characters or parts of characters of minimal models and of lattice VOAs.
Section \ref{sec_newmm_minW} treats the six theories related to the exceptional series from $\mathfrak g_2$ to $\mathfrak e_8$, which give
characters of $M(3,7)$, $M(2,13)$ and lattice VOAs (table \ref{tab_other_obs}).
Section \ref{sec_newmm} treats the other representations of $SU(2)$, which give characters of Virasoro and $\mathcal N=1$ superconformal minimal models
(table \ref{tab_su2_obs}).
Section \ref{sec_newmm_obs} treats $Spin(11)_{-13}$ with $\mathbf{32}$, which gives those of $M(3,26)$ (table \ref{tab_spin11}).

For a pseudo-real $\mathbf R$ the chiral multiplet $Q$ has the field content of a half-hyper\-multiplet, and the scalars of
$Q$ and their conjugates form the representation $(\mathbf2,\mathbf R)$ of $SU(2)\times G$.
For $USp(2n)$ this $SU(2)$ is the R-symmetry $SU(2)_H$, while for the other theories it acts only on the
representation $(\mathbf2,\mathbf R)$ and is not a symmetry of the theory.
By the criterion of \cite{Gaiotto:2008sd}, a Chern-Simons theory with half-hypermultiplets in $\mathbf R$ has $\mathcal N=4$ supersymmetry
when ${\rm Lie}(G)\oplus\mathbf R$ is a Lie superalgebra whose invariant form restricts to the Chern-Simons levels.
For a simple $G$ and an irreducible $\mathbf R$ the only such Lie superalgebra is $\mathfrak{osp}(1|2n)$ \cite{Kac:1977em}.
The theory $SU(2)^3$ with $(\mathbf2,\mathbf2,\mathbf2)$ is the simplest of the theories of $SU(N)^3$ coupled to the 3d $T_N$ theory studied
in \cite{Assel:2022row}.
For it $\mathcal N=4$ supersymmetry requires levels with $\sum_i1/k_i=0$ \cite{Gaiotto:2008sd,Assel:2022row}, which the equal levels $-4$
fixed by the boundary anomaly do not satisfy.
So for these theories, as for real $\mathbf R$, we do not expect $\mathcal N=4$ supersymmetry and use the same H-twist point
\eqref{other_HHm} as in section \ref{sec_other_setup}.

\subsection{The exceptional series}
\label{sec_newmm_minW}
For $USp(2n)$ the scalars of $Q$ and their conjugates in $(\mathbf2,\mathbf{2n})$ complete $\mathfrak{su}(2)\oplus\mathfrak{usp}(2n)$ to
$\mathfrak{usp}(2n+2)=(\mathbf3,\mathbf1)\oplus(\mathbf1,\mathfrak{usp}(2n))\oplus(\mathbf2,\mathbf{2n})$.
The theories of table \ref{tab_other_obs} are those in which the same holds with another simple Lie algebra $\mathfrak g$, which contains
${\rm Lie}(G)$,
\begin{align}
\label{series_decomp}
\mathfrak g=(\mathbf3,\mathbf1)\oplus\bigl(\mathbf1,{\rm Lie}(G)\bigr)\oplus(\mathbf2,\mathbf R).
\end{align}
Here the first factor $\mathfrak{su}(2)$ of $\mathfrak{su}(2)\oplus{\rm Lie}(G)\subset\mathfrak g$ acts on the doublet formed by the scalars of
$Q$ and their conjugates.
These $\mathfrak g$ are
\begin{itemize}
\item $\mathfrak g_2$ for $SU(2)$ with $\mathbf4$,
\item $\mathfrak f_4$ for $USp(6)$ with $\mathbf{14}'$,
\item $\mathfrak{so}(8)$ for $SU(2)^3$ with $(\mathbf2,\mathbf2,\mathbf2)$,
\item $\mathfrak e_6$ for $SU(6)$ with $\mathbf{20}$,
\item $\mathfrak e_7$ for $Spin(12)$ with $\mathbf{32}$,
\item $\mathfrak e_8$ for $E_7$ with $\mathbf{56}$.
\end{itemize}
They are the members beyond $A_2$ of the exceptional series $A_1\subset A_2\subset G_2\subset D_4\subset F_4\subset E_6\subset E_7\subset E_8$
of Deligne \cite{MR1378507,MR1952563}.
Table \ref{tab_series} lists them with $k$ of the Chern-Simons levels $-k$, the indices $\iota$ of ${\rm Lie}(G)\subset\mathfrak g$ and the levels of
$\mathfrak g$,
\begin{align}
\label{series_level}
\frac k\iota-h^\vee_{\mathfrak g}=-\frac{h^\vee_{\mathfrak g}}6-1,
\end{align}
where the equality holds for the six members of the exceptional series.
For $USp(2n)$ the level is $-\frac12$.

\begin{table}[!htbp]
\centering\footnotesize
\renewcommand{\arraystretch}{1.3}\setlength{\tabcolsep}{4pt}
\begin{tabular}{@{}lll>{\raggedright\arraybackslash}p{0.17\textwidth}>{\raggedright\arraybackslash}p{0.24\textwidth}>{\raggedright\arraybackslash}p{0.15\textwidth}@{}}
\toprule
theory & $\mathbf R$ & $\mathfrak g$ & characters of & all modules & which appear\\
\midrule
$SU(2)_{-7}$ & $\mathbf4$ & $\mathfrak g_2$ & $M(3,7)$ & $(1,s)$, $1\le s\le6$ & all\\
$USp(6)_{-\frac{13}2}$ & $\mathbf{14}'$ & $\mathfrak f_4$ & $M(2,13)$ & $(1,s)$, $1\le s\le6$ & all\\
$SU(2)^3_{-4}$ & $(\mathbf2,\mathbf2,\mathbf2)$ & $\mathfrak{so}(8)$ & $V_{\sqrt8\Zb}$ & $b$, $0\le b\le4$ & $b\ne2$\\
$SU(6)_{-9}$ & $\mathbf{20}$ & $\mathfrak e_6$ & $V_{\sqrt{18}\Zb}$ & $b$, $0\le b\le9$ & all\\
$Spin(12)_{-14}$ & $\mathbf{32}$ & $\mathfrak e_7$ & $V_{\sqrt{28}\Zb}$ & $b$, $0\le b\le14$ & all\\
$(E_7)_{-24}$ & $\mathbf{56}$ & $\mathfrak e_8$ & $V_{\sqrt{48}\Zb}$ & $b$, $0\le b\le24$ & all\\
\bottomrule
\end{tabular}
\caption{The theories of section \ref{sec_newmm_minW}, the VOAs whose characters the H-twist limits of their line defect half-indices
give up to monomials, all their modules and those which appear, with $(r,s)\sim(p-r,p'-s)$ for $M(p,p')$.
The characters of the modules $b$ of the lattice VOAs are \eqref{deligne_ch}, the same for $b$ and $-b$, so that their rows list the
distinct characters.
The column $\mathfrak g$ gives the Lie algebra of \eqref{series_decomp} (table \ref{tab_series}), and tables \ref{tab_su2_obs} and
\ref{tab_spin11} treat the other representations of $SU(2)$ and $Spin(11)_{-13}$ with $\mathbf{32}$.}
\label{tab_other_obs}
\end{table}

\begin{table}[!htbp]
\centering\footnotesize
\renewcommand{\arraystretch}{1.3}\setlength{\tabcolsep}{4pt}
\begin{tabular}{@{}lllllllll@{}}
\toprule
$\mathfrak g$ & $\dim\mathfrak g$ & $G$ & $\mathbf R$ & $k$ & $\iota$ & $h^\vee_{\mathfrak g}$ & level of $\mathfrak g$ & characters of\\
\midrule
$\mathfrak{usp}(2n+2)$ & $(n+1)(2n+3)$ & $USp(2n)$ & $\mathbf{2n}$ & $n+\frac32$ & $1$ & $n+2$ & $-\frac12$ & $M(2,2n+3)$\\
\midrule
$\mathfrak g_2$ & $14=3+3+8$ & $SU(2)$ & $\mathbf4$ & $7$ & $3$ & $4$ & $-\frac53$ & $M(3,7)$\\
$\mathfrak f_4$ & $52=3+21+28$ & $USp(6)$ & $\mathbf{14}'$ & $\frac{13}2$ & $1$ & $9$ & $-\frac52$ & $M(2,13)$\\
$\mathfrak{so}(8)$ & $28=3+9+16$ & $SU(2)^3$ & $(\mathbf2,\mathbf2,\mathbf2)$ & $4$ & $1$ & $6$ & $-2$ & $V_{\sqrt8\Zb}$\\
$\mathfrak e_6$ & $78=3+35+40$ & $SU(6)$ & $\mathbf{20}$ & $9$ & $1$ & $12$ & $-3$ & $V_{\sqrt{18}\Zb}$\\
$\mathfrak e_7$ & $133=3+66+64$ & $Spin(12)$ & $\mathbf{32}$ & $14$ & $1$ & $18$ & $-4$ & $V_{\sqrt{28}\Zb}$\\
$\mathfrak e_8$ & $248=3+133+112$ & $E_7$ & $\mathbf{56}$ & $24$ & $1$ & $30$ & $-6$ & $V_{\sqrt{48}\Zb}$\\
\bottomrule
\end{tabular}
\caption{The theories of table \ref{tab_other_obs} related to the exceptional series, with the $USp(2n)$ theories for comparison
(section \ref{sec_newmm_minW}).
The Lie algebra $\mathfrak g$ decomposes as \eqref{series_decomp}, so that $\dim\mathfrak g=3+\dim G+2\dim\mathbf R$.
The theory is $G_{-k}$, $\iota$ is the index of ${\rm Lie}(G)\subset\mathfrak g$ and the level of $\mathfrak g$ is $\frac k\iota-h^\vee_{\mathfrak g}$ \eqref{series_level}.
The last column gives the VOAs whose characters the H-twist limits of the line defect half-indices give.}
\label{tab_series}
\end{table}

The levels of the six members of the exceptional series in table \ref{tab_series} are those of the Deligne series of affine VOAs.
For $\mathfrak{so}(8)$, $\mathfrak e_6$, $\mathfrak e_7$ and $\mathfrak e_8$ these affine VOAs are associated \cite{Beem:2013sza,Beem:2017ooy}
with the 4d $\mathcal N=2$ superconformal field theories of rank one, the $SU(2)$ theory with four flavors \cite{Seiberg:1994aj} and the
Minahan-Nemeschansky theories \cite{Minahan:1996fg,Minahan:1996cj}.
For $\mathfrak g_2$ and $\mathfrak f_4$ the 4d interpretation was left open in \cite{Beem:2017ooy}.

\subsubsection{The general form}
For all the six theories we find that the H-twist limits of the line defect half-indices of all the Wilson lines are
fixed by finitely many of them.
We also find that these finitely many H-twist limits are given by the characters of the VOAs in table \ref{tab_series}.
As for real $\mathbf R$ in \eqref{real_reduce}, we find that the H-twist limit of the line defect half-index of $W_\mu$ vanishes if $\mu+\rho$ lies on a
reflecting hyperplane of $W\ltimes kQ_{\rm long}$.
Otherwise
\begin{align}
\label{series_reduce}
\lin{W_\mu}{G_{-k}}{\Ncal,N,\Gamma}\Big|_{\rm H}=\epsilon(g)\,q^{h_\mu-h_\lambda}\,\lin{W_\lambda}{G_{-k}}{\Ncal,N,\Gamma}\Big|_{\rm H},
\end{align}
where $\lambda$ is the unique weight in the open alcove (section \ref{sec_other_unit_real}) with $g(\lambda+\rho)=\mu+\rho$ for some
$g=t_\beta w\in W\ltimes kQ_{\rm long}$.
Here $h_\mu=\frac{(\mu,\mu+2\rho)}{2k}$ as in \eqref{real_reduce}.
The sign is $\epsilon(g)={\rm sgn}(w)$ for the integral levels $k$, and for $USp(6)_{-\frac{13}2}$ it is given in \eqref{usp6_sign}.
For $\lambda+\rho$ in the open alcove
\begin{align}
\label{series_form}
\lin{W_\lambda}{G_{-k}}{\Ncal,N,\Gamma}\Big|_{\rm H}=\sum_{x\in S_\lambda}\varepsilon_{\lambda,x}\,q^{h_\lambda+h_x-h_{x_0}}\,\chi_x,
\end{align}
where $\chi_x$ is the normalized character of the module $x$ of the VOA of table \ref{tab_series} and $h_x$ is its conformal weight.
Here $x_0$ is the module whose character the H-twist limit of the half-index gives, $S_\lambda$ is a set of modules with at most one element for
$\mathfrak g_2$ and $\mathfrak f_4$ and at most two for the simply-laced members, and $\varepsilon_{\lambda,x}=\pm1$.
The following paragraphs and tables \ref{tab_app_so8}--\ref{tab_app_e8} give $S_\lambda$ and $\varepsilon_{\lambda,x}$.

\subsubsection{\texorpdfstring{$SU(2)_{-7}$ with $\mathbf4$}{SU(2) with 4}}
For $\mathfrak g_2$ the gauge group is $SU(2)$ with the spin-$\frac32$ representation $\mathbf R=\mathbf4$, and \eqref{series_decomp} reads
\begin{align}
\label{g2_decomp}
\mathfrak g_2=(\mathbf3,\mathbf1)\oplus\bigl(\mathbf1,\mathfrak{su}(2)\bigr)\oplus(\mathbf2,\mathbf4),
\end{align}
where ${\rm Lie}(G)=\mathfrak{su}(2)$ has the index $\iota=3$ in $\mathfrak g_2$.
Here $k=7$, and the open alcove contains the Wilson lines $W_{(a)}$ of spin $\frac a2$ with $0\le a\le5$, with
conformal weights $h_{(a)}=\frac{a(a+2)}{28}$.
We find
\begin{align}
\label{su24_table}
\lin{W_{(0)}}{SU(2)_{-7}}{\Ncal,N,\Gamma}\big|_{\rm H}&=\chi^{M(3,7)}_{(1,3)},&
\lin{W_{(1)}}{SU(2)_{-7}}{\Ncal,N,\Gamma}\big|_{\rm H}&=q^{\frac32}\chi^{M(3,7)}_{(1,6)},
\nonumber\\
\lin{W_{(2)}}{SU(2)_{-7}}{\Ncal,N,\Gamma}\big|_{\rm H}&=-q\,\chi^{M(3,7)}_{(1,5)},&
\lin{W_{(3)}}{SU(2)_{-7}}{\Ncal,N,\Gamma}\big|_{\rm H}&=-q^{\frac12}\chi^{M(3,7)}_{(1,2)},
\nonumber\\
\lin{W_{(4)}}{SU(2)_{-7}}{\Ncal,N,\Gamma}\big|_{\rm H}&=q\,\chi^{M(3,7)}_{(1,1)},&
\lin{W_{(5)}}{SU(2)_{-7}}{\Ncal,N,\Gamma}\big|_{\rm H}&=q^{\frac32}\chi^{M(3,7)}_{(1,4)},
\end{align}
the characters of all the six modules of the minimal model $M(3,7)$.
This is \eqref{series_form} with $x_0=(1,3)$ and the conformal weights $h_{1,s}$ of $M(3,7)$.
For the other Wilson lines \eqref{series_reduce} reads, with $m=a+1$,
\begin{align}
\label{su24_rule}
\lin{W_{(a)}}{SU(2)_{-7}}{\Ncal,N,\Gamma}\big|_{\rm H}=
\begin{cases}
0, & m\equiv0,7\bmod14,\\
\pm q^{h_{(a)}-h_{(m_0-1)}}\,\lin{W_{(m_0-1)}}{SU(2)_{-7}}{\Ncal,N,\Gamma}\big|_{\rm H}, & m\equiv\pm m_0\bmod14,
\end{cases}
\end{align}
with $1\le m_0\le6$.

\subsubsection{\texorpdfstring{$USp(6)_{-13/2}$ with $\mathbf{14}'$}{USp(6) with 14'}}
For $\mathfrak f_4$ the gauge group is $USp(6)$ with the representation $\mathbf R=\mathbf{14}'$ in $\Lambda^3\mathbf6=\mathbf{14}'\oplus\mathbf6$, and
\begin{align}
\label{f4_decomp}
\mathfrak f_4=(\mathbf3,\mathbf1)\oplus\bigl(\mathbf1,\mathfrak{usp}(6)\bigr)\oplus(\mathbf2,\mathbf{14}').
\end{align}
Here $k=\frac{13}2$.
With the Young diagrams $\lambda$ of $USp(6)$ and $\lambda+\rho=(\lambda_1+3,\lambda_2+2,\lambda_3+1)$ we have $W\ltimes kQ_{\rm long}=W\ltimes13\Zb^3$,
and the open alcove contains the $20$ diagrams $\lambda\subseteq\ysmall{3,3,3}$.
We find that the H-twist limits of the line defect half-indices of the Wilson lines of the eight
diagrams $\yf$, $\yaa$, $\ysmall{3}$, $\ysmall{3,1}$, $\ysmall{2,2,1}$, $\ysmall{2,2,2}$, $\ysmall{3,3,1}$ and $\ysmall{3,3,2}$ vanish.
Those of the other twelve are
\begin{align}
\label{usp6_table}
\lin{W_\varnothing}{USp(6)_{-13/2}}{\Ncal,N,\Gamma}\big|_{\rm H}&=\chi^{M(2,13)}_{(1,5)},&
\lin{W_{\ysmall{3,2}}}{USp(6)_{-13/2}}{\Ncal,N,\Gamma}\big|_{\rm H}&=-q^{\frac32}\chi^{M(2,13)}_{(1,5)},
\nonumber\\
\lin{W_{\ysmall{1,1,1}}}{USp(6)_{-13/2}}{\Ncal,N,\Gamma}\big|_{\rm H}&=-q^{\frac12}\chi^{M(2,13)}_{(1,6)},&
\lin{W_{\ybbd}}{USp(6)_{-13/2}}{\Ncal,N,\Gamma}\big|_{\rm H}&=q\,\chi^{M(2,13)}_{(1,6)},
\nonumber\\
\lin{W_{\yb}}{USp(6)_{-13/2}}{\Ncal,N,\Gamma}\big|_{\rm H}&=-q\,\chi^{M(2,13)}_{(1,3)},&
\lin{W_{\ysmall{3,2,1}}}{USp(6)_{-13/2}}{\Ncal,N,\Gamma}\big|_{\rm H}&=-q^2\chi^{M(2,13)}_{(1,3)},
\nonumber\\
\lin{W_{\yba}}{USp(6)_{-13/2}}{\Ncal,N,\Gamma}\big|_{\rm H}&=q^{\frac32}\chi^{M(2,13)}_{(1,2)},&
\lin{W_{\ysmall{3,2,2}}}{USp(6)_{-13/2}}{\Ncal,N,\Gamma}\big|_{\rm H}&=q^{\frac52}\chi^{M(2,13)}_{(1,2)},
\nonumber\\
\lin{W_{\ysmall{2,1,1}}}{USp(6)_{-13/2}}{\Ncal,N,\Gamma}\big|_{\rm H}&=-q^2\chi^{M(2,13)}_{(1,1)},&
\lin{W_{\ysmall{3,3,3}}}{USp(6)_{-13/2}}{\Ncal,N,\Gamma}\big|_{\rm H}&=-q^{\frac72}\chi^{M(2,13)}_{(1,1)},
\nonumber\\
\lin{W_{\ysmall{3,1,1}}}{USp(6)_{-13/2}}{\Ncal,N,\Gamma}\big|_{\rm H}&=q^{\frac32}\chi^{M(2,13)}_{(1,4)},&
\lin{W_{\ysmall{3,3}}}{USp(6)_{-13/2}}{\Ncal,N,\Gamma}\big|_{\rm H}&=q^2\chi^{M(2,13)}_{(1,4)},
\end{align}
so that each of the six modules of $M(2,13)$ appears twice.
This is \eqref{series_form} with $x_0=(1,5)$ and the conformal weights $h_{1,s}$ of $M(2,13)$.
For the other Wilson lines the sign in \eqref{series_reduce} is
\begin{align}
\label{usp6_sign}
\epsilon(g)={\rm sgn}(w)\,(-1)^{\frac{\beta_1+\beta_2+\beta_3}{13}},\qquad \beta=(\beta_1,\beta_2,\beta_3)\in13\Zb^3,
\end{align}
and the extra sign is absent at the integral levels.

\subsubsection{The simply-laced members}
For $\mathfrak{so}(8)$, $\mathfrak e_6$, $\mathfrak e_7$ and $\mathfrak e_8$ we have $\iota=1$, and the VOAs of table \ref{tab_series} are the lattice VOAs
$V_{\sqrt{2k}\Zb}$ of section \ref{sec_twist_lat}, with $(\alpha,\alpha)=2k$, whose modules $b$ have the conformal weights $h_b=\frac{b^2}{4k}$.
By \eqref{ramanujan_f} their normalized characters \eqref{lattice_chi} are
\begin{align}
\label{deligne_ch}
\chi^{V_{\sqrt{2k}\Zb}}_b=\frac{f(q^{k+b},q^{k-b})}{f(-q)},\qquad f(q^{k+b},q^{k-b})=\sum_{m\in\Zb}q^{km^2+bm},
\end{align}
which are invariant under $b\to-b$ and are multiplied by $q^{-k-b}$ under $b\to b+2k$, so that up to monomials the labels are
$0\le b\le k$.
The open alcove consists of the dominant weights $\lambda$ with $(\lambda,\theta)\le j_{\mathbf R}$ for the highest root $\theta$ of each simple factor
of $G$.
The H-twist limit of the half-index gives the character of $x_0=b_0=\frac{h^\vee_{\mathfrak g}}6$, and \eqref{series_form} reads
\begin{align}
\label{deligne_gen}
\lin{W_\lambda}{G_{-k}}{\Ncal,N,\Gamma}\Big|_{\rm H}=\sum_{b\in S_\lambda}\varepsilon_{\lambda,b}\,q^{h_\lambda+\frac{b^2-b_0^2}{4k}}\,\chi^{V_{\sqrt{2k}\Zb}}_b,
\qquad
\II^{G_{-k}}_{\Ncal,N,\Gamma}\Big|_{\rm H}=\chi^{V_{\sqrt{2k}\Zb}}_{b_0}.
\end{align}
Of the $27$, $56$, $71$ and $79$ weights in the open alcoves of $SU(2)^3_{-4}$, $SU(6)_{-9}$, $Spin(12)_{-14}$ and $(E_7)_{-24}$, respectively
$9$, $20$, $31$ and $49$ give non-zero H-twist limits of the line defect half-indices.
Tables \ref{tab_app_so8}--\ref{tab_app_e8} list them, and $S_\lambda$ and $\varepsilon_{\lambda,b}$ in \eqref{deligne_gen} can be read off
from these tables.
The labels $b$ of all the modules appear except $b=2$ for $SU(2)^3$.
In the tables the weights are written in terms of the fundamental weights $\omega_i$.
For $SU(2)^3$ the $\omega_i$ are the doublets of the three factors, and for $SU(6)$ they give $\Lambda^i\mathbf6$.
For $Spin(12)$ and $E_7$ they follow Bourbaki \cite{MR1890629} (appendix \ref{app_char}).
For $Spin(12)$ the highest weights $\omega_6$ and $\omega_5$ give the two half-spinors, $\omega_6$ giving $\mathbf R=\mathbf{32}$, and
$\omega_p$ gives $\Lambda^p\mathbf{12}$ for $p\le4$.
For $E_7$ the $\omega_1,\dots,\omega_7$ give the representations with the dimensions $133$, $912$, $8645$, $365750$, $27664$, $1539$ and $56$.
The rows are ordered by $h_\lambda$.
Below we give the H-twist limits of the half-indices and of the line defect half-indices of the Wilson lines in the fundamental representations.

\paragraph{\texorpdfstring{$SU(2)^3_{-4}$ with $(\mathbf2,\mathbf2,\mathbf2)$}{SU(2)**3 with (2,2,2)}}
For $\mathfrak{so}(8)$ the gauge group is $SU(2)^3$ with $\mathbf R=(\mathbf2,\mathbf2,\mathbf2)$, and
\begin{align}
\label{so8_decomp}
\mathfrak{so}(8)=(\mathbf3,\mathbf1)\oplus\bigl(\mathbf1,\mathfrak{su}(2)^{\oplus3}\bigr)\oplus\bigl(\mathbf2,(\mathbf2,\mathbf2,\mathbf2)\bigr).
\end{align}
Here $k=4$ and $b_0=1$, so that the H-twist limit of the half-index is $\chi^{V_{\sqrt8\Zb}}_1=f(q^5,q^3)/f(-q)$.
The H-twist limits of the line defect half-indices of the Wilson lines in the doublets of the three factors vanish.

\begin{table}[!htbp]
\centering\footnotesize
\renewcommand{\arraystretch}{1.25}
\begin{tabular}{@{}l@{\hspace{0.8em}}l@{\hspace{2em}}l@{\hspace{0.8em}}l@{}}
\toprule
$\lambda$ & $f(-q)\lin{W_\lambda}{SU(2)^3_{-4}}{\Ncal,N,\Gamma}\big|_{\rm H}$ & $\lambda$ & $f(-q)\lin{W_\lambda}{SU(2)^3_{-4}}{\Ncal,N,\Gamma}\big|_{\rm H}$\\
\midrule
$0$ & $f(q^{5},q^{3})$ & $2\omega_{2}+2\omega_{3}$ & $qf(q^{5},q^{3})$\\
$2\omega_{3}$ & $-qf(q^{7},q)$ & $2\omega_{1}+2\omega_{3}$ & $qf(q^{5},q^{3})$\\
$2\omega_{2}$ & $-qf(q^{7},q)$ & $2\omega_{1}+2\omega_{2}$ & $qf(q^{5},q^{3})$\\
$2\omega_{1}$ & $-qf(q^{7},q)$ & $2\omega_{1}+2\omega_{2}+2\omega_{3}$ & $-q^{2}f(q^{7},q)$\\
$\omega_{1}+\omega_{2}+\omega_{3}$ & $-q^{\frac{1}{2}}f(q^{4},q^{4})+q^{\frac{3}{2}}f(q^{8},1)$ & &\\
\bottomrule
\end{tabular}
\caption{The non-zero H-twist limits of the line defect half-indices of $SU(2)^3_{-4}$, for $\mathfrak{so}(8)$, with $\lambda$ in the open alcove, $9$ of its $27$
weights.}
\label{tab_app_so8}
\end{table}

\paragraph{\texorpdfstring{$SU(6)_{-9}$ with $\mathbf{20}$}{SU(6) with 20}}
For $\mathfrak e_6$ the gauge group is $SU(6)$ with $\mathbf R=\mathbf{20}=\Lambda^3\mathbf6$, and
\begin{align}
\label{e6_decomp}
\mathfrak e_6=(\mathbf3,\mathbf1)\oplus\bigl(\mathbf1,\mathfrak{su}(6)\bigr)\oplus(\mathbf2,\mathbf{20}).
\end{align}
Here $k=9$ and $b_0=2$, so that the H-twist limit of the half-index is $\chi^{V_{\sqrt{18}\Zb}}_2=f(q^{11},q^7)/f(-q)$.
The H-twist limits of the line defect half-indices of the Wilson lines of the fundamental representations vanish
except that of $W_{\ysmall{1,1,1}}$,
\begin{align}
\label{e6_fund}
\lin{W_{\ysmall{1,1,1}}}{SU(6)_{-9}}{\Ncal,N,\Gamma}\big|_{\rm H}=-q^{\frac12}\,\frac{f(q^{10},q^8)}{f(-q)},
\end{align}
which gives the character of the module $1$.

\begin{table}[!htbp]
\centering\footnotesize
\renewcommand{\arraystretch}{1.25}
\begin{tabular}{@{}l@{\hspace{0.8em}}l@{\hspace{2em}}l@{\hspace{0.8em}}l@{}}
\toprule
$\lambda$ & $f(-q)\lin{W_\lambda}{SU(6)_{-9}}{\Ncal,N,\Gamma}\big|_{\rm H}$ & $\lambda$ & $f(-q)\lin{W_\lambda}{SU(6)_{-9}}{\Ncal,N,\Gamma}\big|_{\rm H}$\\
\midrule
$0$ & $f(q^{11},q^{7})$ & $2\omega_{1}+\omega_{4}$ & $q^{3}f(q^{17},q)$\\
$\omega_{3}$ & $-q^{\frac{1}{2}}f(q^{10},q^{8})$ & $\omega_{1}+\omega_{3}+\omega_{5}$ & $q^{\frac{3}{2}}f(q^{12},q^{6})-q^{\frac{7}{2}}f(q^{18},1)$\\
$\omega_{1}+\omega_{5}$ & $-qf(q^{13},q^{5})$ & $2\omega_{2}+\omega_{5}$ & $-q^{\frac{3}{2}}f(q^{10},q^{8})$\\
$\omega_{4}+\omega_{5}$ & $q^{\frac{3}{2}}f(q^{14},q^{4})$ & $\omega_{1}+2\omega_{4}$ & $-q^{\frac{3}{2}}f(q^{10},q^{8})$\\
$\omega_{1}+\omega_{2}$ & $q^{\frac{3}{2}}f(q^{14},q^{4})$ & $\omega_{3}+\omega_{4}+\omega_{5}$ & $-q^{2}f(q^{13},q^{5})$\\
$\omega_{2}+\omega_{4}$ & $qf(q^{9},q^{9})-q^{2}f(q^{15},q^{3})$ & $\omega_{1}+\omega_{2}+\omega_{3}$ & $-q^{2}f(q^{13},q^{5})$\\
$3\omega_{5}$ & $-q^{\frac{5}{2}}f(q^{16},q^{2})$ & $\omega_{2}+\omega_{3}+\omega_{4}$ & $q^{\frac{5}{2}}f(q^{14},q^{4})$\\
$3\omega_{1}$ & $-q^{\frac{5}{2}}f(q^{16},q^{2})$ & $3\omega_{4}$ & $q^{2}f(q^{11},q^{7})$\\
$2\omega_{3}$ & $q^{3}f(q^{17},q)$ & $3\omega_{2}$ & $q^{2}f(q^{11},q^{7})$\\
$\omega_{2}+2\omega_{5}$ & $q^{3}f(q^{17},q)$ & $3\omega_{3}$ & $-q^{\frac{7}{2}}f(q^{16},q^{2})$\\
\bottomrule
\end{tabular}
\caption{The non-zero H-twist limits of the line defect half-indices of $SU(6)_{-9}$, for $\mathfrak e_6$, with $\lambda$ in the open alcove, $20$ of its $56$
weights.}
\label{tab_app_e6}
\end{table}

\paragraph{\texorpdfstring{$Spin(12)_{-14}$ with $\mathbf{32}$}{Spin(12) with 32}}
For $\mathfrak e_7$ the gauge group is $Spin(12)$ with the half-spinor $\mathbf R=\mathbf{32}$, and
\begin{align}
\label{e7_decomp}
\mathfrak e_7=(\mathbf3,\mathbf1)\oplus\bigl(\mathbf1,\mathfrak{so}(12)\bigr)\oplus(\mathbf2,\mathbf{32}).
\end{align}
Here $k=14$ and $b_0=3$, so that the H-twist limit of the half-index is $\chi^{V_{\sqrt{28}\Zb}}_3=f(q^{17},q^{11})/f(-q)$.
The H-twist limits of the line defect half-indices of the Wilson lines of the fundamental representations vanish
except those of $W_{\omega_6}$, $W_{\omega_2}$ and $W_{\omega_4}$,
\begin{align}
\label{e7_fund}
\lin{W_{\omega_6}}{Spin(12)_{-14}}{\Ncal,N,\Gamma}\big|_{\rm H}&=-q^{\frac12}\,\frac{f(q^{16},q^{12})}{f(-q)},&
\lin{W_{\omega_2}}{Spin(12)_{-14}}{\Ncal,N,\Gamma}\big|_{\rm H}&=-q\,\frac{f(q^{19},q^9)}{f(-q)},
\nonumber\\
\lin{W_{\omega_4}}{Spin(12)_{-14}}{\Ncal,N,\Gamma}\big|_{\rm H}&=q\,\frac{f(q^{15},q^{13})}{f(-q)},
\end{align}
which give the characters of the modules $2$, $5$ and $1$.

\begin{table}[!htbp]
\centering\footnotesize
\renewcommand{\arraystretch}{1.25}
\begin{tabular}{@{}l@{\hspace{0.8em}}l@{\hspace{2em}}l@{\hspace{0.8em}}l@{}}
\toprule
$\lambda$ & $f(-q)\lin{W_\lambda}{Spin(12)_{-14}}{\Ncal,N,\Gamma}\big|_{\rm H}$ & $\lambda$ & $f(-q)\lin{W_\lambda}{Spin(12)_{-14}}{\Ncal,N,\Gamma}\big|_{\rm H}$\\
\midrule
$0$ & $f(q^{17},q^{11})$ & $3\omega_{6}$ & $q^{\frac{9}{2}}f(q^{26},q^{2})$\\
$\omega_{6}$ & $-q^{\frac{1}{2}}f(q^{16},q^{12})$ & $3\omega_{1}+\omega_{5}$ & $q^{\frac{9}{2}}f(q^{26},q^{2})$\\
$\omega_{2}$ & $-qf(q^{19},q^{9})$ & $2\omega_{3}$ & $q^{2}f(q^{15},q^{13})-q^{5}f(q^{27},q)$\\
$\omega_{1}+\omega_{5}$ & $q^{\frac{3}{2}}f(q^{20},q^{8})$ & $\omega_{2}+2\omega_{6}$ & $-q^{5}f(q^{27},q)$\\
$\omega_{4}$ & $qf(q^{15},q^{13})$ & $\omega_{2}+2\omega_{5}$ & $q^{2}f(q^{15},q^{13})$\\
$2\omega_{5}$ & $-q^{2}f(q^{21},q^{7})$ & $2\omega_{1}+\omega_{4}$ & $-q^{5}f(q^{27},q)$\\
$\omega_{2}+\omega_{6}$ & $q^{\frac{3}{2}}f(q^{18},q^{10})$ & $\omega_{1}+\omega_{3}+\omega_{6}$ & $q^{\frac{11}{2}}f(q^{28},1)$\\
$2\omega_{1}+\omega_{6}$ & $-q^{\frac{5}{2}}f(q^{22},q^{6})$ & $2\omega_{1}+2\omega_{6}$ & $q^{3}f(q^{21},q^{7})$\\
$\omega_{3}+\omega_{5}$ & $-q^{\frac{3}{2}}f(q^{14},q^{14})$ & $\omega_{1}+\omega_{4}+\omega_{5}$ & $q^{\frac{5}{2}}f(q^{18},q^{10})$\\
$\omega_{1}+\omega_{5}+\omega_{6}$ & $-q^{2}f(q^{19},q^{9})+q^{3}f(q^{23},q^{5})$ & $\omega_{1}+\omega_{5}+2\omega_{6}$ & $-q^{\frac{7}{2}}f(q^{22},q^{6})$\\
$2\omega_{1}+\omega_{2}$ & $q^{3}f(q^{23},q^{5})$ & $\omega_{1}+3\omega_{5}$ & $-q^{\frac{5}{2}}f(q^{16},q^{12})$\\
$\omega_{4}+\omega_{6}$ & $-q^{\frac{7}{2}}f(q^{24},q^{4})$ & $\omega_{4}+2\omega_{6}$ & $q^{4}f(q^{23},q^{5})$\\
$\omega_{1}+\omega_{2}+\omega_{5}$ & $-q^{\frac{7}{2}}f(q^{24},q^{4})$ & $\omega_{4}+2\omega_{5}$ & $-q^{3}f(q^{19},q^{9})$\\
$\omega_{2}+\omega_{4}$ & $-q^{2}f(q^{17},q^{11})+q^{4}f(q^{25},q^{3})$ & $4\omega_{6}$ & $-q^{5}f(q^{25},q^{3})$\\
$4\omega_{1}$ & $-q^{4}f(q^{25},q^{3})$ & $4\omega_{5}$ & $q^{3}f(q^{17},q^{11})$\\
$2\omega_{5}+\omega_{6}$ & $q^{\frac{5}{2}}f(q^{20},q^{8})$ & &\\
\bottomrule
\end{tabular}
\caption{The non-zero H-twist limits of the line defect half-indices of $Spin(12)_{-14}$, for $\mathfrak e_7$, with $\lambda$ in the open alcove, $31$ of its $71$
weights.}
\label{tab_app_e7}
\end{table}

\begin{table}[!htbp]
\centering\footnotesize
\renewcommand{\arraystretch}{1.25}
\begin{tabular}{@{}l@{\hspace{0.8em}}l@{\hspace{2em}}l@{\hspace{0.8em}}l@{}}
\toprule
$\lambda$ & $f(-q)\lin{W_\lambda}{(E_7)_{-24}}{\Ncal,N,\Gamma}\big|_{\rm H}$ & $\lambda$ & $f(-q)\lin{W_\lambda}{(E_7)_{-24}}{\Ncal,N,\Gamma}\big|_{\rm H}$\\
\midrule
$0$ & $f(q^{29},q^{19})$ & $2\omega_{3}$ & $q^{3}f(q^{25},q^{23})-q^{6}f(q^{41},q^{7})$\\
$\omega_{7}$ & $-q^{\frac{1}{2}}f(q^{28},q^{20})$ & $\omega_{1}+2\omega_{2}$ & $q^{3}f(q^{25},q^{23})$\\
$\omega_{1}$ & $-qf(q^{31},q^{17})$ & $\omega_{2}+3\omega_{7}$ & $q^{6}f(q^{41},q^{7})$\\
$\omega_{2}$ & $q^{\frac{3}{2}}f(q^{32},q^{16})$ & $2\omega_{6}+\omega_{7}$ & $q^{\frac{9}{2}}f(q^{36},q^{12})$\\
$\omega_{6}$ & $qf(q^{27},q^{21})$ & $\omega_{6}+3\omega_{7}$ & $-q^{\frac{13}{2}}f(q^{42},q^{6})$\\
$\omega_{1}+\omega_{7}$ & $q^{\frac{3}{2}}f(q^{30},q^{18})$ & $\omega_{1}+\omega_{2}+2\omega_{7}$ & $-q^{\frac{13}{2}}f(q^{42},q^{6})$\\
$\omega_{5}$ & $-q^{\frac{3}{2}}f(q^{26},q^{22})-q^{\frac{5}{2}}f(q^{34},q^{14})$ & $\omega_{1}+2\omega_{6}$ & $q^{4}f(q^{33},q^{15})$\\
$\omega_{2}+\omega_{7}$ & $-q^{2}f(q^{31},q^{17})$ & $\omega_{2}+\omega_{3}+\omega_{7}$ & $q^{7}f(q^{43},q^{5})$\\
$\omega_{1}+\omega_{6}$ & $-q^{2}f(q^{29},q^{19})+q^{3}f(q^{35},q^{13})$ & $\omega_{1}+\omega_{6}+2\omega_{7}$ & $-q^{5}f(q^{37},q^{11})+q^{7}f(q^{43},q^{5})$\\
$\omega_{4}$ & $q^{2}f(q^{25},q^{23})$ & $\omega_{2}+\omega_{4}$ & $q^{\frac{7}{2}}f(q^{28},q^{20})-q^{\frac{15}{2}}f(q^{44},q^{4})$\\
$2\omega_{1}+\omega_{7}$ & $-q^{\frac{7}{2}}f(q^{36},q^{12})$ & $\omega_{3}+\omega_{6}+\omega_{7}$ & $-q^{\frac{15}{2}}f(q^{44},q^{4})$\\
$\omega_{2}+\omega_{6}$ & $q^{\frac{5}{2}}f(q^{30},q^{18})$ & $5\omega_{7}$ & $q^{\frac{15}{2}}f(q^{44},q^{4})$\\
$\omega_{5}+\omega_{7}$ & $q^{3}f(q^{33},q^{15})$ & $\omega_{4}+\omega_{6}$ & $q^{8}f(q^{45},q^{3})$\\
$2\omega_{6}$ & $-q^{4}f(q^{37},q^{11})$ & $\omega_{1}+4\omega_{7}$ & $-q^{8}f(q^{45},q^{3})$\\
$3\omega_{1}$ & $q^{4}f(q^{37},q^{11})$ & $3\omega_{2}$ & $-q^{\frac{7}{2}}f(q^{26},q^{22})+q^{\frac{17}{2}}f(q^{46},q^{2})$\\
$\omega_{1}+\omega_{5}$ & $q^{\frac{5}{2}}f(q^{28},q^{20})$ & $\omega_{3}+3\omega_{7}$ & $q^{\frac{11}{2}}f(q^{38},q^{10})+q^{\frac{17}{2}}f(q^{46},q^{2})$\\
$\omega_{1}+\omega_{6}+\omega_{7}$ & $-q^{\frac{7}{2}}f(q^{34},q^{14})+q^{\frac{9}{2}}f(q^{38},q^{10})$ & $2\omega_{5}$ & $q^{4}f(q^{31},q^{17})-q^{9}f(q^{47},q)$\\
$\omega_{2}+\omega_{3}$ & $-q^{\frac{5}{2}}f(q^{24},q^{24})$ & $\omega_{4}+2\omega_{7}$ & $-q^{9}f(q^{47},q)$\\
$\omega_{3}+2\omega_{7}$ & $-q^{5}f(q^{39},q^{9})$ & $2\omega_{2}+\omega_{6}$ & $-q^{9}f(q^{47},q)$\\
$2\omega_{1}+\omega_{6}$ & $-q^{5}f(q^{39},q^{9})$ & $\omega_{2}+\omega_{5}+\omega_{7}$ & $q^{\frac{19}{2}}f(q^{48},1)$\\
$\omega_{2}+\omega_{5}$ & $-q^{3}f(q^{29},q^{19})$ & $3\omega_{6}$ & $-q^{5}f(q^{35},q^{13})$\\
$2\omega_{1}+2\omega_{7}$ & $q^{4}f(q^{35},q^{13})$ & $\omega_{2}+4\omega_{7}$ & $-q^{\frac{13}{2}}f(q^{40},q^{8})$\\
$\omega_{5}+\omega_{6}$ & $-q^{\frac{7}{2}}f(q^{32},q^{16})$ & $\omega_{6}+4\omega_{7}$ & $q^{7}f(q^{41},q^{7})$\\
$\omega_{1}+\omega_{3}+\omega_{7}$ & $q^{\frac{11}{2}}f(q^{40},q^{8})$ & $6\omega_{7}$ & $-q^{8}f(q^{43},q^{5})$\\
$\omega_{1}+\omega_{4}$ & $-q^{3}f(q^{27},q^{21})$ & &\\
\bottomrule
\end{tabular}
\caption{The non-zero H-twist limits of the line defect half-indices of $(E_7)_{-24}$, for $\mathfrak e_8$, with $\lambda$ in the open alcove, $49$ of its $79$
weights.}
\label{tab_app_e8}
\end{table}

\paragraph{\texorpdfstring{$(E_7)_{-24}$ with $\mathbf{56}$}{E7 with 56}}
For $\mathfrak e_8$ the gauge group is $E_7$ with $\mathbf R=\mathbf{56}$, and
\begin{align}
\label{e8_decomp}
\mathfrak e_8=(\mathbf3,\mathbf1)\oplus\bigl(\mathbf1,\mathfrak e_7\bigr)\oplus(\mathbf2,\mathbf{56}).
\end{align}
Here $k=24$ and $b_0=5$, so that the H-twist limit of the half-index is $\chi^{V_{\sqrt{48}\Zb}}_5=f(q^{29},q^{19})/f(-q)$.
The H-twist limits of the line defect half-indices of the Wilson lines of the fundamental representations vanish
only for $W_{\mathbf{8645}}$, and those of the other six are
\begin{align}
\label{e8_fund}
&\lin{W_{\mathbf{56}}}{(E_7)_{-24}}{\Ncal,N,\Gamma}\big|_{\rm H}=-q^{\frac12}\,\frac{f(q^{28},q^{20})}{f(-q)},\qquad
\lin{W_{\mathbf{133}}}{(E_7)_{-24}}{\Ncal,N,\Gamma}\big|_{\rm H}=-q\,\frac{f(q^{31},q^{17})}{f(-q)},
\nonumber\\
&\lin{W_{\mathbf{1539}}}{(E_7)_{-24}}{\Ncal,N,\Gamma}\big|_{\rm H}=q\,\frac{f(q^{27},q^{21})}{f(-q)},\qquad
\lin{W_{\mathbf{912}}}{(E_7)_{-24}}{\Ncal,N,\Gamma}\big|_{\rm H}=q^{\frac32}\,\frac{f(q^{32},q^{16})}{f(-q)},
\nonumber\\
&\lin{W_{\mathbf{365750}}}{(E_7)_{-24}}{\Ncal,N,\Gamma}\big|_{\rm H}=q^2\,\frac{f(q^{25},q^{23})}{f(-q)},
\nonumber\\
&\lin{W_{\mathbf{27664}}}{(E_7)_{-24}}{\Ncal,N,\Gamma}\big|_{\rm H}=-\frac{q^{\frac32}f(q^{26},q^{22})+q^{\frac52}f(q^{34},q^{14})}{f(-q)},
\end{align}
which give the characters of the modules $4$, $7$, $3$, $8$ and $1$, and of the two modules $2$ and $10$ for $W_{\mathbf{27664}}$.
For example, $W_{\mathbf{56}}$ has $h_\lambda=\frac{57}{96}$ and $b=4$ in \eqref{deligne_gen}.

\FloatBarrier

\subsection{\texorpdfstring{Other representations of $SU(2)$}{Other representations of SU(2)}}
\label{sec_newmm}
We consider the other representations of $SU(2)$ in table \ref{tab_su2_obs}, with $W_{(a)}$ the Wilson line of spin $\frac a2$ and $m=a+1$.
Here $k=\frac92$, $\frac{15}2$, $\frac{39}2$, $\frac{25}2$ and $44$ for $\mathbf R=\mathbf2\oplus\mathbf3$,
$\mathbf2\oplus\mathbf4$, $\mathbf6$, $\mathbf2\oplus\mathbf5$ and $\mathbf8$, and the conformal weights are $h_{(a)}=\frac{a(a+2)}{4k}$.
As for $\mathfrak g_2$ in \eqref{su24_rule}, $W\ltimes kQ_{\rm long}$ acts by $m\to\pm m+\beta$ with $\beta\in2k\Zb$, and the open alcove is
$1\le m<k$.
Then \eqref{series_reduce} reads
\begin{align}
\label{su2_rule}
\lin{W_{(a)}}{SU(2)_{-k}}{\Ncal,N,\Gamma}\big|_{\rm H}=
\begin{cases}
0, & m\in k\Zb,\\
\pm(-1)^{\beta}\,q^{h_{(a)}-h_{(m_0-1)}}\,\lin{W_{(m_0-1)}}{SU(2)_{-k}}{\Ncal,N,\Gamma}\big|_{\rm H}, & m=\pm m_0+\beta,
\end{cases}
\end{align}
with $1\le m_0<k$ and $\beta\in2k\Zb$.
The sign $(-1)^\beta$ is $-1$ only at the half-integral levels for odd $\frac\beta{2k}$, as in \eqref{usp6_sign} for $USp(6)$.
So the H-twist limits of the line defect half-indices of the Wilson lines with $m<k$ fix those of all the others, and the
following paragraphs give them.

\begin{table}[!htb]
\centering\footnotesize
\renewcommand{\arraystretch}{1.3}\setlength{\tabcolsep}{4pt}
\begin{tabular}{@{}ll>{\raggedright\arraybackslash}p{0.26\textwidth}>{\raggedright\arraybackslash}p{0.3\textwidth}>{\raggedright\arraybackslash}p{0.14\textwidth}@{}}
\toprule
theory & $\mathbf R$ & characters of & all modules & which appear\\
\midrule
$SU(2)_{-\frac92}$ & $\mathbf2\oplus\mathbf3$ & $M(2,9)$ at $q^{\frac12}$ & $(1,s)$, $1\le s\le4$ & all\\
 & & $M(4,9)$, $\Zb_2$ gauged & $(1,s)$, $1\le s\le8$, and $(2,s)$, $1\le s\le4$ & all\\
$SU(2)_{-\frac{15}2}$ & $\mathbf2\oplus\mathbf4$ & $M(2,15)$ & $(1,s)$, $1\le s\le7$ & $s=1,2,4,5,7$\\
$SU(2)_{-\frac{39}2}$ & $\mathbf6$ & $M(2,39)$ & $(1,s)$, $1\le s\le19$ & $s\ne13$\\
$SU(2)_{-\frac{25}2}$ & $\mathbf2\oplus\mathbf5$ & $M(3,25)$ at $q^{\frac14}$ & $(1,s)$, $1\le s\le24$ & even $s$\\
 & & extension of $M(12,25)$, $\Zb_2$ gauged & $(1,s)\oplus(7,s)$, $(5,s)\oplus(11,s)$ and $(4,s)\oplus(8,s)$, $1\le s\le12$ & all\\
$SU(2)_{-44}$ & $\mathbf8$ & $SM(3,11)$ & $(1,s)$, $1\le s\le10$ & all\\
\bottomrule
\end{tabular}
\caption{The other theories of $SU(2)$ (section \ref{sec_newmm}), the VOAs whose characters
the H-twist limits of their line defect half-indices give up to monomials, all their modules and those which appear, with $(r,s)\sim(p-r,p'-s)$ for $M(p,p')$ and
$SM(p,p')$.
For $\mathbf2\oplus\mathbf3$ and $\mathbf2\oplus\mathbf5$ the characters of the first VOAs appear at $q^{\frac12}$, resp.\ $q^{\frac14}$, as in
\eqref{su223_main} and \eqref{su225_main}, and those of the second VOAs, $M(4,9)$ and the extension of $M(12,25)$ of type $(E_6,A_{24})$, appear at
the ${\rm H}^-$-twist point and in the twisted sector with the $\Zb_2$ gauged.
The modules $(1,s)$ of $SM(3,11)$ are Neveu-Schwarz modules for odd $s$ and Ramond modules for even $s$, where the Neveu-Schwarz modules
appear through the characters of their even and odd parts.}
\label{tab_su2_obs}
\end{table}

\subsubsection{\texorpdfstring{$SU(2)_{-9/2}$ with $\mathbf2\oplus\mathbf3$}{SU(2) with 2+3}}
For $\mathbf2\oplus\mathbf3$ and $\mathbf2\oplus\mathbf5$ no element of the center acts as $-1$ on $\mathbf R$, and the $\Zb_2$ which acts by $-1$ on $Q$ and
$\Gamma$ in the real representation $\mathbf3$ or $\mathbf5$ is a global symmetry.
It differs from the $\Zb_2$ which acts on all of $\mathbf R$ and defines the ${\rm H}^-$-twist point \eqref{other_Hm} by the central element
$-\mathbf 1_2$ of $SU(2)$, whose insertion multiplies the line defect half-index of $W_{(a)}$ by $(-1)^a$.
For $\mathbf2\oplus\mathbf3$ we have $k=\frac92$, and the open alcove contains $W_{(a)}$ with $0\le a\le3$.
For all $a$
\begin{align}
\label{su223_main}
\lin{W_{(a)}}{SU(2)_{-9/2}}{\Ncal,N,\Gamma}\big|_{\rm H}&=(-1)^aq^{\frac{a^2}2}\,\frac{f\bigl(-q^{\frac{5-4a}2},-q^{2+2a}\bigr)}{f(-q)}
\nonumber\\
&=\pm q^{h_{(a)}+\frac12(h_{1,s_a}-h_{1,4})}\,\frac{f(-q^{\frac12})\,\chi^{M(2,9)}_{(1,s_a)}(q^{\frac12})}{f(-q)},
\end{align}
with $s_a\equiv\pm4m$ modulo $9$, $1\le s_a\le4$, and the conformal weights $h_{1,s}$ of $M(2,9)$.
The product of the two arguments of $f$ is $q^{\frac92}$, so that by \eqref{ramanujan_f} the theta function is, up to a sign and a power of $q$,
$f(-q^{\frac s2},-q^{\frac{9-s}2})$ with $1\le s\le4$, and $f(-q^{\frac s2},-q^{\frac{9-s}2})=f(-q^{\frac12})\chi^{M(2,9)}_{(1,s)}(q^{\frac12})$ by \eqref{n1_chi}
for $n=3$.
For $0\le a\le3$ the sign is $-1$ when the power of $q$ is half-integral, and the values of $s_a$ and the monomials are
\begin{align*}
\begin{array}{c|cccc}
a & 0 & 1 & 2 & 3\\
\hline
s_a & 4 & 1 & 3 & 2\\
\textrm{monomial} & 1 & -q^{\frac12} & -q^{\frac12} & q
\end{array}.
\end{align*}
For the other $a$ the sign follows from \eqref{su2_rule}.

At the ${\rm H}^-$-twist point \eqref{other_int_m} the line defect half-indices are given by the characters of $M(4,9)$, and we find
for $0\le a\le3$
\begin{align}
\label{su223_Hm}
\lin{W_{(a)}}{SU(2)_{-9/2}}{\Ncal,N,\Gamma}\big|_{{\rm H}^-}=\sum_{x=(1,2m),\,(3,2m)}q^{h_{(a)}+h_x-h_{1,2}}\,\chi^{M(4,9)}_x,
\end{align}
with the conformal weights $h_{r,s}$ of $M(4,9)$.
The power of $q$ is integral for one of the two modules and half-integral for the other, and \eqref{su223_Hm} with
$q^{\frac12}\to-q^{\frac12}$ is \eqref{su223_main}.
Gauging the $\Zb_2$ separates the parts with integral and with half-integral powers of $q$.
So with the $\Zb_2$ gauged, the H-twist limits of the line defect half-indices of the four Wilson lines give the characters of the eight modules $(1,2m)$ and
$(3,2m)=(1,9-2m)$ of $M(4,9)$ individually.
In the twisted sector \eqref{other_tw} of the $\Zb_2$ which acts by $-1$ on $Q$ and $\Gamma$ in $\mathbf3$ we have $\mathbf R_+=\mathbf2$,
$\mathbf R_-=\mathbf3$ and $d_-=3$.
The three fermion zero modes give the spinor $\mathbf2$ of $Spin(3)=SU(2)$, and the matter factor is
\begin{align}
\label{su223_ramond}
q^{\frac3{16}}(-q;q)_\infty\,(s+s^{-1})\,(-qs^2;q)_\infty(-qs^{-2};q)_\infty\,\poch{q^{\frac12}s^{\pm}}.
\end{align}
For $0\le a\le3$ we find
\begin{align}
\label{su223_tw}
\lin{W_{(a)}}{SU(2)_{-9/2}}{\Ncal,N,\Gamma}\big|^{\rm tw}_{\rm H}=(-1)^{a+1}q^{h_{(a)}+h_{2,2m}-h_{1,2}}\,\chi^{M(4,9)}_{(2,2m)},
\end{align}
the characters of the four modules $(2,s)$.
The rule \eqref{su2_rule} holds also in the twisted sector, and at the ${\rm H}^-$-twist point it holds with $q^{\frac12}\to-q^{\frac12}$.

\subsubsection{\texorpdfstring{$SU(2)_{-15/2}$ with $\mathbf2\oplus\mathbf4$}{SU(2) with 2+4}}
For $\mathbf2\oplus\mathbf4$ we have $k=\frac{15}2$, and the open alcove contains $W_{(a)}$ with $0\le a\le6$.
With $\chi_s=\chi^{M(2,15)}_{(1,s)}$ we find
\begin{align}
\label{su224_table}
\lin{W_{(0)}}{SU(2)_{-15/2}}{\Ncal,N,\Gamma}\big|_{\rm H}&=\chi_7+q\,\chi_2,&
\lin{W_{(1)}}{SU(2)_{-15/2}}{\Ncal,N,\Gamma}\big|_{\rm H}&=-q^{\frac12}\chi_4+q^{\frac32}\chi_1,
\nonumber\\
\lin{W_{(2)}}{SU(2)_{-15/2}}{\Ncal,N,\Gamma}\big|_{\rm H}&=0,&
\lin{W_{(3)}}{SU(2)_{-15/2}}{\Ncal,N,\Gamma}\big|_{\rm H}&=-q^{\frac12}\chi_7-q^{\frac32}\chi_2,
\nonumber\\
\lin{W_{(4)}}{SU(2)_{-15/2}}{\Ncal,N,\Gamma}\big|_{\rm H}&=2q\,\chi_5,&
\lin{W_{(5)}}{SU(2)_{-15/2}}{\Ncal,N,\Gamma}\big|_{\rm H}&=0,
\nonumber\\
\lin{W_{(6)}}{SU(2)_{-15/2}}{\Ncal,N,\Gamma}\big|_{\rm H}&=-q^2\chi_4+q^3\chi_1.
\end{align}
This is of the form \eqref{series_form} with $x_0=(1,7)$ and the conformal weights $h_{1,s}$ of $M(2,15)$, with at most two modules in
$S_{(a)}$, except that the H-twist limit of the line defect half-index of $W_{(4)}$ has the coefficient $2$.
The characters of the modules $(1,s)$ with $s=1,2,4,5,7$ appear.

\subsubsection{\texorpdfstring{$SU(2)_{-39/2}$ with $\mathbf6$}{SU(2) with 6}}
For $\mathbf6$ we have $k=\frac{39}2$, and the open alcove contains $W_{(a)}$ with $0\le a\le18$.
The H-twist limit of the half-index is
\begin{align}
\label{su26_vac}
\II^{SU(2)_{-39/2}}_{\Ncal,N,\Gamma}\Big|_{\rm H}=\chi^{M(2,39)}_{(1,18)}=\prod_{j\not\equiv0,\pm18\,(39)}\frac{1}{1-q^j}.
\end{align}
Table \ref{tab_su26} gives the H-twist limits of the line defect half-indices of the Wilson lines in the open alcove.
They are \eqref{series_form} with $x_0=(1,18)$ and the conformal weights $h_{1,s}$ of $M(2,39)$, with at most two modules in $S_{(a)}$.
They give the characters of all the modules $(1,s)$ with $s\neq13$, some of them only in sums of two.

\begin{table}[!htbp]
\centering\footnotesize
\renewcommand{\arraystretch}{1.25}
\begin{tabular}{@{}ll@{\hspace{2em}}ll@{\hspace{2em}}ll@{}}
\toprule
$a$ & $\lin{W_{(a)}}{SU(2)_{-39/2}}{\Ncal,N,\Gamma}\big|_{\rm H}$ & $a$ & $\lin{W_{(a)}}{SU(2)_{-39/2}}{\Ncal,N,\Gamma}\big|_{\rm H}$ & $a$ & $\lin{W_{(a)}}{SU(2)_{-39/2}}{\Ncal,N,\Gamma}\big|_{\rm H}$\\
\midrule
$0$ & $\chi_{18}$ & $7$ & $q^{\frac32}\chi_{12}$ & $14$ & $q^3\chi_{16}-q^4\chi_{10}$\\
$1$ & $-q^{\frac72}\chi_3$ & $8$ & $q\chi_{19}-q^3\chi_7$ & $15$ & $q^{\frac72}\chi_{15}$\\
$2$ & $-q\chi_{11}-q^4\chi_2$ & $9$ & $-q^{\frac32}\chi_{15}$ & $16$ & $-q^6\chi_6$\\
$3$ & $q^{\frac52}\chi_6$ & $10$ & $-q^5\chi_3$ & $17$ & $-q^{\frac92}\chi_{14}+q^{\frac{17}2}\chi_1$\\
$4$ & $q\chi_{12}$ & $11$ & $q^{\frac72}\chi_8+q^{\frac92}\chi_5$ & $18$ & $q^6\chi_9$\\
$5$ & $-q^{\frac12}\chi_{17}+q^{\frac72}\chi_4$ & $12$ & $0$ & & \\
$6$ & $-q^2\chi_9$ & $13$ & $-q^{\frac52}\chi_{18}$ & & \\
\bottomrule
\end{tabular}
\caption{The H-twist limits of the line defect half-indices of $SU(2)_{-\frac{39}2}$ with $\mathbf6$, for the weights in the open alcove, with
$\chi_s=\chi^{M(2,39)}_{(1,s)}$.}
\label{tab_su26}
\end{table}

\subsubsection{\texorpdfstring{$SU(2)_{-25/2}$ with $\mathbf2\oplus\mathbf5$}{SU(2) with 2+5}}
For $\mathbf2\oplus\mathbf5$ we have $k=\frac{25}2$, and the open alcove contains $W_{(a)}$ with $0\le a\le11$.
For these
\begin{align}
\label{su225_main}
\lin{W_{(a)}}{SU(2)_{-25/2}}{\Ncal,N,\Gamma}\big|_{\rm H}=\pm q^{h_{(a)}+\frac14(h_{1,s_a}-h_{1,8})}\,\frac{f(-q^{\frac14})}{f(-q)}\,\chi^{M(3,25)}_{(1,s_a)}(q^{\frac14}),
\end{align}
with $s_a\equiv\pm8m$ modulo $50$, $2\le s_a\le24$ even, the conformal weights $h_{1,s}$ of $M(3,25)$ and the sign $-1$ when the power
of $q$ is half-integral, as for $\mathbf2\oplus\mathbf3$.
The values of $s_a$ and the monomials are
\begin{align*}
\begin{array}{c|cccccccccccc}
a & 0 & 1 & 2 & 3 & 4 & 5 & 6 & 7 & 8 & 9 & 10 & 11\\
\hline
s_a & 8 & 16 & 24 & 18 & 10 & 2 & 6 & 14 & 22 & 20 & 12 & 4\\
\textrm{monomial} & 1 & -q^{\frac12} & q^2 & q & -q^{\frac12} & q & q & -q^{\frac32} & q^3 & q^3 & -q^{\frac52} & q^3
\end{array}.
\end{align*}

As for $\mathbf2\oplus\mathbf3$, the line defect half-indices at the ${\rm H}^-$-twist point and in the twisted sector are given by the
characters of a second VOA, the extension of $M(12,25)$ of type $(E_6,A_{24})$.
Its vacuum module is $(1,1)\oplus(7,1)$, from the modular invariant of type $E_6$ of $\widehat{\mathfrak{su}}(2)_{10}$
\cite{Cappelli:1987xt}, and its modules are $(1,s)\oplus(7,s)$, $(5,s)\oplus(11,s)$ and $(4,s)\oplus(8,s)$ with $1\le s\le12$.
With $r_a\equiv\pm2m$ modulo $25$, $1\le r_a\le12$, and the conformal weights $h_{r,s}$ of $M(12,25)$ we find for $0\le a\le11$
\begin{align}
\label{su225_Hm}
\lin{W_{(a)}}{SU(2)_{-25/2}}{\Ncal,N,\Gamma}\big|_{{\rm H}^-}&=\sum_{r=1,7,5,11}q^{h_{(a)}+h_{r,r_a}-h_{1,2}}\,\chi^{M(12,25)}_{(r,r_a)},
\\
\label{su225_tw}
\lin{W_{(a)}}{SU(2)_{-25/2}}{\Ncal,N,\Gamma}\big|^{\rm tw}_{\rm H}&=(-1)^{a+1}\sum_{r=4,8}q^{h_{(a)}+h_{r,r_a}-h_{1,2}}\,\chi^{M(12,25)}_{(r,r_a)}.
\end{align}
Here \eqref{su225_Hm} with $q^{\frac12}\to-q^{\frac12}$ is \eqref{su225_main}.
In the twisted sector \eqref{other_tw} we have $\mathbf R_-=\mathbf5$ and $d_-=5$.
The matter factor is \eqref{su223_ramond} with the factors of $\mathbf3$ replaced by
\begin{align}
\label{su225_ramond}
q^{\frac5{16}}(-q;q)_\infty\,(s^3+s+s^{-1}+s^{-3})\prod_{\mu=2,4}(-qs^\mu;q)_\infty(-qs^{-\mu};q)_\infty,
\end{align}
in which the five fermion zero modes give the spinor $\mathbf4$ of $Spin(5)$, the representation of spin $\frac32$ of $SU(2)$.
In \eqref{su225_Hm} the powers of $q$ are integral for $r=1,7$ and half-integral for $r=5,11$, or the other way around.
So with the $\Zb_2$ acting on $\mathbf5$ gauged, the parts of \eqref{su225_Hm} with integral and with half-integral powers of
$q$ and the twisted sector \eqref{su225_tw} give the characters of all the $36$ modules of the extension.

\subsubsection{\texorpdfstring{$SU(2)_{-44}$ with $\mathbf8$}{SU(2) with 8}}
For $\mathbf8$ the level $k=44$ is integral, and the open alcove contains $W_{(a)}$ with $0\le a\le42$.
The modules $(1,s)$ of $SM(3,11)$, with the conformal weights $h_{1,s}$, are Neveu-Schwarz modules for odd $s$ and Ramond modules for
even $s$.
For the parts of the Neveu-Schwarz characters with integral and with half-integral powers of $q$ and for the Ramond characters we write
\begin{align}
\label{su28_parts}
\chi^\pm_s=\frac12\Bigl(\chi^{SM(3,11)}_{{\rm NS},(1,s)}\pm\chi^{SM(3,11)}_{\widetilde{\rm NS},(1,s)}\Bigr),\qquad
\chi_s=\chi^{SM(3,11)}_{{\rm R},(1,s)}.
\end{align}
The H-twist limits of the line defect half-indices of $W_{(a)}$ with $a\equiv1$ modulo $4$ and of $W_{(10)}$ and $W_{(32)}$ vanish.
Table \ref{tab_su28} gives the other thirty.
For example, the H-twist limit of the half-index is
\begin{align}
\label{su28_vac}
\II^{SU(2)_{-44}}_{\Ncal,N,\Gamma}\Big|_{\rm H}=\chi^+_3=\frac12\Bigl(\chi^{SM(3,11)}_{{\rm NS},(1,3)}+\chi^{SM(3,11)}_{\widetilde{\rm NS},(1,3)}\Bigr).
\end{align}
In all of them the module is $(1,s)$ with $s\equiv\pm3m$ modulo $22$, which is a Ramond module for odd $a$, and the power of $q$ is
$h_{(a)}+h_{1,s}-h_{1,3}$.
So they are \eqref{series_form} with $x_0=(1,3)$ and with $\chi^\pm_s$ and $\chi_s$ of \eqref{su28_parts} as $\chi_x$.
They give both parts of the five Neveu-Schwarz characters and the five Ramond characters.

\begin{table}[!htbp]
\centering\footnotesize
\renewcommand{\arraystretch}{1.25}
\begin{tabular}{@{}ll@{\hspace{2em}}ll@{\hspace{2em}}ll@{}}
\toprule
$a$ & $\lin{W_{(a)}}{SU(2)_{-44}}{\Ncal,N,\Gamma}\big|_{\rm H}$ & $a$ & $\lin{W_{(a)}}{SU(2)_{-44}}{\Ncal,N,\Gamma}\big|_{\rm H}$ & $a$ & $\lin{W_{(a)}}{SU(2)_{-44}}{\Ncal,N,\Gamma}\big|_{\rm H}$\\
\midrule
$0$ & $\chi^+_{3}$ & $15$ & $-q^{\frac32}\chi_{4}$ & $28$ & $q^{5}\chi^+_{1}$\\
$2$ & $-q\chi^+_{9}$ & $16$ & $q^{2}\chi^+_{7}$ & $30$ & $q^{\frac{11}2}\chi^-_{5}$\\
$3$ & $-q^{\frac32}\chi_{10}$ & $18$ & $q^{3}\chi^+_{9}$ & $31$ & $-q^{\frac{13}2}\chi_{8}$\\
$4$ & $q^{\frac12}\chi^-_{7}$ & $19$ & $-q^{\frac52}\chi_{6}$ & $34$ & $q^{7}\chi^+_{5}$\\
$6$ & $-q^{\frac12}\chi^-_{1}$ & $20$ & $q^{\frac52}\chi^-_{3}$ & $35$ & $-q^{\frac{15}2}\chi_{2}$\\
$7$ & $-q^{\frac12}\chi_{2}$ & $22$ & $q^{3}\chi^+_{3}$ & $36$ & $-q^{8}\chi^+_{1}$\\
$8$ & $q^{\frac12}\chi^-_{5}$ & $23$ & $-q^{\frac72}\chi_{6}$ & $38$ & $q^{9}\chi^+_{7}$\\
$11$ & $-q^{\frac32}\chi_{8}$ & $24$ & $q^{\frac92}\chi^-_{9}$ & $39$ & $-q^{\frac{21}2}\chi_{10}$\\
$12$ & $q\chi^+_{5}$ & $26$ & $q^{\frac92}\chi^-_{7}$ & $40$ & $-q^{\frac{21}2}\chi^-_{9}$\\
$14$ & $q^{\frac32}\chi^-_{1}$ & $27$ & $-q^{\frac92}\chi_{4}$ & $42$ & $q^{\frac{21}2}\chi^-_{3}$\\
\bottomrule
\end{tabular}
\caption{The non-zero H-twist limits of the line defect half-indices of $SU(2)_{-44}$ with $\mathbf8$, for the weights in the open alcove, with $\chi^\pm_s$ and $\chi_s$ of \eqref{su28_parts}.}
\label{tab_su28}
\end{table}

\FloatBarrier

\subsection{\texorpdfstring{$Spin(11)_{-13}$}{Spin(11)}}
\label{sec_newmm_obs}
For $Spin(11)_{-13}$ with the pseudo-real spinor $\mathbf{32}$ the level $k=13$ is integral and $j_{\mathbf R}=4$.
We write $\lambda=\sum_i\lambda_i\mathbf e_i$ as for $Spin(7)$, so that $(\frac12^5)=(\frac12,\dots,\frac12)$ is the spinor and $(1)$ the vector.
The open alcove contains the $39$ dominant weights with $\lambda_1+\lambda_2\le4$, and we find that the H-twist
limits of the line defect half-indices of the other Wilson lines are given by \eqref{series_reduce} with $\epsilon(g)={\rm sgn}(w)$.
We also find that those of the six weights $(2)$, $(2^2,1)$, $(2^3,1)$, $(2^4)$, $(2,1^4)$ and $(2^3,1^2)$ vanish.
The other $33$ are \eqref{series_form} with the characters $\chi_s=\chi^{M(3,26)}_{(1,s)}$, $x_0=(1,7)$ and the conformal weights $h_{1,s}$
of $M(3,26)$, and table \ref{tab_spin11} lists them.
For example,
\begin{align}
\label{spin11_obs}
&\II^{Spin(11)_{-13}}_{\Ncal,N,\Gamma}\Big|_{\rm H}=\chi^{M(3,26)}_{(1,7)},\qquad
\lin{W_{(\frac12^5)}}{Spin(11)_{-13}}{\Ncal,N,\Gamma}\big|_{\rm H}=-q^{\frac12}\chi^{M(3,26)}_{(1,10)},
\nonumber\\
&\lin{W_{(1)}}{Spin(11)_{-13}}{\Ncal,N,\Gamma}\big|_{\rm H}=q^8\chi^{M(3,26)}_{(1,25)}.
\end{align}
Six of them contain the characters of two modules, $(1,s)$ and $(1,26-s)=(2,s)$ with odd $s$.
Together the H-twist limits of the $33$ line defect half-indices give the characters of all the $25$ modules $(1,s)$ of $M(3,26)$.

\begin{table}[!htb]
\centering\footnotesize
\renewcommand{\arraystretch}{1.25}
\begin{tabular}{@{}ll@{\hspace{2.5em}}ll@{}}
\toprule
$\lambda$ & $\lin{W_\lambda}{Spin(11)_{-13}}{\Ncal,N,\Gamma}\big|_{\rm H}$ & $\lambda$ & $\lin{W_\lambda}{Spin(11)_{-13}}{\Ncal,N,\Gamma}\big|_{\rm H}$\\
\midrule
$\varnothing$ & $\chi_{7}$ & $(\frac32^4,\frac12)$ & $q^{\frac52}\chi_{14}$\\
$(1)$ & $q^{8}\chi_{25}$ & $(\frac52,\frac32,\frac12^3)$ & $q^{\frac{17}2}\chi_{24}$\\
$(\frac12^5)$ & $-q^{\frac12}\chi_{10}$ & $(\frac32^5)$ & $-q^{\frac{11}2}\chi_{20}$\\
$(1^2)$ & $-q\chi_{5}$ & $(2^2,1^2)$ & $-q^{2}\chi_{11}-q^{3}\chi_{15}$\\
$(1^3)$ & $q\chi_{11}$ & $(3,1^2)$ & $q^{3}\chi_{15}$\\
$(\frac32,\frac12^4)$ & $q^{\frac32}\chi_{4}$ & $(2^2,1^3)$ & $q^{2}\chi_{7}+q^{5}\chi_{19}$\\
$(1^4)$ & $q\chi_{9}$ & $(4)$ & $q^{5}\chi_{19}$\\
$(1^5)$ & $-q^{2}\chi_{3}$ & $(\frac72,\frac12^4)$ & $-q^{\frac72}\chi_{16}$\\
$(2,1)$ & $-q^{2}\chi_{3}-q^{7}\chi_{23}$ & $(\frac52,\frac32^2,\frac12^2)$ & $-q^{\frac92}\chi_{18}$\\
$(\frac32^2,\frac12^3)$ & $q^{\frac52}\chi_{2}$ & $(2^3)$ & $q^{2}\chi_{9}+q^{4}\chi_{17}$\\
$(2,1^2)$ & $-q^{3}\chi_{1}-q^{9}\chi_{25}$ & $(3,1^3)$ & $q^{4}\chi_{17}$\\
$(3)$ & $q^{3}\chi_{1}$ & $(3,1^4)$ & $-q^{8}\chi_{23}$\\
$(\frac52,\frac12^4)$ & $q^{\frac{13}2}\chi_{22}$ & $(\frac52,\frac32^3,\frac12)$ & $q^{\frac52}\chi_{12}$\\
$(2,1^3)$ & $-q^{2}\chi_{13}$ & $(\frac52,\frac32^4)$ & $-q^{\frac52}\chi_{6}$\\
$(2^2)$ & $q^{2}\chi_{13}$ & $(2^4,1)$ & $-q^{3}\chi_{13}$\\
$(\frac32^3,\frac12^2)$ & $-q^{\frac32}\chi_{8}$ & $(2^5)$ & $q^{3}\chi_{5}+q^{7}\chi_{21}$\\
$(3,1)$ & $-q^{6}\chi_{21}$ & & \\
\bottomrule
\end{tabular}
\caption{The non-zero H-twist limits of the line defect half-indices of $Spin(11)_{-13}$ with $\mathbf{32}$, for the weights in the open alcove, with $\chi_s=\chi^{M(3,26)}_{(1,s)}$ and $\lambda=\sum_i\lambda_i\mathbf e_i$, in the order of $h_\lambda$.
All the $25$ modules $(1,s)$, $1\le s\le25$, of $M(3,26)$ appear, with $(r,s)\sim(3-r,26-s)$.}
\label{tab_spin11}
\end{table}

\FloatBarrier
\section{Character formulae}
\label{app_char}
This appendix lists, for the gauge groups of section \ref{sec_other} and appendix \ref{app_pseudo}, the data which enter
\eqref{other_general}.
These are the coordinates $s_i$ of the maximal torus, the roots $\Delta$, the weights of $\mathbf R$ and the characters $\chi_\lambda(s)$ of the
Wilson lines.
A weight $\mu=\sum_i\mu_i\mathbf e_i$ gives the monomial $s^\mu=\prod_is_i^{\mu_i}$.
For every $G$ the character of the irreducible representation with highest weight $\lambda$ is given by the Weyl character formula
\cite{MR1153249}
\begin{align}
\label{app_weyl}
\chi_\lambda(s)=\frac{\sum_{w\in W}{\rm sgn}(w)\,s^{w(\lambda+\rho)}}{\sum_{w\in W}{\rm sgn}(w)\,s^{w\rho}},
\end{align}
where $\rho$ is half the sum of the positive roots.
For the classical groups \eqref{app_weyl} is a ratio of determinants.
For the exceptional groups we give the simple roots $\alpha_i$, numbered as in \cite{MR1890629}, and write $\lambda=\sum_i\lambda_i\omega_i$
with the fundamental weights $\omega_i$.
The coordinates of $E_6$ and $E_7$ are those of the subgroups $SU(3)^3$ and $SU(8)$, whose branching rules are listed in
\cite{Slansky:1981yr}.

\paragraph{$SU(N)$}
We use $s_1,\dots,s_N$ with $s_1\cdots s_N=1$, the roots $\Delta=\{\mathbf e_i-\mathbf e_j\ (i\neq j)\}$ and the weights $\mathbf e_i$ of $\mathbf N$, and the Weyl
group has order $|W|=N!$.
For a Young diagram $\lambda$ with at most $N-1$ rows the character is the Schur polynomial
\begin{align}
\label{app_su}
\begin{split}
&\chi^{SU(N)}_\lambda(s)=\frac{\det\bigl(s_j^{\lambda_i+N-i}\bigr)_{1\le i,j\le N}}{\det\bigl(s_j^{N-i}\bigr)_{1\le i,j\le N}},\\
&\chi^{SU(N)}_{(1)}(s)=\sum_{i=1}^Ns_i,\qquad
\chi^{SU(N)}_{(1^p)}(s)=e_p(s_1,\dots,s_N),
\end{split}
\end{align}
with the elementary symmetric polynomial $e_p$, e.g.\ $\mathbf{20}=\ysmall{1,1,1}$ of $SU(6)$.
For $SU(2)$ we write $s_1=s$ and $s_2=s^{-1}$.
Then $\chi^{SU(2)}_{(a)}(s)=\sum_{t=0}^as^{a-2t}$ is the character of the representation of spin $\frac a2$, as in
section \ref{sec_characters}, and for $SU(2)^3$ the characters are products, e.g.\
\begin{align}
\chi^{SU(2)^3}_{(\mathbf2,\mathbf2,\mathbf2)}(s)=\prod_{I=1}^3\bigl(s^{(I)}+(s^{(I)})^{-1}\bigr)
\end{align}
with the fugacities $s^{(I)}$ of the three factors, as for $E_6$ below.

\paragraph{$USp(2n)$}
We use $s_1,\dots,s_n$, the roots $\Delta=\{\pm \mathbf e_i\pm \mathbf e_j\ (i<j),\ \pm2\mathbf e_i\}$ and the weights $\pm \mathbf e_i$ of $\mathbf{2n}$, as in appendix
\ref{app_proof}, and the Weyl group has order $|W|=2^nn!$.
Besides \eqref{detchar}, the character of the Young diagram $\lambda$ is
\begin{align}
\label{app_usp}
\chi^{USp(2n)}_\lambda(s)=\frac{\det\bigl(s_j^{\lambda_i+n-i+1}-s_j^{-(\lambda_i+n-i+1)}\bigr)_{1\le i,j\le n}}
{\det\bigl(s_j^{n-i+1}-s_j^{-(n-i+1)}\bigr)_{1\le i,j\le n}},\qquad
\chi^{USp(2n)}_{(1)}(s)=\sum_{i=1}^n(s_i+s_i^{-1}),
\end{align}
and $\mathbf{14}'$ of $USp(6)$ is $\ysmall{1,1,1}$, with $\chi_{\ysmall{1,1,1}}=E_3-E_1$ by \eqref{antisym_sym}.

\paragraph{$Spin(2n+1)$}
We use $s_1,\dots,s_n$ and the roots $\Delta=\{\pm \mathbf e_i\pm \mathbf e_j\ (i<j),\ \pm \mathbf e_i\}$, and the Weyl group has order $|W|=2^nn!$.
We label the representations by $\lambda=\sum_i\lambda_i\mathbf e_i$ with $\lambda_1\ge\dots\ge\lambda_n\ge0$, all integral or all half-integral.
Then
\begin{align}
\label{app_so_odd}
\begin{split}
&\chi^{Spin(2n+1)}_\lambda(s)=\frac{\det\bigl(s_j^{\lambda_i+n-i+\frac12}-s_j^{-(\lambda_i+n-i+\frac12)}\bigr)_{1\le i,j\le n}}
{\det\bigl(s_j^{n-i+\frac12}-s_j^{-(n-i+\frac12)}\bigr)_{1\le i,j\le n}},\\
&\chi^{Spin(2n+1)}_{(1)}(s)=1+\sum_{i=1}^n(s_i+s_i^{-1}),\qquad
\chi^{Spin(2n+1)}_{(\frac12^n)}(s)=\prod_{i=1}^n\bigl(s_i^{\frac12}+s_i^{-\frac12}\bigr),
\end{split}
\end{align}
for the vector and the spinor, which is $\mathbf R$ for $Spin(7)$, $Spin(9)$ and $Spin(11)$.

\paragraph{$Spin(2n)$}
We use $s_1,\dots,s_n$ and the roots $\Delta=\{\pm \mathbf e_i\pm \mathbf e_j\ (i<j)\}$, and the Weyl group has order $|W|=2^{n-1}n!$.
We label the representations by $\lambda=\sum_i\lambda_i\mathbf e_i$ with $\lambda_1\ge\dots\ge\lambda_{n-1}\ge|\lambda_n|$, all integral or all half-integral.
For table \ref{tab_app_e7} we use the simple roots $\alpha_i=\mathbf e_i-\mathbf e_{i+1}$, $1\le i\le n-1$, and $\alpha_n=\mathbf e_{n-1}+\mathbf e_n$, numbered as in
\cite{MR1890629}, with the fundamental weights $\omega_p=\mathbf e_1+\dots+\mathbf e_p$ for $p\le n-2$, $\omega_{n-1}=\frac12(\mathbf e_1+\dots+\mathbf e_{n-1}-\mathbf e_n)$ and
$\omega_n=\frac12(\mathbf e_1+\dots+\mathbf e_n)$.
With $l_i=\lambda_i+n-i$
\begin{align}
\label{app_so_even}
\begin{split}
&\chi^{Spin(2n)}_\lambda(s)=\frac{\det\bigl(s_j^{l_i}+s_j^{-l_i}\bigr)_{1\le i,j\le n}+\det\bigl(s_j^{l_i}-s_j^{-l_i}\bigr)_{1\le i,j\le n}}
{\det\bigl(s_j^{n-i}+s_j^{-(n-i)}\bigr)_{1\le i,j\le n}},\\
&\chi^{Spin(2n)}_{(1)}(s)=\sum_{i=1}^n(s_i+s_i^{-1}),\qquad
\chi^{Spin(2n)}_{(\frac12^{n-1},\pm\frac12)}(s)=\frac12\Bigl[\prod_{i=1}^n\bigl(s_i^{\frac12}+s_i^{-\frac12}\bigr)\pm\prod_{i=1}^n\bigl(s_i^{\frac12}-s_i^{-\frac12}\bigr)\Bigr],
\end{split}
\end{align}
for the vector and the two half-spinors, e.g.\ $\mathbf{32}=(\frac12^6)$ of $Spin(12)$.
For $SO(N)$ the Young diagrams $\lambda$ give the tensor representations, whose characters are given by \eqref{app_so_odd} and
\eqref{app_so_even}.
For $N=2n$ and $\lambda_n>0$ the tensor representation is the sum of $\lambda$ and of $\lambda$ with $\lambda_n$ replaced by $-\lambda_n$, e.g.\
for $(1^n)$ the sum of the self-dual and anti-self-dual parts.
For the discrete fugacity $\chi=-1$ of the charge conjugation $\Zb_2^{\mathcal C}$ (section \ref{sec_voa_lattice}) the integral runs over the
component $O(N)^-$ of $O(N)$ with determinant $-1$ \cite{Okazaki:2021pnc,Okazaki:2024kzo}.
The Haar measure on $O(N)^-$ is normalized to unit volume.
For odd $N$ the element $-\mathbf 1_N$ has determinant $-1$, so that $O(N)^-=-SO(N)=\{-g\,|\,g\in SO(N)\}$.
Since $-\mathbf 1_N$ is central, the vector multiplet contributes as for $SO(N)$.
The integrand is that of $SO(N)$ with the eigenvalues $s^\mu$ of the vector representation replaced by $-s^\mu$, i.e.\ with the substitution
$(a,x)\to(-a,-x)$ of section \ref{sec_other_setup}, and $\chi_\lambda(-s)=(-1)^{|\lambda|}\chi_\lambda(s)$ for the Young diagram $\lambda$ with
$|\lambda|$ boxes.
For $N=2n$ the element $-\mathbf 1_N$ lies in $SO(N)$, and every element of $O(N)^-$ is conjugate to $\sigma_0s$, where
$\sigma_0={\rm diag}(1,\dots,1,-1)$ and $s$ is in the maximal torus of $SO(N-1)$.
The eigenvalues of $\sigma_0s$ are $s_i^{\pm1}$, $1\le i\le n-1$, $1$ and $-1$, so that $\chi_{(1^p)}(\sigma_0s)=E_p-E_{p-2}$ with the $E_p$
of $s_1,\dots,s_{n-1}$.
This is the character of $(1^p)$ of $USp(2n-2)$ in \eqref{antisym_sym} for $p\le n-1$, and it vanishes for $p=n$.
The vector multiplet contributes $(q)_\infty^{n-1}\prod_e\poch{e}$ with the eigenvalues $e\neq1$ of $\sigma_0s$ in $\Lambda^2\mathbf N$.

\paragraph{$G_2$}
We use the fugacities $s_1,s_2,s_3$ of $SU(3)\subset G_2$ with $s_1s_2s_3=1$, i.e.\ the weights in the plane $\sum_ix_i=0$ of $\mathbb R^3$, where $\mathbf e_i$ stands for
its projection $\mathbf e_i-\frac13\sum_j\mathbf e_j$.
The long roots $\mathbf e_i-\mathbf e_j$, $i\neq j$, are those of $SU(3)$, and the short roots are $\pm \mathbf e_i$.
The simple roots are $\alpha_1=\mathbf e_2$ and $\alpha_2=\mathbf e_1-\mathbf e_2$, and the Weyl group has order $|W|=12$.
The representation $\mathbf7=\mathbf1\oplus\mathbf3\oplus\overline{\mathbf3}$ of $\omega_1$, which is $(1)$ in section \ref{sec_voa_g2}, and the
adjoint $\mathbf{14}=\mathbf8\oplus\mathbf3\oplus\overline{\mathbf3}$ of $\omega_2$ have
\begin{align}
\label{app_g2}
\chi^{G_2}_{\mathbf7}(s)=1+\sum_{i=1}^3(s_i+s_i^{-1}),\qquad
\chi^{G_2}_{\mathbf{14}}(s)=2+\sum_{i\neq j}s_is_j^{-1}+\sum_{i=1}^3(s_i+s_i^{-1}),
\end{align}
and the other characters are given by \eqref{app_weyl}.

\paragraph{$E_6$}
We use the fugacities $s^{(I)}_i$, $I,i=1,2,3$, of $SU(3)^3\subset E_6$, with $s^{(I)}_1s^{(I)}_2s^{(I)}_3=1$, and write $\mathbf e^{(I)}_i$ for the
corresponding weights $\mathbf e_i$ of the factor $I$, as for $G_2$, so that $\mu=\sum_{I,i}\mu^{(I)}_i\mathbf e^{(I)}_i$ gives $s^\mu=\prod_{I,i}(s^{(I)}_i)^{\mu^{(I)}_i}$.
The $72$ roots are the roots $\mathbf e^{(I)}_i-\mathbf e^{(I)}_j$, $i\neq j$, of $SU(3)^3$ and $\pm(\mathbf e^{(1)}_i+\mathbf e^{(2)}_j+\mathbf e^{(3)}_k)$, i.e.\
$\mathbf{78}=(\mathbf8,\mathbf1,\mathbf1)\oplus(\mathbf1,\mathbf8,\mathbf1)\oplus(\mathbf1,\mathbf1,\mathbf8)\oplus(\mathbf3,\mathbf3,\mathbf3)\oplus
(\overline{\mathbf3},\overline{\mathbf3},\overline{\mathbf3})$, the simple roots are
\begin{align}
\begin{split}
&\alpha_1=\mathbf e^{(1)}_1-\mathbf e^{(1)}_2,\qquad\alpha_2=\mathbf e^{(2)}_1-\mathbf e^{(2)}_2,\qquad\alpha_3=\mathbf e^{(1)}_2-\mathbf e^{(1)}_3,\\
&\alpha_4=\mathbf e^{(1)}_3+\mathbf e^{(2)}_2+\mathbf e^{(3)}_3,\qquad\alpha_5=\mathbf e^{(3)}_2-\mathbf e^{(3)}_3,\qquad\alpha_6=\mathbf e^{(3)}_1-\mathbf e^{(3)}_2,
\end{split}
\end{align}
and $|W|=51840$.
The representation $\mathbf{27}=(\mathbf3,\overline{\mathbf3},\mathbf1)\oplus(\mathbf1,\mathbf3,\overline{\mathbf3})\oplus(\overline{\mathbf3},\mathbf1,\mathbf3)$ of $\omega_1$ has
\begin{align}
\label{app_e6}
\chi^{E_6}_{\mathbf{27}}(s)=\sum_{i,j=1}^3\Bigl(\frac{s^{(1)}_i}{s^{(2)}_j}+\frac{s^{(2)}_i}{s^{(3)}_j}+\frac{s^{(3)}_i}{s^{(1)}_j}\Bigr),
\end{align}
$\overline{\mathbf{27}}$, of $\omega_6$, is its conjugate, and $\mathbf{351}=\Lambda^2\mathbf{27}$ and $\mathbf{2925}=\Lambda^3\mathbf{27}$ of section
\ref{sec_voa_exc} are the representations of $\omega_3$ and $\omega_4$.

\paragraph{$E_7$}
We use the fugacities $s_1,\dots,s_8$ of $SU(8)\subset E_7$ with $s_1\cdots s_8=1$, i.e.\ the weights in the hyperplane $\sum_ix_i=0$ of $\mathbb R^8$, where $\mathbf e_i$ stands
for its projection, as for $G_2$.
The $126$ roots are the $56$ roots $\mathbf e_i-\mathbf e_j$, $i\neq j$, of $SU(8)$ and the $70$ weights $\mathbf e_i+\mathbf e_j+\mathbf e_k+\mathbf e_l$, $i<j<k<l$, of $\Lambda^4\mathbf8$, i.e.\
$\mathbf{133}=\mathbf{63}\oplus\mathbf{70}$, the simple roots are
\begin{align}
\alpha_1=\mathbf e_2-\mathbf e_3,\qquad\alpha_2=\mathbf e_5+\mathbf e_6+\mathbf e_7+\mathbf e_8,\qquad\alpha_i=\mathbf e_i-\mathbf e_{i+1},\quad3\le i\le7,
\end{align}
and $|W|=2903040$.
The representation $\mathbf{56}=\mathbf{28}\oplus\overline{\mathbf{28}}$ of $\omega_7$ and the adjoint $\mathbf{133}$ of $\omega_1$ have
\begin{align}
\label{app_e7}
\chi^{E_7}_{\mathbf{56}}(s)=\sum_{i<j}\bigl(s_is_j+s_i^{-1}s_j^{-1}\bigr),\qquad
\chi^{E_7}_{\mathbf{133}}(s)=7+\sum_{i\neq j}s_is_j^{-1}+\sum_{i<j<k<l}s_is_js_ks_l,
\end{align}
and $\omega_2,\dots,\omega_6$ give the representations $\mathbf{912}$, $\mathbf{8645}$, $\mathbf{365750}$, $\mathbf{27664}$ and $\mathbf{1539}$ of
section \ref{sec_newmm_minW} and table \ref{tab_app_e8}.

\FloatBarrier
\bibliographystyle{utphys}
{\small\renewcommand{\baselinestretch}{1.07}\selectfont
\makeatletter\def\@listi{\leftmargin\leftmargini\topsep 4\p@\parsep\z@\itemsep\z@}\makeatother
\bibliography{ref}}

\end{document}